\documentclass[12pt]{iopart}
\usepackage{graphicx,color}

\def\am{angular momentum }
\def\amd{angular momentum}
\def\momi{moment of inertia }
\def\momis{moments of inertia }

\def\momid{moment of inertia}
\def\momisd{moments of inertia}

\def\J{{\cal J}}
\def\Jr{{\cal J}_{rig}}
\def\J2{{\cal J}^{(2)}}
\def\conf{configuration }
\def\confs{configurations }
\def\qp{quasiparticle }
\def\qps{quasiparticles }
\def\qn{quasineutron }
\def\qpr{quasiproton }
\def\qprs{quasiprotons }
\def\qns{quasineutrons }
\def\mr{magnetic rotation }

\def\confd{configuration}
\def\confsd{configurations}
\def\qpd{quasiparticle}
\def\qpsd{quasiparticles}
\def\qnd{quasineutron}

\def\qnsd{quasineutrons}

\def\tw{tidal wave }
\def\twd{tidal wave}
\newcommand{\Om}{\Omega}
\newcommand{\vth}{\vartheta}
\newcommand{\al}{\alpha}
\newcommand{\De}{\Delta}
\newcommand{\ga}{\gamma}
\newcommand{\la}{\lambda}

\newcommand{\de}{\delta}
\newcommand{\f}{\varphi}
\newcommand{\eps}{\varepsilon}
\newcommand{\om}{\omega}
\newcommand{\ka}{\kappa}
\newcommand{\barr}{\begin{array}}
\newcommand{\bea}{\begin{eqnarray}}
\newcommand{\beq}{\begin{equation}}
\newcommand{\ear}{\end{array}}
\newcommand{\eea}{\end{eqnarray}}
\newcommand{\eeq}{\end{equation}}
\newcommand{\bef}{\begin{figure}}
\newcommand{\enf}{\end{figure}}
\begin{document}
\title{Beyond the Unified Model}

\author{S. Frauendorf}

\address{Department of Physics, University Notre Dame, Indiana 46557, USA}

\ead{sfrauend@nd.edu}

\begin{abstract}
The key elements of the Unified Model are reviewed. 
The microscopic derivation of the Bohr Hamiltonian by means of adiabatic time-dependent mean
 field theory is presented. By checking against experimental data the limitations of the Unified Model are delineated.
 The description of the strong coupling between the rotational and intrinsic degrees of freedom in framework of the rotating mean field 
 is presented from a conceptual point of view. 
 The classification of rotational bands as configurations of rotating quasiparticles is introduced.
 The occurrence of uniform rotation about an axis  that differs from the principle axes of the
 nuclear density distribution is discussed. The physics behind this tilted-axis rotation, unknown in molecular physics, is explained on a basic level.  
  The new symmetries of the rotating mean field that arise from the various orientations of the \am vector with respect to the triaxial 
  nuclear density distribution  and their manifestation by the level sequence of rotational bands are discussed. Resulting phenomena, as  
 transverse wobbling, rotational chirality, magnetic rotation and band termination are discussed.        
 Using the concept of spontaneous symmetry breaking the  microscopic underpinning of the  rotational degrees is refined. 
\end{abstract}
\maketitle

\section{Introduction}

This Focused Issue commemorates  the 40-year Anniversary of the Nobel Prize for
A. Bohr, B. R. Mottelson and L. Rainwater, which was awarded 
"for the discovery of the
connection between collective motion and particle motion in atomic nucleus and the
development of the theory of the structure of the atomic nucleus based on this
connection." 
Before 1952, two apparently incompatible models coexisted. The Liquid Drop Model \cite{LD} describes the nucleus as 
a droplet of incompressible nuclear liquid, where  the shape parameters are the degrees of freedom. 
As major achievements, it  determined the stability of nuclei against 
spontaneous fission and explained  the phenomenon of induced fission. 
   The shell model considers the nucleus as a system of independent 
protons  and neutrons confined to the spherical nuclear potential \cite{SM1,SM2}. 
It succeeded in explaining existence of special  ("magic") numbers of protons and neutrons for which nuclei 
are particularly strongly bound and it well accounted for excitation energies of these nuclei.        

The experimental evidence for a substantial deformation of the nuclear charge deformation and
for the existence of low-energy rotational excitations inspired A. Bohr and B. R. Mottelson  to combine the two
models to the Unified Model. In their pioneering papers \cite{UM1,UM2}, they introduced the innovative concept 
of the coexistence and the interplay of collective and single particle degrees of freedom, which has become  a 
fundamental of understanding low-energy nuclear structure. They realized that the collective motion of the density 
is accompanied by  a corresponding motion of shell-model potential, which modifies the motion of the nucleons in the potential.
The response of the nucleon ensemble  determines the inertial parameters and the potential of the collective Hamiltonian.
In addition, they assumed that the collective motion is slow compared to the single particle motion.  

The Unified Model was 
extremely successful in accounting for the observation of collective and particle-like phenomena in one and the same nucleus. The concept
has developed considerably during the succeeding years. 
 In their contributions to this Focus Issue, D. R. Bes \cite{NCBes} and K. Heyde and J. L.Wood \cite{NCHeydeWood} 
 review the history that lead to birth of the Unified Model and its subsequent development in a comprehensive way.
  A. Bohr and B. R. Mottelson expose in great detail the mature version of the Unified Model  in their famous monograph {\it Nuclear Structure Vol. II: Nuclear
Deformations} \cite{BMII}. 
For a first encounter with the Unified Model, the excellent textbook {\it Nuclear Collective Motion: Models and Theory} by D. J. Rowe is
recommended. A extended exposure of models describing the collective and single particle modes 
 is given in {\it Fundamentals of Nuclear Models} by D. J. Rowe and J. L. Wood \cite{RW}.

In the following I will refer to the "Unified Model" as presented in these books.  A central element of the theory is the presumption that the collective 
motion is slow as compared to the single particle motion, which simplifies the treatment of the  coupling between collective and single particle degrees of freedom.
The  nucleonic orbitals adjust  "adiabatically" to the changing potential, which means, they are determined  by the potential at given instant of time.
Additional coupling terms due to the finite velocity of the changing potential are either neglected or taken into account in lowest order perturbation theory.
The "adiabatic approximation" shapes the interpretation of data in a fundamental way.  For this reason  it seems appropriate using the name "Unified Model"
when the adiabatic approximation is applicable.  Clearly the concept of nucleons moving in a time-dependent potential extends beyond the realm of the
Unified Model specified in this way.  However, when the coupling between collective and single particle degrees of freedom becomes strong
new phenomena emerge and  it may  be useful changing the perspective of looking on the data. Some of these new aspects  beyond the Unified Model
will be presented in my contribution.

The advances of experimental techniques after 1970, in particular the combination heavy ion accelerators with arrays of $\gamma$-detectors,
 provided new results that demonstrated the limitations of the Unified Model and stimulated the  development of  new theoretical approaches.
 My contribution will report on the progress in understanding the behavior of rapidly rotating nuclei,  
 where the focus is on novel concepts and not  on a full scale presentation of the theoretical approaches. Section 2  
 summarizes the essentials of the Unified Model and its microscopic foundation.
 The range of applicability is delineated by comparing its predictions with data.  The non-adiabtic
 regime has been furthest explored in in the framework of the semiclassical rotating mean field approach. Section 3 lays out its 
 simplest version based on the schematic pairing + quadrupole - quadrupole interaction, analyzes the symmetries and their
 observable consequences and introduces the Cranked Shell Model classification of the multi-band spectra. The central role of
 rotational frequency for the interpretation of the data and the appearance of uniform rotation about a tilted axis  are discussed.    Based on the
 concept of spontaneous symmetry breaking, 
 the emergence and disappearance of collective degrees of freedom is analyzed in section 4.  The definition  collective angles is refined 
 by changing the focus from the orientation of the deformed  nuclear potential to the orientation in space of the nodal structure 
 of the mean field many-body state. The new perspective accounts for band termination and magnetic rotation, which are
 phenomena beyond the Unified Model. The yrast states of spherical and 
 weakly deformed nuclei are described as "tidal waves" running over the nuclear surface.
 Section 5 lists some challenging phenomena  beyond the Unified Model, for which the appropriate theoretical approaches are yet to be developed. 
Appendix A provides a semiclassical analysis of the interaction between the high-j particle orbitals and the deformed nuclear potential.
Appendix B contains a table for navigating the paper.

The paper is meant as an introduction to structure of rotating  nuclei  on 
the  graduate student level. I apologize to colleagues working in the field for repeating   
too many well known things and to newly interested ones for not having well enough explained certain things. 
 To be self-contained, the present  contribution repeats some  material that I have reviewed before. In particular,
 Sections    \ref{sec:SCCM}, \ref{sec:crdiabatic}, \ref{sec:AppTAC}, \ref{sec:axialTAC}, \ref{sec:approxTAC},
\ref{sec:ChangeSym}, \ref{sec:TriaxTAC} (paragraph {\it Chirality}), \ref{sec:termination}, \ref{sec:MR}, \ref{sec:cohl}, 
\ref{sec:reg} contain excerpts from  Ref.\cite{RMP},
where I reviewed  the  development of the  cranking model  until 2000 focusing on the symmetries.

\section{The Unified Model: virtues  and limits}\label{sec:UM}
The structure of Unified Model   is  analog  to molecules, in case of which there are two classes of degrees of freedom,
the positions of the nuclei and of the electrons. The electrons move about one thousand times faster than the heavy nuclei. Their wave functions
adiabatically adjust to the slow re-arrangement of the nuclei. Adiabaticity allows one to find the molecular states in two steps (Born-Oppenheimer approximation).
 The electronic wave functions are calculated for fixed positions of the nuclei, which are varied. The energy of a certain electronic configuration 
 as function of the nuclear positions represents the potential energy for the Hamiltonian that describes the motion of the nuclei. Its eigenstates are found in the second step. 
 The eigenstates of the nuclear Hamiltonian are combinations of the rotation of the molecule as a whole and vibrations of the atoms relative to each other. 
 For low angular momentum and excitation energy the vibrations is typically a factor 10-100 larger than the energy difference $\hbar \omega_R$ between the 
 adjacent rotational levels, such that a rotational band is built on each vibrational configuration, which is  defined by the number of vibrational phonons in each eigenmode $\hbar \omega_V$.
 With increasing angular momentum  the rotational and vibrational modes progressively couple with each other, forming the so-called "rovib" states. The  
 adiabatic separation between electronic and nuclear degrees of freedom holds as long as the energy difference between two electronic configurations remains
 large compared with the rotational and vibrational frequencies $\hbar \omega_R$ and $\hbar \omega_V$. In this context, "electronic configuration" means a certain electronic
 state that continuously changes with gradually moving the positions of the nuclei and "energy difference" concerns the minimum with respect to all position explored by the    
 nuclear motion. Sometimes the energy difference between two electronic configurations becomes small for a certain arrangement of the nuclei, which causes a coupling between 
 the two configurations and the nuclear degrees of freedom.
 
 In analogy, the Unified Model invokes  two classes of degrees of freedom. The "collective" degrees of freedom describe motion  of the nuclear surface 
 and the "intrinsic" ones the motion of the nucleons in a fixed average potential that reflects the  nuclear shape.  It is assumed that collective motion
 is slow compared to the intrinsic motion such that the adiabatic approximation holds. The Unified Model considers the collective and intrinsic degrees of freedom as independent.
 It disregards the fact that the collective coordinates emerge as a consequence of correlations between the nucleons, that is, they merely describe a coherent motion overlaid to the
 intrinsic one.  In this respect the Unified Model differs from molecules where the two kinds of degrees of freedom refer 
 to different constituents (electrons and nuclei). The over-counting  of degrees of freedom is unproblematic for many applications of the Unified Model.

 In contrast to molecules, the collective motion of nuclei is only marginally slower than the intrinsic motion. 
For low spin and well deformed nuclei, rotational  transition energies are about a factor 10 lower than the energies of intrinsic excitations at the best.
 For vibrational excitations the ratio is  only 3 at the best. Exciting the nucleus, one early encounters the non-adiabatic regimes. 
 Bohr and Mottelson studied two important special solutions of the collective Hamiltonian. One case is deformed  nuclei that execute small
 vibrations around  the axial shape, which I will discuss   in sections \ref{sec:UMdSM} and \ref{sec:UMdef}
 for the example of the Er isotopes.  The microscopic basis of the Unified Model for this case will be discussed in section \ref{sec:UMmicro}. 
 The second case are spherical nuclei that execute oscillations around the equilibrium shape, which will be discussed in section \ref{sec:UMsph}. 
General solutions of the Bohr Hamiltonian for even-even nuclei have been studied by many authors. They apply 
 to even-even nuclei only. For these nuclei  the pairing  correlation generate a gap of about 2 MeV between   the intrinsic ground state and the
 lowest two-quasiparticle excitations, which ensures a reasonable adiabatic separation between the collective and intrinsic degrees of freedom.
 They will be briefly reviewed in section {\ref{sec:BHgen}.  Complementary discussions of the material of this section can be found in Refs. \cite{Frauendorf15,Frauendorf16}.

 \subsection{The Bohr Hamiltonian}\label{sec:BH}
 The collective motion is described by the Bohr Hamiltonian (BH), which describes the surface motion of a droplet of  liquid \cite{UM1}. 
  The droplet has a well localized surface because the 
 liquid is hard to compress. The shape is described by a multipole expansion of the distance of the surface from the center of gravity 
\beq\label{eq:LDR}
 R(\theta,\phi)=R_0\left[1+\sum\limits_{\lambda=2,\mu=-\lambda}\limits^{\infty,\mu=\lambda}\alpha_{\la\mu} Y_{\lambda\mu}(\theta,\phi)\right], ~~~R_0=1.2A^{1/3} fm.
 \eeq  
  The collective dynamics has been mainly  studied for the quadrupole mode. The five coefficients  $\alpha_{2\mu} \equiv\alpha_\mu$ represent the collective shape coordinates.
  They are re-expressed in terms of the two deformation variables $\beta$ and $\gamma$, which describe the lengths of the principle axes of 
  the triaxial shape \cite{UM1}, 
  \bea
  R_1=R_0\left[1+\sqrt{\frac{5}{16\pi}}\beta\left(-\cos\gamma+\sqrt{3}\sin\gamma\right) \right],\\
  R_2=R_0\left[1+\sqrt{\frac{5}{16\pi}}\beta\left(-\cos\gamma-\sqrt{3}\sin\gamma\right) \right],\\
  R_3=R_0\left[1+\sqrt{\frac{5}{16\pi}}2\beta\cos\gamma \right], 
   \eea
   and the Euler angles
  $\Om=\{\psi,\vth,\f\}$, which specify the orientation of the shape,
 \begin{equation}\label{eqn-bohr-q}
\alpha_\mu = \beta \left[ \cos\gamma\,D^2_{0,M}(\Omega) +
\frac{1}{\sqrt{2}} \sin\gamma \left[ D^2_{2,M}(\Omega) +
D^2_{-2,M}(\Omega) \right] \right].
\end{equation}
The Bohr Hamiltonian takes the generic form
\beq\label{eq:BH}
H_{BH}=T+V(\beta,\ga), ~~T=T_{\beta\beta}+T_{\ga\ga}+T_{\beta\ga}+H_{ROT}.
\eeq
It is composed of the rotational energy $H_{ROT}$, the kinetic energies $ T_{\beta\beta}$, $T_{\ga\ga}$ of the two 
deformation parameters $\beta$ and $\ga$, a kinetic coupling term $T_{\beta\ga}$,
 and the deformation potential $V(\beta,\ga)$.
The rotational part is the Hamiltonian of the triaxial rotor
\beq\label{eq:HROT}
H_{ROT}=\sum\limits_{i=1,2,3}\frac{\hat R_i^2}{2{\cal J}_i(\beta,\ga)}.
\eeq
The angular momentum components $R_i$ expressed in terms of the Euler angles are given in standard 
 texts on angular momentum (e. g. Ref.  \cite{Rose57},  see also section 1A in Ref. \cite{BMI} and Appendix B in Ref. \cite{R70}).
  It is common to assume that the three moments of inertia 
 depend on the deformation parameters as expected for irrotational flow of an ideal liquid (see Ref. \cite{BMII})
 \beq\label{eq:momiIF}
{ \cal J}_i(\beta,\ga)=4B\beta^2\sin^2(\gamma-\frac{2\pi i}{3}),
  \eeq
  with $B$ being a free constant to be adjusted to the experimental energies. 
 
\begin{figure}[t]
 \begin{center}
 \includegraphics[width=\linewidth,angle=-90]{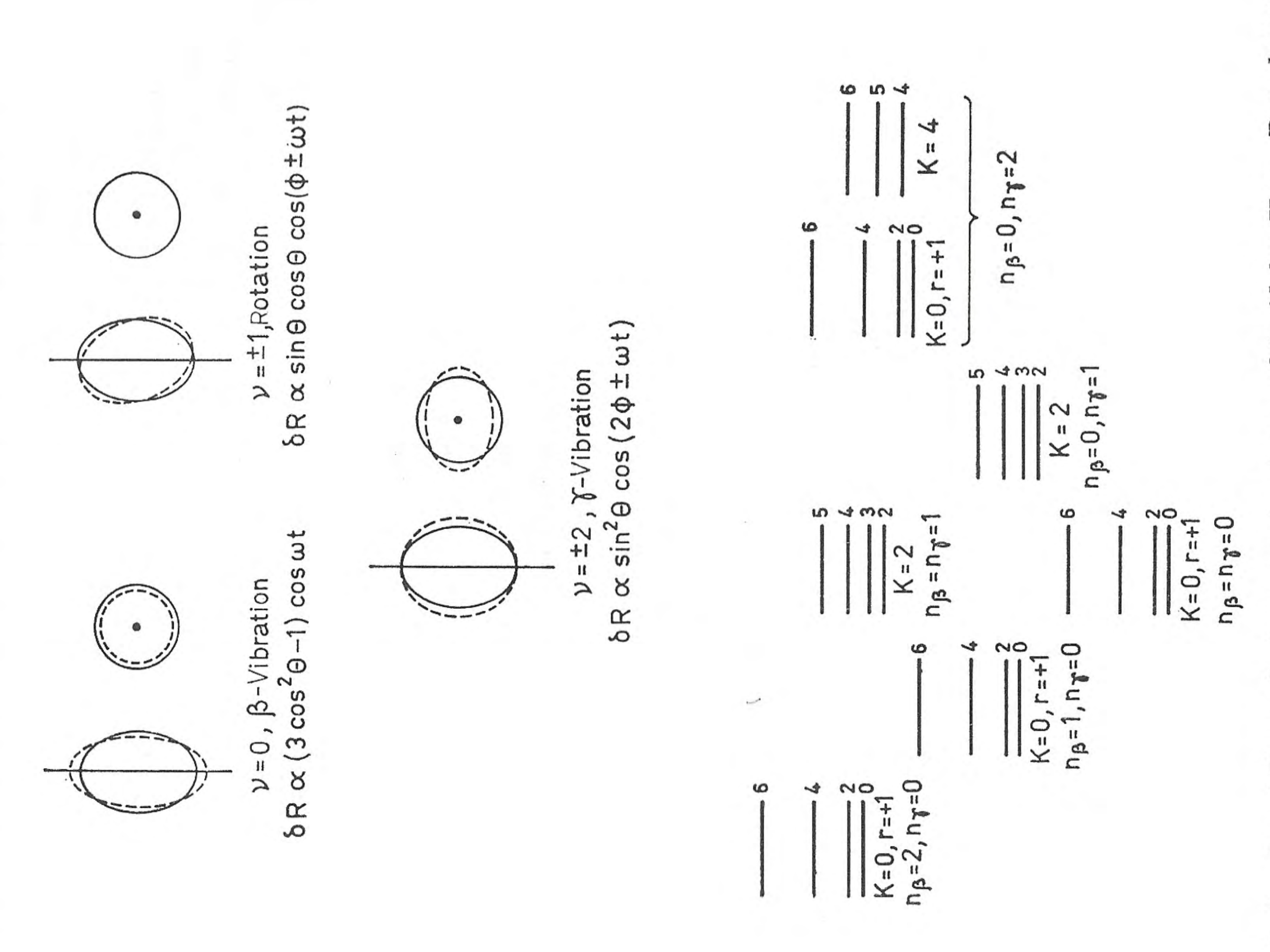}
 \caption{ \label{f:RoVib} Various quadrupole shape oscillations in a prolate nucleus. The upper part shows projections of the nuclear shape 
 perpendicular and parallel to the symmetry axis. The lower part shows the spectra associated with the excitation of one or two
 vibrational quanta including the  specific values of the various oscillation energies $\hbar \om_\beta$ and $\hbar \om_\ga$. 
 The corresponding classical motion of the deviation of the surface from a sphere, $\delta R(\vth=\theta,\f=\phi,t)$ is indicated. Reproduced from \cite{BMIIG}.
 }
  \end{center}
  \end{figure}
  
   \subsection{The Deformed Shell Model}\label{sec:UMdSM}
According to the concept of adiabaticity, the Unified Model assumes that the nucleons move in a
 deformed potential $v(\beta,\ga)$ with a shape that corresponds to the instantaneous values 
 of the slowly changing deformation variables $\beta$ and $\ga$.   The deformed single particle Hamiltonian 
 \beq
 h_{def}\phi_i=\left(t+U(\beta,\ga)\right)\phi_i=e_i\phi_i,
 \eeq
generates the single particle energies $e_i$ and the single particle wave functions $\phi_i$. 
 The most successful 
 versions are the modified oscillator or Nilsson potential \cite {Nilsson55,MOglobal}, the Woods Saxon potential 
 \cite{WSglobal} and the folded Yukawa potential \cite{FYglobal}. 
Because the single particle Hamiltonian   $h(\beta,\ga)$ is invariant under time reversal, it has twofold 
 eigenstates $(i,\bar i)$ (orbitals)  with the energy $e_i$, 
which are related by time reversal. For axial potentials it has become custom to label the single particle by the 
Nilsson quantum numbers $[Nn_z\Lambda]\Omega$ of the modified oscillator potential (see \cite{Nilsson55,BMII}).
 Respectively, they indicate the total number of oscillator quanta, the nodal number along the symmetry axis, 
 the orbital \am and the total \am projections on the symmetry axis.  For axial potentials the pair of time-reversed orbitals are 
 the states with two projections $\pm\Omega$ 
 and the same other quantum numbers.       

\setcounter{footnote}{0}

The pair correlations are taken into account in the framework of the  BCS theory 
\footnote{BCS is the acronym of Bardeen, Cooper and Schrieffer, who invented it for describing superconductivity in metals \cite{BCS}.}
  by introducing the monopole pair 
potential $\Delta P^\dagger =\De\sum\limits_i c^\dagger_ic^\dagger_{\bar i}$, which generates pairs
in the time reverse orbitals $(i,\bar i)$, 
and $\lambda \hat N$, which is adjusted such that the  expectation value of the particle number $\langle \hat N \rangle$ agrees with the actual
particle number $N$.  
The quasiparticle Hamiltonian
\beq\label{eq:hqp}
h_{qp}=h_{def}(\beta,\ga)+\Delta(P^\dagger+P)-\lambda \hat N
\eeq
is diagonalized by introducing quasiparticles with the energy 
\beq\label{eq:Eqp}
E_i=\sqrt{\left(e_i-\lambda\right)^2+\Delta^2}.
\eeq 
The intrinsic states are configurations of excited  quasiparticles with an energy equal to the sum of the quasiparticle energies,
where  excited states in even-$N$ nuclei have  even numbers of quasiparticles  and excited states in odd-$N$ nuclei have  
odd numbers of quasiparticles.  
The strength $\De$  of the pair potential is called the pairing gap, because it is the lowest possible 
quasiparticle energy. The lowest two-quasiparticle energy in even-N nuclei is larger than $2\De$, which generates a gap in 
the excitation spectrum.  The lowest one-quasiparticle state, which is
 the ground state of the odd-$N$ neighbor, is equal or slightly larger than $\De$. For this reason $\De$ is approximately
 given by the energy difference between
the ground state energy of the odd-$N$ nucleus and the average of the ground state energies of its two even-$N$ neighbors,
which is called the even-odd mass difference $\De_{eo}$ (three-point even-odd mass difference). 
The preceding is only a sketch of  the necessary rudiments of the BCS theory,
which is well exposed in many textbooks, as for example in Refs.\cite{BMII,R70,NR,Ringbook,blaizotbook}  
and will be presented in more detail in section \ref{sec:SCCM}.

 In a phenomenological approach to the Unified Model, the
 experimental value of the even-odd mass difference $\De_{eo}$ is taken  to fix the pair 
 gap $\De$, the deformation of the potential is determined from the 
 experimental value of the reduced transition probability  $B(E2,2^+_1\rightarrow0^+_1)$.  
 Usually, the requirement $\langle \hat N \rangle=N$ places the chemical potential $\lambda$  between the last occupied and the first free level.
 One may allow for some adjustment of $\lambda$ to improve the agreement  with the observed energies of the one-quasiparticle levels.
 In the following we will denote the quasiparticle \confs by $\phi(\xi)$, where $\xi$ is the short-hand notation for the nucleon coordinates 
 with respect to principal axes of the deformed potential.

\begin{figure}[t]
 \begin{center}
 \includegraphics[width=16cm]{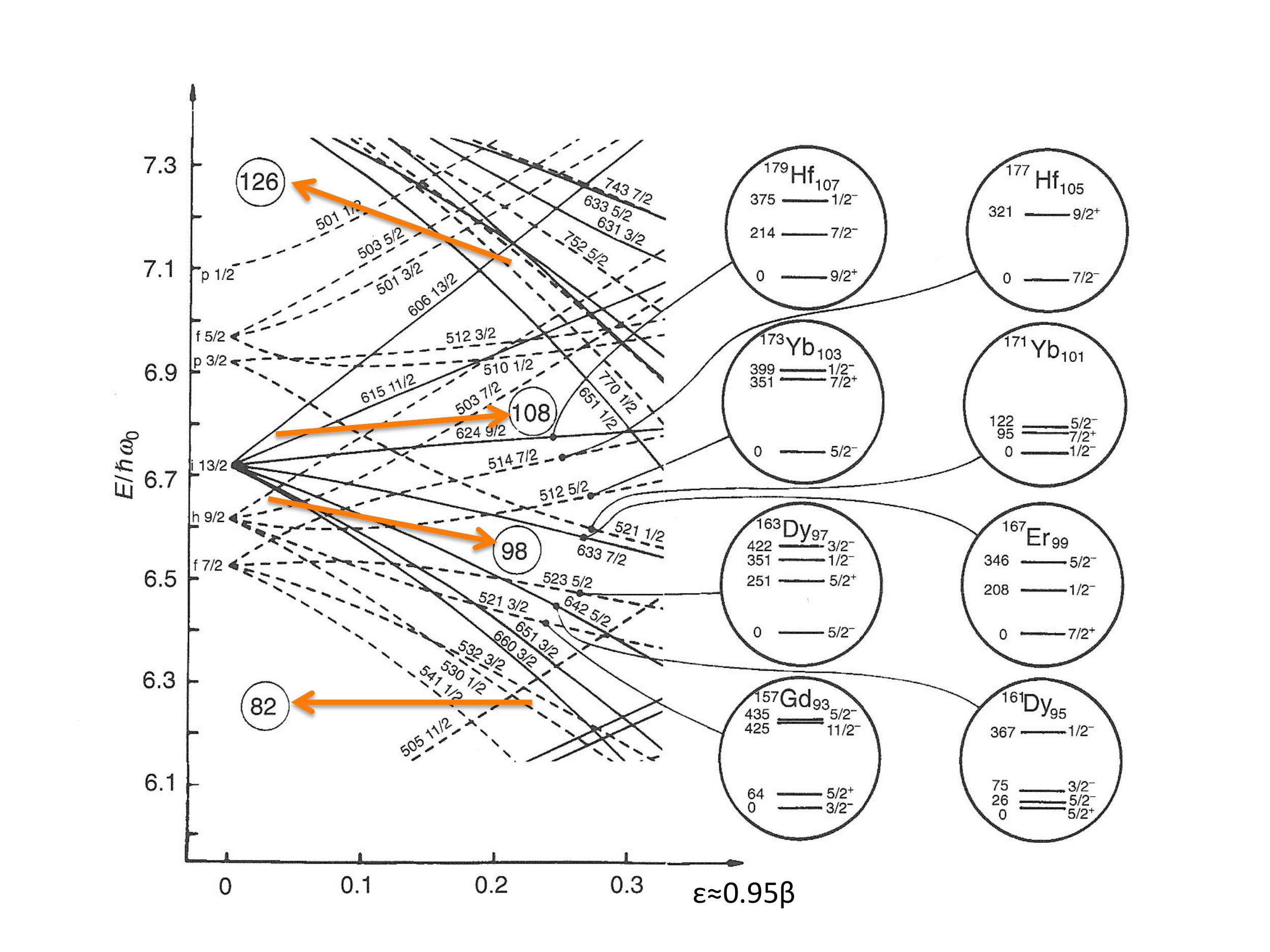}
 \caption{ \label{f:nilsson}  Single neutron orbitals in the rare earth region calculated by means of the modified oscillator potential. The circles show the position of the Fermi level
 for various neutron numbers.
 The deformation parameter $\varepsilon$ differs slightly from $\beta$ (see \cite{BMII,NR}). For  small deformation they are related by the indicated proportionality.
 The energy scale is \mbox{$\hbar \om_0=41 A^{-1/3}$ MeV=7.5 MeV.}
 The experimental band head energies of odd-$N$ rare-earth nuclei are shown to the right, which represent the one-\qn excitations. Each circle is  
  connected with the  Nilsson level pertinent to the ground state rotational band. The red arrows show how the nucleus changes its equilibrium deformation 
 by avoiding high level density.
  Reproduced and modified from Ref. \cite{NR}. 
 }
  \end{center}
  \end{figure}

\setcounter{footnote}{0}

 \subsection{Deformed prolate nuclei}\label{sec:UMdef}

 When the potential of the collective motion $V(\beta,\gamma)$ has a deep minimum at $\beta_0$ and  $\ga=0$, the deformed nuclear surface
  rotates and executes small-amplitude
 oscillations around its equilibrium shape. 
 The resulting harmonic rotation-vibration spectrum spectrum is illustrated by the well known Fig. \ref{f:RoVib}   from 
 the monograph \cite{BMII}. It is assumed that, like in molecules, the rotational frequency $\omega_R$ is substantially smaller than the vibrational frequencies
 $\omega_\beta$ and $\omega_\ga$ and much smaller than  the frequencies of the nucleons moving in the deformed potential.
 The different nucleonic \confs and the vibrational excitations are called the "intrinsic states". 
 The mutual coupling between the intrinsic modes and the rotational mode is neglected which gives the total energy
 as the sum of the energies of the individual  modes,
 \bea\label{eq:Eint}
 E=E_{ROT}+E_{int},~~E_{int}=E_{qp}+E_{V}, \\
 E_{V}=(n_\beta+1/2)\hbar\om_\beta+(n_\ga+1)\hbar\om_\ga.\nonumber
  \eea 
  A rotational band is built an each intrinsic state, where
 sequence of rotational levels is given by the eigenvalues of the rotor Hamiltonian  (\ref{eq:HROT}).
 The lowest state of  the band is called the band head.  
 
 The sequence of states with lowest energy at 
 a given \am is called the yrast \footnote{"yrast": Swedish "most dizzy".  These are the states with the highest
 \am for given energy, which are the lowest states for given \am.
 Sometimes the state above the yrast state is referred to as " yrare": Swedish "more dizzy". } line. 
 The energy range up to  about few MeV above the yrast line  is called the yrast region.
 The rotational bands  in this region are built on different \confs of the nucleons in the deformed rotation potential, while the  
 vibrational modes are in the  ground  or the one-phonon state.

Molecules have also three different modes: rotation, vibrations of the atomic nuclei relative to each other and the motion of the valence electrons. 
 There is a clear separation of the energy scales of the motion of the electrons  and of the nuclei, 
 which is due to the difference of their masses. As both constituents are confined to  the size of the molecule, 
 \beq
 \frac{\hbar \om_{\mathrm{electronic}}}{\hbar\om_{\mathrm{nuclear}}}
 \sim\frac{ m_{\mathrm{nucleus}}c^2}{m_{\mathrm{electron}}c^2}\sim \frac{50000\mathrm{MeV}}{0.5\mathrm{MeV}}=10^5.
 \eeq 
 The electronic states are determined by the instantaneous potential generated by the slowly moving nuclei. The time dependence of
 this potential is neglected, which is called the lowest order adiabatic approximation. 
 
    The Unified Model is based on the adiabatic approximation, which is also called the strong coupling limit. It is 
     assumed that the rotational frequency is small as compared to the typical frequencies of the nucleons in the rotating
 potential such that the reaction of the nucleons to the inertial forces can be neglected 
 (or taken into account in low-order perturbation theory, see Ref. \cite{BMII} 4A-3). 
 The scale ratio for normal deformed nuclei is given  by the ratio between the average splitting of the single particle levels in the deformed 
 potential $\hbar \om_{intrinsic}\sim\beta \hbar\om_0$ (see Fig. \ref{f:nilsson}) and the typical rotational transition energies $\hbar\om_{R}= R/{\cal J}$, which is
 \beq
 \frac{\hbar \om_{intrinsic}}{\hbar\om_{R}}\sim\beta \hbar\om_0 \frac{{\cal J}}{R}
 \sim 0.3\times 7.4~\mathrm{MeV}\frac{30~~~}{\mathrm{MeV}~R}\approx\frac{ 60}{R}
 \eeq
 at the best. The reason for the rather moderate scale ratio is that nuclei are composed of protons and 
 neutrons that have nearly the same mass.

\subsubsection{The strong coupling limit}\label{sec:StrongCoupling}

The adiabatic approximation implies that the Hamiltonian of the Unified Model is the sum of the
rotor Hamiltonian (\ref{eq:HROT}) and the quasiparticle Hamiltonian (\ref{eq:hqp}).  
\beq\label{eq:HUM}
H_{UM}=H_{ROT}+H_{int},~~H_{int}=h_{qp}+H_{V},
\eeq
  where $H_{V}=T_{\beta\beta}+T_{\ga\ga}+C_\beta/2(\beta-\beta_0)^2+C_\ga/2\ga^2$ is
   the vibrational part of the Bohr-Hamiltonian (\ref{eq:BH}) for a  potential $V(\beta,\ga)$ that is quadratic
   around its minimum at $\beta_0$ and $\ga=0$.
The eigenstates are the product of the collective rotor wave functions  and the intrinsic wave functions $\phi=\phi_{qp}\phi_{V}$ represented by
 the \qp \confs in the deformed potential $\phi_{qp}$ and the vibrational states $\phi_{V}=\phi_{n_\beta}(\beta)\phi_{n_\ga}(\ga)$.

In case of molecules the well localized positions of the different atomic nuclei allows one to 
define the orientation of the molecule in  a straightforward way.
In contrast, nuclei are composed of protons and neutrons, 
which are identical Fermions concerning the symmetry of their
 many-body mean-field state in the deformed potential (see \ref{sec:UMdSM}).  
 The indistinguishability of the constituents has cutting
 consequences for the rotational motion, which are illustrated in Fig. \ref{f:UMrot} for the case of an axial rotor.

 The collective degrees of freedom are the Euler angles which determine the orientation of 
 the quasiparticle state $\phi(\xi)$. 
 \setcounter{footnote}{0}
  As illustrated by Fig. \ref{f:UMrot} (a) this intrinsic wave function is not changed by a rotation about the symmetry axis (3).
 The intrinsic wave function $\phi$ is an eigenfunction of $\hat J_3$, the projection of the \am on the symmetry axis,
 with the eigenvalue $\Om$. 
 \footnote{There is an unfortunate ambiguity in notation. Conventionally $\Om$
 is used as a short-hand symbol for the Euler angles. Within the Unified Model 
 $\Om$ denotes the eigenvalue of the \am projection on the symmetry axis. In oder to keep with established notation
 in the literature I use $\Om$ for both. Their meaning should become clear from the context.},
 The rotation operation
 \beq
 {\cal R}_3(\f)\phi=\exp [-i\f\frac{ \hat J_3}{\hbar}] \phi=\exp [-i\f \Om] \phi
  \eeq   
  does not reorient 
 $\phi$, because   a state is only defined up to an arbitrary phase of the wave function. 
 This means the nucleus cannot collectively rotate about the 3-axis and
   the \momi   ${\cal J}_3=0$. The other two \momis are equal, ${\cal J}_1={\cal J}_2={\cal J}$.

  \begin{figure}[t] 
  \begin{center}
 \includegraphics[width=10cm]{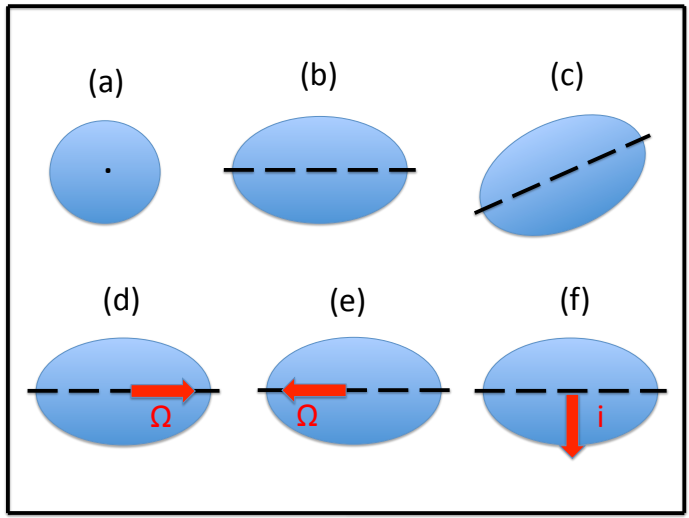}
 \caption{\label{f:UMrot} Rotating a prolate ensemble  composed of indistinguishable particles. The red arrow indicates the angular momentum of the intrinsic state.
 (a): Rotating about the symmetry axis does not change the system because the constituents are
 indistinguishable. (b) and (c): Rotating perpendicular to the symmetry axis changes the system. (b) and (f): Rotation by $\pi$  perpendicular
 to the symmetry axis does not change the system. The orientation angle is restricted to one hemisphere. 
 (d) and (e): Rotation by $\pi$  perpendicular
 to the symmetry axis changes the system. The orientation angles covers the full unit sphere. }
  \end{center}
 \end{figure}

 \setcounter{footnote}{0} 
 The eigenvalues and eigenfunctions of the axial limit of the rotor Hamiltonian (\ref{eq:HROT}) are discussed in
  many textbooks, as  e. g. \cite{BMII,R70,Ringbook,blaizotbook}, which provide citations of the original work. They are given by
   \beq\label{eq:ErotR}
 E(I)=\frac{I(I+1)-K^2}{2{\cal J}},~~~\Psi_{IMK}(\Om)=\left(\frac{2I+1}{8\pi^2}\right)^{1/2}D^I_{M,K}(\Om).
  \eeq
The eigenfunction are the Wigner D-functions $D^I_{M,K}$ which are exposed in all standard treatments of 
   \am in quantum mechanics   (see e. g. Ref. \cite{Rose57},  see also section 1A in Ref. \cite{BMI} and Appendix B in Ref. \cite{R70})
   \footnote{We adopt the definition of the D-functions and phase conventions used by Bohr and Mottelson \cite{BMII} and Rowe \cite{R70}.}.
   Their indices $I$, $M$, $K$, respectively, indicate the absolute \amd, the \am projection on the laboratory z-axis and the \am projection on
   the symmetry axis 3 in units of $\hbar$.  
     
  First we consider bands in even-even nuclei who's 
  intrinsic wave functions do not  carry individual angular momentum along the 3-axis, i.e. $\Om=0$.
     The total \am agrees with the collective \am of the rotor $\hat J_i=\hat R_i$ and $I=R$.
  The energy and the wave function of the Unified Model Hamiltonian (\ref{eq:HUM}) are given by   
   \beq\label{eq:ErotR}
   E(I)=E_{int}+\frac{I(I+1)}{2{\cal J}},~~~\Psi_{IM}(\Om)=\left(\frac{2I+1}{8\pi^2}\right)^{1/2}D^I_{M,0}(\Om)\phi_{K=0}(\xi). 
   \eeq
    It is important keeping in mind that the 
    \qp coordinates $\xi$ are defined relative to the body fixed frame of reference, that is, they change with the Euler angles $\Om$. 
   
   As illustrated by Fig. \ref{f:UMrot} (b), rotating the intrinsic wave 
   function $\phi$ by the angle $\pi$ about the 1-axis brings it back into an indistinguishable position, which  
    means that it can only differ from the original one by a phase factor,
   \beq
    {\cal R}_1(\pi)=\exp [-i\f \frac{\hat J_1}{\hbar}\pi] \phi=r\phi=\exp(-i\alpha \pi)\phi.
        \eeq
 The phase factor $r$ or its exponent $\alpha$ are called the signature of the intrinsic state.  Changing the 
 Euler angles such that they correspond  to the same rotation ${\cal R}_1(\pi)$ multiplies
 the $D^I_{M,0}$ functions  by the phase factor  $\exp(iI\pi)$, which is
 the phase change generated by collective rotation. To have an identical state, this phase  must compensate
 with the phase generated by the direct rotation  of $\phi$,
    which means that only states with $I=\alpha+2n$, $n$ integer are possible.

    Each second value of $I$ is forbidden, because
     the symmetry of the intrinsic wave functions permits specifying the orientation of the symmetry axis only within a hemisphere.
 The ground state of even-even nuclei is even with respect to ${\cal R}_1(\pi)$, because the orbitals with $\Om_i$ and $-\Om_i$ are occupied 
 with equal probability in the BCS ground state and the ground states of the vibrational modes are even as well.
 As a consequence, 
$\alpha=0$,  $r=+1$ and the ground state band has only even spins. The $\beta$ vibrations have $r=1$ as well, which means that the 
one-phonon $\beta$ band comprises only even spins. 
The $K=0$ collective octupole vibration is an oscillation of a pear-shaped
deformation overlaying the prolate shape. 
The one-phonon state has $r=-1$, which implies that the octupole band comprises only odd values of $I$.

Figs. \ref{f:UMrot} (d) and (e) illustrate the case when
  the intrinsic wave function carries its own \am $\hbar\Om$, which for nucleons in a non-rotating  axial potential must have the direction of the symmetry axis. 
 Since there is no collective rotation about the symmetry axis possible, the projection of the total \am on the symmetry axis $K$ must be equal to the intrinsic one,
 $K=\Om$. Using $\hat J_1=\hat R_1,~\hat J_2=\hat R_2,$ eigenvalues and eigenfunctions of the Unified Model Hamiltonian (\ref{eq:HUM}) are
  \beq\label{eq:ErotI}
 E(I)=E_{int}+\frac{I(I+1)-K^2}{2{\cal J}},~~~\Psi_{IMK}(\Om)=\left(\frac{2I+1}{8\pi^2}\right)^{1/2}D^I_{M,K}(\Om)\phi_K(\xi),
  \eeq
 where  $I\geq K>0$. 
 As seen in Fig. \ref{f:UMrot} (d,e), the presence of intrinsic \am
 along the symmetry  axis (arrow with a tip)  makes it possible to specify the orientation for the whole $\Om$ sphere, which 
 has the consequence that there is no 
 signature selection rule that restricts the values of $I$. The \qp \confs in the axial potential are always two-fold degenerate, corresponding 
 to the two \am projections (d) and (e) in  Fig. \ref{f:UMrot}. They cannot be considered as two different intrinsic states because 
 the full sphere of the possible orientations of the \am arrow  includes both projections in, respectively, the eastern and western hemispheres.
 Thus only $K>0$ should  be considered as an individual intrinsic state.
   Care has to be taken when evaluating matrix elements with the wave functions (\ref{eq:ErotI}) for  $K=1/2$. Operators like $J_1$ may
   flip the \am projection on the symmetry axis which generates matrix elements that connect the two hemispheres.  
  In order to account for them in an explicit way, Bohr and Mottelson introduced a symmetrized  version  of the wave function,
  \bea\label{eq:PsiRotSym}
  \Psi_{IMK}(\Om)=\nonumber\\
  =\left(\frac{2I+1}{16\pi^2}\right)^{1/2}\left(D^I_{M,K}(\Om)\phi_K(\xi)+(-1)^{I+K}D^I_{M,-K}(\Om){\cal R}_2(-\pi)\phi_K(\xi) \right).  
   \eea
  Its form is suggested by the consideration that instead of constructing  the wave function from the intrinsic \conf with the \am
  projection pointing in the direction of the positive 3-axis, as in Eq.  (\ref{eq:ErotI}), one may also construct it from the intrinsic \conf with the \am
  projection pointing in the direction of the negative  3-axis, which   is the second term. The phase factor is derived from the requirement that 
  the two forms must be the same wave function, including the phase. The detailed derivation can be found in the textbooks \cite{BMII} (section 4.2c)
  and \cite{R70} (sections 6.1-6.3).
 As discussed in section \ref{sec:QTR} below, taking into account the coupling between rotation and the \qp degrees of freedom in first order
 perturbation theory modifies the energy of the $K=1/2$ bands. The extended energy expression
 for all values of $K$ becomes   
\beq\label{eq:ErotIext}
E(I)=E_{int}+\frac{1}{2{\cal J}}\left(I(I+1)-K^2+\de_{K,1/2}a(-1)^{I+1/2}(I+1/2)\right).
\eeq
The additional term splits the $K=1/2$ bands into two branches with the signatures $\alpha=\pm1/2$, where  the splitting is determined by
the "decoupling parameter" $a$.

Evaluating electromagnetic transition matrix elements, the transition operator must be transformed to the body fixed frame of reference
where the integration over the intrinsic coordinates $\xi$ can be carried out. Multipole operators transform under the rotation from the 
laboratory coordinate system to the body-fixed system  by
\beq\label{eq:Mbody2lab}
 {\cal M}_{\lambda,\mu}=\sum\limits_\nu D^\lambda_{\mu,\nu}(\Om){\cal M}_{\lambda,\nu}(\xi).
  \eeq 
  The $\Om$ integration over the product of the three $D$ - functions  results in the product of two Clebsch-Gordan coefficients
   (see e. g. Ref. \cite{Rose57},  see also section 1A in Ref. \cite{BMI} and Appendix B in Ref. \cite{R70}). That is,
 the ratios of the electromagnetic transition matrix elements between the  
 different members of the bands are determined by the rotational wave functions (\ref{eq:ErotI}) only in form of the two Clebsch-Gordan coefficients.
  Accordingly, the reduced transition matrix elements between two bands  based on the  intrinsic  states $\phi_{K_f}$ and  $\phi_{K_i}$
 are given by the Alaga rules \cite{Alaga55o}
\bea\label{eq:ME2AR}
\langle I_fK_f\vert\vert {\cal M}_{\lambda}\vert \vert I_iK_i\rangle=\left[ \frac{2I_i+1}{(1+\de_{K_i0})(1+\de_{K_f0})}\right]^{1/2}\times \nonumber \\
\{ \langle I_iK_i\lambda\nu\vert I_fK_f\rangle\langle \phi_{K_f}\vert {\cal M}_{\lambda,\nu}\vert\phi_{K_i}\rangle \nonumber \\
+(-1)^{I_i+K_i} \langle I_i-K_i\lambda\mu\vert I_fK_f\rangle\langle \phi_{K_f}\vert {\cal M}_{\lambda,\mu}\vert {\cal R}_2(-\pi)\phi_{K_i}\rangle \} \nonumber \\
\nu=K_f-K_i,~~\mu=K_f+K_i.
\eea     
 For transition within a band
 one has to use the diagonal matrix element  $\langle I_fK_i\vert\vert {\cal M}_{\lambda}\vert \vert I_iK_i\rangle $.
  D. Rowe lists  explicit expressions for the experimentally important electromagnetic and weak- interaction transition matrix elements in his textbook \cite{R70}
   (Eqs. (6.32) - (6.52) in section 6.7).

  \begin{figure} 
   \begin{center}
   \vspace*{-4cm}
 \includegraphics[width=\linewidth]{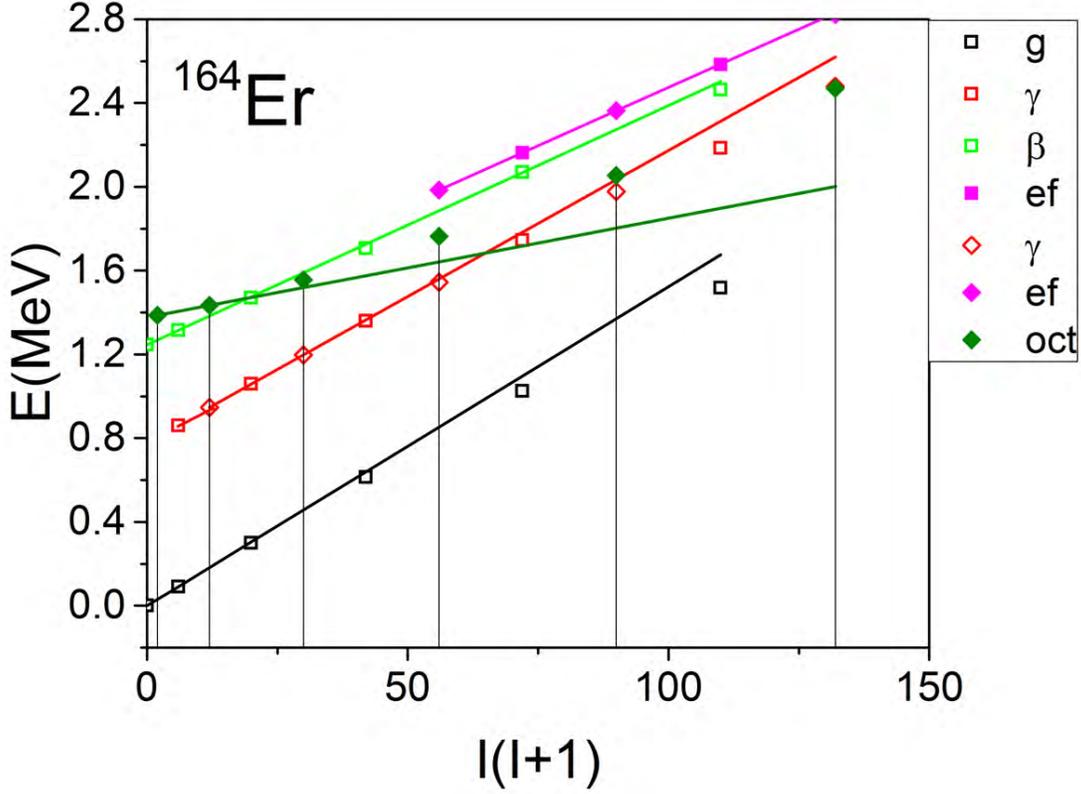}
  \vspace*{-4cm}
   \caption{\label{f:Er164Elow} Experimental rotational bands in $^{164}$Er. The labels indicate the bands built on the ground state  (g),  and on 
    the  one-phonon $\beta$, one-phonon $\ga$,
 two-quasiproton (ae, af) and one-phonon octupole (oct) excitations. The  convention for  $(\pi,\alpha)$ is: $\pi=+$ open symbols, $\pi=-$  full symbols,
 $\alpha=0$ side up, $\alpha=1$ angle up.
 The drop lines indicate the spin value $I$. Data from Ref. \cite{ENSDF}.  }
  \end{center}
  \end{figure}
  
  \subsubsection{Comparison with experiment}
   Fig. \ref{f:Er164Elow} shows the low-spin part of the rotational bands of the well-deformed  nuclide $^{164}$Er. The ground state band (g) has only even $I$ because the
 all even-even nuclei have $r=+1$. The energies follow the rigid rotor values (\ref{eq:ErotI}) $\propto I(I+1)$ only approximately. By $I=10$ they deviate by  200 keV. 
 Band (ef) is built on the intrinsic state that carries two excited quasiprotons, denoted by e and f. They generate an \am projection of $\hbar\Om=7\hbar$ along the symmetry axis.
 Even and odd values of $I$ merge into a smooth $\Delta I=1$ sequence, because the large \am projection strongly breaks the ${\cal R}_1(\pi)$ symmetry. 
 The \momi (slope) of the band is larger than the one of the g-band. The increase is caused by a combination of weakening of pair correlations, change of deformation and
 modification of the microstructure of the intrinsic state.   
 
The $\ga$ vibration represents a deviation of the nuclear shape from axial symmetry, which travels as a wave over the nuclear surface (see Fig. \ref{f:RoVib}). It
 carries an \am of $J_3=\hbar K=2 \hbar$ along the symmetry axis.  Accordingly, the $\gamma$ band is a $\Delta I=1$ sequence. 
 Its \momi is close to the one of the g-band, which is expected for a harmonic 
 shape vibration. This is an example for the general observation that the one-phonon $\gamma$ excitations  come closest to the shape-vibrations of the Bohr Hamiltonian (see section  \ref{sec:BHgen}).  There is no clear evidence for harmonic the two-phonon  $\ga$ excitations. 
 
 The 
 band denoted by $\beta$ is traditionally interpreted as the one-phonon excitation of the axial $\beta$ shape vibration, which does not carry \am of its own and has $\alpha=0$. 
 However,  its structure is likely  a more complex combination of the collective shape oscillation and a two-quasiparticle excitation (see section \ref{sec:BHgen}, Sec.\ref{sec:reg},
 the contribution to this Focus Issue by K. Heyde and J. L.  Wood \cite{NCHeydeWood} and e. g. Ref. \cite{Kulp08}). 
 Accordingly, its \momi differs from the one of the g-band being close to the one of the two-quasiproton band ef. 
 
 The band  (oct) is based on the $K=0$ octupole vibration.
 It is the oscillation   of a pear-shape distortion, which is odd under space inversion and odd under a rotation by $\pi$ perpendicular to the symmetry axis.
 Accordingly,  the intrinsic state has odd parity and $\alpha=1$ and  the band consists of
 the sequence of states with $I^\pi=1^-,~3^-,~5^-,~ ...$. The energies strongly deviate from the rigid rotor values, which indicates that the structure of the intrinsic state 
 changes with \amd.

  \begin{figure} 
  \begin{center}
   \vspace*{-4cm}
    \includegraphics[width=\linewidth]{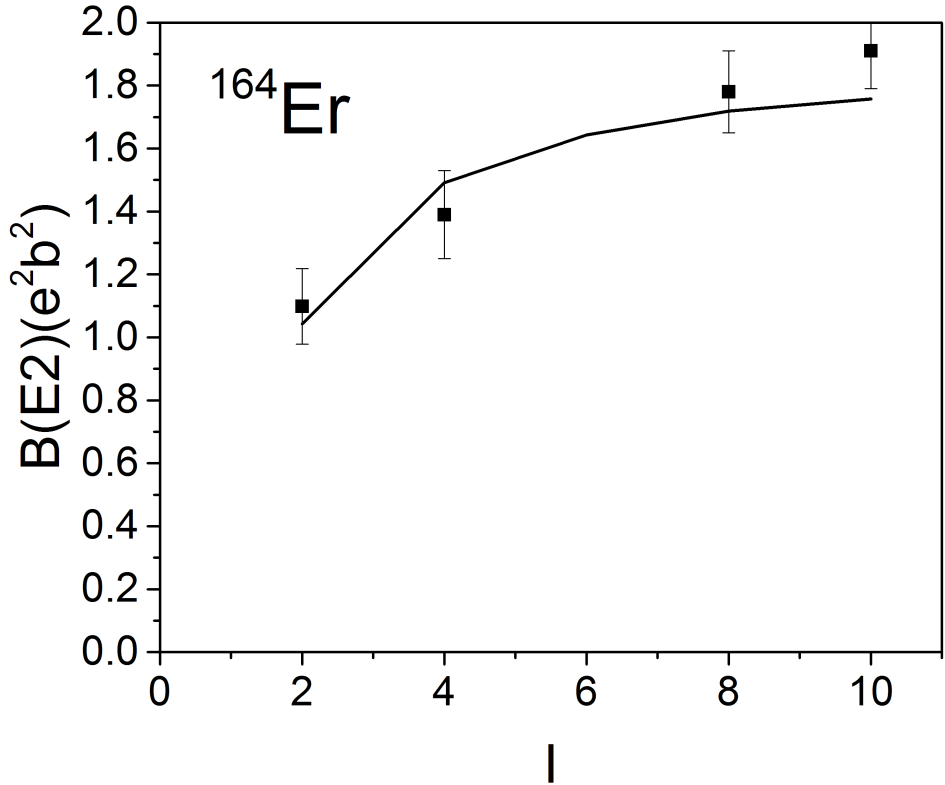}
 \vspace*{-4cm}
 \caption{\label{f:Er164BE2} Reduced transition probabilities $B(E2, I \rightarrow I-2)$ for the ground state band of $^{164}$Er. The line shows the axial rotor values 
 (\ref{eq:BE2AR}) calculated with the intrinsic charge quadrupole moment  $Q_0$=7.24 eb. Data from Ref. \cite{Er164BE2}}
  \end{center}
  \end{figure}

  \begin{figure} [t]
   \begin{center}
 \vspace*{-4cm}
 \includegraphics[width=\linewidth]{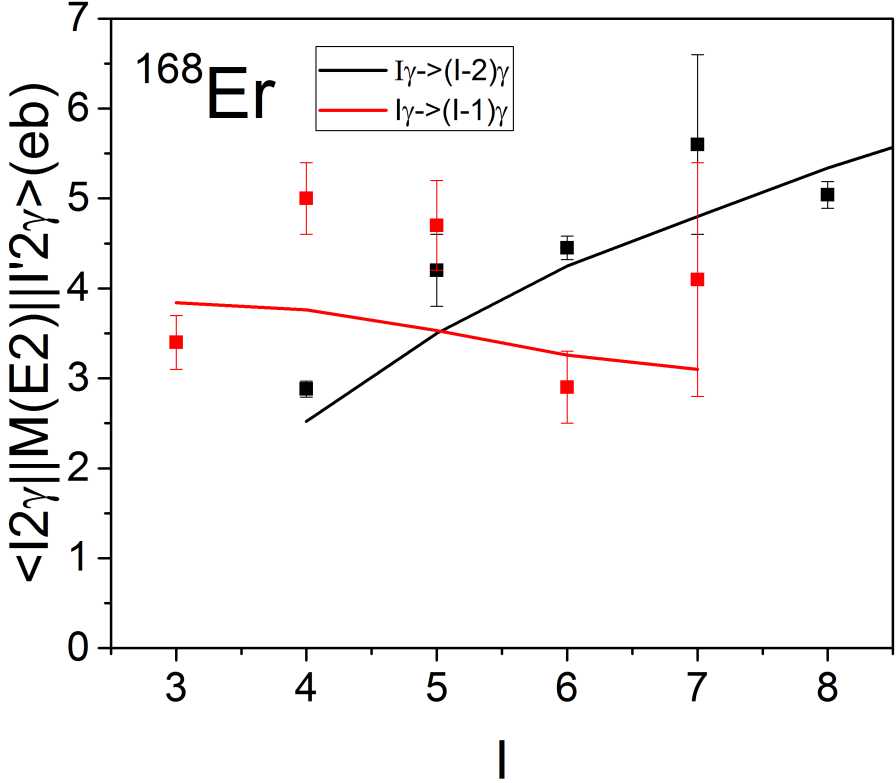}
 \vspace*{-4cm}
 \caption{\label{f:Er168ME2gaga} Reduced transition matrix elements $\langle I2,\ga\vert\vert {\cal M}_{\lambda}\vert \vert I'2,\ga\rangle$ 
 for transitions within the $\ga$ band  of $^{168}$Er. 
 The dashed lines (not visible)  show the axial rotor values 
 (\ref{eq:ME2AR}) calculated with the intrinsic charge quadrupole moment  $Q_0$= 7.7 eb. 
 The full lines show the triaxial rotor calculations. 
 The dashed lines are not visible because two calculations are too close.
   Data from Ref.\cite{Er168ME2}.}
  \end{center}
 \end{figure}
  \begin{figure}[t]
   \begin{center}
 \vspace*{-4cm}
 \includegraphics[width=\linewidth]{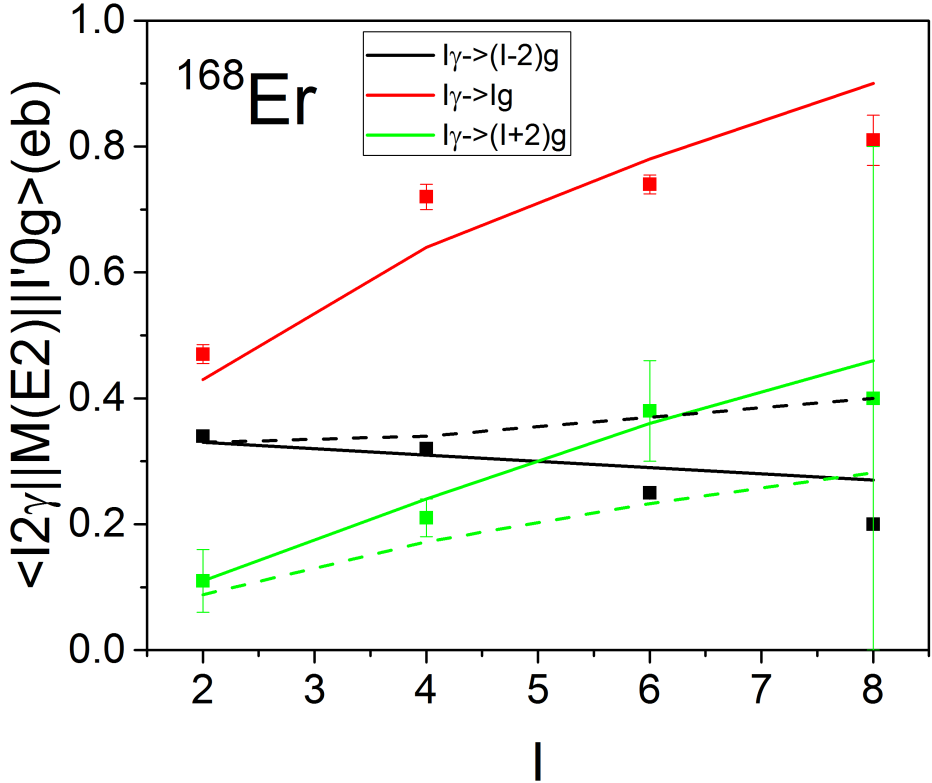} 
  \vspace*{-4cm}
 \caption{\label{f:Er168ME2gag} Reduced transition matrix elements $\langle I2,\ga\vert\vert {\cal M}_{\lambda}\vert \vert I'0,g\rangle$ 
 for transitions between the $\ga$ band and the ground band of $^{168}$Er. The full lines show the the triaxial rotor calculations, the dashed lines the axial rotor values 
 (\ref{eq:ME2AR}) calculated with the transition charge quadrupole moment  $Q_2$= 0.76  eb.   For $I\rightarrow I$ transitions the dashed lines are not
 visible because the two calculations are too close. Data from Ref. \cite{Er168ME2}.}
  \end{center}
 \end{figure}
\begin{figure}
   \begin{center}
 \vspace*{-4cm}
 \includegraphics[width=\linewidth]{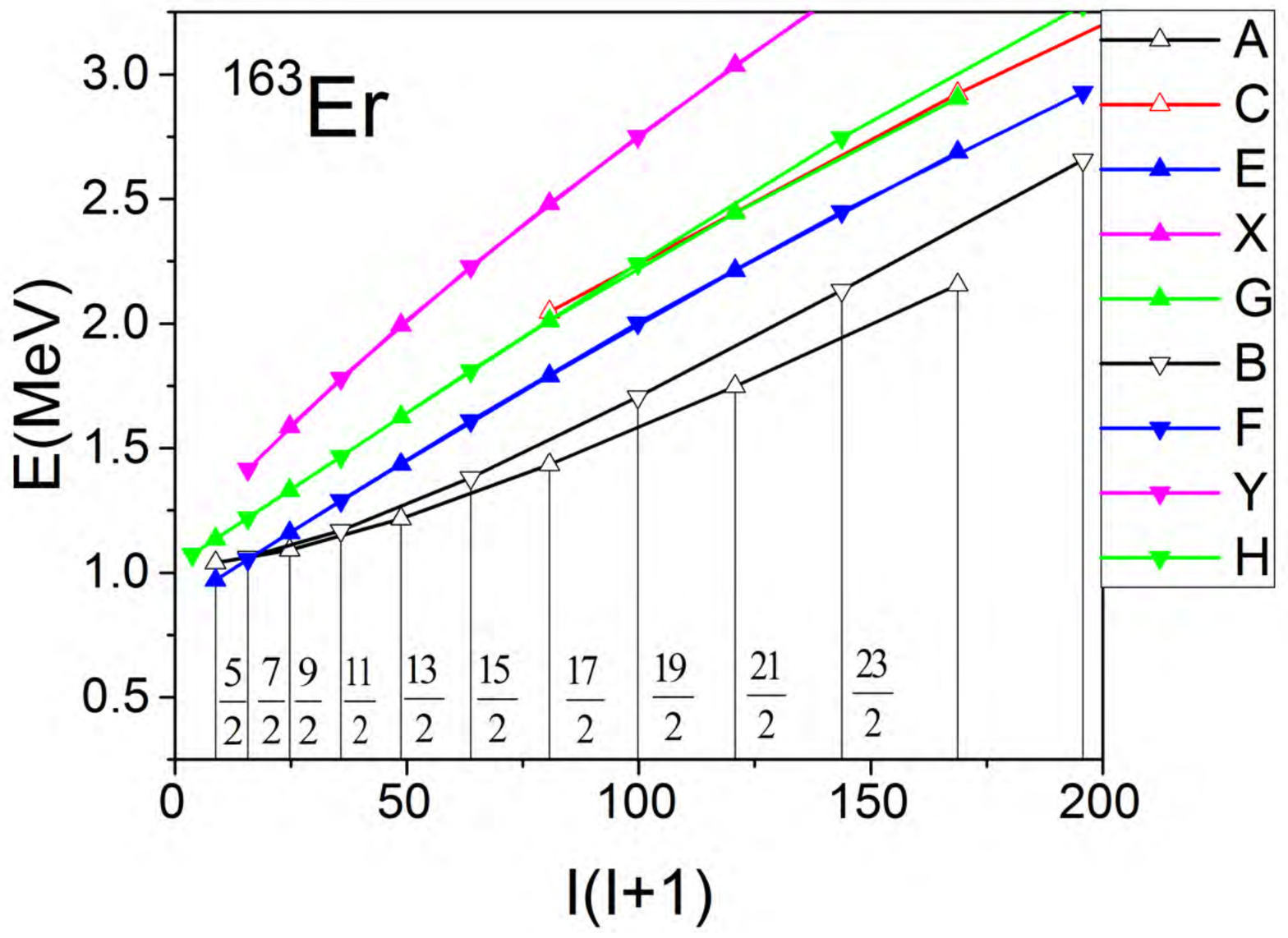} 
  \vspace*{-4cm}
 \caption{\label{f:Er163Elow} Experimental rotational bands in $^{163}$Er. The labels indicate the bands built on various one-quasineutron states that are 
  are labeled by letters. The associated Nilsson labels are quoted in Tab. \ref{t:code}. The  convention for  $(\pi,\alpha)$ is: $\pi=+$ open symbols, $\pi=-$  full symbols,
 $\alpha=1/2$ angle up, $\alpha=-1/2$ angle down.
  For consistency with later discussions the ground state has assigned an energy of 0.92 MeV,
   which takes into account the difference between its 
  binding energy and the mean of the binding  energies of its even-N neighbors. It represents one-quasineutron excitation energy. Seven more bands are identified with $I<7/2$ and $E^* <$ 0.8 MeV,
  which are not shown.  Data from Ref. \cite{ENSDF}. }
  \end{center}
 \end{figure}

 In case of the g-band of even-even nuclei the Alaga rules (\ref{eq:ME2AR}) give for the intraband transitions $I\rightarrow I-2$ the reduced transition probability
 \beq\label{eq:BE2AR}
 B(E2, I\rightarrow I-2)=\frac{5}{16\pi}\langle I~0~2~0\vert I-2~0\rangle^2e^2Q_0^2,
 \eeq
 where $eQ_0$ is the quadrupole moment of the prolate charge distribution as defined in \cite{BMII}. 
Fig. \ref{f:Er164BE2} demonstrates that the experimental values follow closely the rigid rotor ratios

The other transition matrix elements of $^{164}$Er are not well known. A good set has been extracted from the COULEX work by Kotili\'nski
{\it et al.} \cite{Er168ME2} on $^{168}$Er, which
is presented in Figs. \ref{f:Er168ME2gaga} and \ref{f:Er168ME2gag}. 
(The energies follow the $I(I+1)$ rule with about the same accuracy as in $^{164}$Er.)
The figures  show the reduced transition matrix elements (\ref{eq:ME2AR}), where 
the intrinsic transition matrix elements are taken as  $ \langle 0, g\vert {\cal M}_{2,0}\vert 0,g\rangle=\sqrt{5/16 \pi}Q_0$ and
  $ \langle 2, \ga\vert {\cal M}_{2,2}\vert 0,g\rangle=\sqrt{5/16 \pi}Q_2$ with $Q_0=7.7$ eb and $Q_2=0.76$ eb. The Alaga  rules (\ref{eq:ME2AR}) reproduce the experiment fairly well.
The $I\rightarrow I\pm2$ transition matrix elements show systematic deviations.  The detailed analysis of the similar 
observation in $^{166}$Er by Bohr and Mottelson \cite{BMII} demonstrated that the deviations are due to the  coupling between the $\gamma$ and the g- band. 
Such a coupling can be taken into account  by assuming a slight triaxiality $\gamma >0$ for the rotor Hamiltonian (\ref{eq:HROT}). The rotor states for small triaxiality are given in
Ref. \cite{BMII} (section 4.5c). Assuming that the moments of inertia depend on $\gamma$ as expected for irrotational flow (\ref{eq:momiIF}) and that the ratio between the intrisic quadrupole moments is
$Q_2/Q_0=\tan \gamma\sqrt{2}$, the $I\rightarrow I\pm 2$ transition matrix elements are well described  with the choice of $\gamma=8^\circ$ \cite{Er168ME2}. 
The modification  of the other matrix elements by the coupling is too small to be visible in Figs. \ref{f:Er168ME2gaga} and \ref{f:Er168ME2gag}.
There are noticeable deviations from the triaxial rotor calculations for the $\De I=1$ transitions, which have a mixed E2/M1character. 
The M1 component is disregarded in the calculations and cannot be well extracted from the COULEX data. 

The examples illustrate the general observation that the adiabatic pattern of Fig. \ref{f:UMrot}  is realized in well deformed even-even nuclei for $I<10,~ A\approx 170$  and $I< 20,~A\approx 240$.
The large gap of about 2 MeV between the ground state and the first two-quasiparticle excitations as well as the rigid shape are the reason for the adiabatic behavior.
The situations is less favorite in odd mass nuclei. Fig. \ref{f:Er163Elow} shows the low-spin part of the rotational bands in $^{163}$Er, which are based on different intrinsic 
one-\qn states. Fig. \ref{f:nilsson} shows that the band heads  of the $N=95$ isoton $^{161}$Dy  have quite similar energies. 
 There is no energy gap between the ground and excited bands, because the distances between lowest quasiparticle energies  $\sqrt{\left(e_i-\lambda\right)^2+\Delta^2}$ 
are strongly reduced by the pair correlations ($\De \sim $ 1 MeV, $\vert e_i-\lambda\vert \sim $0.3 MeV).
The rotational bands are extracted from the dense spectrum using the fact that the intraband E2 transitions are much stronger than transitions between different bands, 
and  that the transition energies change in a smooth way with $I$. 
 As seen in Fig. \ref{f:Er163Elow}, the deviations of several bands from the adiabatic 
$I(I+1)$ rule are substantial for some of the bands.

There is no evidence for the existence of a collective $\gamma$ vibration built on  the intrinsic ground state (F). In contrast to 
the even-even neighbor, the collective excitation, which is expected at about 0.8 MeV, is situated among   quasineutron states to which 
it couples. There are seven more bands identified  with band heads  below 0.8 MeV, which are not shown in the figure. 
The coupling fragments the collective mode among the states to which it couples.  
Studying the quasiparticle-phonon coupling experimentally requires the measurement pertinent  transition matrix elements, which does not exist.
Conceptually, one has reached the limit of the Unified Model.


\begin{table}
\hspace*{2.2cm}
\begin{tabular}{c c c c c c c c c c c   }
A &$\nu [642]\frac{5}{2}$&$(+,+\frac{1}{2}$)&B &$\nu [642]\frac{5}{2}$&$(+,-\frac{1}{2}$)& C &$\nu[651] \frac{3}{2}$&$(+,+\frac{1}{2}$) \\
 E&$\nu [523]\frac{5}{2}$&$(-,+\frac{1}{2}$)&F &$\nu[523] \frac{5}{2}$&$(-,-\frac{1}{2}$)&G&$\nu [521]\frac{3}{2}$&$(-,+\frac{1}{2}$)\\
 H&$\nu [521]\frac{3}{2}$&$(-,-\frac{1}{2}$)&X &$\nu [505]\frac{11}{2}$&$(-,+\frac{1}{2}$)&Y&$\nu[505] \frac{11}{2}$&$(-,-\frac{1}{2}$)\\
 a&$\pi [404]\frac{7}{2}$&$(+,+\frac{1}{2}$)&e&$\pi [523]\frac{7}{2}$&$(-,+\frac{1}{2}$)&f&$\pi [523]\frac{7}{2}$&$(-,-\frac{1}{2}$)\\
\end{tabular}
\caption{\label{t:code}Letter code used to label quasiparticle orbitals.  The Nilsson labels are quoted after the letter to be followed by the parity and signature. }
\end{table}
\begin{figure} 
\begin{center}
 \hspace*{1cm}
\includegraphics[width=\linewidth]{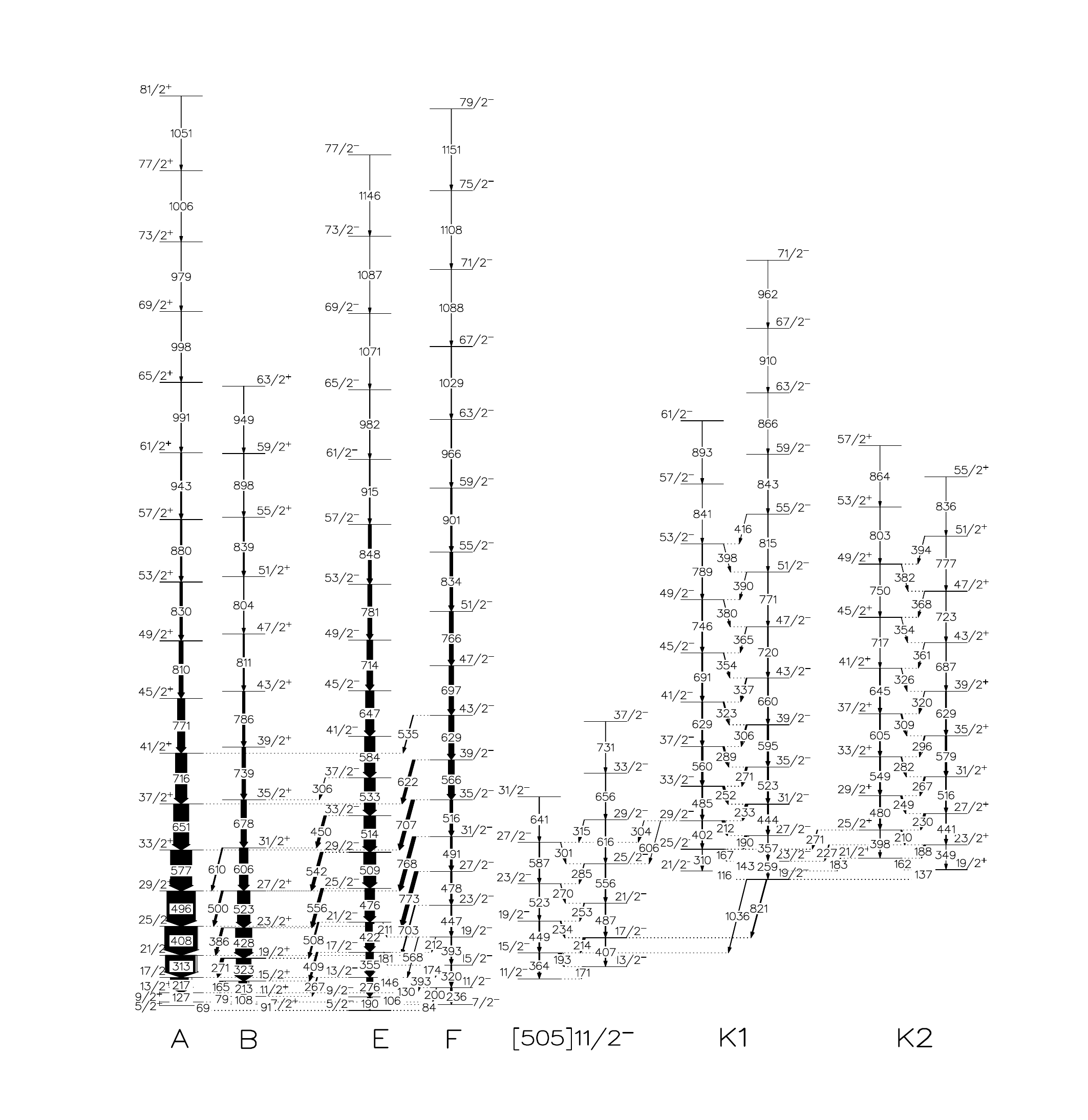} 
\caption{\label{f:Er163spec} Partial level scheme of the nucleus $^{163}$Er
 presented by  Hagemann {\it et al.} \cite{Er163}.
The energies are given in keV. Each level is labeled by the spin 
and parity $I^\pi$.
The bands are labelled  by the  quasiparticle configurations, which are
discussed in  Sects.  \ref{sec:CSM} and  \ref{sec:TAC}.
}
  \end{center}
  \end{figure}

 \begin{figure}
   \begin{center}
   \vspace*{-4cm}
 \includegraphics[width=\linewidth]{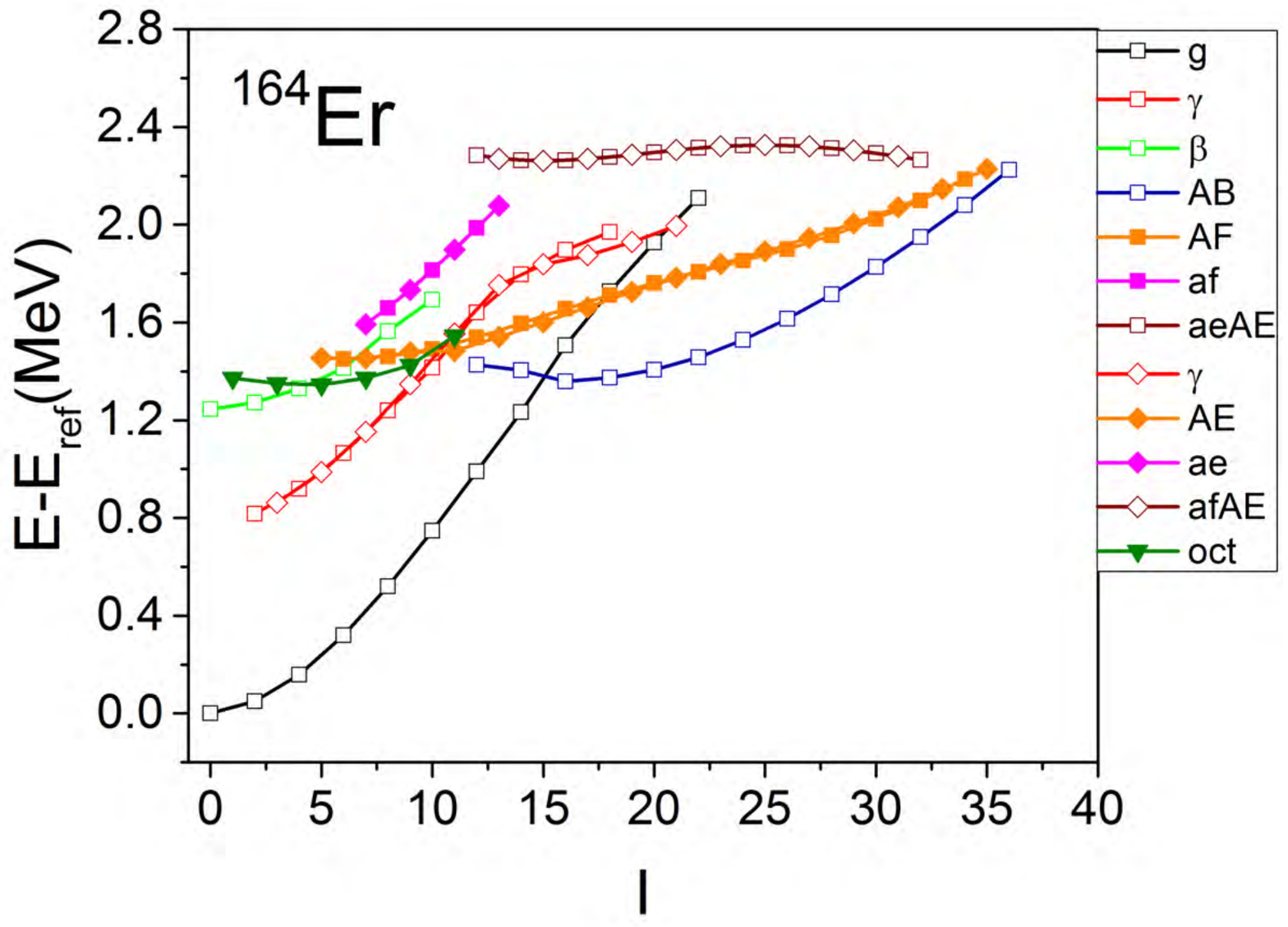} 
  \vspace*{-4cm}
 \caption{\label{f:Er164Ehigh} Experimental rotational bands in $^{164}$Er. A reference energy \mbox{$E_{ref}=0.007\times I(I+1)$ MeV} is subtracted. The labels indicate the configuration
  of rotating quasiparticles. Symbol convention as in Fg. \ref{f:Er164Elow}. Data from Ref. \cite{ENSDF}. }
  \end{center}
 \end{figure}
  \begin{figure}
   \begin{center}
 \vspace*{-4cm}
 \includegraphics[width=\linewidth]{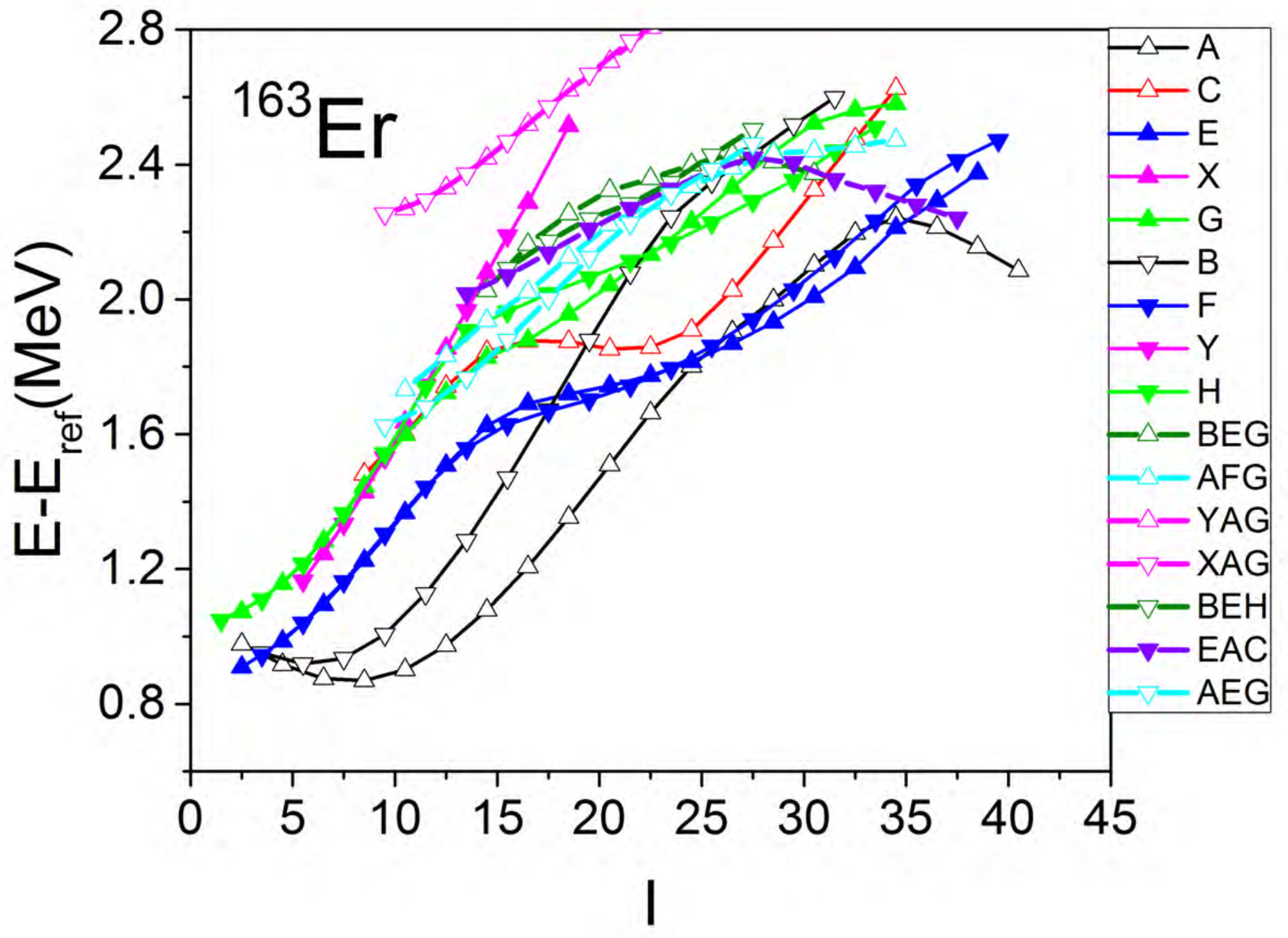} 
  \vspace*{-4cm}
 \caption{\label{f:Er163Ehigh} Experimental rotational bands in $^{163}$Er. A reference energy \mbox{$E_{ref}=0.007\times I(I+1)$ MeV} is subtracted.   
The labels indicate the configuration
  of rotating quasiparticles. Symbol convention as in Fg. \ref{f:Er163Elow}. Data from Ref. \cite{ENSDF}. }
  \end{center}
 \end{figure}
 
  Fig. \ref{f:Er163spec} shows an
example of a high-spin level scheme.
The transition energies between the members of a band are comparable with
the energy differences between the bands, which are given by the distances
between levels of the same spin.
Thus differences in the energy
scales cannot be used to group the levels into bands.
 The experimentalists 
 arrange the measured $\ga$ lines 
into a rotational spectrum like Fig. \ref{f:Er163spec}  as follows.\\
1) The states  of a band are connected by fast 
electromagnetic transitions of low multipolarity
($E2$,  $M1$, $E1$, $E3$).\\
2) The transition energy grows with the angular momentum $I$ in a smooth way.\\
3) The transition matrix elements connecting the states gradually 
 change with $I$.\\
These criteria are quite handy tools for systematizing the data.
However they also reflect   those features  of bands 
that  remain valid at high spin.
The first criterion states that the 
nucleus has large electromagnetic multipole moments, which are carried along
with the rotation. They are the source of the radiation, which manifests 
itself as a cascades of
sequential fast transitions.
The multi-coincidence $\gamma$-detector arrays are very good filters for
such cascades. The second and third  criterion state that
the intrinsic nuclear structure  changes only gradually along a band.
Figs. \ref{f:Er164Ehigh} and \ref{f:Er163Ehigh} display the energies of rotational bands up to the highest spins reached in experiment. 
The  bands show kinks and bends which indicate major intrinsic rearrangements.

The figures illustrate the structure of the  high-spin yrast region  about 2 MeV above yrast line,  which is
 the sequence of levels with minimal energy for given \amd.  Although the energy increases by 10 MeV in the shown range of spin, the 
 level density remains about the same in the  region of about 1 MeV above the yrast line.  There one can experimentally identify rotational bands and assign to them an intrinsic structure, which 
 however is substantially modified by the inertial forces. In contrast, 
  at zero \am the same excitation energy leads into a region of very high level density (level distance eV),  where one has to change to a statistical description in terms of average quantities (see Fig \ref{f:YrastRadiation} in section \ref{sec:beyond}).  
  
   \begin{figure}
  \begin{center}
  \hspace*{2cm}
 \includegraphics[width=14cm]{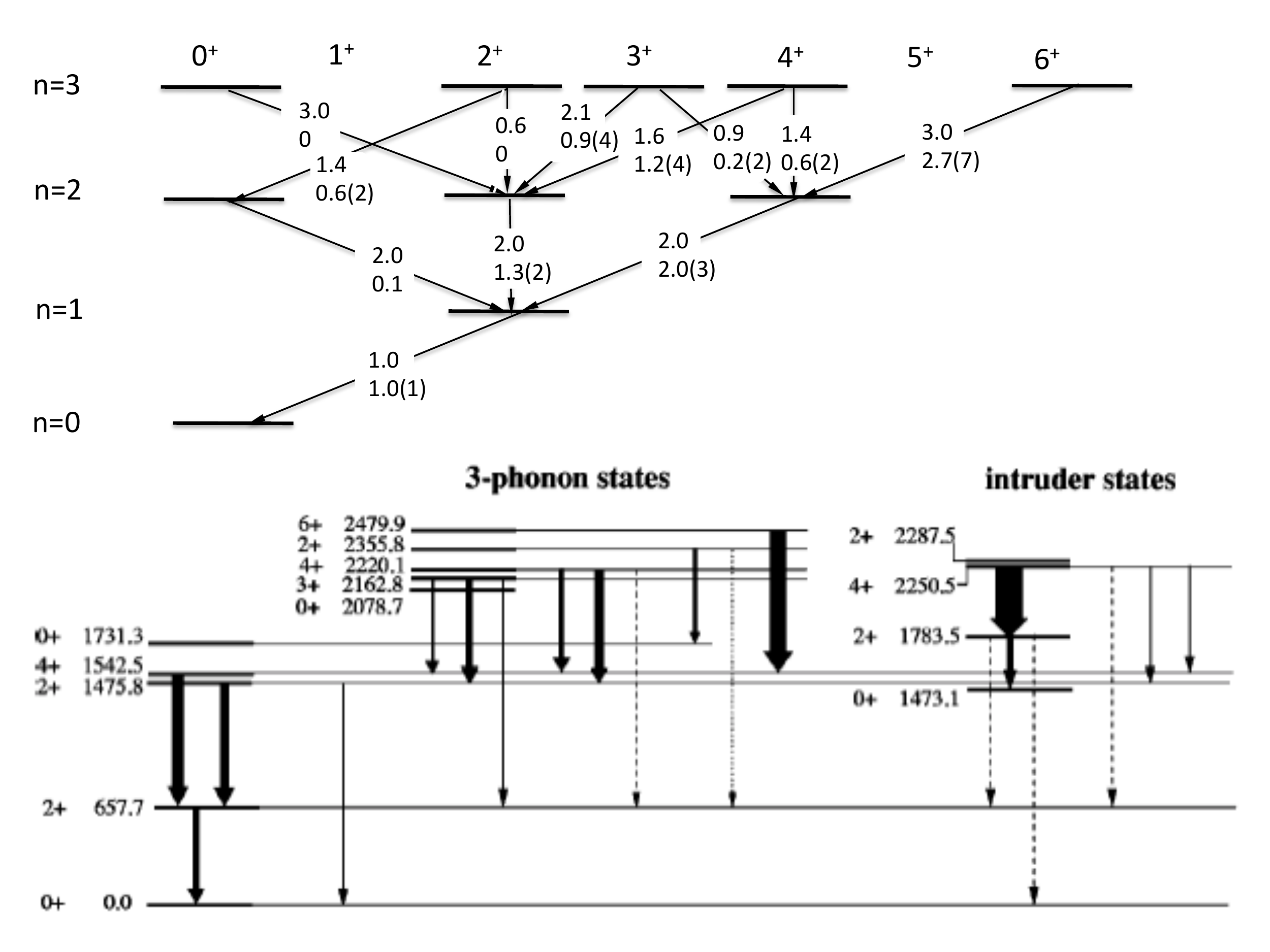} 
  \caption{\label{f:Cd110Elow} Upper part: The harmonic vibrator excitation spectrum. The allowed transitions are indicated by arrows. All other are forbidden. The numbers on the arrows quote the ratios of the reduced E2 transition probabilities.
   The upper numbers are the harmonic vibrator ratios and the lower are the experimental ratios $B(E2, I,n\rightarrow I',n-1)/B(E2,2,1\rightarrow0,0)$, which are taken   
  from \cite{Cd110BE2NDS,Cd110BE2Corminboeuf,Cd110BE2Piiparinen}. The value $B(E2,2,1\rightarrow0,0)$=27(15) Wu from \cite{Cd110BE2Piiparinen} is used. 
  Lower part: The experimental levels associated with the phonon multiplets and the rotational band based on the $0_2^+$ "intruder level". The width of the arrows correspond to the magnitude of the 
  experimental $B(E2)$ values. Figure is taken from Ref. \cite{Cd110BE2Corminboeuf}. }
  \end{center}
 \end{figure}
  
\subsection{Spherical nuclei}\label{sec:UMsph} 
The minimum of the collective potential $V(\alpha_\mu)$  lies at spherical shape $\alpha_\mu=0$.  The shape executes harmonic quadrupole vibrations,
which are described by the harmonic vibrator Hamiltonian
\beq
H_{HV}=\sum\limits_{\mu=-2}\limits^{2} (-1)^\mu\left[ -\frac{\hbar^2}{2B}\frac{\partial^2}{\partial \alpha_\mu\partial\alpha_{-\mu}}+\frac{C}{2}\alpha_\mu\alpha_{-\mu}\right],~~\om_V=\sqrt{\frac{C}{B}}.
\eeq
It has the harmonic vibration spectrum 
\beq
E(n)=\hbar\om_V\left(n+\frac{5}{2}\right),
\eeq
 which is generated by exciting $n$ of the  five degenerate phonons. The phonons are described by the operators 
\beq
a^\dagger_\mu=\frac{1}{\sqrt{2}}\left(-b\frac{\partial}{\partial \alpha_\mu}+\frac{\alpha_{\mu}}{b}\right),~~a_\mu=\frac{1}{\sqrt{2}}\left(b\frac{\partial}{\partial \alpha_\mu}+\frac{\alpha_{\mu}}{b}\right),
~~b=\left(\frac{\hbar^2}{4CB}\right)^{1/4},
\eeq
where $b$ is the zero point amplitude. The phonon carries the \am  of 2$\hbar$ with the projection of $\hbar\mu$.  
The spectrum is composed of equidistant multiplets of $n$ phonons, which couple to  the total angular momenta of
$R=0$ to $R=2n\hbar$. The upper panel of Fig.  \ref{f:Cd110Elow} illustrates the spectrum and indicates the ratios between the 
reduced transition probabilities.  For the harmonic vibrator the $B(E2)$ values increase linearly with
 the phonon number $n$ as long as the phonons are independently excited. Coupling them to good \am redistributes the transition strength. 
 The coefficients of fractional parentage, which determine
 the redistribution, are discussed in Ref. \cite{BMII} (appendix 6b). The lower panel of Fig. \ref{f:Cd110Elow} shows the lowest positive 
 parity excitations in $^{110}$Cd, which is considered as one of the nuclides 
 that come as close as possible to the harmonic vibrator limit.  The $2^+$ one-phonon excitation lies at $\hbar \om_V$=658 keV. 
 Around 1500 keV  are the three states $0^+,~2^+,~4^+$, which are interpreted as the two-phonon triplet.
 Their energy is about 200 keV higher than $2\hbar\om_V$.   The group $0^+,~2^+,~3^+,~4^+,~6^+$ has been associated 
 with the three-phonon quintuplet. The experimental ratios of the yrast energies $E(I)/E(2)$=  
 2.3, 3.7, 5.0 for $I$=4, 6, 8 are larger than the harmonic vibrator values  2, 3, 4;  their ratios $B(E2, I,n\rightarrow I',n-1)/B(E2,2,1\rightarrow0,0)=$1.8,  2.3
 for $I=$4, 6 are smaller then the harmonic vibrator values of 2, 3 (see section \ref{sec:tidal}). The energies of the 
 non-yrast levels deviate from the harmonic vibrator values by comparable amounts. Their $B(E2)$ values strongly deviate from 
 the harmonic vibrator, where the deviations increase with the distance from the yrast line. One notices
  that transitions that go parallel to the yrast line come closer to the harmonic vibrator values than the other. The transitions from the 0$^+$ states are completely off.

\begin{figure}
 \begin{center}
  \hspace*{2cm}
 \includegraphics[width=13cm]{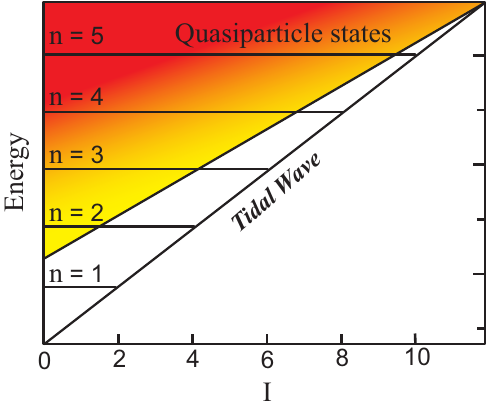} 
  \caption{\label{f:HV-qp}  Schematic representation of the location of the collective quadrupole vibrational excitations relative to the quasiparticle excitations. 
  The darker shades approximately indicate higher densities of quasiparticle states. Figure from Ref. \cite{TWPd102}.}
  \end{center}
 \end{figure}
  \begin{figure}
  \begin{center}
 \vspace*{-4cm}
 \includegraphics[width=\linewidth]{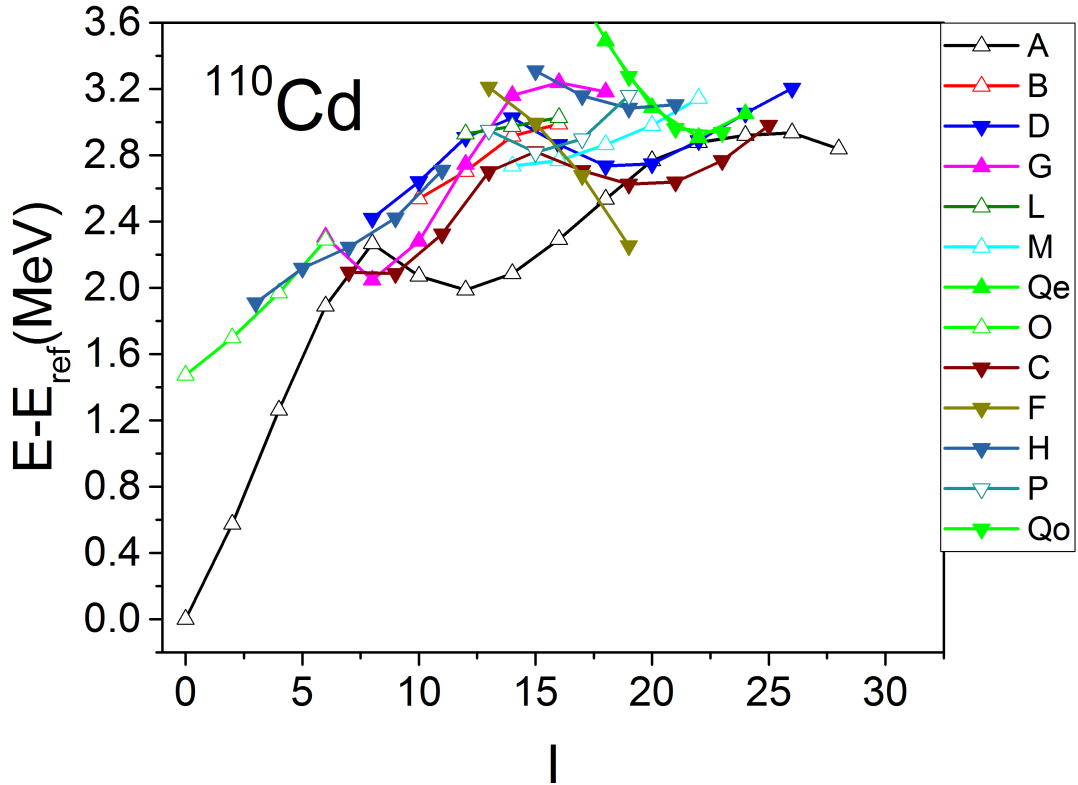} 
  \vspace*{-4cm}
 \caption{\label{f:Cd110Ehigh}  Experimental rotational bands in $^{110}$Cd. A reference energy \mbox{$E_{ref}=0.014\times I(I+1)$ MeV} is subtracted. 
 The  convention for  $(\pi,\alpha)$ is: $\pi=+$ open symbols, $\pi=-$  full symbols,
 $\alpha=0$ angle up, $\alpha=1$ angle down.
  The parity of Qe and Qo is not known.  Data from \cite{ENSDF},
  from which the band labels are adopted.    }
  \end{center}
 \end{figure}

  The experimental spectrum contains an additional
  0$^+$ state upon which a rotational sequence is built. Because it does not fit into the harmonic vibrator scenario, it is often called "intruder band". It is the rotational band built on a deformed four quasiparticle
  excitation, which coexist with the harmonic vibrator spectrum built on the spherical ground state. The coexistence of the spherical shape with a deformed one is common for nuclei classified as harmonic vibrators. K. Heyde and J. L. Wood \cite{NCHeydeWood} discuss the shape coexistence phenomenon in great detail in their contribution to this Focus Issue.
  
  The example describes how close real nuclei come to the  harmonic vibrator spectral characteristics in general. 
  Fig. \ref{f:HV-qp} illustrates the limits of the Unified Model description of spherical nuclei in a schematic way. 
  With increasing phonon number $n$, the zero \am members of the vibrational multiplets move into the region of high density of quasiparticle excitations. The coupling to intrinsic states causes
  a fragmentation of the collective transition strength among an increasing number of states. The fragmentation already sets in for the $0^+$ state of the two-phonon triplet. 
  The level density remains low  for the yrast states of the multiplets, because of \am conservation. As a consequence, the harmonic vibrator characteristics persist to higher phonon numbers. 
  The harmonic vibrator pattern erodes with decreasing spin within a multiplet (see section \ref{sec:tidal}).
 
 Fig. \ref{f:Cd110Ehigh} shows the yrast region of $^{110}$Cd up to the highest spins measured. The states can be grouped into quasi rotational sequences using the same 
 experimental criteria as for well deformed nuclei: Strong E2 transitions connecting adjacent band members and a regular increase of $E(I)$ with $I$. 
 The quasi rotational sequences are shorter than the rotational bands in  well deformed nuclei, but the band pattern is clearly discernible. In Figs. \ref{f:Er164Ehigh}, \ref{f:Er163Ehigh} and 
 \ref{f:Cd110Ehigh} an average  reference energy $E_{ref}(I)=I(I+1)/2{\cal J}_{ref}$ is subtracted. The values ${\cal J}_{ref}=71 \hbar^2$/MeV and $36  \hbar^2$/MeV
 are close to the \momis of rigid spheres, ${\cal J}_{rig}=2mr_0^2A^{5/3}/5$, with $A=$ 164 and 110, which are $72 \hbar^2$/MeV and $37 \hbar^2$/MeV, respectively.
 This comes not unexpected, because the yrast energy of the nucleonic Fermi gas is equal to $I^2 / 2{\cal J}_{rig}$. However, gross shell structure  causes deviations of the average \momi from the rigid body value \cite{Deleplanque04}.

\subsection{Microscopic basis of the Unified Model}\label{sec:UMmicro}
The preceding discussions considered the Unified Model from a phenomenological perspective. The parameters of the 
Unified Model, $\beta$, $\ga$, $\De$, $Q_0$, ${\cal J}$, ... were considered as 
adjustable  to describe the experiment. The parameters can be  calculated in the framework  time-dependent and time-independent mean field approaches 
 of increasing sophistication.   
Let us start with the potential $V(\beta,\ga)$, which is given by the time-independent mean field.  
The underlying Hartree-Fock-Bogolyubov  mean field theory is presented in standard textbooks on nuclear theory, 
 as for example \cite{R70,RW,Ringbook,blaizotbook}. Nowadays, there are four major approaches in practice, which have been refined over the last four decades such that
 they allow us to predict with good accuracy  static properties as binding energies, radii and deformation parameters. These approaches are the shell correction method,
 the Skyrme energy density functional, the Gogny effective interaction and the relativistic mean field theory. A review of the immense work invested into
 developing these tools is beyond the focus of my contribution.  Bender, Heenen and Reinhard  excellently reviewed the work in 2003 \cite{Bender03}.   
 For more recent developments see the literature cited in the contributions to this Focus Issue by Reinhard  \cite{NCReinhard}, Satula and Nazarewicz  \cite{NCSatula},
 Egido  \cite{NCEgido}, Meng and Zhao \cite{NCMengZhao} and Zhou \cite{NCZhou}.  
 
 Here I will only discuss the shell correction method \cite{Strutinsky67,Brack72}, which is also called the micro-macro approach.
It  over arcs  the two columns  of the Unified Model, the liquid drop model and the deformed shell model, in a way that 
allows one to calculate  $V(\beta,\ga)$ with considerable precision.    It exposes the mechanism behind the appearance of stable deformation  most directly and 
it is a time-proven simple method with predictive power that is  
comparable with the other selfconsistent mean field approaches being used at present. 
The new phenomena beyond the realm of the Unified Model, which will be discussed in this contribution, have been described in a 
quantitative way by the shell correction method.

Selfconsistency is the central element of the mean field approaches. The single particle orbitals, which are the eigenstates the single particle Hamiltonian $h_{def}$,
generate the density matrix $\rho$. The  deformed potential   of $h_{def}$ is calculated from $\rho$ in different ways, which depend on the
approach: Skyrme energy density functionals, the Gogny effective interaction and the relativistic mean field theory. 
The shell correction  method uses selfconsistency only in a restricted form, which 
accounts for the short range of the effective nucleon-nucleon interaction. It assumes that the nuclear matter density distribution corresponds to the one of a deformed droplet, which is constant inside and drops to 
zero in a thin surface layer. The central single particle potential generated by the short-range interaction will have a similar profile and shape, 
\beq
U\vec r~)=\int \rho(\vec r~') v(\vec r, \vec r~')d^3r' ~~~\mathrm{if} ~~~ v(\vec r, \vec r~')=v_0\delta(\vec r- \vec r~') \rightarrow U(\vec r~)=v_0\rho(\vec r~).
\eeq
Its depth and surface thickness are parameters.
The profile of the spin-orbit potential is taken as the gradient of the profile of the central potential with the strength as another parameter. The total energy is the sum of the "macroscopic"
energy of a liquid drop of nuclear matter $E_{LD}(\beta,\ga)$ and a "microscopic" shell correction $E_{SC}(\beta,\ga)$,
\beq\label{eq:SC}
E(\beta,\ga)=E_{LD}(\beta,\ga)+E_{SC}(\beta,\ga).
\eeq
Both depend on the shape, which is parametrized by a set of deformation parameters.
Here we only expose the  case of pure quadrupole deformation, but higher multipoles are taken into account when needed. 
The  macroscopic liquid drop energy is the sum of a volume term, a surface term and a Coulomb term, respectively,
\bea\label{eq:Eld}
E_{LD}(\beta,\ga)=-a_V\left(1-\kappa_VI^2\right)A+a_S\left(1-\kappa_SI^2\right)A^{2/3}B_S(\beta,\ga)\nonumber \\
+a_C\frac{Z^2}{A^{1/3}}B_C(\beta,\ga), ~~I=\frac{N-Z}{A},
\eea
 where $B_S$ is the ratio of the surfaces areas of the deformed and the spherical droplet and $B_C$ is the ratio of the Coulomb energies of a homogeneously charged  
   deformed and the spherical droplet \cite{MeyersSwiatecki}. The microscopic shell correction energy takes into account
    the deviations of the total energy from the macroscopic liquid drop energy
   due to the quantization of the nucleonic motion in the nuclear potential, which generates the shell structure. 
   It is taken as the difference between the sum of the energies of the occupied
   levels $e_i$ of $h_{def}$ and of the levels $\tilde e_i$ of a fictitious nucleus  that does not show shell structure,  
   \beq\label{eq:Esc}
   E_{SC}=\sum\limits_{i\leq N}e_i-\sum\limits_{i\leq N}\tilde e_i.
   \eeq   
Here, $e_N$ is the Fermi level, i. e. the $N^{th}$ state counting time reversed states $i$ and $\bar i$ explicitly. The smooth spectrum $\tilde e_i$ is obtained from the 
real spectrum $e_i$ by means of Strutinsky's averaging procedure.  It is described in Refs. \cite{Strutinsky67,Brack72}, which also derive the simple expression (\ref{eq:SC}) for the 
total energy   (see also Ref. \cite{NR}). 

 \begin{figure}
 \center{\includegraphics[width=0.5\linewidth]{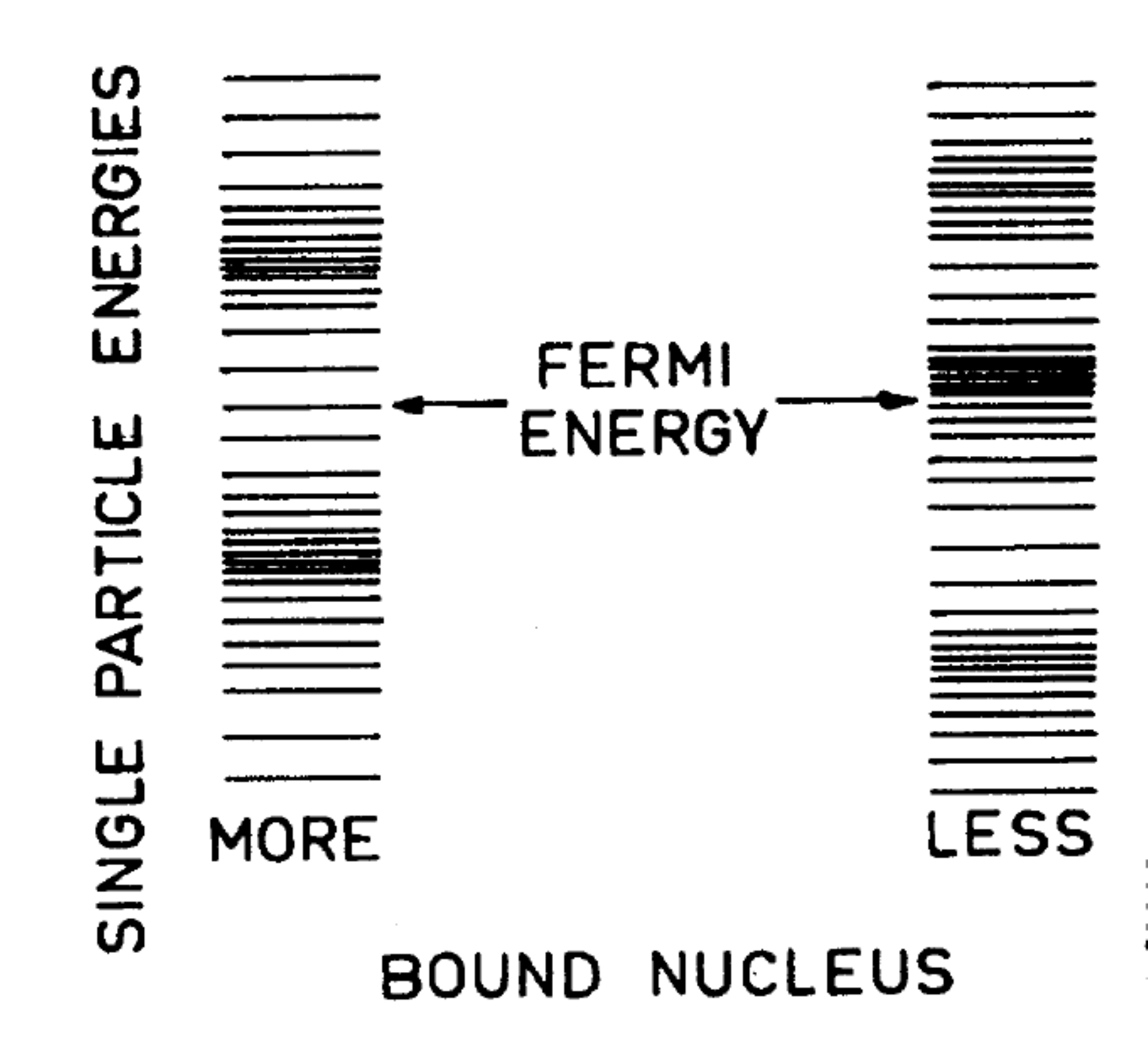} }
 \caption{\label{f:stability} Qualitative illustration of the connection between the level density at the Fermi energy and the binding energy of the nucleus. 
 Left: The occupied levels just below the Fermi energy have a smaller energy than equidistant levels with the average distance. 
 Their contribution reduces the sum energy as compared to the sum energy of the equidistant levels.  
  Right:  The occupied levels just below the Fermi energy have a larger energy than equidistant levels with the average distance. 
 Their contribution increases  the sum energy as as compared to the sum energy of the equidistant levels. 
 When all levels are occupied the modulation of the level density  averages out.  From Ref. \cite{Brack72} }
 \end{figure}

The microscopic potential of the Bohr Hamiltonian is then given as 
\beq
V(\beta,\ga)=E_{LD}(\beta,\ga)+E_{SC}(\beta,\ga).
\eeq 
For $\ga=0$ and up to quadratic order in the deformation parameter $\eps\approx0.95\beta$,
\beq
E_{LD}(\eps)-E_{LD}(\eps=0)=\eps^2\left[\frac{2}{5}a_S\left(1-\kappa_SI^2\right)A^{2/3}-\frac{1}{5}a_C\frac{Z^2}{A^{1/3}}\right].
\eeq
Since the liquid drop energy always prefers spherical shape, the shell energy determines the  form of the potential. The shell energy is the difference of between the sum
of  the energies of the levels near the Fermi level and the same sum of fictitious levels with a constant average spacing. This difference is positive in regions with a level 
density higher than the average and negative in  regions with a level density smaller than the average (see Fig. \ref{f:stability}).
That is, the maxima of $V(\beta,\ga)$ will lie in regions of high level density 
and the minima in regions of low level density. This allows one to easily obtain an estimate of the equilibrium shape from a single particle energy diagram
like Fig. \ref{f:nilsson}.  By changing its shape the  nucleus tries to avoid high level density near the Fermi energy and tries  to approach regions of low level density
near the Fermi energy. As indicated by the red arrows in Fig. \ref{f:nilsson}, the level density is low for spherical shape
    near the magic number $N=82$ and 126 but high for deformed shape.  Somewhat away from these magic numbers the level density becomes rapidly high at spherical shape, 
the shell energy develops a pronounced maximum, which results in a maximum of $V(\beta,\ga)$.  There must be a minimum at deformed shape, because
at large deformation the liquid drop energy prevails. The deformed minimum is stabilized by the relatively low 
level density for $N$ around 100, which has been called a "deformed shell closure" in analogy to a spherical shell closure \cite{Brack72}.

Let us now turn to the kinetic energy of the Bohr Hamiltonian.
In the case of molecules, the kinetic energy of the nuclear motion is simply the sum of the kinetic energies of a set of point masses. In the case of the nuclear Unified Model one calculates 
the increase of the energy when the shape  of the potential in  the single particle Hamiltonian $h_{def}\left(\alpha(t)_\mu\right)$ slowly changes  as a function  of the five quadruple
deformation parameters $\alpha(t)_\mu$.
  The kinetic energy for the Bohr Hamiltonian 
 for the zero-quasiparticle state is obtained by solving the Schr\"odinger equation for the time-dependent quasiparticle Hamiltonian (\ref{eq:hqp})
 by means of time-dependent  perturbation theory, which explicitly brings in the adiabatic approximation. 
 The energy increase caused by the collective motion is
 \beq\label{eq:mass}
 T=\frac{1}{2}\sum\limits_{\mu,\nu}B_{\mu\nu}\dot\alpha_\mu \dot\alpha_\nu, ~~ 
 B_{\mu\nu}=2 \sum\limits_{i,j}\frac{\vert\langle 0\vert\partial h_{def}/\partial \alpha_\mu\vert ij\rangle\vert\langle 0\vert\partial h_{def}/\partial \alpha_\nu\vert ij\rangle\vert}{\left(E_i+E_j\right)^3},
  \eeq   
  where the sum runs over all two-quasiparticle excitations $\vert ij\rangle$. Eq. (\ref{eq:mass}) provides the generic cranking (Belyayev) expressions 
for the inertial parameters $B_{\mu\nu}$ used in the microscopic Bohr Hamiltonian \cite{Belyayev59}. Re-expressed in terms of the intrinsic shape 
parameters $\alpha_0=\beta \cos\ga$ and $\alpha_2=\beta \sin\ga/\sqrt{2}$   and the orientation angles,
 $T$ takes the form  (\ref{eq:BH}),  where the three Inglis-Belyaev  \momis  \cite{Inglis,Belyayev59}
  \beq\label{eq:IBmomi}
 {\cal J}_{1,2,3}=2 \sum\limits_{i,j}\frac{\vert\langle 0\vert j_{1,2,3}\vert ij\rangle\vert^2}{E_i+E_j}  
 \eeq 
  appear as the inertial parameters for the angle degrees of freedom. 
  
  Kumar and Baranger \cite{KumarBaranger} pioneered the microscopic calculation of the Bohr Hamiltonian and its collective wave functions in the framework of 
  the pairing+quadrupole-quadrupole model (see section \ref{sec:SCCM}).  Subsequent approaches based on  the various selfconsistent mean field theories 
  have been reviewed in Refs. \cite{Bender03,Frauendorf15,Niksic11,NCMatsuyanagi}.   The Inglis-Belyaev expressions (\ref{eq:mass},\ref{eq:IBmomi}) well reproduce the deformation dependence of the inertial parameters.
 Overall, they are too small. The suppression depends on the approach but is typically of the order of 20-30\%. Different ad-hoc prescriptions have been suggested to bring the mass parameters to the experimental scale, which have been used in practical calculations.  
  When selfconsistent time-odd mean fields are taken into account, the rotational \momis are changed to the Thouless-Valatin form \cite{Thouless62}, 
  which re-produce very well the experimental values. State-dependent pairing (adding quadruple pairing to the monopole pairing as the simplest variant) generates an important contribution.
 This seems to indicate that using the Thouless-Valatin inertial parameters for the deformation 
   degrees of freedom will bring them to the experimental scale. 
   Taking the time-odd selfconsistent fields into account is still a challenging task. The contribution by K. Matsuyanagi, M. Matsuo, T. Nakatsukasa,
K. Yoshida, N. Hinohara and K. Sato to this Focus Issue \cite{NCMatsuyanagi} reports progress meeting the challenge. Further developments along
these lines are reviewed in Refs. \cite{Frauendorf15,Niksic11}. 
    
 The quasiparticle random phase approximation (QRPA) is an alternative method to describe the vibrational excitations in a microscopic way. It can be seen as 
 an approximation of the time-dependent mean field approach. The oscillating mean field is assumed deviating only slightly from the static equilibrium field. This allows one
 to linearize the equations, which then describe a harmonic vibrator.  The analogy is the small-amplitude limit for a physical pendulum, which is the harmonic  
 mathematical pendulum. The QRPA does not use the adiabatic approximation. It describes the coupling of the one-phonon vibrational excitations
  with the surrounding two-quasiparticle states. The QRPA is well exposed in standard textbooks on nuclear theory, as e. g. \cite{R70,RW,Ringbook,blaizotbook}.
  The present status of describing the quadrupole vibrations in the framework of the QRPA will not be reviewed. 
  Only the QRPA work on wobbling and chiral vibrations will be touched in section \ref{sec:ChangeSym}.

 \subsection{Transitional nuclei}\label{sec:BHgen}
  The collective excitations of the nuclei in-between the well deformed and spherical limits have are described by the general solutions of the Bohr Hamiltonian (\ref{eq:BH}). 
  On the phenomenological level,
  the potential is parametrized by few terms of appropriate symmetry with constants that are adjusted to the 
  experiment. The kinetic energy term is assumed to be of irrotational flow type with the overall scale as a  
  adjustable parameter. The details have recently been reviewed  by Frauendorf \cite{Frauendorf15}.
   The virtues and limits of the Unified Model discussed for 
  deformed and spherical nuclei appear in the transitional nuclei 
  in the same way. The collective yrast states are best accounted for and the excited $0^+$ states least. Only the first excited $0^+$ states in nuclei located near the transition
  between spherical and deformed shape show the collective properties predicted by the Bohr Hamiltonian. They are particularly low in energy and have an extended wave function 
  because of the $\beta$-flatness of the potential  around the transition.  
  
  As discussed in section \ref{sec:UMmicro}, the Bohr Hamiltonian has been derived microscopically in the framework of  the
  various versions of adiabatic time-dependent mean field theory or the equivalent generator method with Gaussian approximation (see Ref. \cite{Bender03}). The
 application of the different approaches to specific nuclei
has been recently reviewed Frauendorf \cite{Frauendorf15}.  The method, often referred to as the "Five-Dimensional Bohr Hamiltonian" 
  (5DBH), has considerable predictive power. Of course, it is constraint by  the adiabatic approximation or the
  Gaussian overlap approximation.   
  Large scale calculations of the lowest collective excitations have been carried out. 
  In a bench mark study, Delaroche {\it et al.} \cite{Delaroche10} calculated all even-even
with  $10\leq Z\leq 100$ and $20 \leq A\leq 200$. The results for the energies and E2 and E0
matrix elements for the yrast levels with $ I \leq 6$, the lowest excited $0^+$ states and the
two next yrare $2^+$ states are accessible in the form of a table as supplemental material
to the publication. The authors carried out a thorough statistical analysis of the merits of performance of the method and state: 
"We assess its accuracy by comparison with
experiments on all applicable nuclei where the systematic tabulations of the data are
available. We find that the predicted radii have an accuracy of 0.6\%, much better
than the one that can be achieved with a smooth phenomenological description.
The correlation energy obtained from the collective Hamiltonian gives a significant
improvement to the accuracy of the two-particle separation energies and to their
differences, the two-particle gaps. Many of the properties depend strongly on the
intrinsic deformation and we find that the theory is especially reliable for strongly
deformed nuclei. The distribution of values of the collective structure indicator
$R42 = E(4^+_1 )/E(2^+_1 )$ has a very sharp peak at the value 10/3, in agreement with the
existing data. On average, the predicted excitation energy and transition strength
of the first $2^+$ excitation are 12\% and 22\% higher than experiment, respectively,
with variances of the order of 40-50\%. The theory gives a good qualitative account
of the range of variation of the excitation energy of the first excited $0^+$ state, but
the predicted energies are systematically 50\% high. The calculated yrare $2^+$ states
show a clear separation between $\ga$ and $\beta$ excitations and the energies of the $2^+,~ \ga$
vibrations accord well with experiment. The character of the $0^+_2$ state is interpreted
as shape coexistence or $\beta$-vibrational excitations on the basis of relative quadrupole
transition strengths. Bands are predicted with the properties of $\beta$ vibrations for
many nuclei having R42 values corresponding to axial rotors, but the shape coexistence
phenomenon is more prevalent.Ó"
In addition the authors  observe that  the theory describes the $0^+_2$ states 
generally as Òtoo vibrationalÓ.

\subsection{ Quasiparticle triaxial rotor model}\label{sec:QTR}

The quasiparticle triaxial rotor model studies a  triaxial rotor to which one or more \qps are coupled.
 Using the body-fixed frame of reference, the Hamiltonian of the coupled system is
\bea\label{eq:HQR}
H_{QTR}=H_{ROT}+h_{qp}=\sum\limits_{i=1,2,3}\frac{\left(\hat J_i-\hat j_i\right)^2}{2{\cal J}_i(\beta,\ga)}+h_{qp}(\beta,\ga),\\
=\sum\limits_{i=1,2,3}\frac{1}{2{\cal J}_i(\beta,\ga)}\left(\hat J_i^2-2\hat J_i\hat j_i+\hat j_i^2\right)+h_{qp}(\beta,\ga),
\eea   
where $\hat J_i=\hat R_i+\hat j_i$ is the total \am and $j_i$ the quasiparticle \amd. 
The rotor represents the collective motion of all nucleons but the explicitly treated \qpsd.  
The intrinsic states are the configurations  generated by combining the quasiparticles
that belong to the Hamiltonian $h_{qp}$, Eq. (\ref{eq:hqp}). 
The model does not resort to the adiabatic approximation. Rather it takes fully
 into account  the impact of the inertial forces on the \qpsd. The 
  coupling between the \qp degrees of freedom and the rotational motion is facilitated by  the "Coriolis coupling"
  terms \hspace*{0.2cm} \mbox{$-\sum_{i=1,2,3}\hat J_i\hat j_i/{\cal J}_i(\beta,\ga)$}.

 The axial rotor limit of the Unified Model, discussed in section \ref{sec:UMdef}, is recovered by setting ${\cal J}_3=0$ 
 and neglecting the terms originating from $j_1$ and $j_2$, which describe action of the inertial forces on the \qp
 motion.  The wave functions  (\ref{eq:PsiRotSym}) are the eigenfunctions and 
  the strong coupling energies (\ref{eq:ErotI}) are the eigenvalues  of the truncated Hamiltonian.
  In first order perturbation theory with respect to the \qp - rotation coupling the expectation value of the full Hamiltonian is taken
   with the unperturbed wave functions  (\ref{eq:PsiRotSym}), which gives the 
   energy expression (\ref{eq:ErotIext}). The signature-dependent term of the  $K=1/2$ bands is the expectation value of the 
   Coriolis coupling. The resulting expression for the decoupling parameter is
   \beq
   a=  -\sum\limits_i\langle \phi_{K=1/2}\vert ( j_{1,i}+ij_{2,i}){\cal R}_2(-\pi)\vert \phi_{K=1/2}\rangle.
   \eeq
The so called "recoil term"   \mbox{$\sum_{i=1,2,3}\hat j_i^2/2{\cal J}_i(\beta,\ga)$} is usually considered to be already
taken into account by $h_{def}$ when fitting  the deformed shell model potential to the \qp levels in odd-A nuclei.

  The step beyond the adiabatic limit is to diagonalize the full quasiparticle triaxial rotor model Hamiltonian   (\ref{eq:HQR}) in the basis
of the Unified Model wave functions (\ref{eq:PsiRotSym}) constructed from the various quasiparticle configurations. 
The details are laid out in the textbooks \cite{BMII,R70,NR}.  
Limits to the number of the coupled  \qps are set by the dimensions 
of the Hamiltonian matrix  but also by the Pauli Principle between the explicitly treated \qps and the ones in the rotor core.  

\subsection{Approaches beyond the Unified Model}\label{sec:beyond}
The  territory beyond the applicability of the Unified Model depends on the direction one crosses its borderlines,
and so the methods and concepts that have been developed. Fig. \ref{f:HV-qp} illustrates the point for spherical nuclei
and Fig. \ref{f:YrastRadiation} for deformed. 
One possibility is to increase the \am but stay close to the yrast line. 
In the yrast region, the level density
remains small enough that one can study individual quantum states. 
The reason is \am conservation,
which limits the ways the excitation energy can be distributed among the nucleons. There is only one way at the yrast line, where
the nucleus has "zero temperature". This corresponds to the well known observation that increasing the energy of a body by setting
it into rotational motion does not 
increase its temperature (see e. g. Landau and Lifshitz {\it Statistical Physics} \cite{landauSP}).
 In the yrast region up to about 1 MeV above the yrast line, experimentalists can identify the transitions between individual states, which    
organize into rotational bands, or may not do so, depending on the nuclide. 
The rotational bands are identified by measuring multi coincidence events  in large arrays of $\gamma$ ray detectors. The method and important results 
are discussed in the contribution to this Speccial Edition by M. A. Riley, J. Simpson and E. S. Paul \cite{NCRiley}.

The density of intrinsic states increases exponentially with the excitation energy above the yrast line.  A one-to-one association
of individual experimental and calculated states loses sense. One has to resort to statistical concepts as averages of energy and transition rates,
their fluctuations and level densities, which implicitly invokes a certain degree of randomness that the models cannot account for. 
The methods of statistical mechanics have long be used for the region of excitation energies at and above   the nucleon binding energy ($\sim $ 8 MeV).
The computational power of modern computers has opened new avenues. The possibility to diagonalize matrices of dimension 10$^6$ and more 
 makes  Shell-Model-like approaches new powerful tools. 
In their contribution to this Focus Issue \cite{NCLeoni}, S. Leon and A. Lopez-Martens  discuss 
"warm nuclei", which is the region of 2-3 MeV above the yrast line. The large density of bands 
   and their mutual coupling require  special techniques for analyzing the data from large arrays of $\gamma$ ray detectors
 and new concepts (rotational damping) for interpreting the
$\ga$ ray spectra emitted by these warm nuclei.

The latter  path is not taken in this contribution.  Rotational 
bands in the yrast region will be analyzed in the traditional way by comparing spectroscopic data of individual quantum states with   
theory. I will present work exploring the yrast region at high \amd, which is based  on  the rotating mean field approach (section \ref{sec:RMF})  
and the quasiparticle triaxial rotor model (section \ref{sec:QTR}).

Important  alternative developments beyond the Unified Model are  left away. One example is
 the extensive work in the framework of the \qp random phase approximation (QRPA).
 In their contributions to this Focus Issue, T. Nakatsukasa, K. Matsuyanagi, M. Matsuzaki and Y. R. Shimizu \cite{NCNakatsukasa} discuss the 
 description of vibrational excitations in rotating nuclei by means of the QRPA starting from the rotating mean field.  
 D. R. Bes \cite{NCBes} and  R. A. Broglia, P. F. Bortignon, F. Barranco, E. Vigezzi, A. Idini and G. Potel \cite{NCBroglia} present the Nuclear Field Theory.
  J. L. Egido  \cite{NCEgido} exposes the Generator Coordinate Method. The reviews \cite{Frauendorf15,Niksic11,Bender03} provide good citations 
  of work on beyond-mean-field approaches.

\section{Rotating mean field}\label{sec:RMF}

The rotating mean field  approach bases on the assumption that the nucleus rotates {\it uniformly} about a body-fixed axis. The time dependence is removed
by transforming the theory to the frame of reference that rotates with the angular velocity $\vec \om$, within which the nucleus stands still. 
It is known from Classical Mechanics that the Hamiltonian $H'$ in the rotating frame is related to the Hamiltonian $H$ in the laboratory frame 
by the simple transformation 
\beq\label{eq:2bodyRouthian}
H'=H-\vec \om \cdot \vec J,      
\eeq
where $\vec J$ is the total \amd. The transformed  Hamiltonian has been called the two-body routhian \cite{BF79}
\setcounter{footnote}{0}\footnote{
The name adopts the terminology of classical hamiltonian mechanics (see e. g. Landau and Lifshitz, 
{\it Classical Mechanics} \cite{landauCM}).
 There, Eq. (\ref{eq:2bodyRouthian})
is a partial canonical transformation  from a Hamiltonian operating with the canonical  \am $\vec J$  to a Lagrangian  
operating with the  angular velocity $\vec \om$. routhian is called a combination that is a Hamiltonian for one
part of the degrees of freedom and Lagrangian for the remaining part.}.
The fact that $\vec J$ is a one-body operator 
 leads to a dramatic simplification
because it allows one to apply the mean field approximation in the same way as for the ground state.
The new term $-\vec \om \cdot \vec J$ just modifies the mean field Hamiltonian by "cranking" it with the angular velocity $\vec \om$.   
For this reason applying the mean field approximation to the two-body routhian 
 (\ref{eq:2bodyRouthian}) is also called the selfconsistent cranking model, which will be exposed in sections  \ref{sec:SCCM}, \ref{sec:mfshape}. 
 
  \begin{figure}
 \begin{center}
 \includegraphics[width=0.48\linewidth]{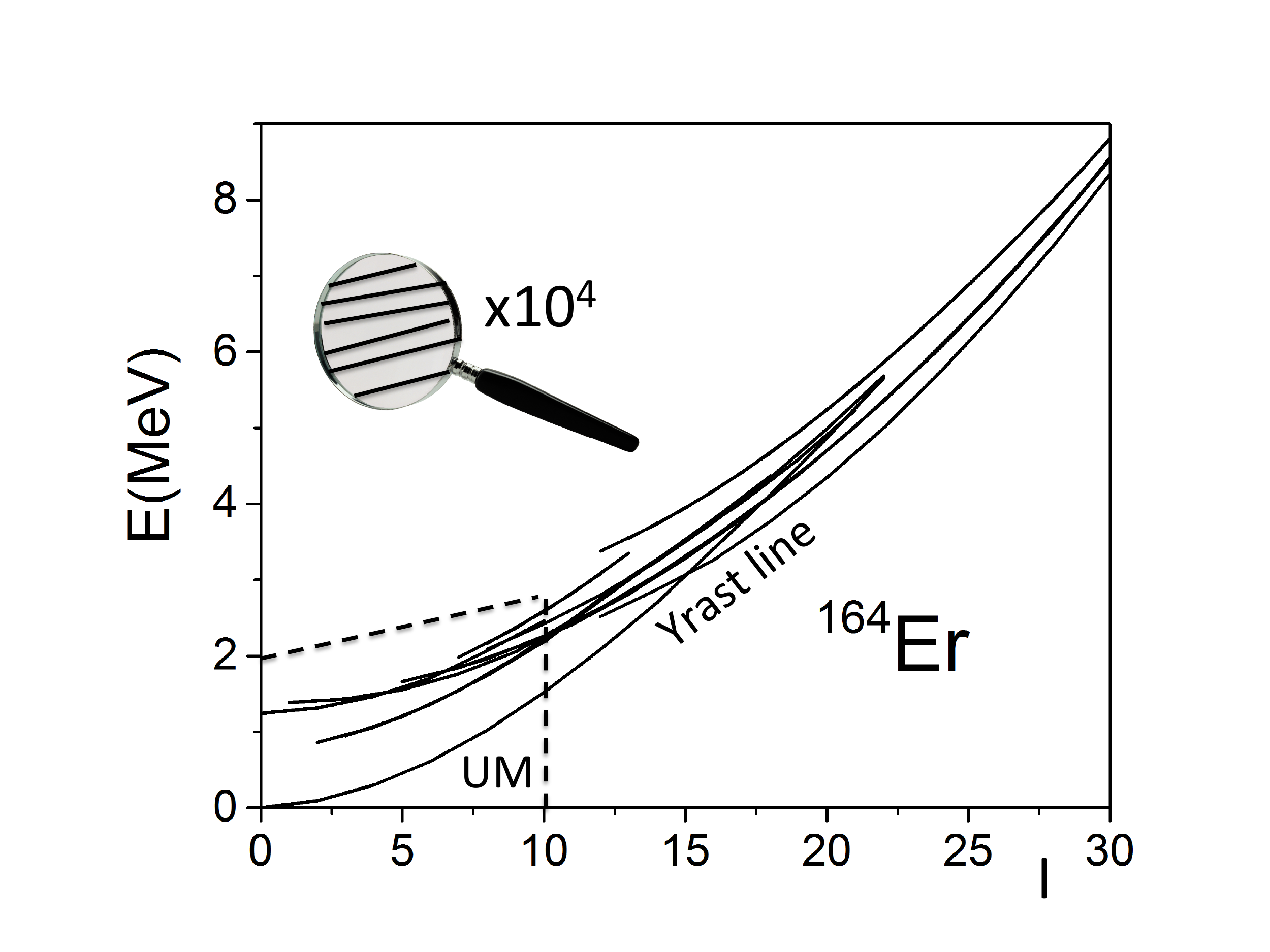} 
\includegraphics[width=0.48\linewidth]{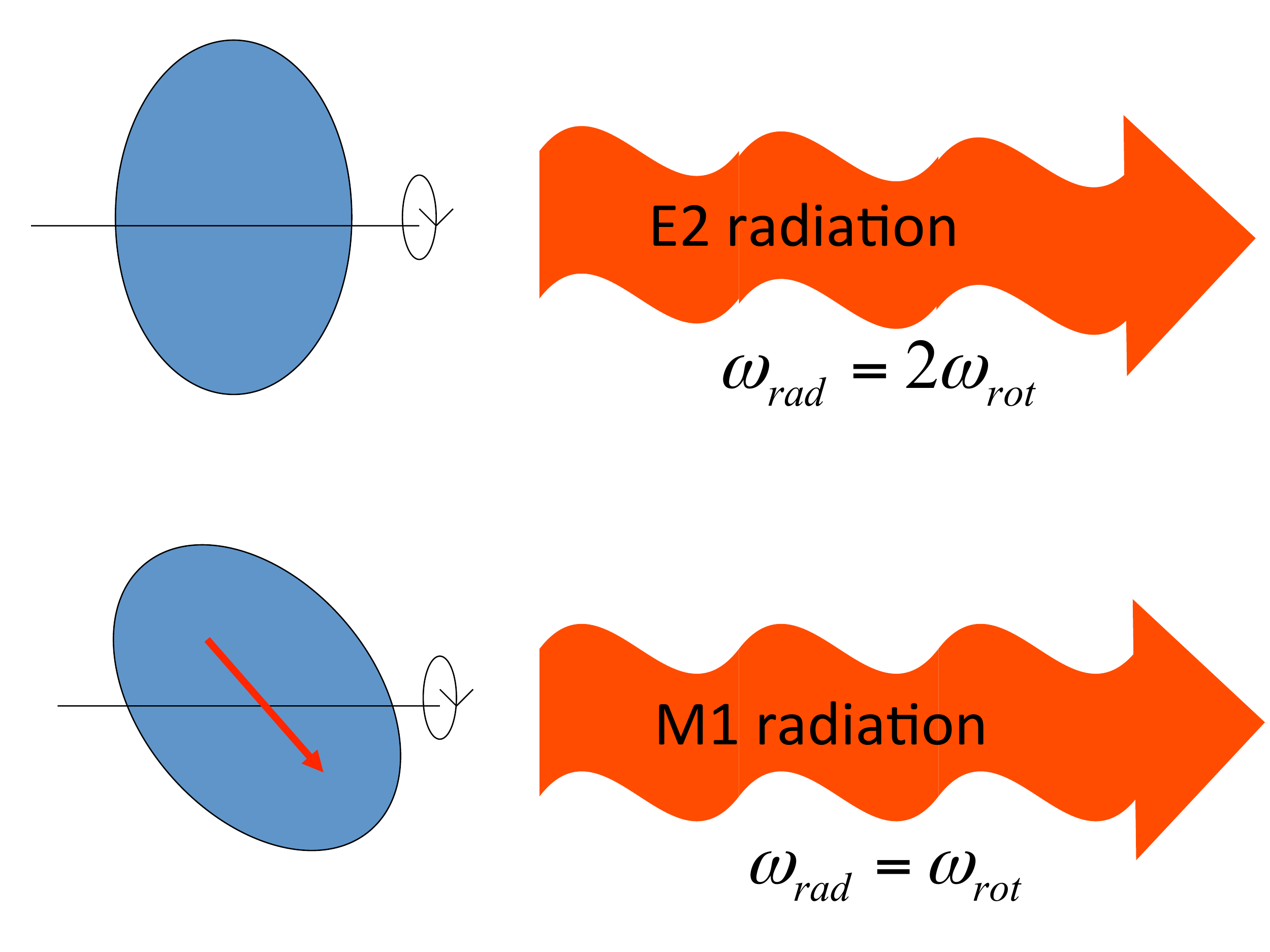} 
 \end{center}
 \caption{\label{f:YrastRadiation}  Left panel: Experimental energies of rotational bands in the normal deformed nucleus 
 $^{164}$Er. There are no states below the yrast line. The dashed linens delineate the realm of the Unified Model.
 The spacing between the
   rotational bands decreases exponentially with the excitation energy above the yrast line, which is illustrated by the magnifying glass.
   The indicated magnification factor corresponds to the location of the glass. \\ Right panel:
   Classical radiation from rotating nuclei.  Upper case: The rotating charge distribution generates  E2 radiation with twice the rotational frequency, 
   because already after a half turn it returns to its original position. 
 Lower case: The rotating magnetic moment generates M1 with the rotational frequency.  The rotating charge distribution generates two kinds of E2
  radiation, one with  and another with $\om_{rad}=2\om_{rot}$. }
 \end{figure}

Our focus beyond the Unified Model is on large \amd, for which uniform rotation about  a body fixed axis can be described with good accuracy
 in a semiclassical way. The expressions from classical electrodynamics  for radiation emission  from rotating charged and magnetized bodies 
 provide the transition matrix elements in semiclassical approximation. As illustrated by the right panel of Fig. \ref{f:YrastRadiation},
 the frequency of the emitted radiation provides directly the rotational frequency $\om$ of the nucleus. In this way, the rotational frequency can be measured by detecting
 the energy $\hbar \om(E,M1)$ of the photons (cf. section \ref{sec:intfreq}). The interpretation of high-spin phenomena  becomes more transparent 
 when the angular frequency,  instead of the \amd, is used to label the  sequence of rotational states. 
 It is useful to "dequantize"   the data, that is, to interpolate between the discrete values of the energies of the transitions connecting 
 the rotational levels.  The resulting continuous
 functions are directly compared with the rotating mean field calculations. This way, the multi band spectrum of the yrast region
 is classified in terms  of \confs of \qps in the rotating potential, where  their response  to the inertial forces is fully taken into account 
 (sections \ref{sec:symmetry}, \ref{sec:intfreq}, \ref{sec:orbits}, \ref{sec:CSM}). 

There is a complimentary perspective on  the rotating mean field approach.    
Applying the mean field approximation to the two-body routhian (\ref{eq:2bodyRouthian} is equivalent with  minimizing its expectation value 
with with respect to all possible mean field states. The variation 
\beq
\de \left(\langle H \rangle-\vec\om\cdot\langle \vec J\rangle\right)=0
\eeq
with respect to the \qp states provides the  pertinent  \qp routhian.  
The derivation is given by Ring and Schuck \cite{Ringbook} and Blaizot and Ripka \cite{blaizotbook}.
According to the Ritz Variational Principle one determines the best possible mean field state under 
the constraint that its expectation value of the angular momentum operator is different from zero.   The mean field solutions found
may have a lower symmetry than the two-body routhian, which is called spontaneous symmetry breaking. Breaking the rotational symmetry 
with respect to $\vec\om$ axis is the prerequisite for the appearance of a rotational band. Breaking the symmetry with respect to additional discrete 
symmetries of the two-body routhian is reflected by different sequences of spin and parity of the band members. The discussion of the 
mean field symmetries in section \ref{sec:symmetry} extends the  discussion of the $\De=1$ and $\De=2$ bands in the framework of the Unified Model
in section \ref{sec:StrongCoupling}. 
 The rotating mean field  reveals the microscopic underpinning of the rotational degrees of freedom and has extended  our view of what these are (section \ref{sec:emergence}). 
 Important new aspects are the insight that the rotating mean field  may uniformly rotate about an axis titled away from the principal axes of the density distribution (section \ref{sec:TAC}), 
  a quantitative description of band termination (section \ref{sec:termination}) and the discovery magnetic rotation ((section \ref{sec:MR}), a new mode not anticipated in the Unified Model framework.

  \begin{figure}
  \begin{center}
  \includegraphics[width=\linewidth,angle=0.6]{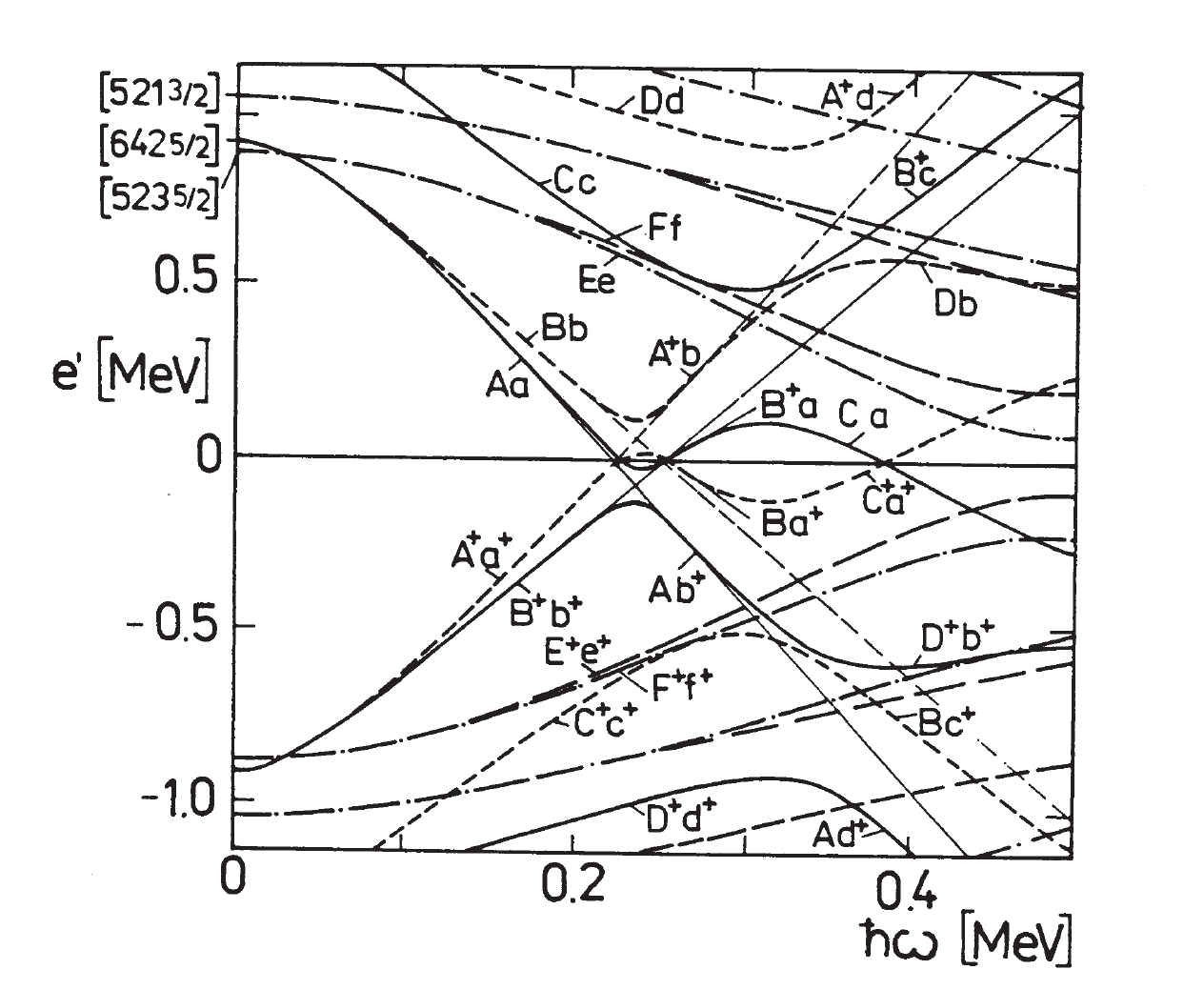} 
  \caption{\label{f:spagN96} Quasineutron routhian trajectories for $N = 96$. 
  The figure shows the trajectories as obtained by diagonalizing the \qp routhian  (\ref{eq:hpqp}) , labeled with lowercase letters, as well as
the diabatic trajectories constructed from them, labeled with capital letters. The diabatic
trajectories are extended through the crossings only for the levels
$A,~ B,~ A^+,~ B^+$, where $+$ indicates the conjugate quasiparticle. Note, the frequency $\om$ is the frequency  $\om_1$ discussed in the text.
  The figure is constructed using the parameters $\beta= 0.258,~ \De = 0.86$ MeV.
 The line types indicate parity and signature $(\pi,\alpha)$:
  full $(+, +1/2 )$, short-dashed $(+, -1/2)$, dot-dashed $(-, 1/2)$, long-dashed $(-, -1/2)$.  Reproduced from \cite{BFM86},
  from which the band labels are adopted.    }
  \end{center}
 \end{figure}

The key assumptions -uniform rotation and semiclassics-  are also the  limitations of the selfconsistent cranking model. Certain phenomena are missing, like wobbling 
 as a manifestation of non-uniform rotation with respect to the body fixed frame. Angular momentum is not conserved by the rotating mean field,  which, among many other issues,
 leads  to inaccuracies of the semiclassical transition rates or causes  problems at band crossings. The quasiparticle triaxial rotor model, often used  in combination with the selfconsistent cranking model, 
complements for these deficiencies of the rotating mean field  to some extend (section \ref{sec:TriaxTAC}).   
The  Projected Shell Model (see the contributions by Y. Sun \cite{NCSun}, P. M. Walker and F. R. Xu \cite{NCWalker},
 and J. A. Sheikh, G. H.  Bhat, W. A. Dar,  S. Jehangir and P. A. Ganai  \cite{NCSheikh} to this Focus Issue) 
is another approach that has been very  successful in reproducing  numerous spectroscopic data from the non-adiabatic regime. 
It starts from a set of \qp configurations generated by  the non-rotating mean field  of the pairing + quadrupole-quadrupole Hamiltonian,
  generates rotational bands by projecting a sequence of states of good angular momentum  from  each,
and finally diagonalizes a rotationally invariant Hamiltonian in the non-orthogonal basis generated this way.

\subsection{Selfconsistent cranking model}\label{sec:SCCM}

 The selfconsistent cranking model  extends the mean field approaches from low spin to the whole yrast region. 
 The rotational mode is treated semiclassically, which make the rotational frequency $\om$ a central concept. 
  The cranking term $\vec \om \cdot \vec {\hat J}$ is subtracted from the 
 deformed quasiparticle Hamiltonian (\ref{eq:hqp}),
 \beq\label{eq:hpqp}
 h'_{qp}(\beta,\ga,\De,\lambda,\om_1)=h_{qp}(\beta,\ga,\De,\lambda)-\vec \om \cdot \vec {\hat J},
 \eeq 
 where $\vec {\hat J}$ is the total \am operator.  
 The cranking term describes the transformation of the \qp Hamiltonian from the non-rotating frame of reference to 
 the frame  that uniformly rotates with the angular velocity $\om$. We will refer to Hamiltonian (\ref{eq:hpqp}) 
 as  the quasiparticle routhian.
    Like $h_{qp}(\beta,\ga)$, it describes the motion of independent quasiparticles, because $\vec {\hat J}$ is a one-body operator.
 The motion in the rotating frame is subject to the inertial forces.  Their strength is directly determined by the angular velocity, not the angular momentum,
 which is the reason for its central role of $\om$.  Compared to the microscopic Unified Model discussed in section \ref{sec:UMmicro},  the new step is to take the cranking term
 fully into account instead of using perturbation theory.  Fig. \ref{f:spagN96} shows  the \qn routhians of an axial nucleus as  functions of the 
 frequency, which  rotates about  an axis  perpendicular to its symmetry axis.

 Conventionally (see  \cite{NR,Ringbook,blaizotbook,Szymanski}) it was assumed that the rotational axis agrees with one of the principal axes of the deformed density distribution. 
 Such principal axis cranking (PAC) solutions always exist.  Frauendorf
 \cite{tac} demonstrated the existence of stable selfconsistent rotating mean field  solution that     
 uniformly rotate  about an axis that is titled with respect to  one of the principal axes of the deformed potential. These tilted axis cranking (TAC) solutions 
 were only reluctantly accepted by the nuclear community, because  their existence  seemed to contradict basic classical mechanics, which states that a rigid body
 can only uniformly rotate about one of its principal axes.  The origin of the tilt is discussed in section  \ref{sec:TAC}.

  The elements of  the selfconsistent cranking model are most transparently presented for  the schematic
   pairing+quadrupole-quadrupole interaction \cite{PQQ}.
  The starting point is two-body routhian, which has the form 
\bea \label{eq:HpPQQ}
H'=H_{sph}-\frac{\chi}{2}\sum\limits_{\mu =-2}^{2} Q^+_\mu Q_\mu - GP^+P-\lambda \hat N - \om \hat J_z,
\eea
where the z-axis is chosen as rotational axis. As discussed above, applying the mean field approximation to it can be seen 
from two complementary points of view. Interpreting  expression (\ref{eq:HpPQQ}) as the Hamiltonian transformed to the  
rotation frame of reference, its stationary mean field solutions uniformly rotate in  the laboratory system, 
which straightforwardly leads to the semiclassical expressions for electromagnetic transition probabilities given below.
Interpreting the cranking term as a constraint,  the mean field solutions
are the  best approximation for a given  finite value of the \am expectation value. This complementary viewpoint is more
appropriate when discussing symmetries and the nature of the rotational degrees of freedom in section \ref{sec:symmetry}.

The  pairing+quadrupole-quadrupole  model \cite{PQQ} incorporates three important aspects of the nuclear many-body system.
The nucleons move  in a spherical potential $U_{sph}$ with a strong spin-orbit
term as the Nilsson \cite{MOglobal} or the Woods-Saxon \cite{WSglobal} potentials. 
Sometimes  the energies $e_k$ of the levels in the spherical potential  are directly adjusted to the experiment \cite{KumarBaranger}.
In second quantization spherical single particle Hamiltonian reads
\beq \label{eq:hsph}
h_{sph}=t+U_{sph}=\sum\limits_k e_kc^+_kc_k,
\eeq
where $k$ labels the single particle states.
\setcounter{footnote}{0}
\footnote{To keep the notation compact, only one letter is used as index. It is understood that it stands for the full set of single particle
quantum numbers, which includes the \am projection in case of spherical or axial potentials. The states with opposite sign of
the \am projection, denoted by $k$ and $\bar k$, are related by the time reversal operation.   If not explicitly indicated, as 
in Eq. (\ref{eq:hsph}), the sum runs over the positive and negative \am projections. The restriction  $k>0$, as in Eq. (\ref{eq:P}),  
means that the sum runs only over the positive \am projections.}  

 The long range particle-hole correlations are taken into account by the
second term, the quadrupole interaction. It assumed
to be separable in form of a product of the quadrupole operators
\beq\label{eq:Q}
 Q_\mu=\sum\limits_{kk'}\sqrt{\frac{4\pi}{5}}\langle k|r^2Y_{2\mu}|k'\rangle c^+_kc_{k'}.
\eeq
This part of the interaction is responsible for the quadrupole deformation
of the mean field.

The short range particle-particle pair correlations are taken into account
by the third term, the  pairing interaction. It is a
 product of the   operators of the monopole pair field
\beq\label{eq:P}
P^+=\sum\limits_{k>0} c^+_kc^+_{\bar k},
\eeq
were $\bar k$ is the time reversed state of $k$. The monopole pair field
consists of Cooper pairs of protons or neutrons
 coupled to angular momentum zero.

The term $-\lambda \hat N $ controls the expectation value of the particle number $\langle \hat N\rangle$. To simplify the
notation, the routhian (\ref{eq:HpPQQ}) is written only for one
kind of  particles. The terms $h_{sph}$, $Q_\mu$ and $\hat J_z$ must be understood
as sums of a proton and a neutron part and there are   terms
$- GP^+P $   and $ -\lambda \hat N $ for both protons and neutrons.

The Hartree--Bogoliubov   
approximation (see \cite{Ringbook,blaizotbook}) is used for the state vector $|\rangle $,
 which is an eigenstate of the mean field routhian $h'$,
which is given by
\bea \label{eq:hhfb}
h'=h_{sph}-\chi\sum\limits_{\mu=-2}^{2} q_\mu Q^+_\mu
 - \De(P^++P)-\lambda \hat N -\om \hat J_z.
\eea
The selfconsistency equations determine the 
deformed part of the potential
\beq \label{eq:scq}
q_\mu=\langle Q_\mu \rangle,
\eeq
the pair potential
\beq \label{eq:scp}
\De =G\langle P \rangle,
\eeq
and implicitly the chemical potential $\la$ by
\beq \label{eq:n}
N=\langle \hat N \rangle.
\eeq
The quasiparticle operators
\beq \label{eq:qp}
\al_i^+=\sum\limits_kU_{ki}c^+_k+V_{ki}c_k
\eeq
obey the equations of motion
\beq \label{eq:eom}
[h',\al_i^+]=e_i'\al_i^+,
\eeq
which define the   eigenvalue equations  for the
quasiparticle amplitudes $U_{ki}$ and $V_{ki}$. The explicit
form of these Hartree--Bogoliubov 
equations are given by Ring and Schuck \cite{Ringbook} and Blaizot and Ripka \cite{blaizotbook}. 
The eigenvalues $e'_i$ are called the quasiparticle routhians.  Examples are shown in  Fig. \ref{f:spagN96}.

The quasiparticle operators refer to the vacuum state $|0\rangle$
\beq \label{eq:vacuum}
\al_i|0\rangle=0~~ \forall~ i
\eeq
and define the excited quasiparticle configurations
\beq \label{eq:qpconf}
|i_1,i_2,...\rangle=\al_{i_1}^+\al_{i_2}^+...|0\rangle.
\eeq
The construction of a configuration for a sequence of $\om$ values, which
represents a rotational band, will be discussed in section \ref{sec:crdiabatic}.  

The set of Hartree-Fock-Bogoliubov 
equations (\ref{eq:hhfb})-(\ref{eq:qpconf}) can be
solved for any configuration $|\rangle=|i_1,i_2,...\rangle$.
For such a selfconsistent solution, the total routhian
\beq \label{eq:e'total}
E'=\langle H'\rangle
\eeq
 has an extremum
\beq
\frac{\partial E'}{\partial q_\mu}\vert_\om=0,
~~ \frac{\partial E'}{\partial \De}\vert_\om=0
\eeq
for the values of $q_\mu$ and $\De$ determined from the selfconsistency requirements (\ref{eq:scq},\ref{eq:scp}).
The total energy as function of the angular momentum is given by
\bea \label{eq:etotal}
E(J)=E'(\om)+\om J(\om),\\
\label{eq:J}
J(\om)=\langle \hat J_z\rangle,
\eea
where Eq. (\ref{eq:J}) implicitly determines $\om(J)$. The total
 energy is  extremal for a fixed value of $J$,
\beq
\frac{\partial E}{\partial q_\mu}\vert_J=0,
~~ \frac{\partial E}{\partial \De}\vert_J=0.
\eeq
For a family of selfconsistent solutions $|\om \rangle$
found for different values of $\om$, 
the following canonical relations hold
\bea \label{eq:canonical}
\frac{dE}{dJ}=\om,~~\frac{dE'}{d\om}=-J,~~E'=E-\om J.
\eea

\begin{figure}[t]
 \begin{center}
 \includegraphics[width=0.7\linewidth]{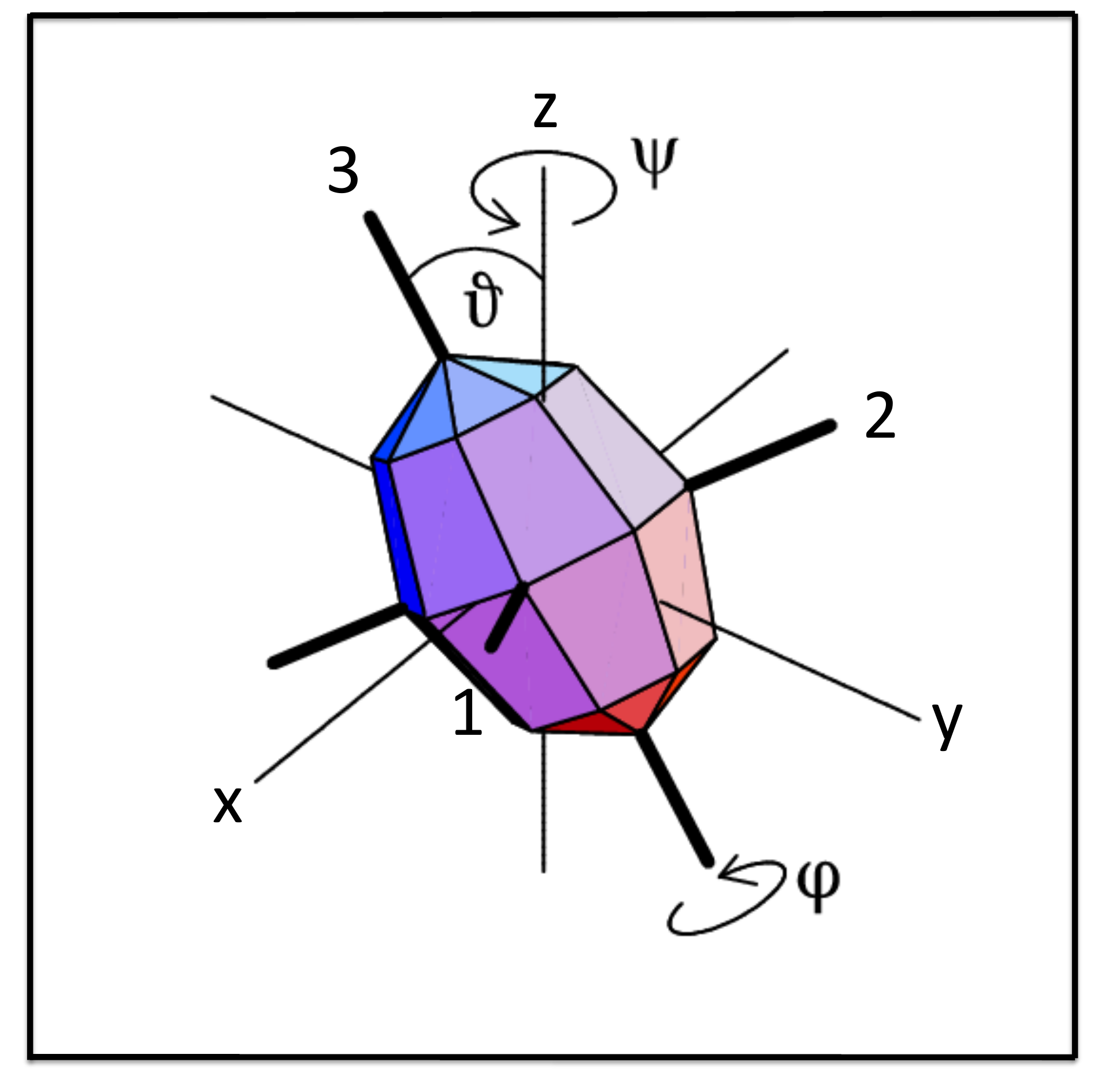}
  \caption{\label{f:angles}
Euler angles specifying the orientation of the triaxial reflection symmetric
density distribution. Reproduced from \cite{RMP}.
 }
  \end{center}
  \end{figure}

The calculations are most conveniently  carried out in the body-fixed  frame.
It is the frame of  the  principal axes 1, 2, 3, within which the
quadrupole tensor takes the simple form $q'_{-1}=q'_1=0$ and $q'_{-2}=q'_2$. Its
 orientation  with respect to the laboratory frame is fixed by the
three Euler angles $\psi$, $\vartheta$ and $\f$. 
Fig. \ref{f:angles} illustrates the definition of these  angles.
In our convention, $\psi(=\om t)$ is the angle that  grows 
as the nucleus rotates uniformly about the z-axis.
The angles $\vartheta$ and and $\f$ are the orientation angles of $\vec J$ 
(i.e. of the  z-axis) with respect to the principal axes
\setcounter{footnote}{0} 
\footnote{This convention has been used in the tilted axis cranking  literature.  
 The meaning of the angles $\psi$ and $\f$ is 
inverted as compared to Ref. \cite{BMII}.}.
The two intrinsic quadrupole moments $q'_0$ and $q'_2$
specify the deformation of the potential.
The quadrupole moments in the laboratory
frame are related to them   by 
\beq\label{eq:intlab}
q_{\mu}=D^{2}_{\mu 0}(\psi,\vartheta,\f)q'_{0}+(D^{2}_{\mu 2}(\psi,\vartheta,\f)+
D^{2}_{\mu -2}(\psi,\vth,\f))q'_{2},
\eeq
where $D^{2}_{\nu \mu}(\psi,\vth,\f)$ are the Wigner $D$-functions
\footnote{We use the definition of the $D$-functions
 by Ref. \cite{BMII}}.  
 
 In the body-fixed frame, the \qp routhian (\ref{eq:hhfb}) takes the form 
 \beq\label{eq:hpbf}
h'= h_{def}-\om(\sin\vth\cos\f\hat J_1+\sin\vth\sin\f\hat J_2+\cos\vth\hat J_3)
 \eeq
 with
 \beq\label{eq:hMO}
 h_{def}=h_{sph}- q'_0Q'_0-q'_2(Q'_2 +Q'_{-2}), 
 \eeq
 which becomes the modified oscillator Hamiltonian when one
 expresses  quadrupole moments  by the 
deformation parameters
\setcounter{footnote}{0} 
 \footnote{The sign of $q'_2$ is taken according to the  "Lund convention" which is 
 opposite to the "Copenhagen convention" of the Bohr Hamiltonian of Ref. \cite{BMII}. This unfortunate inconsistency persist in the literature and
 between section \ref{sec:UM} and the following as well.}   
\beq\label{eq:deflund}
q'_0=K\beta\cos \ga,~~ q'_2=-K\beta\sin \ga/\sqrt{2},
\eeq
where $K=\hbar\omega_0/\sqrt{4\pi/5}$, $\hbar\omega_0=41\mathrm{MeV}/A^{1/3}$ sets the energy scale for the deformed potential.
The modified oscillator Hamiltonian involves a careful parametrization of the spherical single particle energies \cite{MOglobal}.

The five selfconsistency equations (\ref{eq:scq}) are reduced to two
\beq \label{scQintr}
q'_0=\kappa\langle Q'_0\rangle ,\,q'_2=\kappa\langle Q'_2\rangle ,
\eeq
which determine $\beta$ and $\ga$. They are complemented by
the condition   that 
$\vec J=\langle \hat{\vec J}\rangle$ must be parallel to $\vec \om$ at the point of selfconsistency,
which is used to determine  the angles $\vth$ and $\f$. 
The routhian $E'(\vth,~\f$) has an extremum for this orientation.

The stationarity of the eigenvalues with respect to parameter changes  ensues that the negative slope
of the quasiparticle routhians with respect to $\om$,
\beq\label{eq:qpalignment}
i=-\frac{de'(\om)}{d\om}, 
\eeq
 is the projection of the quasiparticle \am on the rotational axis $\vec\om$, which is called the aligned \am or simply alignment. 
 For example, the routhian A in Fig. \ref{f:spagN96} has the large alignment of 4.1$\hbar$.

The possible ratios between the lengths of three axes of the deformed potential are given by the values of the triaxiality parameter 
 restricted to the interval $0\leq \ga\leq 60^o$. Extending its range  repeats
the family of shapes with a different association of the principal axes long (l), medium (m) ,
short (m) with the axes labels 1, 2, 3  to which the Euler angles are attached. Tab. \ref{t:axes}
lists the association of the 1- and 3-axes. As long as the rotational axis lies in one of the principal planes
one may chose the 1-3 plane and extend the $\ga$ interval to  $-60^o\leq \ga\leq 120^o$
to include the  possible combinations m-l, s-l, s-m.  
Frauendorf \cite{tacdic}    introduced this practical convention, which  extends the 
long-practiced convention that the nucleus rotates about the 1-axis in case that the rotational
axis agrees with one of the three principal axes.
More  details and further illustrations  
are given by Nilsson and Ragnasson  \cite{NR} and Szymanski \cite{Szymanski}.

\begin{table}[h]
\begin{minipage}{0.48\linewidth}
\begin{tabular}{|cccc|}
$\gamma$&shape&1-axis&3-axis\\
\hline
$-240^o$&prolate&short&short\\
        &triaxial&medium&short\\
$-180^o$&oblate&long&short$^*$\\
        &triaxial&long&short\\
$-120^o$&prolate&long$^*$&short\\
        &triaxial&long&medium\\
$-60^o$ &oblate&long&long\\
        &triaxial&medium&long\\
$0^o$   &prolate&short&long$^*$\\
        &triaxial&short&long\\
$60^o$  &oblate&short$^*$&long\\
        &triaxial&short&medium\\
$120^o$ &prolate&short&short
\end{tabular}
\end{minipage}
\hfill
\begin{minipage}{0.45\linewidth}
\includegraphics[width=\textwidth,angle=-90]{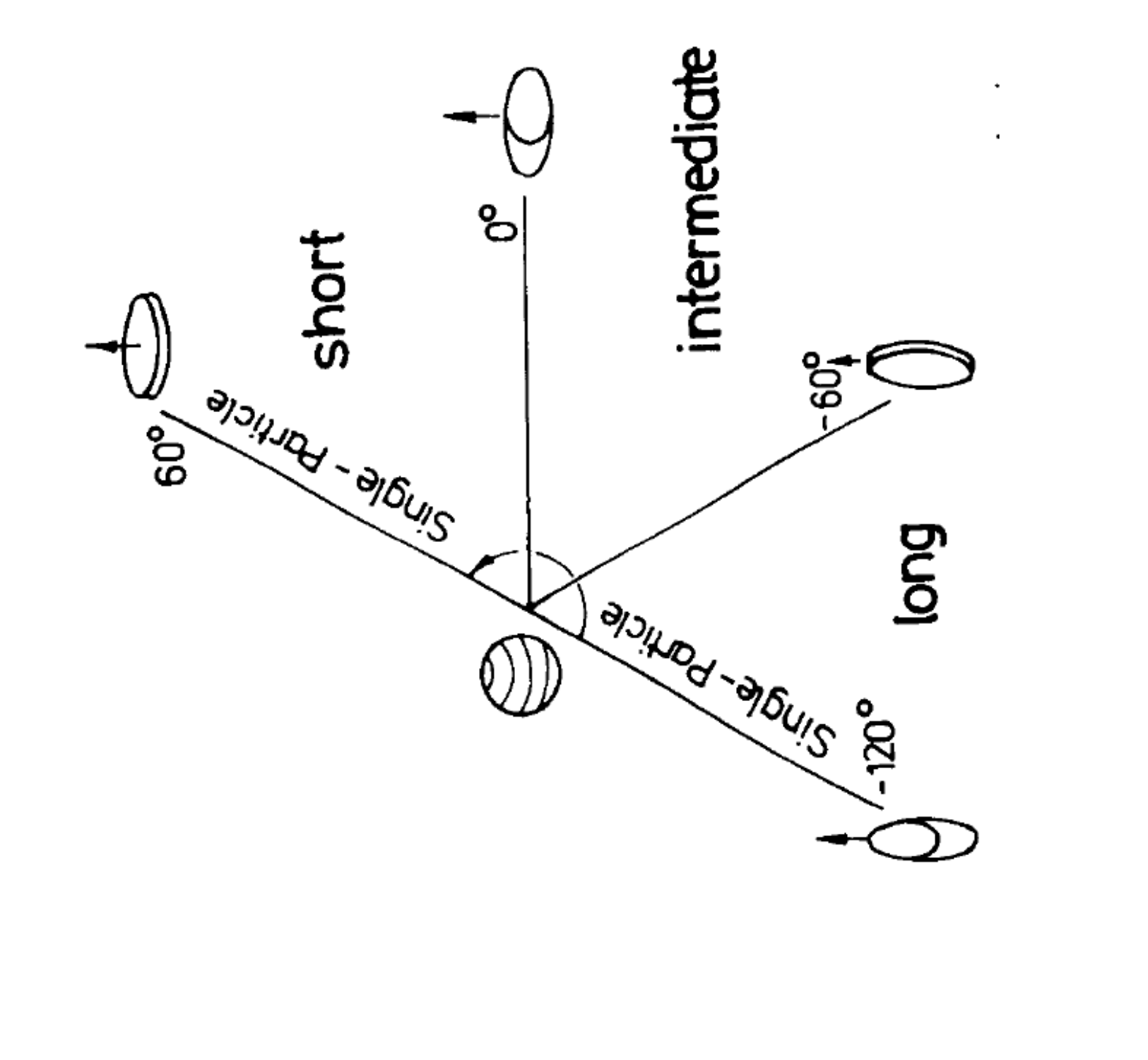} 
\end{minipage}
\caption{\label{t:axes} Association of
the semi axes of the triaxial potential with the principal axes 1 and 3.
The star indicates a symmetry axis.  The figure on the right illustrates the association by displaying the deformation
parameters as polar coordinates, where $\beta$ is the radius and $\ga$ the angle.
It is quite common to visualize the potential energy $V(\beta,\ga)$ and other quantities 
using this kind of polar coordinates, which are called the "$\beta-\ga-$plane".
 The vertical arrows in 
the drawing indicate the direction of  the rotational 1-axis for axial shape.
For triaxial shape the axis associated with 1 is indicated by its name (the medium axis is called intermediate).  }
\end{table}

 Semiclassically, the probabilities for electromagnetic 
transitions are given by the classical radiative power divided by the photon energy. 
The expressions for uniformly rotating magnetic dipoles and electric quadrupoles 
can be found in textbooks on classical electrodynamics, as
 e. g. Landau and Lifshitz {\it Classical Fields} \cite{landauEM}. 
 The semiclassical transition amplitudes are proportional to the E2 and M1 multipole moments in the laboratory frame, which are related
to the time-independent intrinsic moments by the transformation (\ref{eq:Mbody2lab}).  For uniform rotation
the angle $\psi=\om t$ and  for planar geometry $\f=0$ if rotational lies in the 1-3- principal plane (see  Fig. \ref{f:angles}).
Using that $D^\lambda_{\mu\nu}(\psi,\vth,\f)=e^{i\mu\psi}d^\lambda_{\mu\nu}(\vth)e^{i\nu\f}$, the transformation  
(\ref{eq:Mbody2lab}) becomes
 \beq
 {\cal M}^{lab}_{\lambda,\mu}(t)=\sum\limits_\nu e^{i\om\mu t}d^\lambda_{\mu,\nu}(\vth){\cal M}^{int}_{\lambda,\nu}.
\eeq
The expression has a straightforward meaning in classical electrodynamics, which is illustrated 
by the right panel of Fig. \ref{f:YrastRadiation}. The static multipole moments
are given by ${\cal M}^{lab}_{1,0}(M1)$ and ${\cal M}^{lab}_{2,0}(E2)$. The multipole moments 
${\cal M}^{lab}_{1,1}(M1,t)$ and ${\cal M}^{lab}_{2,1}(E2,t)$ are sources of radiation with the frequency $\om$.
 The multipole moment  ${\cal M}^{lab}_{2,2}(E2,t)$ is a source of radiation with the frequency $2\om$.
Using the explicit expressions for the reduced $d$ - functions given in standard textbooks on 
\am  (e. g. Ref.  \cite{Rose57},  see also section 1A in Ref. \cite{BMI} and Appendix B in Ref. \cite{R70}), Frauendorf 
 \cite{tacdic} derived the semiclassical transition amplitudes  for the case of a planar geometry ($\f=0$).

 The  intra band M1 - transition matrix elements are
\bea\label{eq:m1}
<I-1I-1|{\cal M}_{-1}(M1)|II>=
<{\cal M}_{-1}(M1)>\nonumber \\
=\sqrt{\frac{3}{8\pi }}\left[\mu_3\sin\,\vartheta -\mu_1\cos\,\vartheta \right].
\eea
The components of the transition operator ${\cal M}_{\nu}$
refer the laboratory  system. The expectation value is taken with the 
tilted axis cranking \conf $|>$. In the second line  ${\cal M}_{\nu}$ is expressed by
the components of the magnetic moment in the intrinsic frame for the  case that 
the rotational axis lies  in the 1-3 principal plane. (The other three planes can be reached by
extending the range of the triaxiality parameter to $-120^\circ\leq\ga\leq60^\circ$.)   The components  are related  
by the rotation about the Euler angle $\vartheta$ ( cf. Eq. (\ref{eq:intlab}), $\varphi=0$). 
The reduced M1-transition probability is 
\beq\label{eq:bm1}
B(M1,\,I\rightarrow I-1)=<{\cal M}_{-1}(M1)>^2.
\eeq
The spectroscopic magnetic moment is given by
\bea\label{eq:mu}
\mu_s=<II|\mu_z|II>=\frac{I}{I+1/2}<\mu_z>\nonumber \\
=\frac{I}{I+1/2}[\mu_1\sin\,\vartheta+\mu_3\cos\,\vartheta].
\eea
The factor $I/(I+1/2)$ is a quantal correction which is close to one
for high spin. 
The components of the magnetic moment with respect to the principal axes are calculated by
means of
\bea
\mu_1=\mu_N(J_{1,p}+(\eta 5.58-1)\,S_{1,p}-\eta 3.82\,S_{1,n}),\nonumber\\
\mu_3=\mu_N(J_{3,p}+(\eta 5.58-1)\,S_{3,p}-
\eta 3.82 \,S_{3,n}),
\eea     
where the components 
of the vectors of angular momentum $\vec J=\langle  \hat{\vec J}\rangle$ and of the spin 
$\vec S=\langle \hat {\vec S}\rangle$ are the expectation values  with the tilted axis cranking
\conf $\vert\rangle$. The free spin magnetic moments are attenuated by a factor
$\eta \sim 0.5-0.7$ depending on the mass region.  The components of the 
magnetic moment can expressed as the sum of the contributions of the 
vacuum and the excited \qps (labelled by $i$),   
\bea
\mu_1=g_RR_1+\sum\limits_i g_i j_{1,i},~~
\mu_3=g_RR_3+\sum\limits_i g_i j_{3,i},
\eea
where $R_{1,3}$, $j_{1,3i}$ and $g_R$, $g_i$ are the respective \am components and the $g$ factors of the vacuum
and of the excited \qpsd. This form allows one to use estimates of the ingredients that are taken from experiment:
Experimental  $g$ factors  are often known for a mass region of interest and \qp \am components can be derived from the spectra.
Examples are the semi empirical vector models \cite{vector,vectors} and \cite{magrevCM}, 
which will be discussed in sections \ref{sec:CSM}, \ref{sec:approxTAC} (see Eqs. (\ref{eq:DF})) and \ref{sec:MR}, respectively.   

The intra band E2  transition matrix elements are
 \bea\label{eq:m2}
<I-2I-2|{\cal M}_{-2}(E2)|II>=
<{\cal M}_{-2}(E2)>=&\nonumber\\
=\sqrt{{5\over4\pi}}\left(\frac{eZ}{A}\right)
\Bigl[\sqrt{3\over8}<Q'_0>(\sin\,\vartheta)^2\nonumber\\
+{1\over4}<Q'_2+Q'_{-2}>\left( 1+(\cos\vartheta)^2\right) \Bigr],&\nonumber\\
<I-1I-1|{\cal M}_{-1}(E2)|II>=
<{\cal M}_{-1}(E2)>=\nonumber\\
=\sqrt{{5\over4\pi}}\left(\frac{eZ}{A}\right)
\Bigl[\sin\,\vartheta \cos\vartheta( \sqrt{3\over2}<Q'_0>
-{1\over2}<Q'_2+Q'_{-2}>)\Bigr],
\eea
and the spectroscopic quadrupole moment is 
\bea\label{eq:Qstat}
&Q_s=<II|Q_0^{(BM)}|II>=\frac{I}{I+3/2}<Q_0^{(BM)}>=&
\nonumber\\
&\frac{I}{I+2/3} \frac{2eZ}{A}\Bigl[<Q'_0>\left((\cos\vartheta)^2
-{1\over2}(\sin\,\vartheta)^2\right)&\nonumber\\
&+\sqrt{3\over8}<Q'_2+Q'_{-2}>(\sin\,\vartheta)^2\Bigr].&
\eea
 The reduced E2 transition probabilities are
\beq\label{eq:be22}
B(E2,\,I\rightarrow I-2)=<{\cal M}_{-2}(E2)>^2
\eeq
and
\beq\label{eq:be21}
B(E2,\,I\rightarrow I-1)=<{\cal M}_{-1}(E2)>^2.
\eeq
The mixing ratio is
\beq\label{eq:delta}
\delta=\frac{<{\cal M}_{-1}(E2)>}{<{\cal M}_{-1}(M1)>}.
\eeq
If one uses the high-spin approximation
 for the Clebsch-Gordan coefficients
in  the Unified Model one obtains  
the semiclassical form (\ref{eq:m1}-\ref{eq:delta}).

 \subsection{Symmetries}\label{sec:symmetry}
 
 The symmetries of the two-body routhian (\ref{eq:HpPQQ}) and of the
mean field routhian (\ref{eq:hpqp})
play a central role in the interpretation of the rotating
mean field solutions. The symmetries are not necessarily the same. If the
mean field routhian has a lower symmetry one speaks of  "spontaneous
symmetry breaking".
 The concept of  spontaneous
symmetry breaking is discussed in the textbooks   \cite{Ringbook,blaizotbook} on many-body theory of finite
quantum systems.   It has been conceived for phase transitions in infinite systems.
When a phase has a lower symmetry than the Hamiltonian there are 
more than one symmetry-breaking  mean field  states, which all have the same energy.  Their number 
equals the number of elements of the group of the broken symmetry. 
In the thermodynamic limit the symmetry breaking states become
exact many-body solutions. For finite systems the exact solutions are
 the eigenstates of the  two-body routhian, which have  the same symmetry.
Only approximate solutions may have a lower  symmetry, which are the rotating mean field   solutions.
The exact eigenstates are approximated by superpositions of these degenerate rotating mean field   solutions,
which restore the broken symmetry
 \setcounter{footnote}{0}
\footnote{Working with such superpositions is a special case of the Generator Coordinate Method \cite{Bender03,NCEgido},
which amounts to project states of good \am from the mean field solution. In applications the \am projection often is restricted to non-rotating
mean field states because it is technically simpler. An example is the triaxial projected shell model \cite{NCSun,NCSheikh}.  }.
The amplitudes of such a superposition represent the 
collective wave function.  
Classifying rotating mean field  solutions according to their symmetry is a
fruitful concept, because the different symmetries manifest themselves
 clearly as the different types of rotational bands.

As discussed  in the texbooks \cite{Ringbook,blaizotbook} and section \ref{sec:StrongCoupling},
 breaking of the rotational symmetry with respect to
the angular momentum axis leads to the appearance of a rotational band. The breaking of a
 discrete twofold symmetry results in the appearance of pairs of degenerate states. In a molecule 
 the  rigid localization  of the atomic nuclei and the corresponding electronic wave functions 
 break spontaneously the symmetry of the total Hamiltonian that describes the interaction between its constituents, which leads 
 to specific degeneracies of the rotational levels.  In nuclei the breaking of a two-fold symmetry by the mean field is weaker and 
 the two states are close but not degenerate.

The  order parameter is a related concept that is used for the description of
phase transitions in infinite systems. An order parameter is some physical
quantity that is zero in the symmetry-conserving phase and
is finite in the symmetry-breaking phase. Its values distinguish between the symmetry-conserving and the  symmetry-breaking states.
\footnote{A more profound discussion can be found in the 
textbook {\it Statistical Physics} by Landau and Lifshitz \cite{landauSP}.}.
The spontaneous magnetization of a
ferromagnet  is a way to specify the different orientations of a ferromagnet, which   
are the degenerate symmetry-breaking states.  
The magnetization disappears when the temperature exceeds the critical
value, where the phase transition into  the isotropic
paramagnetic state takes place. 
For finite nuclei  the order parameters  should
 specify the different orientation of the degenerate rotating mean field  solutions.
 Suitable quantities are the magnetic dipole and electric  moments, if 
they are large  compared with a typical single-particle matrix element.

\begin{figure}
 \includegraphics[width=\linewidth]{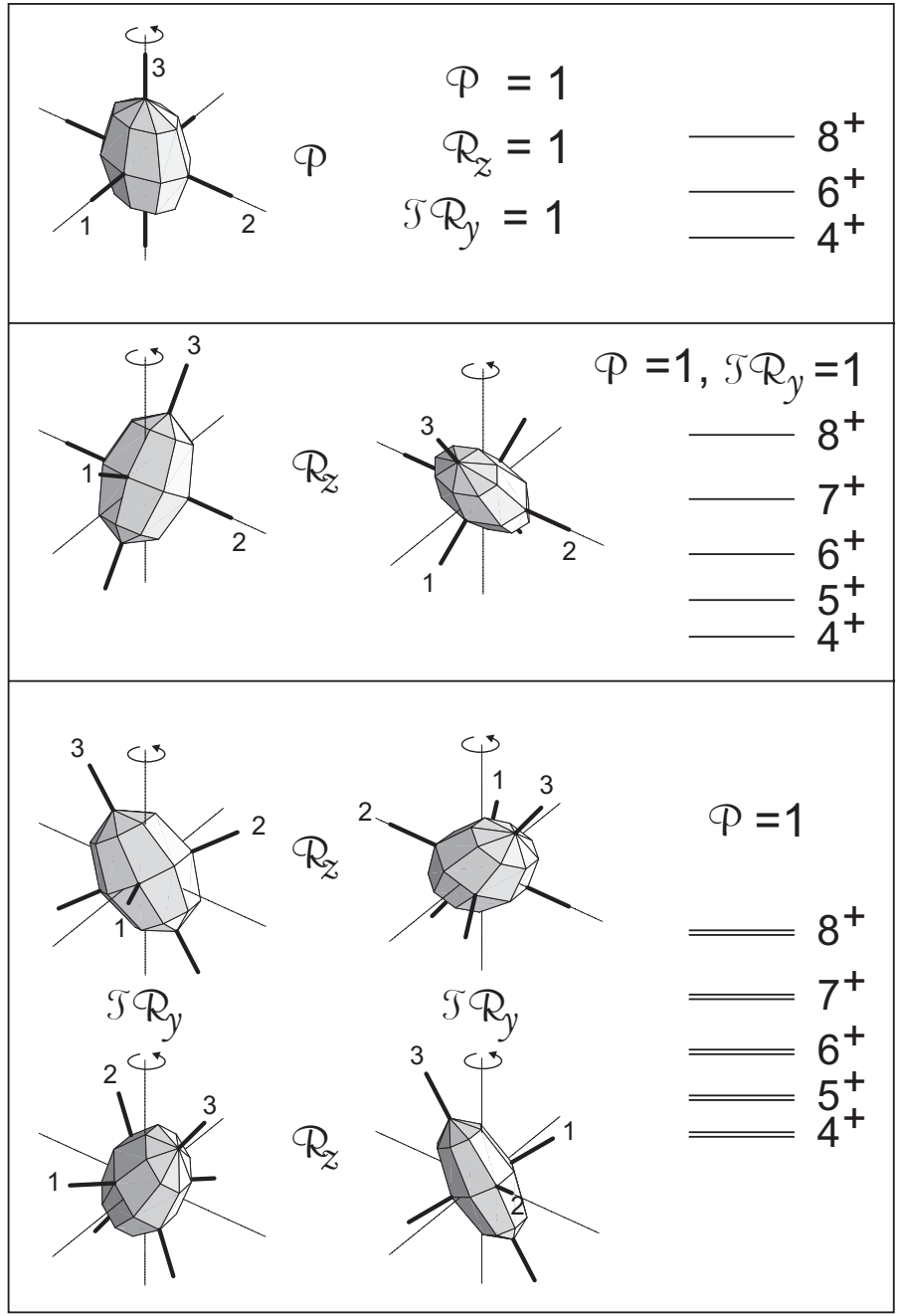}
\caption{\label{f:tacsym1}Discrete symmetries of the mean field 
of a rotating triaxial reflection symmetric  nucleus (three symmetry
planes).  
The mean field is represented by its density distribution. 
A polyhedron  is used to make  the symmetries better visible.  
The axis of rotation
(z) is marked by the circular arrow. It coincides with the angular momentum
$\vec J$.  The figure illustrates the 
reorientation of the density distribution under the three symmetry
 operations that leave the two-body routhian invariant. 
The structure  of the rotational
 bands associated with each
symmetry type is illustrated on the right side.}
\end{figure}

The two-body routhian (\ref{eq:HpPQQ}) is invariant with respect to
\begin{enumerate}
\item ${\cal R}_z(\psi)$,  rotation about the z-axis,
\item ${\cal P}$, space inversion,
\item ${\cal R}_z(\pi)$, rotation about the z-axis by an angle of
$\pi$,
\item  ${\cal  T R}_y(\pi)$,  rotation about the x-axis by an angle of
$\pi$ \\ combined with  the time reversal ${\cal T}$.
\end{enumerate}
The symmetry (iv) is a consequence of $\vec J$ being odd under the time
reversal operation  ${\cal T}$. We are interested in rotational bands, which means that symmetry (i) is broken.
The symmetry operations (ii-iv) are two-fold and commute.  The  pairing+quadrupole-quadrupole  interaction does not generate correlations that  
spontaneously break the space inversion symmetry. In the following it is assumed  that the mean field conserves parity.  The
consequences of breaking the space inversion symmetry are discussed in Ref. \cite{RMP}.

Fig. \ref{f:tacsym1} illustrates the case that
the mean field routhian (\ref{eq:hhfb}) is invariant with respect
to space inversion (indicated by ${\cal P}=1$).
The parity
$\pi$ is a good quantum number of the mean field solutions,
\beq
{\cal P}|\rangle=\pi |\rangle,
\eeq
and the rotational bands are characterized by a fixed parity $\pi$.
  The deformed density distribution is reflection symmetric 
with respect to the three
planes spanned by the principal axes of the quadrupole tensor (see  Eq. ({\ref{eq:intlab})). 
 We refer to them as the principal planes (PP)
and to their intersections as the principal axes (PA).
The different possible orientations of angular momentum vector $\vec J$ with respect to 
the principal axes obey the symmetries (iii) and/or (iv) or break them, which is reflected by
the sequence of states in the rotational bands. 
The following discussion of the symmetries is not restricted to
quadrupole deformations. It is valid for all shapes that have  three
symmetry planes (D$_{2h}$ symmetry). Hexadecapole deformation is the most import reflection-symmetric
deviation from quadrupole deformation. It appears in the lower region of a shell (e. g. $N\approx 90$), where nuclei
look lemon-like and in the upper region (e. g. $N\approx110$), where they look barrel-like (see e. g. \cite{FYglobal}).

 The symmetry operations  ${\cal R}_z(\pi)$ and ${\cal T R}_y(\pi)$ are
 twofold. If the mean field
is not symmetric  with respect to one of them,
 there is for each  selfconsistent 
solution another one 
with the same energy, which
 is  generated by the symmetry operation. 
This follows from the invariance
of the two-body routhian with respect to the symmetry operation.
As a consequence, there will be two rotational bands with
the same energy represented by the linear combinations of the two mean
field solutions that restores the broken symmetry.

\subsubsection{Axis of rotation is a principal axis}\label{sec:rmfspac}
The upper panel of Fig. \ref{f:tacsym1} shows the case when
 $\vec J$ has the direction of one of the
principal axes  (principal axis cranking - PAC). The mean field routhian $h'$ 
is invariant with respect to ${\cal R}_z(\pi)$ and, as a consequence, 
\beq\label{eq:rzal}
{\cal R}_z(\pi)|\rangle=e^{-i\al \pi}|\rangle,
\eeq
where  $\al$ is the signature exponent or shortly the signature of the state
\setcounter{footnote}{0}
\footnote{We use the notation introduced by Ref. \cite{BF79}.
Another convention, 
which follows Ref. \cite{BMII}  uses  
the signature quantum number $r=e^{-i\al \pi}$. In order to avoid long winded
formulations we shall adopt the somewhat loose but 
common terminology calling both $\alpha$ and
$r$ signature.}.
Since
${\cal R}_z(\pi)$ is a subgroup of the full rotational group, invariance
of $h'$ with
respect to it leads to
a selection rule for the total angular momentum,
\beq\label{eq:alI}
I=\al + 2n, ~~n=0,\pm 1,\pm 2,....~ .
\eeq 
The relation between the
total angular momentum and $\al$ follows directly from the decomposition of $|\rangle$
into states of good angular momentum $|I,M=I\rangle$
\beq\label{eq:rz}
{\cal R}_z(\pi)|\rangle=\sum C_I{\cal R}_z(\pi)|I,M=
I\rangle=e^{-i I \pi}|\rangle,
\eeq
which implies by virtue of (\ref{eq:rzal}) that the sum contains only
those values of $I$ that obey  (\ref{eq:alI}).
Naturally such a wave packet is associated with a $\De I=2$ band according
 the selection rule (\ref{eq:alI}).

 Each
quasiparticle configuration corresponds to a band of given
parity and signature $(\pi,\al)$, i. e.
to sequence of states of given parity $\pi$
and angular momentum $I$ that changes in steps of 2, in accordance with
(\ref{eq:alI}).
The  configurations with different signature have different energy.
 This is seen best  if 
 one connects
the $\De I =2 $ sequences of  data points in a smooth way.
As examples, Fig. \ref{f:Er163spec} shows 
the two bands A and B with  $(\pi,\al)=(+,1/2)$ and $(+,-1/2)$, 
which constitute two well separated $\De=2$ bands at large spin.

 The members of the $\De I=2$ bands are interconnected by
fast $E2$-transitions. 
The $\ga$ transitions within  the bands A and B 
 in Fig. \ref{f:Er163spec} are examples.
The transitional quadrupole moment 
\beq\label{eq:Qt}
Q_t\equiv Q_{22}=\sqrt{\frac{4\pi}{5} B(E2,I\rightarrow I-2)},
\eeq
which measures the asymmetry of the
density distribution with respect to the rotational axis $\vec J$
can be considered as the order parameter. 
\footnote{The definition of the transition quadrupole  moment deviates from the commonly used 
 \mbox{ $Q_t=\sqrt{\frac{16\pi}{5} \frac{B(E2,I\rightarrow I-2)}{\langle I020\vert I-20\rangle^2}}$}, which provides
  the intrinsic quadrupole moment of an $K=0$ band of an axial nucleus.}


\subsubsection{Axis of rotation  in a principal plane}
The middle panel of Fig. \ref{f:tacsym1} shows  the case  when the rotational
axis $\vec J$ is tilted away from the principal axes but still lies in one 
of the three principal planes (planar tilted axis cranking - planar TAC). This is always the
 case for axial shape. 
The mean field routhian  is no longer invariant with respect
the rotation ${\cal R}_z(\pi)$. Since
\beq\label{eq:norzal}
{\cal R}_z(\pi)|\rangle\not=e^{-i\al\pi}|\rangle,
\eeq
 there is no restriction  of $I$. The rotational
bands correspond to sequences of states of all possible values of $I$
and fixed parity $\pi$.
If one plots experimental energies or routhians of
the states with $I=I_o + 2n$ and $I=I_o + 1+ 2n$
and connects them smoothly, the two $\De I=2$ sequences are degenerate,
which means they combine
to a $\De I=1$ band. In this way, the expected doubling
of the number of states due to the breaking of the ${\cal R}_z(\pi)$
symmetry shows up.
The bands  K1 and K2 in Fig. \ref{f:Er163spec} 
are examples of broken ${\cal R}_z(\pi)$ symmetry.   
 
The members of the $\De I =1$ bands are linked by strong
$E2$-transitions between the states $I$ and $I-2$ as
well as by   $M1$- and $E2$-transitions between the states $I$ and $I-1$.
The latter are caused by  the finite transitional
 magnetic moment $\mu_{11}\left(=\sqrt{4\pi/3 B(M1,I\rightarrow I-1)}\right)$ and transitional quadrupole moment $Q_{21}\left(=\sqrt{4\pi/5 B(E2,I\rightarrow I-1)}\right)$.
Since both are equal to zero for ${\cal R}_z(\pi)$ symmetry, they may be
considered as order parameters, measuring the tilt of the rotational axis or,
equivalently, how strongly the ${\cal R}_z(\pi)$ symmetry is broken.

The preceding discussion generalizes the symmetry considerations in section \ref{sec:StrongCoupling}.
Within the Unified Model signature appears  as a special quantum number for $K=0$ bands, because  
axial shape and the validity of the adiabatic approximation are assumed. More generally, the signature quantum number
appears as a consequence of the symmetry mean field state under the
rotation  ${\cal R}_i(\pi)$ about one of the principal axes  (see case (f) in Fig. \ref{f:UMrot}).
The bands A and B in Fig.  \ref{f:Er163spec} constitue an example.  At low spin the sequences 
   A and B combine to a $\De I=1$ based on $K=5/2$ intrinsic state.  
   With increasing  spin the inertial forces modify the \qp state such that the $\De=1$ sequence 
   splits into the two $\De=2$ branches with the signature $\alpha\pm 1/2$.  
The corresponding 
 transition  from broken to good  ${\cal R}_z(\pi)$ symmetry will be discussed in more detail in section \ref{sec:ChangeSym}.
 
 \subsubsection{Axis of rotation out of the principal planes}\label{sec:planarTAC}
The lower panel of Fig. \ref{f:tacsym1} displays the aplanar tilted axis cranking  case when
$\vec J$ does not lie in one of the principal planes.
Then the mean field routhian $h'$ is no longer invariant with respect to
$ {\cal T R}_y(\pi)$, because this operation
 leads to a new combination of
$\vec J$ with the system of principal axes. 
The two combinations have opposite chirality.
To see this, first realize that the angular momentum vector $\vec J$ selects three
principal half axes (the ones with a positive projection on $\vec J$).
Look from the arrow head of $\vec J$ into this octant.
Call the principal axes system right-handed  if the short (1), 
medium (2) and
long (3)  axes are ordered  counter clockwise (upper pair in
Fig. \ref{f:tacsym1}) and left-handed if they are ordered  clockwise
(lower pair in Fig. \ref{f:tacsym1}). 
It is not possible to change the
chirality by a rotation. Only the combination
$ {\cal T R}_y(\pi)$ which includes the time reversal
operation ${\cal T}$ reverses the
chirality. 

The breaking of the
 $ {\cal T R}_y(\pi)$ symmetry causes a doubling of the rotational
levels. There are two identical $\De I =1 $ sequences with the same parity,
which are the even and odd linear combinations of the 
left- and right-handed mean field solutions. These linear combinations
restore the broken $ {\cal T R}_y(\pi)$ symmetry.
The members of each band are connected by enhanced $E2$- and $M1$-transition
like the planar tilted axis cranking  solutions. 

\subsection{Rotational frequency}\label{sec:intfreq}
 When the angular momentum is large it behaves  almost like a classical
 quantity. As a consequence, 
the angular frequency $\om$ also becomes  a well defined observable.
This presumes that we can construct a wave packet out of states with different
$I$ that have  otherwise similar structure, i. e. out of the
 members of a rotational band.  
At high spin it is often preferable  to choose the angular frequency
as the  parameter that changes along the band and not the angular momentum. 
 The modifications of the nucleonic motion are directly controlled
by $\om$,   because   the inertial forces depend
 on the angular velocity. The central role of the angular frequency
in the analysis of the experimental spectra 
was  appreciated by Bentsson and Frauendorf \cite{BF79}. They
introduced the experimental routhian
$E'(\om)$, which is  energy in the rotating frame of reference. It is the
appropriate quantity  to consider when referring to the frequency $\om$.
In classical mechanics there are the canonical relations (\ref{eq:canonical}) between the 
energy in the space-fixed and rotating frames of reference,
which hold in semiclassical quantum mechanics as well.
As illustrated by the right panel of Fig. \ref{f:YrastRadiation}, the rotational frequency $\om$ is directly related 
to the frequency of the electromagnetic 
radiation which is equal to the  transition energy. 
One may define $E'$  and $\om$ on the basis of the stretched quadrupole
($\De I =2$)  or dipole ($\De I =1$)
 transitions. The choice depends on the symmetry (cf. section \ref{sec:TAC}).

If  states  differing by one unit of angular momentum arrange into a
 $\De I =1$ band (e. g. the band  K1
in Fig. \ref{f:Er163spec}) one can use
\bea \label{eq:om1exp}
J  = I\hbar,   ~~ \hbar\om(I) = E(I)- E(I-1), \\
E'(I)=\frac{1}{2}[E(I)+E(I-1)]-\om J(I).
\eea
Here we introduced the "classical value" of angular momentum $J$, which can be directly
compared with the results of the mean field theory.

Due to the leading order quantal correction
one must associate the classical angular momentum $J$ with quantal value
$I+\frac{1}{2}$. The rotational frequency $\om$, which is defined by
a transition between two rotational levels, is assigned to
 the mean value of $J$ of the  two levels.

If  states differing by two units of angular momentum arrange into a
 $\De I =2$ band (e. g. band A in Fig. \ref{f:Er163spec})
one must use 
 \bea
\label{eq:om2exp}
J(I)  = (I - \frac 1 2)\hbar,   ~~ \hbar\om(I) = \frac {1} {2} [ E(I)- E(I-2) ],\\
E'(I)=\frac{1}{2}[E(I)+E(I-2)]-\om(I) J(I).
\eea

For $\De I=1$ bands the rotational axis, which has the direction 
of the  \am vector $\vec J$, deviates from the principal axes of the
nuclear shape. A majority of rotating mean field  calculations assumes that the rotational axis is a principal axis.
Conventionally the 1-axis is chosen.
For comparison with these calculations one uses 
the projection $\om_1$ of $\vec \om$ on the principal axis 1.
For axial nuclei with a substantial deformation a quite good approximation is keeping \am projection on the symmetry axis $J_3$ 
constant equal to  its value $K$ at the band head. The relations of classical vector geometry 
\beq
\om_1=\frac{J_1}{J}\om=\frac{dJ}{dJ_1}\frac{dE}{dJ}=\frac{dE}{dJ_1},~~J_1=\sqrt{J^2-J_3^2}
\eeq
are approximated  by  finite differences 
\bea\label{eq:om1EpCSM}
J(I)  = (I - \frac 1 2)\hbar,  ~~J_1=\sqrt{J^2-(\hbar K)^2}, ~~\om_1(I)=\frac{E(I)-E(I-2)}{J_1(I)-J_1(I-2)}, \label{eq:om1}\\
E'(I)=\frac{1}{2}[E(I)+E(I-2)]-\om_1(I) J_1(I).\label{eq:Epom1}
\eea
Figs. \ref{f:Er163Eplow} and \ref{f:Er163Ephigh} show the experimental routhians derived by means of Eqs. (\ref{eq:om1},\ref{eq:Epom1}) from the
spectrum in Fig. \ref{f:Er163spec} 
\setcounter{footnote}{0} 
\footnote{ In the literature $\om_1$ is often denoted by simply $\om$, like in Fig. \ref{f:spagN96}.}.

The rotational  frequency is a more direct experimental observable than the angular momentum, which
must be constructed from the multi polarities of all the $\ga$ transitions that
form the   cascade. If a certain sequence of coincident transitions is
not connected with the rest of the spectrum, 
then one knows the frequency $\om$  but the angular momentum  only up to a constant.

Since $\om $ changes  in many small steps along a band,
it is a convenient measure for
 studying the reaction of the nucleonic motion
to the inertial forces. Choosing the rotational frequency as the band parameter 
 facilitates the comparison with 
the cranking mean field theory because the latter 
is formulated for a given frequency.  The relation  (\ref{eq:canonical}) between the energies and the frequency,
which holds exactly for the selfconsistent cranking  model, ensures that the frequency used in the calculations  can be identified with the
experimental frequency.

Using the frequency as a rotational parameter, one distinguishes between the kinematical \momi ${\cal J}^{(1)}(\om)$ and the
 dynamical \momi ${\cal J}^{(2)}(\om)$ defined as
\beq \label{momi}
{\cal J}^{(1)}=\frac{J}{\om},~~ {\cal J}^{(2)}=\frac{dJ}{d\om}.
\eeq
Authors often show experimental  \momis calculated from the measured  energies 
$\De I=1$ transitions by means of Eqs. (\ref{eq:om1exp}) and $\De I=2$ transitions by means of Eqs. (\ref{eq:om2exp}) and, respectively,
\bea
{\cal J}^{(1)}\left(\om(I)\right)=\frac{J(I)}{\om(I)},\label{eq:J1J2expJ1}\\
{\cal J}^{(2)}\left(\om(I)\right)=\frac{1\hbar}{\om(I)-\om(I-1)}~~~\De I=1,\label{eq:J1J2expJ21}\\
{\cal J}^{(2)}\left(\om(I)\right)=\frac{2\hbar}{\om(I)-\om(I-2)}~~~\De I=2\label{eq:J1J2expJ22}.
\eea

\begin{figure}
 \includegraphics[width=0.48\linewidth]{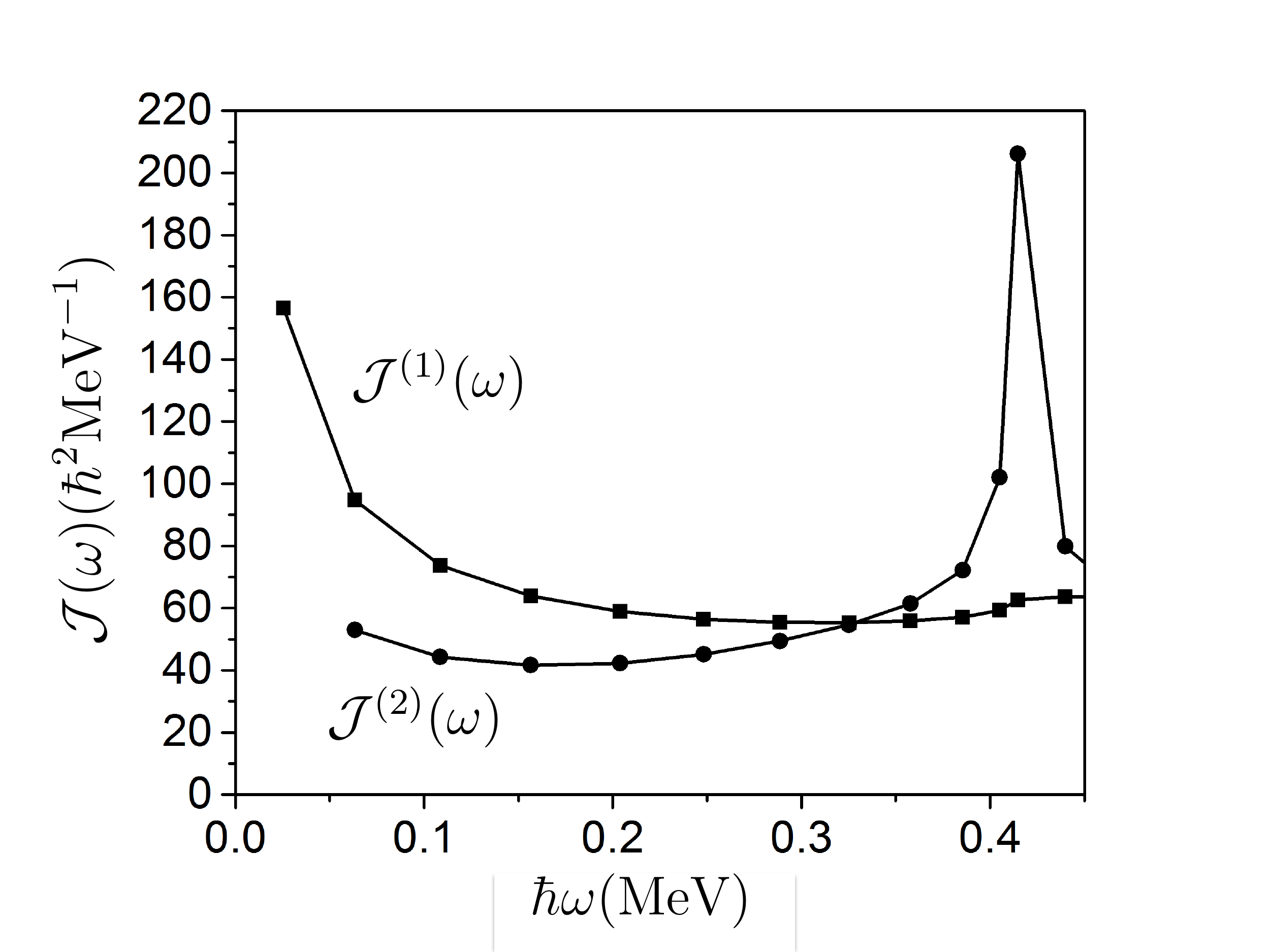} 
 \includegraphics[width=0.48\linewidth]{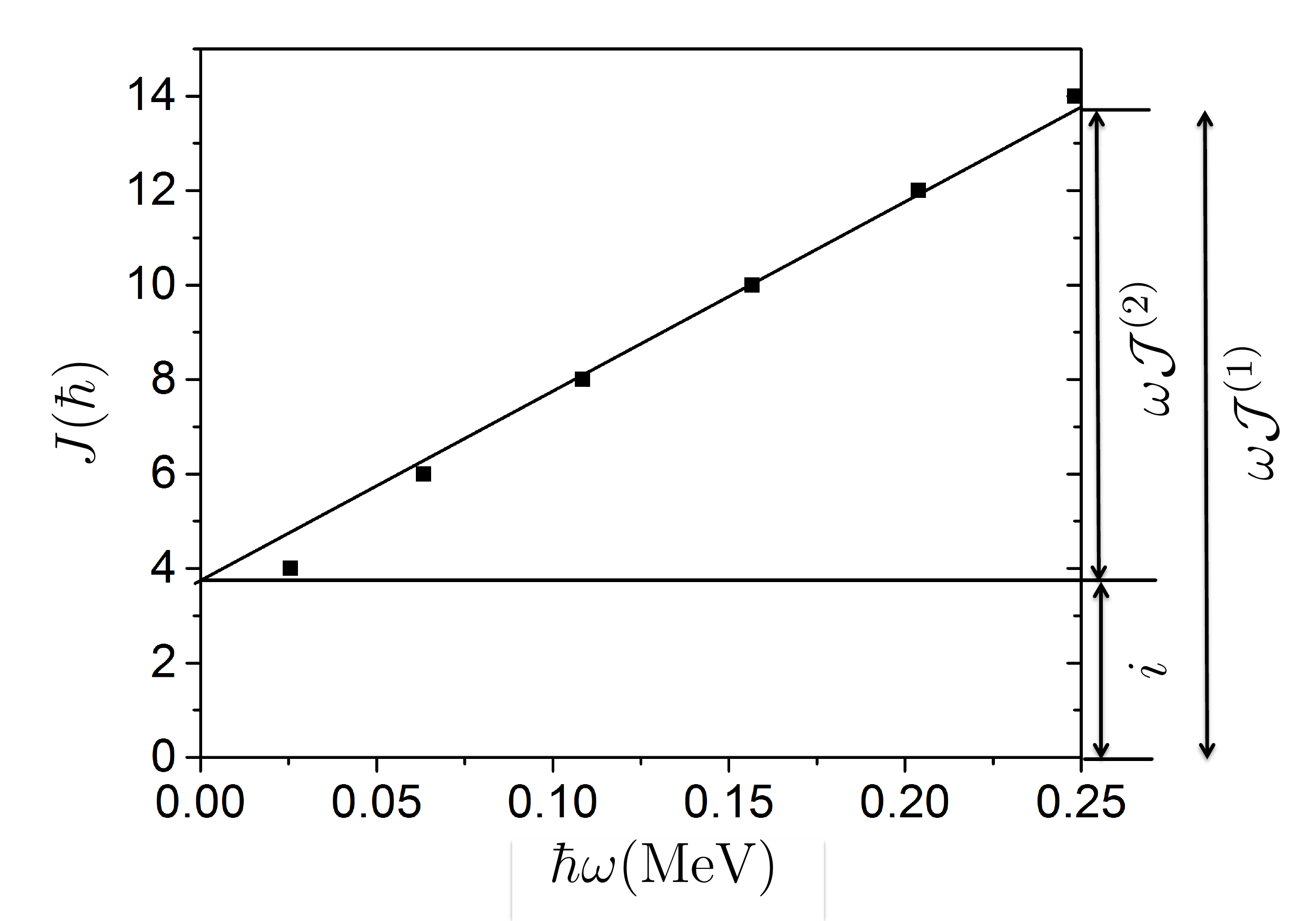} 
 \caption{\label{f:J1J2}The experimental kinematic \momi ${\cal J}^{(1)}$ and dynamic \momi ${\cal J}^{(2)}$ (left) and the \am $J$ (right)  of band A in $^{163}$Er
 calculated by means of Eqs. (\ref{eq:om2exp},\ref{eq:J1J2expJ1},\ref{eq:J1J2expJ22}). The right panel illustrates Eq. (\ref{eq:J1J2}), where the slight curvature of the 
 sequence experimental points is disregarded. Data from \cite{ENSDF}.  } 
 \end{figure}

The dynamical moment of inertia ${\cal J}^{(2)}$ measures the  increment of the angular momentum with the frequency $\om$
for a given \qp configuration. It characterizes the coherent quantal states that form a rotational band. 
The kinematical \momi ${\cal J}^{(1)}$ is a more appropriate measure for the overall increment of the energy of the states
in the yrast region with \amd, $E(J)\approx J^2/(2{\cal J}^{(1)})$, irrespective of structural rearrangements. 

When the adiabatic approximation is valid, the kinematic \momi ${\cal J}^{(1)}$ is equal to the dynamic \momi ${\cal J}^{(2)}$, 
and both coincide with the \momi of the Unified Model.
This is no longer the case when there is \qp \am  $i$ aligned with the rotational axis. Then   
\beq\label{eq:J1J2}
{\cal J}^{(1)}=\frac{J}{\om}=\frac{i+{\cal J}^{(2)}\om}{\om}=\frac{i}{\om}+{\cal J}^{(2)}.
\eeq
 Fig. \ref{f:J1J2} shows the band A in $^{163}$Er as an example.   Below $\hbar\om$=0.3 MeV  the dynamic \momi ${\cal J}^{(2)}$ is nearly constant,
 because it represents the rotational response of all \qps but A, which contributes an approximately constant \am of $i$.  The latter generates the down-slope 
  of the in the kinematic \momi ${\cal J}^{(1)}$ (see Eq. (\ref{eq:J1J2}}). A sudden increase of  ${\cal J}^{(1)}$ followed by a down-slope
   indicate the alignment of \qp \am with the
 rotational axis.  The spike of ${\cal J}^{(2)}$ at $\hbar\om$=0.4 MeV is another signal of rapid \qp alignment, which leads to a gain of $J$ without much
 increase of $\om$ (see the discussion of the back bending effect  in section \ref{sec:crdiabatic}). The irregularites of ${\cal J}^{(2)}$ are used as 
sensitive indicators of \qp alignment processes. They are better visible at high frequency than the up-down in ${\cal J}^{(1)}$ (compare the feeble feature 
 of  ${\cal J}^{(1)}$ at $\hbar\om$=0.4 MeV with the spike of  ${\cal J}^{(2)}$).
 
  Often, a more direct and practical graph is $J(\om)$, as shown in the right panel of Fig. \ref{f:J1J2}.  It shows $i$ as a constant shift of the line  ${\cal J}^{(2)}\om$.
  As seen, the aligned \am $i=3.9\hbar$ is already present at the band head. As a consequence  there is no spike of  ${\cal J}^{(2)}(\om)$ that indicates
  the {\it process} of \qp re-alignment. When the bands in the yrast region have approximately the same dynamic \momi ${\cal J}^{(2)}(\om)$ 
  the distances between of the curves $J(\om)$ directly provide the relative amount of \am aligned with the rotational 
  axis, which is an important information about the   \qp composition of the bands. The discussion of Fig. \ref{f:Erjom} in section \ref{sec:csmclass}
  is an example.   

\subsection{Calculation of the mean field shape}\label{sec:mfshape}
 
  The selfconsistent cranking approach is applied on different levels of sophistication. 
  The simplest version is the Cranked Shell Model introduced by Bengtsson and Frauendorf \cite{BF79}. It assumes that 
 the rotational axis is the principal axis of the deformed potential with the maximal \momid, which is conventially taken as the 1-axis. 
 Its  shape parameters $\beta,~ \ga$ and the pair potential $\De$  are assumed to be  
 known. The rotational bands shown in Figs. \ref{f:Er164Ehigh} and \ref{f:Er163Ehigh} are interpreted as configurations of quasiparticles in a potential rotating with the 
 angular velocity $\om$. The Cranked Shell Model has become an indispensable tool for classifying the 
 wealth of rotational bands measured with  $\ga$ ray detector arrays. 
 It will be discussed in section \ref{sec:CSM}.
 
 The next level of sophistication is to calculate the quasiparticle routhian using selfconsistency.  The majority of these studies assumes that the axis of rotation is a principal axis.  Systematic  calculations have been carried out for all versions of mean field  theory. They are quite successful in reproducing rotational bands 
 up to the highest spins. A review of this work is beyond the scope of this contribution. 
 Only the principal axis cranking  based on the shell correction 
 method (SCPAC) introduced in Refs. \cite{Neergard76,Andersson76} will be presented. 
 From the conceptual standpoint, it provides  the essential new phenomena that  principal axis cranking  describes. 
 The presentation is postponed to section \ref{sec:SCPAC} after discussing
 the properties of the \qp routhians and the relation between \qp \confs and rotational bands in sections \ref{sec:orbits} and \ref{sec:CSM}.

 Finally, the assumption that the rotational axis coincides with one of the principal axes of the density distribution is dropped. 
 Frauendorf demonstrated the existence of selfconsistent solutions of this type, 
 called tilted axis cranking (TAC) solutions, for the schematic  pairing+quadrupole-quadrupole  interaction and gave their interpretation as $\Delta I=1$ bands \cite{tac}. 
  The tilted axis cranking  approach has been laid out above for the  pairing+quadrupole-quadrupole  interaction, because its simplicity brings the new aspects into focus. 
  Frauendorf  \cite{tacdic} introduced the shell correction version
  of tilted axis cranking  (SCTAC) and described in detail how to apply the method. Olbratowski {\it et al.}  \cite{tacdft1,tacdft2} applied tilted axis cranking  to the Skyrme density functional approach. Peng {\it et al.}  \cite{tacrmf1} and Zhao {\it et al.} \cite{tacrmf2}
  developed tilted axis cranking  for the Relativistic Mean Field approach. In their contribution to this Focus Issue, J. Meng and P. Zhao  \cite{NCMengZhao} 
  describe the application of the latter to Magnetic Rotation. 
 Tilted axis cranking  will be discussed in section \ref{sec:TAC} with focus on its physical interpretation and symmetry aspects. 
 A more comprehensive review, which cites the work preceding  Ref. \cite{tac} and early applications, can be found in Ref. \cite{RMP}. 
 
\subsection{Geometry and rotational response of \qp orbitals}\label{sec:orbits}

The high-spin phenomena are governed by the balance between the cranking term $-\vec \om \cdot \vec {\hat J}$, which 
favors the  alignment of the \qp \am with the rotational axis $\vec \om$ (inertial forces) and the deformed potential, which favors maximal overlap 
with the density distribution of a particle (nuclear interaction is short-range attractive). This interplay is particularly transparent for the "high-j intruder orbitals" 
$f_{7/2},~g_{9/2},~h_{11/2},~i_{13/2},~j_{15/2}$, which 
are the key players in high-spin phenomena. As illustrated in in the left panel of Fig. \ref{f:qpcouplings}, they are relatively pure spherical orbitals because the reflection-symmetric potential  cannot mix 
them with the neighboring orbitals of opposite parity. They act like gyroscopes carrying  \am fixed by quantization. They favor maximal overlap of their 
torodial density distribution with the  deformed potential. Away from this orientation,  the torque causes a precession motion. For axial shape  a particle tends to align its \am with
the axis perpendicular to the symmetry axis, because this orientation corresponds to maximal overlap with the attractive potential.
 A hole prefers an orientation that minimizes the overlap of its density with the 
deformed potential, which is  repulsive for a hole. Accordingly it tends to align with the symmetry axis. For triaxial shape a  particle tends to align with the short axis, because this 
orientation maximizes the overlap, whereas a hole tends align with the long axis the long axis,  minimizing the overlap.  

 The interplay between the torque excerted by the deformed potential  and  the inertial forces result in different 
coupling schemes, which are illustrated in Fig. \ref{f:qpcouplings} by means of the precession cone carried out by the \am vector of the \qp \cite{FR81}. 
The figure also shows the fingerprint of the coupling scheme in a quasiparticle diagram like Fig. \ref{f:spagN96}.

At low frequency the torque exerted by the deformed potential causes $\vec j$ to precess around the symmetry axis 3, i. e. $\langle j_1\rangle=i\approx 0$
 and $\langle j_3\rangle\approx \hbar\Omega$, the quantized \am projection on the symmetry axis without rotation. 
 The trajectories of the routhians are horizontal and degenerate with respect to the signature $\alpha$.
 The coupling scheme is called deformation aligned (DAL) because the precession cone is aligned with the symmetry axis of the potential. In this context 
 "aligned with the axis"
  means that the axis is located within the cone. Of course, DAL is the coupling scheme of the adiabatic limit of the Unified Model.   
  
  At large frequency the inertial forces prevail, which cause $\vec j$  
  precessing about the 1-axis, i. e. $\langle j_1\rangle=i\approx const$ and  $\langle j_3\rangle\approx 0$. The precession angle is given by the quantization of 
  $j_1=\hbar m$. The trajectories of the routhians have a constant slope $-\hbar m$  and are ordered according to the energy  $-\hbar\om m$ 
  with alternating signature $\alpha=m + 2n$. 
  This coupling scheme is called rotation aligned (RAL)  because the precession cone is aligned with the rotational axis.
  
    For the lowest \qps the pair field generates a tendency toward RAL. The superposition of particle and hole components, which have opposite quadrupole moments, 
    reduces the net quadrupole moment of the \qp and thus its coupling to the deformed potential.  A new 
    coupling scheme appears. It corresponds to precession about an axis $\vec j_\lambda$, which has the angle $\arccos[j_{3\lambda}/j]$ with the 3-axis, 
    where $j_{3\lambda}$
    is the \am projection of the Fermi level on the symmetry axis. Both $j_{3\lambda}$ and $i$ are approximately constant. 
    The fingerprint corresponds to constantly sloping trajectories, which are degenerate with 
    respect to the signature $\alpha$.  The scheme has been called Fermi aligned (FAL) because the precession cone is aligned with the $\vec j_\lambda$ axis that
     lies on the precession cone of the Fermi level \cite{FR81}. 
     \begin{figure}[t]
  \begin{center}
  \includegraphics[width=7cm]{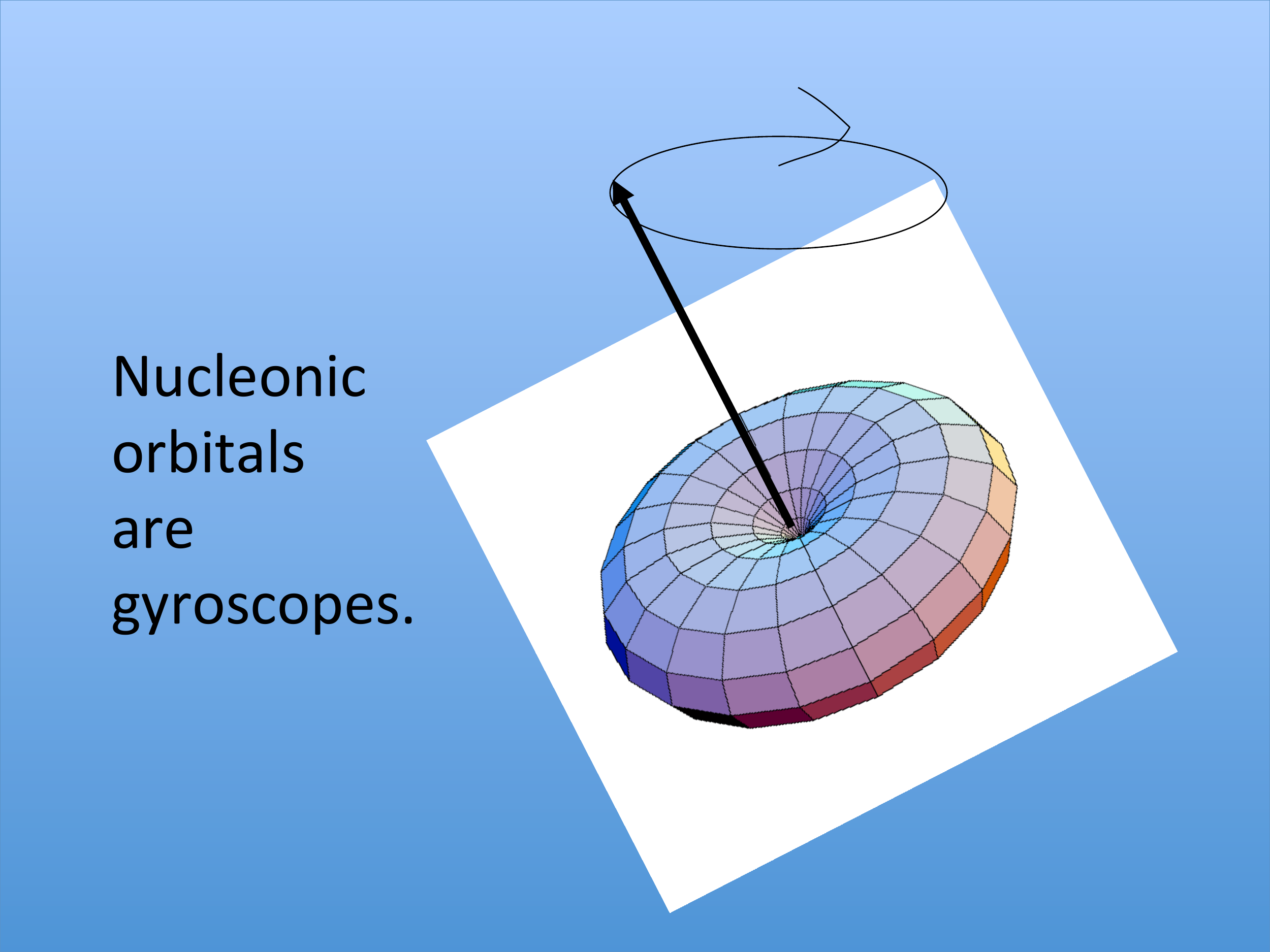}  
 \includegraphics[width=8cm]{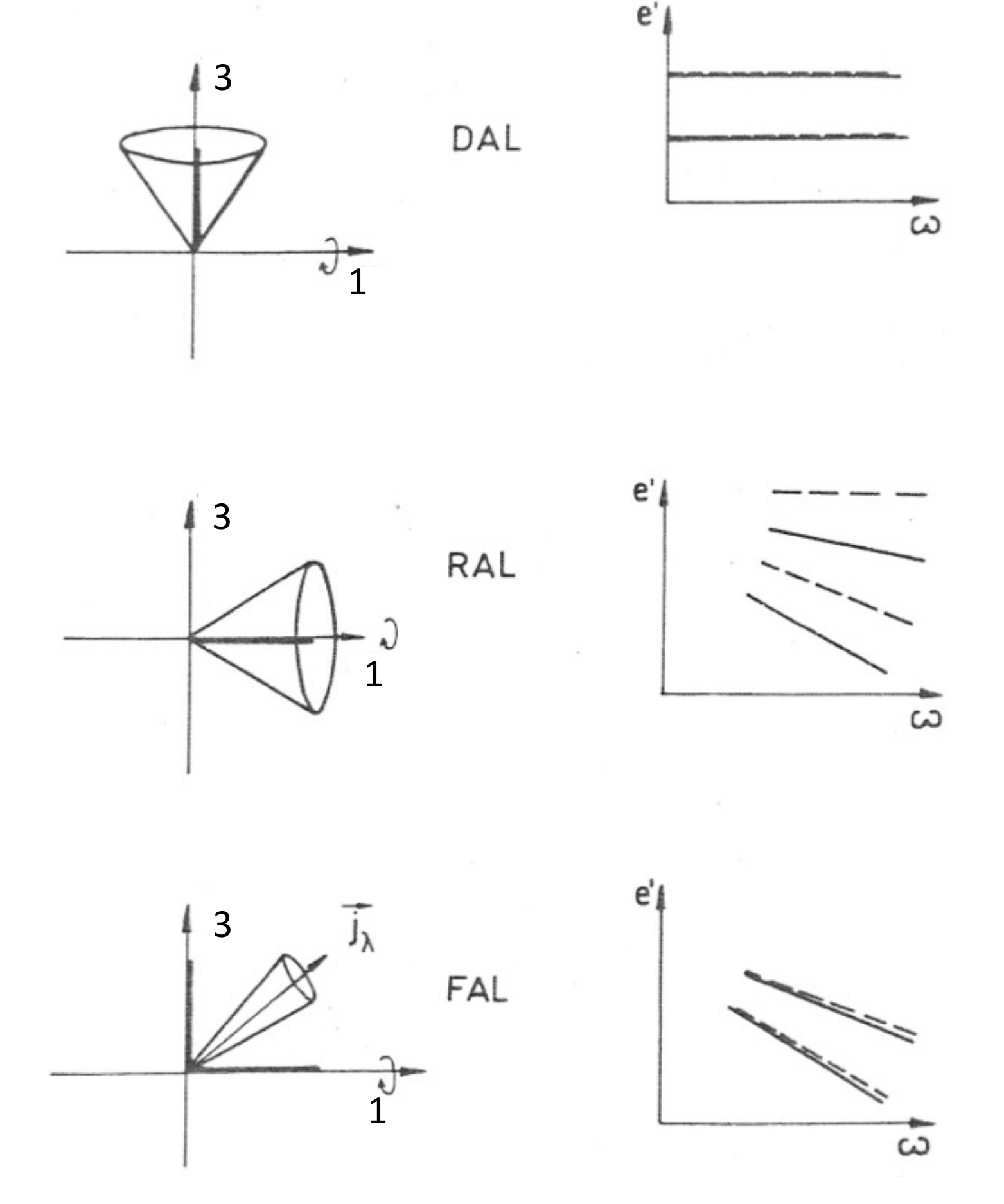} 
    \caption{\label{f:qpcouplings} Left panel: Density distribution  and \am of a high-j orbital (f$_{7/2}$, g$_{9/2}$, h$_{11/2}$, i$_{13/2}$, j$_{15/2}$).
   They remain relatively pure spherical orbitals because  the deformed reflection-symmetric nuclear potential does not  mix them with the surrounding orbitals of opposite parity.
    Its torque  causes a precession motion.
    Right panel: The three coupling schemes for a nucleus rotating about an axis perpendicular to its symmetry axis and their fingerprints in the \qp routhian trajectories.}
  \end{center}
 \end{figure}

The  semiclassical analysis in  \ref{sec:semi}   provides a more detailed picture of
the coupling of high-j orbitals with the triaxial potential, which extends the work in Refs.  \cite{MotANL,FrauDrexel,FR81,HamMot83,FraSob86}.

Frauendorf \cite{tacdic} also analyzed  how  the  normal parity  quasiparticles
 couple to the deformed rotating potential. In the case of 
a prolate axial deformation
the orbitals with a large projection $j_3$ are strongly coupled to the
deformed potential. They have the same  $-j_3 \om \cos \vth$  dependence discussed above for the hole type high-$j$ orbitals.
The states originating from low-$K$ Nilsson levels show a complex behavior,
which cannot be explained in a simple way. An exception are 
the  pseudo spin  singlets, which will be briefly discussed in section \ref{sec:axialTAC}.

  In summary,  the fingerprint for RAL coupling is  a well-split sequence of states 
     of alternating signature with approximately constant aligned \am $i=\langle j_1\rangle=-de'/d\om$, which
     decreases with energy.  The fingerprints of DAL coupling are signature doublets 
     with finite \am aligned with the deformation axis and no (or small negative) \am aligned with the rotation axis.
     The fingerprint of FAL coupling are signature doublets with   with finite \am aligned with the deformation axis and substantial  \am aligned with the rotation axis.

 \begin{figure}[t]
  \begin{center}
 \vspace*{-4cm}
 \includegraphics[width=\linewidth]{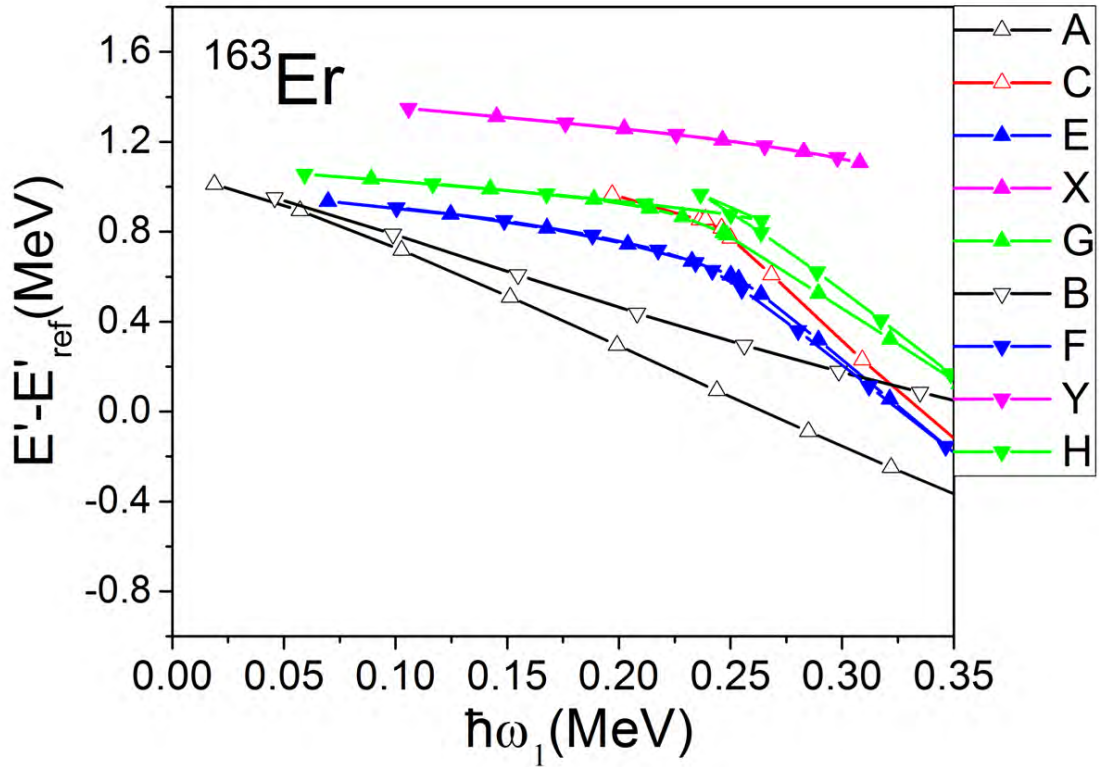} 
 \vspace*{-4cm}
  \caption{\label{f:Er163Eplow}  Experimental routhians in $^{163}$Er relative to the g-band reference routhian $E_{ref}$. 
  The symbol convention is the same as in Fig. \ref{f:Er164Elow}.
  The routhians are calculated by means of Eqs. (\ref{eq:om1},\ref{eq:Epom1}), where $K=5/2$ is used for A, B, E, F, 
  and $K=11/2$ for X, Y.
  Configurations involving quasiprotons and  several three-quasineutron configurations are left away for clarity.
   Data from Refs. \cite{ENSDF} and \cite{Er163}, from which the band labels are adopted. }
  \end{center}
 \end{figure} 
 \begin{figure}
  \begin{center}
 \vspace*{-4cm}
 \includegraphics[width=\linewidth]{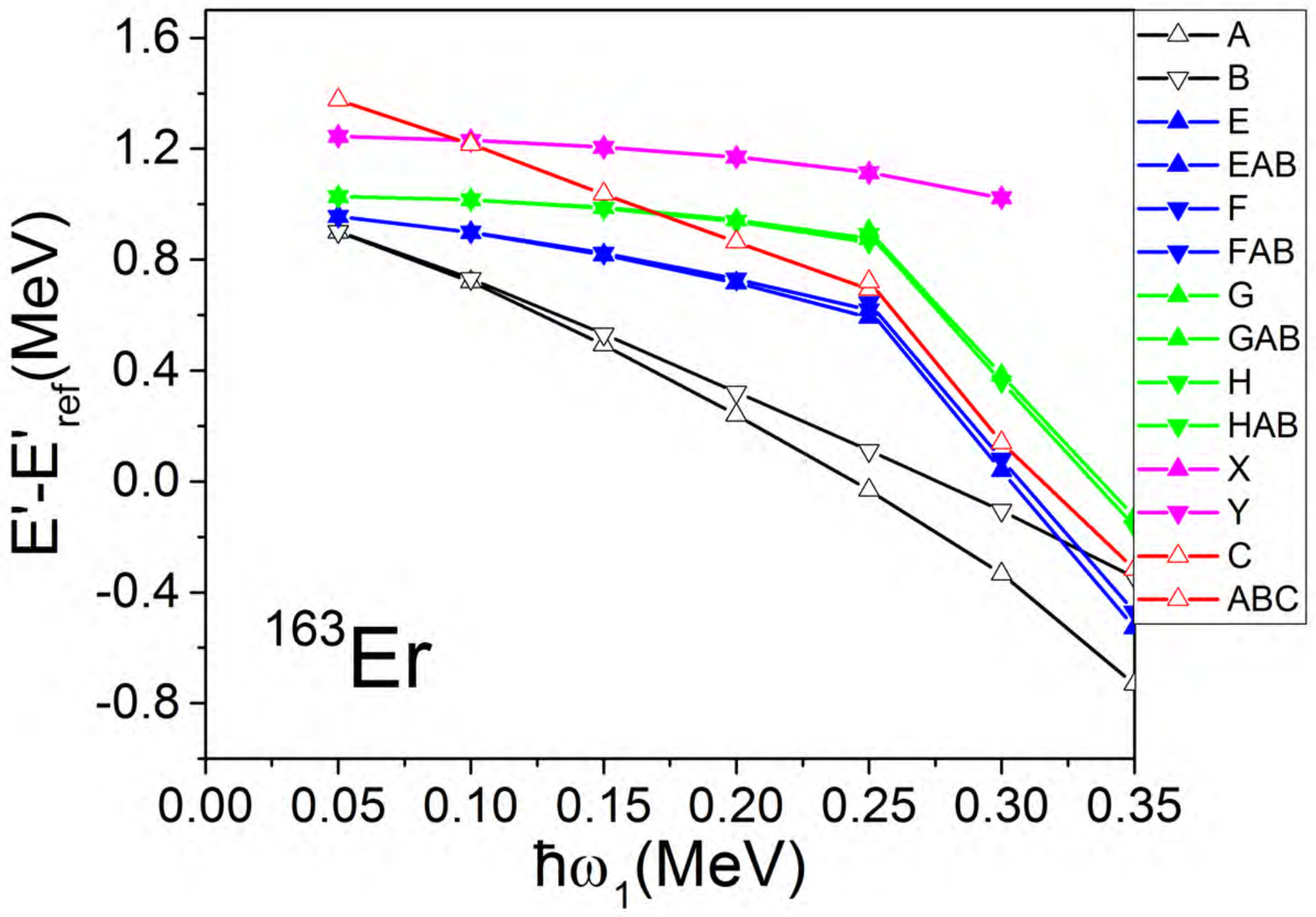} 
 \vspace*{-4cm}
  \caption{\label{f:Er163CSMlow}  Calculated  routhians in $^{163}$Er relative to the calculated g-band reference routhian $E_{ref}$. Symbol convention as in Fig. \ref{f:Er164Elow}. 
  Configurations involving two-quasiproton and  several three-quasineutron configurations are left away for clarity.  }
  \end{center}
 \end{figure} 
\begin{figure}
 \begin{center}
 \vspace*{-4cm}
 \includegraphics[width=\linewidth]{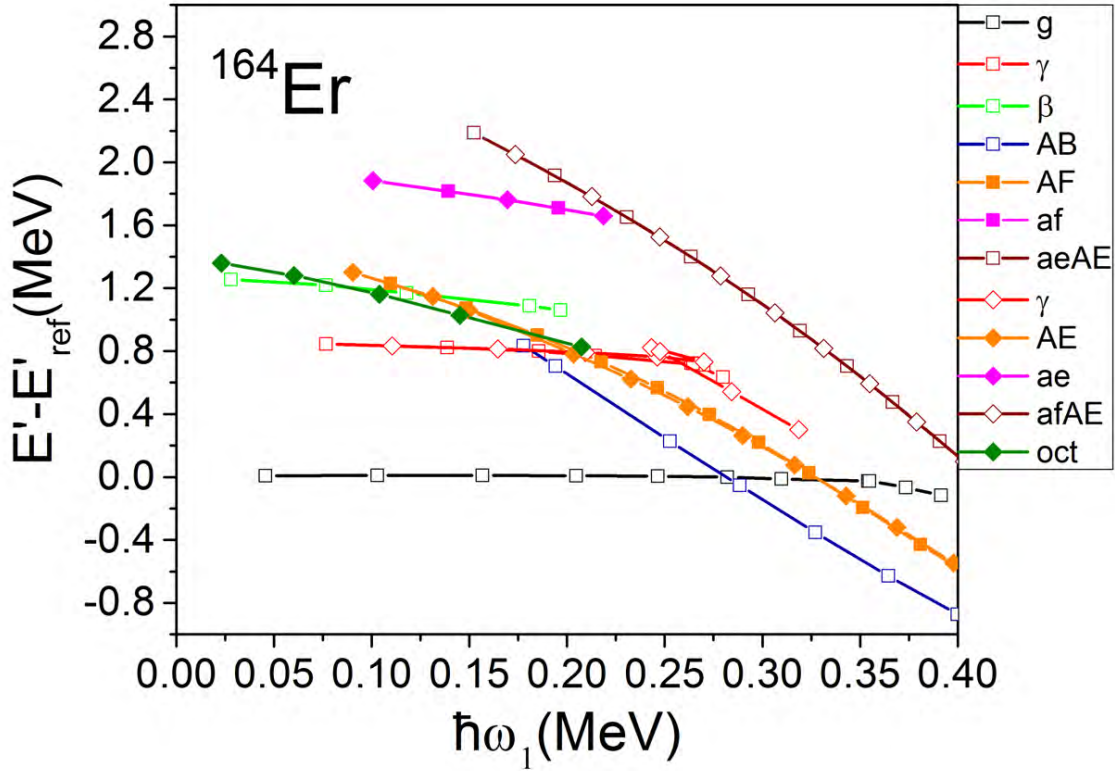} 
 \vspace*{-4cm}
  \caption{\label{f:Er164Ep}  Experimental routhians in $^{164}$Er relative to the g-band reference routhian $E_{ref}$. 
  The symbol convention is the same as in Fig. \ref{f:Er164Elow}.
  Data from Ref. \cite{ENSDF} from which the band labels are adopted.  
 The routhians are calculated by means of Eqs. (\ref{eq:om1},\ref{eq:Epom1}), where $K=0$ is used for g, AB; $\beta$, oct; 
   $K=2$ for $\ga$; $K=5$ for AE, AF; $K=7$ for ae, af; and $K=12$ for aeAE, afAE.
  }
  \end{center}
 \end{figure} 
 \begin{figure}
  \begin{center}
 \vspace*{-4cm}
 \includegraphics[width=\linewidth]{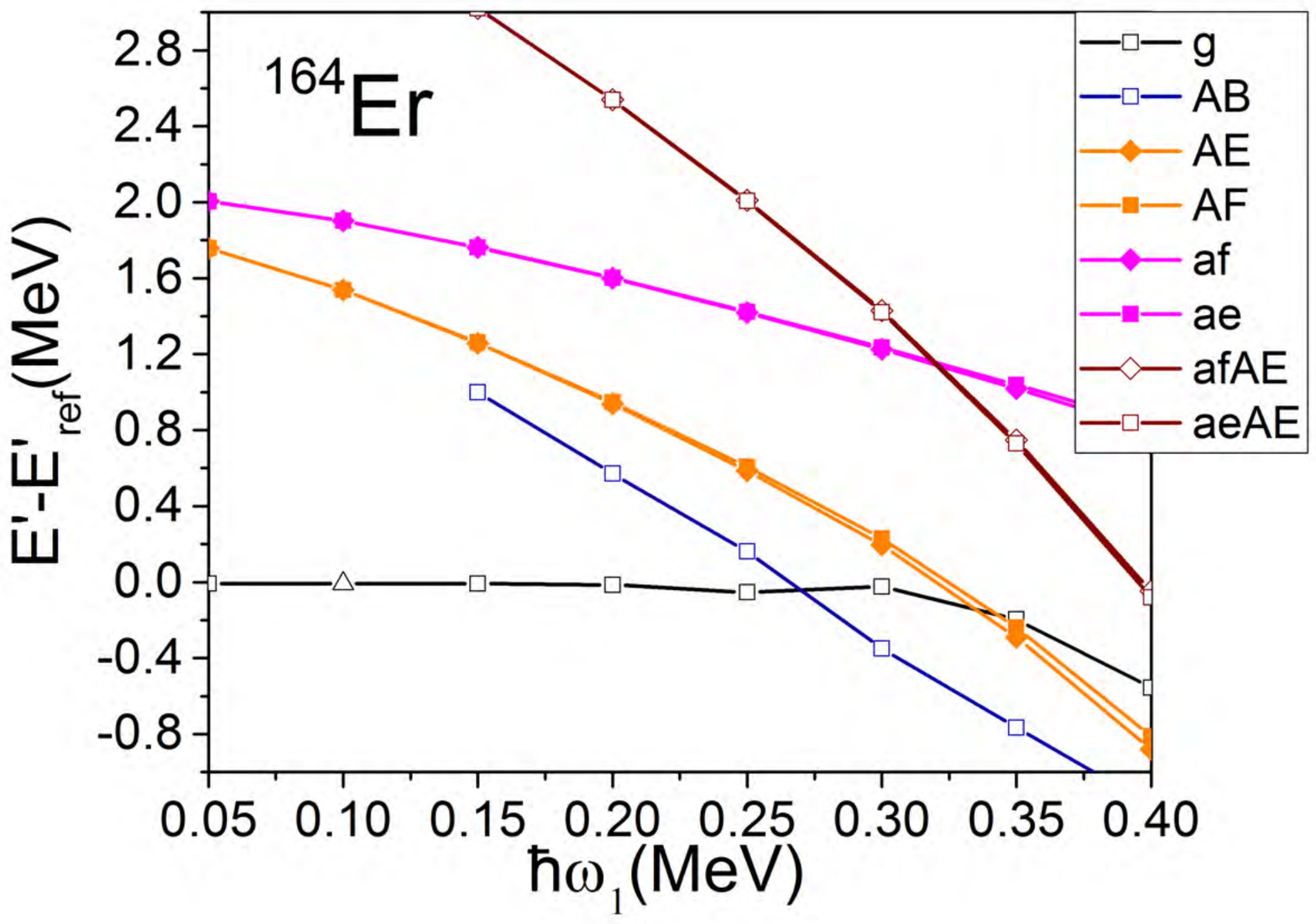} 
 \vspace*{-4cm}
  \caption{\label{f:Er164CSM}  Calculated  routhians in $^{164}$Er relative to the calculated g-band reference routhian $E_{ref}$. 
  The symbol convention is the same as in Fig. \ref{f:Er164Elow}.   }
  \end{center}
 \end{figure}

 \subsection{Cranked Shell Model}\label{sec:CSM}
 The concept of the Cranked Shell Model  was established by Bengtsson and Frauendorf \cite{BF79}. A detailed presentation was 
 given by Bengtsson, Frauendorf and May \cite{BFM86}, which contains a collection of quasiparticle routhians.  
 In the spirit of the conventional Shell Model it considers configurations of independent nucleons which occupy 
 the eigenstates of the single particle routhian 
 \beq\label{eq:h'sp} 
 h'(\beta,\ga,\om_1)\phi_i=\eps'_i(\beta,\ga,\om_1)\phi_i, ~~~ h'(\beta,\ga,\om)= h(\beta,\ga) -\om _1j_1
  \eeq 
   according to the Pauli principle. The energies in the rotating frame of reference, $\eps'_i(\beta,\ga,\om)$, are called the single particle routhians.  
    The single particle routhian operator (\ref{eq:h'sp})  conserves parity. In addition it is invariant with respect to a rotation by $\pi$  about
   the 1-axis, which implies that its eigenstates carry the signature quantum number $\alpha$,
   \beq
   {\cal R}_1(\pi)\phi_i=e^{-i\pi \alpha_i}\phi_i, ~~~\alpha=\pm1/2.
   \eeq
 The total parity is the product of the single particle parities.  The total signature $\alpha$ of a configuration is the sum of the single particle signatures, which 
 is related to the \am of the band by $I=\alpha +2n$ for the same reason as discussed for the Unified Model above. 
 As usual, the relative  energies of the various configurations
   are equal to the differences between the sums of the routhians of all occupied states, provided the deformation parameters {\it and the angular frequency} $\om_1$ are the same.
   This means the bands can be classified as the particle-hole configurations   in the same way as for the non-rotating deformed potential of the Unified Model. 
 The simplicity only reveals when one compares routhians at a given frequency $\om_1$.  
     
   The  particle-hole scheme works only at sufficiently high spin, where the pair correlations can be neglected.
   In their contribution to this Focus Issue \cite{NCRiley}, M. A. Riley, J. Simpson and E. S. Paul
   present the Cranked Shell Model interpretation of the rotational bands  $^{159,160,161,162}$Er. For $\hbar\om>0.4$MeV the experimental routhians can be
   surprisingly well reproduced as particle-hole configurations of single particle routhians.
   
    At moderate spin, where  the pairing correlations must be taken into account, one
   must change to \qp \confsd.   
   The quasiparticle routhians are found by diagonalizing  the double-dimensional quasiparticle routhian (\ref{eq:hpqp}). 
   The details are given e. g. in Ref. \cite{BF79} or the in the textbooks nuclear many body theory
   \cite{Ringbook,blaizotbook}. The quasiparticle routhians $e_i'$ appear in the symmetric way shown in Fig.  \ref{f:spagN96}. At low spin the quasiparticles simply occupy the  
   solutions $e_i>0$.  When the quasiparticle routhians  cross with zero the interpretation simplifies  using an extended occupation scheme that includes the 
   $e_i<0$ part.   The modified rules state that 
   \begin{itemize}
   \item One half of {\it all} quasiparticle routhians is occupied and the conjugate half is free. 
   \item Out of a conjugate pair {\it only one} of the re-occupied states
contributes to the change of energy, aligned angular momentum, signature, parity and other quantities. 
   \end{itemize}
 The second rule is meant for the Cranked Shell Model only, which calculates {\it energy differences} between quasiparticle configurations for {\it fixed} mean-field parameters.
 The excitation energies are taken relative to some {\it reference} configuration, which may be chosen differently depending on the spin range.    
 
 The reference routhian and \am can be well approximated by the Harris expressions
\beq\label{eq:ref}
J_{1ref}=i_r+\om \Theta_0+\om^3\Theta_1, ~~E_{ref}=e_r+i_r\om+\frac{\om^2}{2}\Theta_0+\frac{\om^4}{4}\Theta_1+\frac{\hbar^2}{8\Theta_0}.
\eeq
For low and moderate spin the natural choice is the ground state (g-) band of the even-even nucleus, which has $i_r=e_r=0$.
For our example $^{164}$Er, the experimental energies of the g-band give $\Theta_0=32\hbar^{2}$MeV$^{-1}$ and $\Theta_1=32\hbar^{4}$MeV $^{-3}$.  
This reference is subtracted in Figs. \ref{f:Er163Eplow}, \ref{f:Er164Ep} and \ref{f:Erjom}. The experimental relative routhians are compared 
with the calculated ones  in Figs. \ref{f:Er163CSMlow} and \ref{f:Er164CSM}
\setcounter{footnote}{0} \footnote{The Cranked Shell Model parameters are quoted in the caption of Fig. \ref{f:spagN96}.}.
The parameters of the Cranked Shell Model reference are adjusted to the calculated routhian of the g-configuration.    
   
 \subsubsection{Bands as quasiparticle configurations}\label{sec:qpconfs}
 
Fig. \ref{f:spagN96} shows the quasineutron routhians for $N\approx 96$.
 A collection of such quasiparticle diagrams for the rare earth region can be found in Ref. \cite{BFM86}.
The quasiparticle trajectories are classified
by means of the parity and signature, $(\pi,\al)$. The different rotational
bands correspond to the various possibilities of occupying the quasiparticle levels, i. e. to quasiparticle configurations.
The large number
of bands identified with multi $\ga$ ray detector arrays necessitates a compact notation. A letter is assigned to each \qpd, where
the convention is used that the first letters of the alphabet A, B, C, D are used for the high-j intruder states (i$_{13/2}$ neutron in our example)
and  E, F, G, .... for the normal parity states.  For the discussion we use the letter code of  Fig. \ref{f:spagN96} and its extension  in Table. \ref{t:code}. 

The vacuum configuration [0] has $(\pi,\al)=(0,0)$. It is the even-$I$ ground state
rotational band of $^{164}_{68}$Er$_{96}$.
Exciting one quasineutron to the levels A or B generates the configurations [A] and [B] with
$(\pi,\al)=(+,1/2)$ and $(+,-1/2)$, respectively. They represent  bands
with $I=1/2+2n$ and $I=-1/2+2n$ in $^{163}_{68}$Er$_{95}$, which are
in Figs. \ref{f:Er163Eplow}
and   \ref{f:Er163CSMlow} denoted  by  A and B, respectively. Likewise,
the excitation of quasineutrons to the levels E, F, F, G generates the negative parity bands. 
These quasineutron routhians combine to the two-quasineutron configurations AE, AF in $^{164}$Er
shown in Figs. \ref{f:Er164Ep} and \ref{f:Er164CSM}. Shown are also the two-quasiproton excitations
ae and af, as well as the two-quasiproton-two-quasineutron configurations  afAE and aeAE.
Several three-quasiparticle excitations are left away in  Figs. \ref{f:Er163Eplow} and   \ref{f:Er163CSMlow} for clarity.

    \subsubsection{Bandcrossings}\label{sec:crdiabatic}

At $\hbar\om=0.23$ MeV, the levels A and  B$^+$ "quasicross", 
i. e. they exchange their character in the  narrow region where they
repel each other. 
The yrast \conf    changes its character from 
the vacuum [0] to what was the two-quasineutron configuration
[AB] before the quasicrossing. Such a rapid structural change is in conflict
with the concept of a band. The appropriate point of view is
to consider the \confs 
[0] and [AB] as two separate bands,  called  
the g- and s-bands (super-), which cross each other (AB-crossing).
The crossing is observed as a "back bending" (S-shape) of the function $J(\om)$
 constructed from  the yrast levels. 
As first relized by Stephens and Simon \cite{backbend}, the back bend is caused 
by the sudden alignment of the angular momentum of 
the two  quasiparticles A and B with the 1-axis.
At given $I$,  the  frequency  in the g-band  
is larger than in the s-band
\beq\label{eq:backbend}
\om_g=(I+1/2)/{\cal J}, ~~\om_s=(I+1/2-j)/{\cal J},~~j=j_{1,A}+j_{1,B},
\eeq  
which results in the
the decrease of $\om$ when the yrast levels change from the g- to the s-band.   
More about the discovery of back bending can be found in the contribution by  
M. A. Riley, J. Simpson and E. S. Paul to this Focus Issue \cite{NCRiley}.
A beautiful mechanical simulation of the effect is available online \cite{bbMovie}.

\begin{figure}[t]
 \begin{center}
 \includegraphics[width=\linewidth]{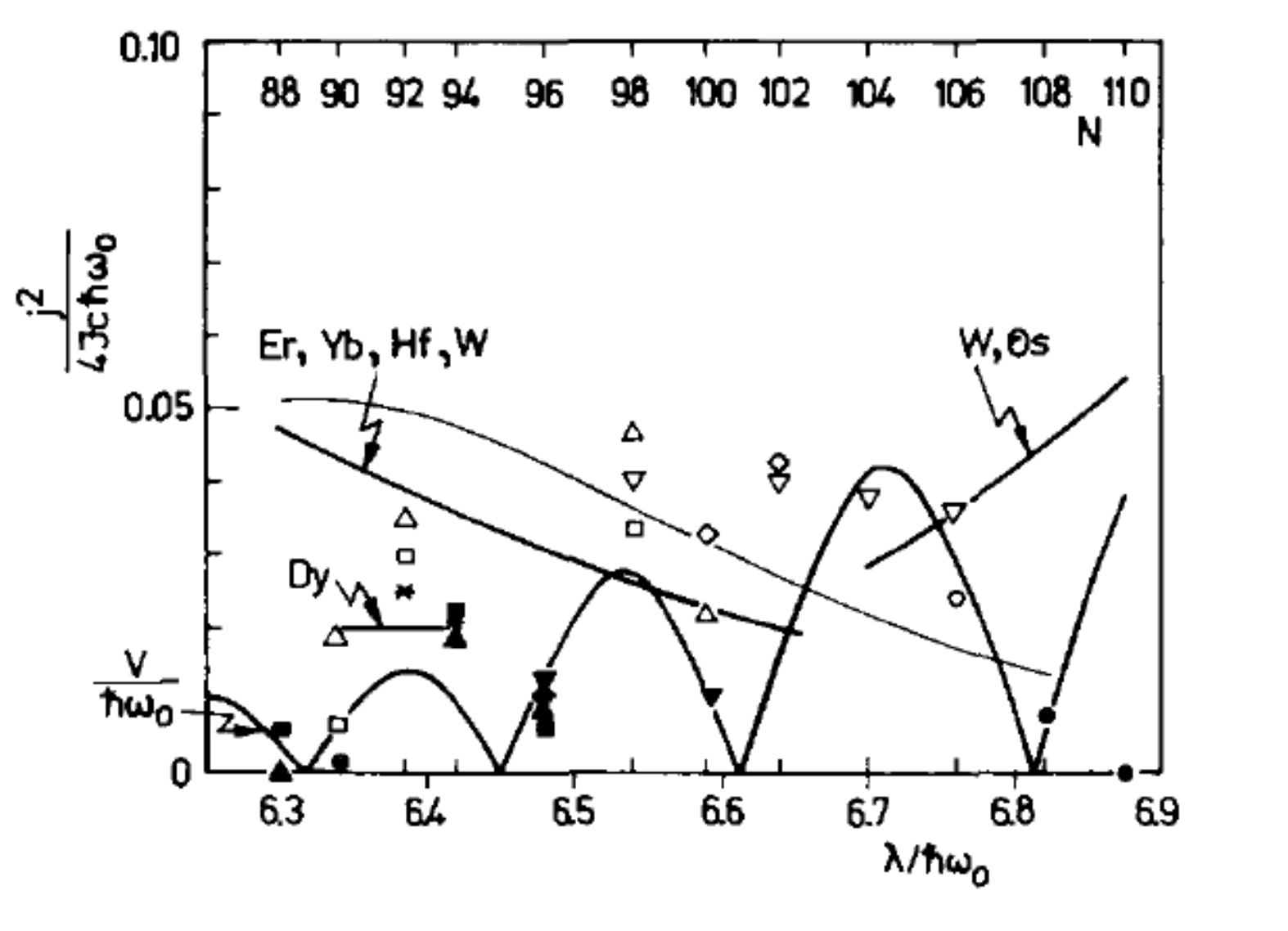}
\caption{\label{f:VABoscillation}
The interaction matrix element $\vert V\vert$ between the g- and s-band. The experimental values are shown
by the scattered symbols. The theoretical values
 correspond to the oscillating curve denoted by $V/\hbar\om_0$ with $\hbar\om_0=41A^{-1/3}$ MeV.
The experimental values  are derived from the experimental function $J_1(\om)$ by an expression 
given in Ref. \cite{BF79a} that relates $\vert V\vert$ to the rapidness of the back bend. The scale of this relation
is determined by  the quantity  $j^2/4{\cal J}_c$ (cf. Eq. (\ref{eq:backbend}), which is
shown by the smooth curves (thick lines - experimental, thin lines CSM). Reproduced from \cite{BF79a}.

}
  \end{center}
  \end{figure}
 \begin{figure}
  \begin{center}
 \vspace*{-4cm}
 \includegraphics[width=\linewidth]{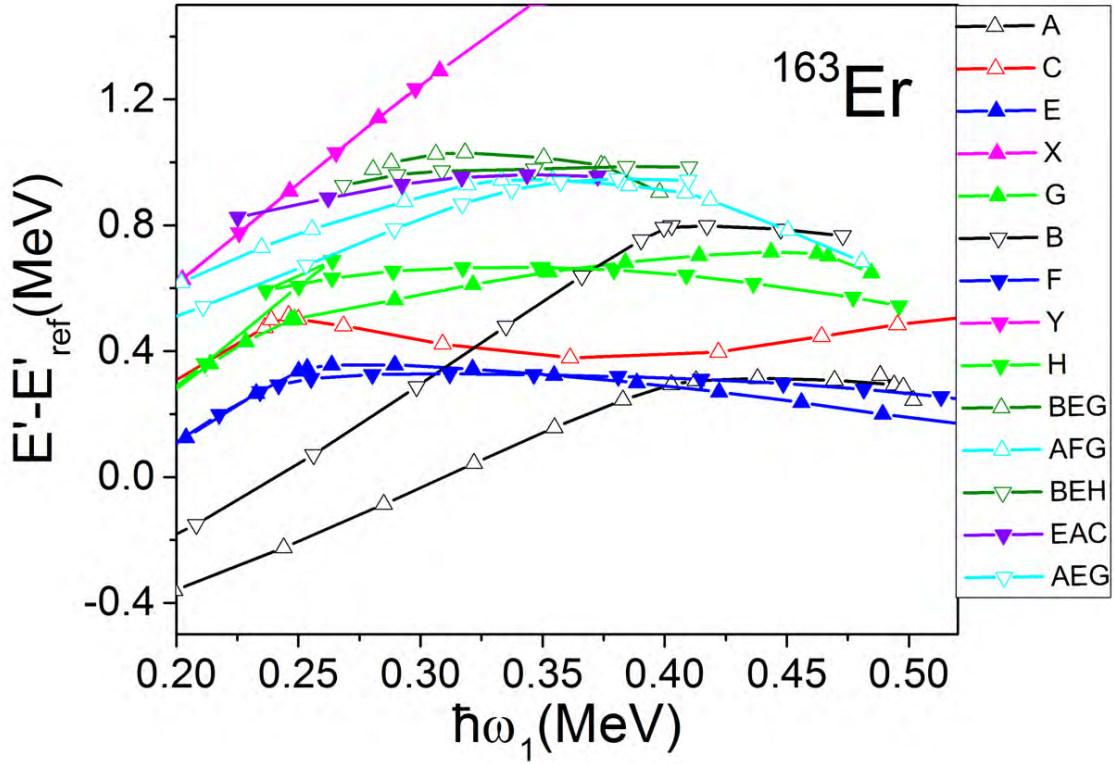} 
 \vspace*{-4cm}
  \caption{\label{f:Er163Ephigh}  Experimental routhians in $^{163}$Er relative
   to the s-band reference routhian $E_{ref}$. The symbol convention is the same as in Fig. \ref{f:Er164Elow}.
   Configurations involving two-quasiproton configurations are left away for clarity.
  The routhians are calculated by means of 
   Eqs. (\ref{eq:om1},\ref{eq:Epom1}), where $K=3/2$ is used for g, H; $K=5/2$  for A, B, E, F; $K=9/2$ for BEG, AFG, EAC, AEG;
  $K=11/2$ for X, Y (cf. \cite{Er163}). Data from Refs. \cite{Er163,ENSDF},
  from which the band labels are adopted.
     }
  \end{center}
 \end{figure} 
 \begin{figure}
  \begin{center}
 \vspace*{-4cm}
 \includegraphics[width=\linewidth]{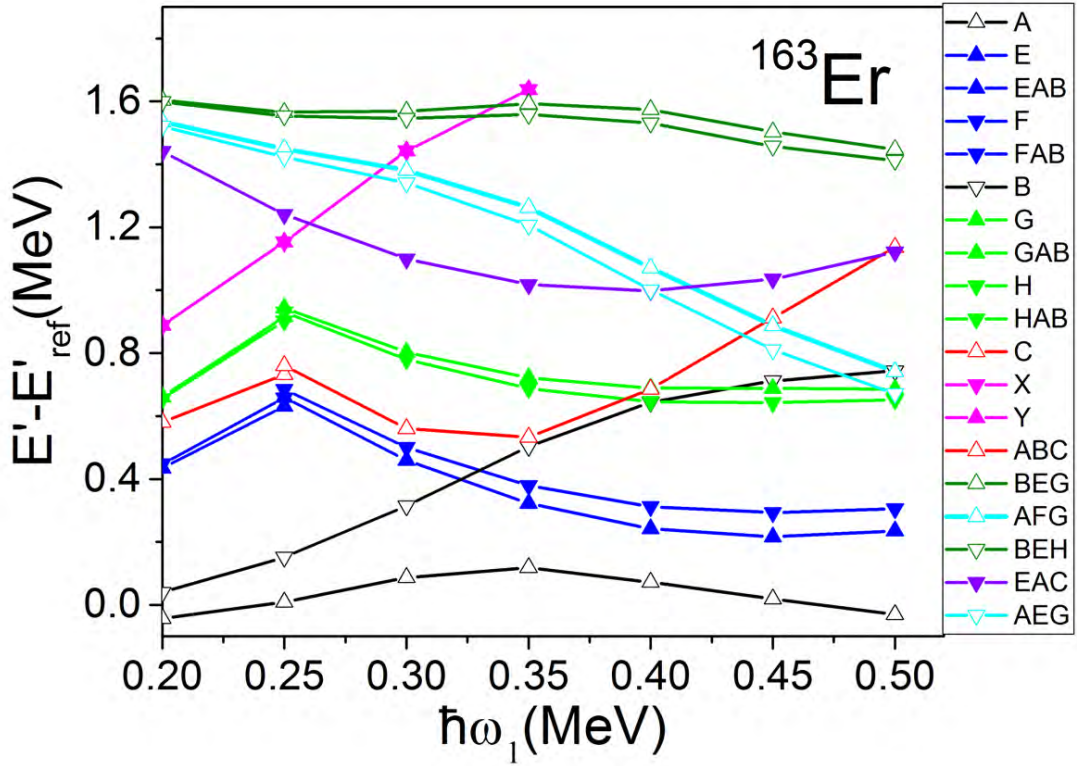} 
 \vspace*{-4cm}
  \caption{\label{f:Er163CSMhigh}  Calculated  routhians in $^{163}$Er relative to the calculated g-band reference routhian $E_{ref}$. Symbol convention as in Fig. \ref{f:Er164Elow}. 
  Configurations involving quasiprotons and  several three-quasineutron configurations are left away for clarity.  }
  \end{center}
 \end{figure} 
 The AB-crossing
appears in yrast lines of the even-$N$
nuclei $^{162,164}$Er.  In Figs. \ref{f:Er163Eplow} and \ref{f:Er163CSMlow}, it is  
seen at the same frequency as the sudden increase of the negative slope 
of the  bands E, F, G, H  in $^{163}$Er.  These quasineutrons
are  just spectators of 
the reaction of A and B to the inertial forces. According to this interpretation the
[AB] content after the band crossing is explicitly indicated in Fig, \ref{f:Er163CSMlow}.
The  configuration [A] is not disturbed by the AB-crossing. It is
``blocked'' because both A and B$^+$ are occupied. The same holds for [B].

For  the systematic analysis of the band crossings
 it is useful to construct "diabatic"
trajectories 
\setcounter{footnote}{0}
\footnote{The concept of diabatic routhians was introduced in Ref. \cite{BF79}. The detailed use of diabatic vs. unmodified 
 routhians is discussed in Ref. \cite{BFM86}. The authors introduced the terminus  "diabatic" and referred to the unmodified quasiparticle routhians as
 "adiabatic". In this contribution I do not use the latter in order to avoid confusion with the adiabatic approximation of the Unified Model.
 The terminology was adopted from the discussion of time dependent processes near a quasi crossing of two quantum states.
 In context of the Cranked Shell Model it just means that the mixing between the quasipartice states near the quasi crossing is switched off
 for the diabatic routhians and not for the adiabatic ones.
  }. 
These are the  thin lines in Fig. \ref{f:spagN96}
obtained by "switching off" the
interaction, which causes the
repulsion between A and B$^+$ near the crossing.
In this way problems   near the crossing  are avoided.  
The rotating mean field  becomes a poor approximation
there  because the basic presumption   
that the dispersion in 
angular momentum depends  weakly on $\om$   is 
violated \cite{fluct1}. This point will be further discussed in section \ref{sec:coherence}.

Diabatic tracing is the simplest way to construct diabatic routhians. 
The  \qp routhian (\ref{eq:hpqp}) is diagonalized for a discrete set of $\om_n$ values. 
The overlaps between the \qp amplitudes of two adjacent steps, which are given by $[\al_i^\dagger(\om_{n-1}),\al_j(\om_n)]$, 
are calculated. The \qp $j$ at $\om_n$ that has the maximal overlap with \qp $i$ at $\om_{n-1}$ continues the diabatic trajectory.
The method depends on the step size. A step of $\Delta\hbar \om$=0.05 MeV has proven a good choice for medium and heavy nuclei.   
The tracing  sometimes fails when a grid point falls close to the center of a quasicrossing.  Usually such problems are easily mended by slightly changing the step size. 

Of course, the \qp trajectories found by tracing are still perturbed by the level repulsion near  the quasicrossing. 
Bengtsson an Frauendorf \cite{BF79a} introduced an interaction removal procedure. The routhians in the $\om$ interval where the level repulsion is substantial
are replaced by a low-order interpolation adjusted to the branches above and below the quasicrossing.  
The deviation of the calculated routhians from the interpolation provides an estimate of the  interaction strength that causes  the level repulsion.
They also suggested an analog analysis of experimental data in terms of
 two crossing bands, which provided  an estimate of the interaction strength that mixes the crossing bands. 
 The two estimates are related in a semiquantitative way.
 The observed degree of band mixing correlates closely with the mixing degree of the quasiparticle configurations. 
A striking example of this correlation are the oscillations of the AB-crossing interaction strength shown in Fig.  \ref{f:VABoscillation}.
Nikam {\it et al.} \cite{berry} suggested that these oscillations are a nuclear manifestation of  Berry's Phase encountered in quantum chemistry and solid state physics.

\subsubsection{Cranked Shell Model classification of bands}\label{sec:csmclass}

The Cranked Shell Model provides  indispensable guidance of which \qp \confs are expected in the yrast region  to be associated with the observed
multitude of rotational bands. As an example, 
Fig. \ref{f:Er163Ephigh} shows the routhians in the yrast region of $^{163}$Er at high spin, which are  associated with the Cranked Shell Model \qn  \confs in Fig. \ref{f:Er163CSMhigh}. 
For the high spin region it is more appropriate to use the [AB] configuration, which is yrast in the $N=96$ system, as a reference. It is well approximated by 
the Harris expression (\ref{eq:ref}) with $i_r=3.4\hbar,~\Theta_0=55.8~\hbar^2$MeV$^{-1},~\Theta_1=0$. 
The experimental routhians clearly correlate with the Cranked Shell Model pattern. 
Of course, accurate agreement cannot be expected because the Cranked Shell Model assumes independent \qps and a configuration-independent rotating mean field. Like the time-honored 
work horses of  the  spherical 
Shell Model and the Nilsson Model for deformed nuclei, which are based on the same simplifications, it represents the starting point for a more sophisticated analysis. 

A specific feature of the Cranked Shell Model is the appearance of characteristic
frequencies of band crossings, which belong to one and the same crossing between two \qp trajectories. It is expected
 to be seen as a rapid increase of $J_1(\om)$
in all  \confs with one of the trajectories occupied and the other free. The crossing shows up as a kink or down bend in the routhian.
If both trajectories are occupied or free, no irregularity appears, the crossing is said to be "blocked". There are three pronounced crossings seen in Fig. \ref{f:spagN96}, which
occur between the high-j i$_{13/2}$ \qn orbitals: the AB-crossing between A$^\dagger$ and B at $\hbar\om=$0.25 MeV, the BC-crossing between B$^\dagger$  and C 
 at $\hbar\om=$0.34 MeV,
and the AD-crossing between A$^\dagger$  and D at at $\hbar\om=$0.38 MeV.
  The AB-crossing  is active in the  \confs [E], [F], [G], [H], [C] and seen in Fig. \ref{f:Er163Ephigh} as kinks in the respective bands. It is blocked in the \confs 
[A] and [B], which are regular around $\hbar\om=$0.25 MeV.  The AD-crossing is active in [B]. It appears as the gradual down bend of the routhian B in Fig.
\ref{f:Er163CSMhigh} and the kink of B in Fig. \ref{f:Er163Ephigh} at $\hbar\om=$0.38 MeV, where [B] changes to [ABD]. The BC-crossing is active in [A].
It appears  as the gradual down bend of the routhian A  and the gradual up bend of ABC in Fig.
\ref{f:Er163CSMhigh} around  $\hbar\om=$0.34 MeV, where the two \confs interchange character: [A] $\rightarrow$ [ABC] and [ABC]$\rightarrow$ [A].  The interchange corresponds 
to the  crossing of the bands A and ABC seen in Fig. \ref{f:Er163Ephigh}  at   $\hbar\om=$0.39 MeV. 
The BC and AD crossings show analog oscillations as AB as function of the location of the chemical potential $\lambda$ within the i$_{13/2}$ shell (cf. Ref. \cite{FR81}).
Therefore, the smoothness of the crossings is sensitive to details of the rotating mean field  and may not be accurately accounted for by the Cranked Shell Model, 
as it is the case in our example.

The Cranked Shell Model approximation implies additivity. That is, the relative routhian $E'(\om)-E'_{ref}(\om)$ of a complex configuration is just the sum the routhians of the constituent \qpsd.
 The same holds for the relative \am $J_1(\om)-J_{1ref}$, which is often called the aligned \am or simply the alignment. Plots of the latter have
  turned out to be a very useful classification tool.
 Following familiar Shell Model practice, the quasiparticle alignments can be directly derived from experiment.  Then they are added up to the total alignment of a multi-\qp \confd.  
Fig. \ref{f:Erjom} shows an example of such "alignment plot". The gain in alignment by the AB crossing is clearly seen as the group of up-bends clustering around \mbox{$\hbar \om_1=$0.25 MeV}.
Summing the experimental \qn alignments somewhat overestimates the experimental alignments. The discrepancy can be traced back to a reduction of the pair field $\De_n$.
Using a Cranked Shell Model \qp diagram like Fig. \ref{f:spagN96} for guidance and  combining the information from the alignment plot 
with the signature and parity quantum numbers (determined by the combining the respective quantum numbers of the constituent
\qpsd), usually allows one to identify the \qp configurations of the lowest bands.
 The analysis of $^{163}$Er in Ref.\cite{Er163} is a good example.

 \begin{figure}
  \begin{center}
 \vspace*{-4cm}
 \includegraphics[width=\linewidth]{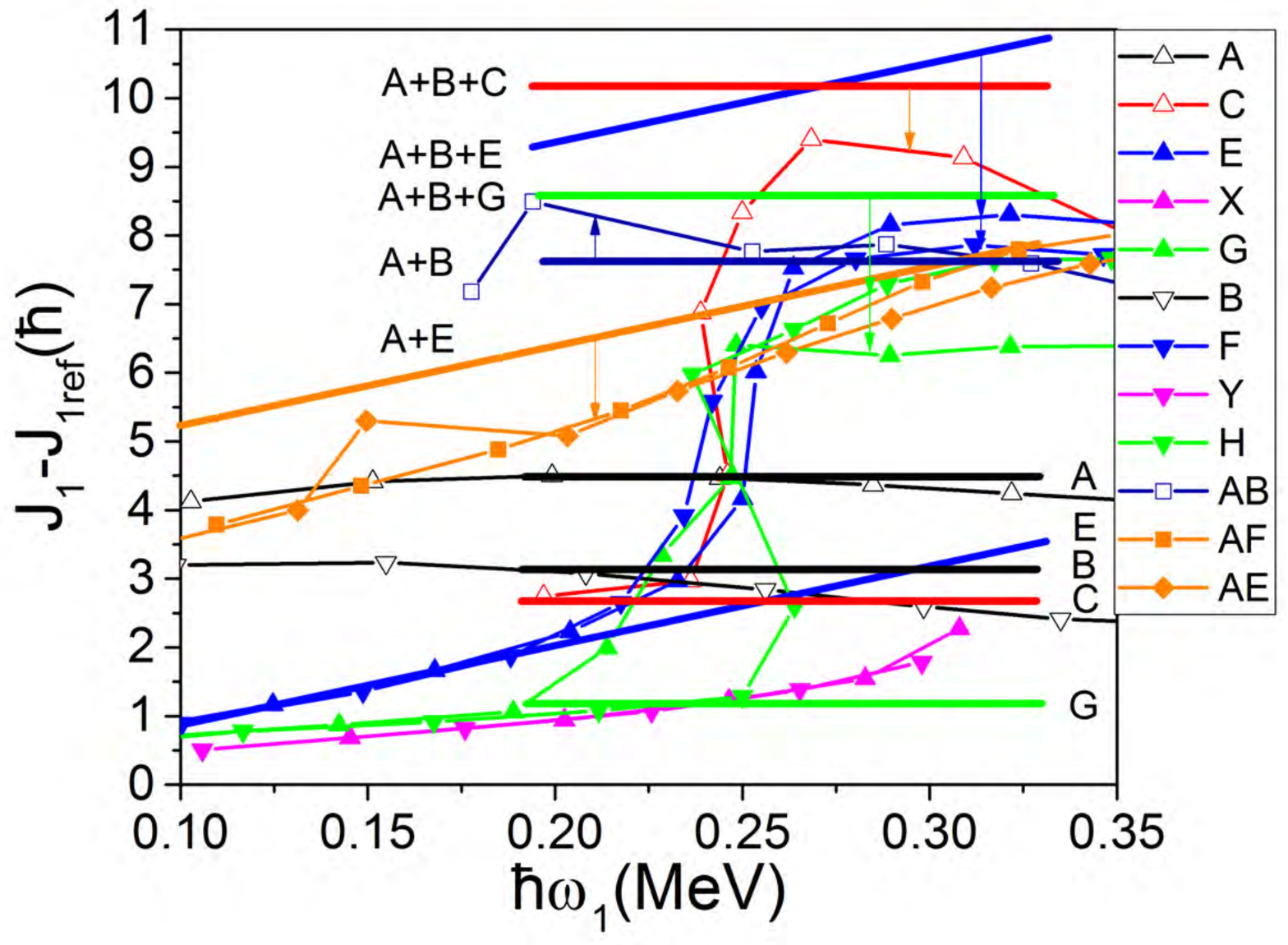} 
 \vspace*{-4cm}
  \caption{\label{f:Erjom}  Experimental \am projection on the 1-axis for  $^{163,164}$Er relative to the g-band reference $J_{1ref}$. 
  Thin lines with symbols display the data from Data from \cite{Er163,ENSDF}, where symbol convention is as in Fig. \ref{f:Er164Elow}.   
  The thick lines show some average of the aligned \am in case of one-\qn \confs and their sums in case of the multi-\qn \confs. 
  The arrows associate the  sums with the respective data. Only \qn configurations are shown
  for clarity.   
  }
  \end{center}
 \end{figure}
 
   \begin{figure}
  \begin{center}
 \includegraphics[width=0.94\linewidth,angle=-0.1]{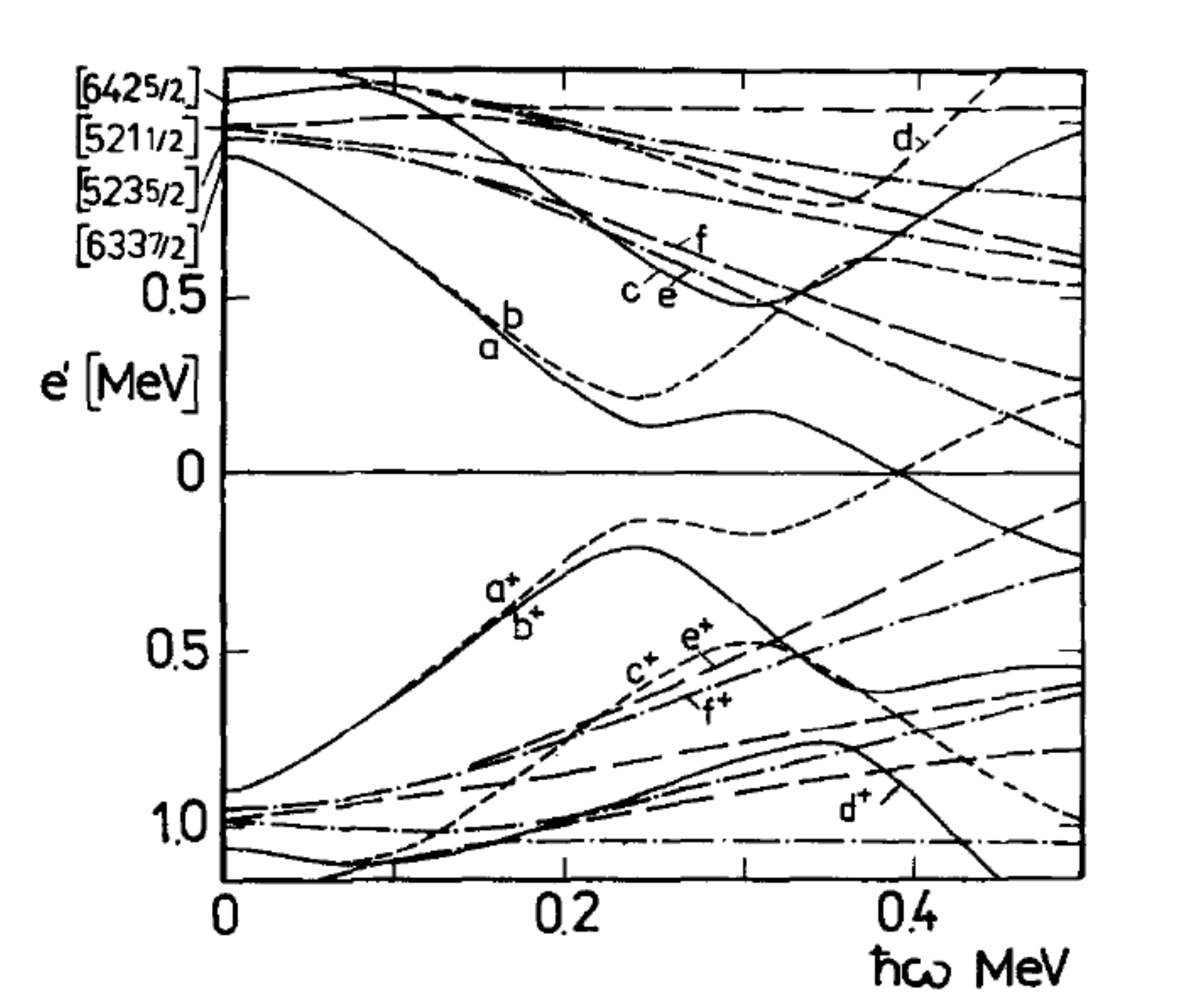} 
  \caption{\label{f:spagN98} Quasineutron routhian trajectories for $N = 98$. 
  The figure shows the trajectories as obtained by diagonalizing the \qp routhian  (\ref{eq:hpqp}) , labeled with lowercase letters, 
  where $+$ indicates the conjugate quasiparticle. Note, the frequency $\om$ is the frequency  $\om_1$ discussed in the text.
  The figure is constructed using the parameters $\beta= 0.267,~ \De = 0.79$ MeV.
 The line types indicate parity and signature $(\pi,\alpha)$:
  full $(+, +1/2 )$, short-dashed $(+, -1/2)$, dot-dashed $(-, 1/2)$, long-dashed $(-, -1/2)$.  Reproduced from \cite{BFM86},
  from which the band labels are adopted.    }
  \end{center}
 \end{figure}

   \begin{figure}
  \begin{center}
 \includegraphics[width=0.7\linewidth,angle=0]{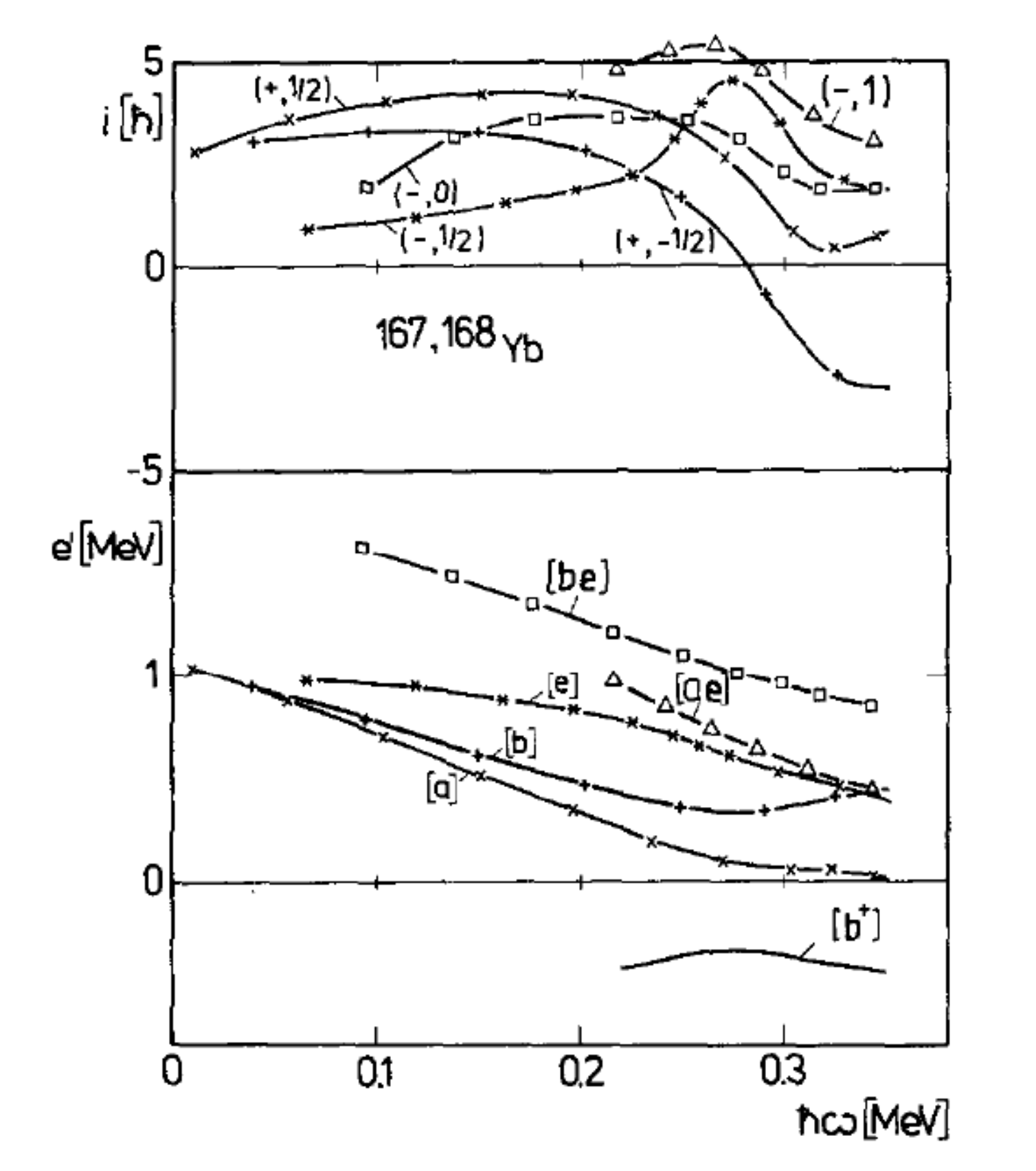} 
  \caption{\label{f:176Ybexp}  The relative angular momenta $i(\om_1)$ and routhians $e'(\om_1)$ of 
  the experimental rotational bands in $^{167}_{70}$Yb$_{97}$ and  $^{168}_{70}$Yb$_{98}$ calculated using 
  the experimental yrast band of $^{168}$Yb as reference. In the upper panel the bands are labeled by 
  parity and signature $(\pi,\alpha)$ in the lower panel by the letters used in Fig. \ref{f:spagN98}. The same symbols
  are used  for the bands in both panels. The conjugate routhian b$^+$ is added as $-e'_b(\om_1)$.  The figure is constructed 
  adding the experimental even-odd mass difference $\De_{eo}$ to all rotational energies of $^{167}$Yb.
  Note, the frequency $\om$ is the frequency  $\om_1$ discussed in the text. 
   From Ref. \cite{BFM86}.
   }
  \end{center}
 \end{figure}

 The quasicrossings between high-j routhians are not always as narrow as between a$^+$ and b$^+$
  in Fig. \ref{f:spagN96}. Examples are the neutron numbers  with a large value of $V$ in Fig. \ref{f:VABoscillation}, which indicates a smooth
  up bend of the \qp routhian a instead of a quasikink.
  Fig. \ref{f:spagN98} shows  the quasineutron routhians of $^{178}_{70}$Yb$_{98}$.  
   For such cases it may be more appropriate to refer to the original quasiparticle routhians instead of
  the diabetic ones constructed from them. 
The  (perturbed) yrast  band in 
the $N=98$ system is assigned to the configuration [0] (all $e'_i(\om_1) <0$ occupied),  and [a] is the lowest 
$(+,1/2)$ band in the $N=97$ neighbor. The difference between their total routhians, which is equal to the quasiparticle routhian $e'_a(\om_1)$,
can  be directly  compared with the difference between the experimental total routhian of the $(+,1/2)$ band in the $N=97$ nucleus
and the total routhian of the yrast band of the $N=98$ neighbor. The same holds for the other one-quasineutron configurations.
The lower panel of Fig. \ref{f:176Ybexp} displays the routhian differences between $^{177}_{70}$Yb$_{97}$ and $^{178}_{70}$Yb$_{98}$, 
which very much look like the quasineutron diagram Fig. \ref{f:spagN98}. The upper panel shows the differences between the 
pertinent angular momenta $i(\om_1)$, which are 
 to  be compared with the negative derivatives of the quasineutron routhians in   Fig. \ref{f:spagN98}.
 Of course, the up bend of $e'_a(\om_1)$ in the lower panel of Fig. \ref{f:176Ybexp} is caused
 by the down bend of the total routhian  of  [0] yrast band  in the the $N=98$ system while 
 the total routhian of the \conf [a]  in the the $N=97$ system continues smoothly. This blocking feature is more transparently
 exposed when referring to the diabatic routhian trajectories and a Harris reference. 
 The  \am difference for $(-,1/2)$ band [e] shows a bump at $\hbar \om_1=0.28$ MeV, which is absent in the calculated slope of e.
 The reason is that the presence of the  quasiparticle e reduces the neutron pair correlations, which shifts the 
 down bends of quasineutron routhians a$^+$ and b$^+$ to lower frequency than in the [0] configuration. As a consequence,  such  they do not completely
 cancel in the difference, as it is the case for the Cranked Shell Model assuming the same  pair gap $\De$ for all \confsd.  
 Using diabetic routhians, the reduction of the pair correlations caused by the     
 presence of the   quasiparticle e shows up as a lowering of the [AB] \confd. 
 The interpretation of the rotational spectra  in terms of the Cranked Shell Model quasiparticle \confs
 without invoking the  diabetic  trajectories is advantageous when the pair correlations are weak.   
 The studies of  nuclei around $N=108$ in Refs. \cite{yb178,yb173}
 are examples for this kind of analysis.

  \subsection{ Cranked shell correction approach}\label{sec:SCPAC}
  
The selfconsistent cranking model based on the  pairing+quadrupole-quadrupole  model Hamltonian presented in 
section \ref{sec:SCCM} has limitations when applied to realistic nuclei.
The equilibrium shape turns out to be very sensitive to the coupling constant $\chi$, such that local adjustments in various mass regions 
 are necessary for reproducing the
experimental information on the nuclear shape. Large changes of deformation within  one nucleus, like the appearance of superdeformation at
high spin, are also problematic because the  pairing+quadrupole-quadrupole  interaction does not conserve volume.  
The cranked 
shell correction method is a very efficient method to generate a map of the total routhian $E'(\beta,\ga,\om)$
or the total energy $E(\beta,\ga,J)$, the minima of which represent the different equilibrium shapes of the rotating nucleus.

Neergard {\it et al.} \cite{Neergard76} and Andersson {\it et al.} \cite{Andersson76} generalized the shell correction method  to the rotating mean field. The shell correction
 is calculated from the single particle routhians and it is added to the routhian of the rotating liquid drop, that is Eq. 
 (\ref{eq:Esc}) is replaced by
 \beq
 E'_{SC}(\om_1)=\sum\limits_{i\leq N}e'_i(\om_1)-\sum\limits_{i\leq N}\tilde e'_i(\om_1)
  \eeq
 and Eq.(\ref{eq:SC}) by
 \beq
 E'(def,\om_1)=E_{LD}(def)-\frac{\om_1^2}{2}{\cal J}_{rig}(def)+E'_{SC}(def,\om_1),  
  \eeq
  where ${\cal J}_{rig}(def)$ is the classical moment of inertia of rigid rotation \cite{Neergard76}
  \setcounter{footnote}{0} \footnote{Andersson {\it et al.} \cite{Andersson76}  renormalize the rotational energy at a given \am $J$.}.
  The shape of the rotating nucleus is found by minimizing the total routhian $ E'(def,\om_1)$ with respect to the deformation parameters at a given value $\om_1$.  
  The prediction of the existence of superdeformed nuclei at high spin in these papers, which were found a decade later \cite{sd} was a tremendous success of the theory. 
  As an example, Fig. \ref{f:SDTRS} left shows the prediction of superdeformed bands in $^{150}$Gd in Ref. \cite{Neergard76}, 
  which were identified by Fallon {\it et al.} \cite{Fallon89}. In their contribution to this Focus Issue, 
  M. A. Riley, J. Simpson and E. S. Paul discuss the phenomenon of super deformation and the history of its discovery 
  in more detail.

 The "total routhian surface" (TRS) represents 
$E'(\beta,\ga,\om)$ of the lowest  \conf for a given
combination of parity and signature $(\pi,\al)$ calculated by means of the cranked shell correction method.
 It is useful for a
global survey of the shapes expected near the yrast line.
TRS calculations 
 have become quite popular for the interpretation of high-spin data.
They are easily
accessible as a public domain computer code \cite{trs}, which is based on the Woods-Saxon potential
and includes pairing in a parametrized form.  As an example, Fig. \ref{f:SDTRS} right shows the TRS 
for the configuration [$\pi$h$_{11/2},~\nu$h$_{11/2}$] in $^{134}$Pr.

 \begin{figure}
  \begin{center}
 \includegraphics[width=0.48\linewidth]{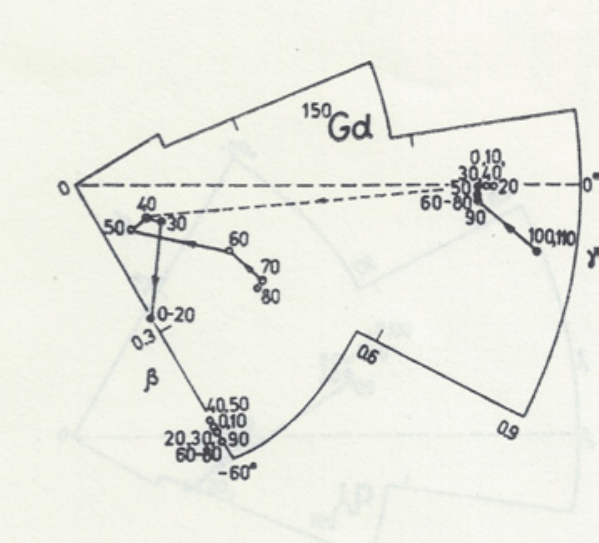} 
  \includegraphics[width=0.48\linewidth]{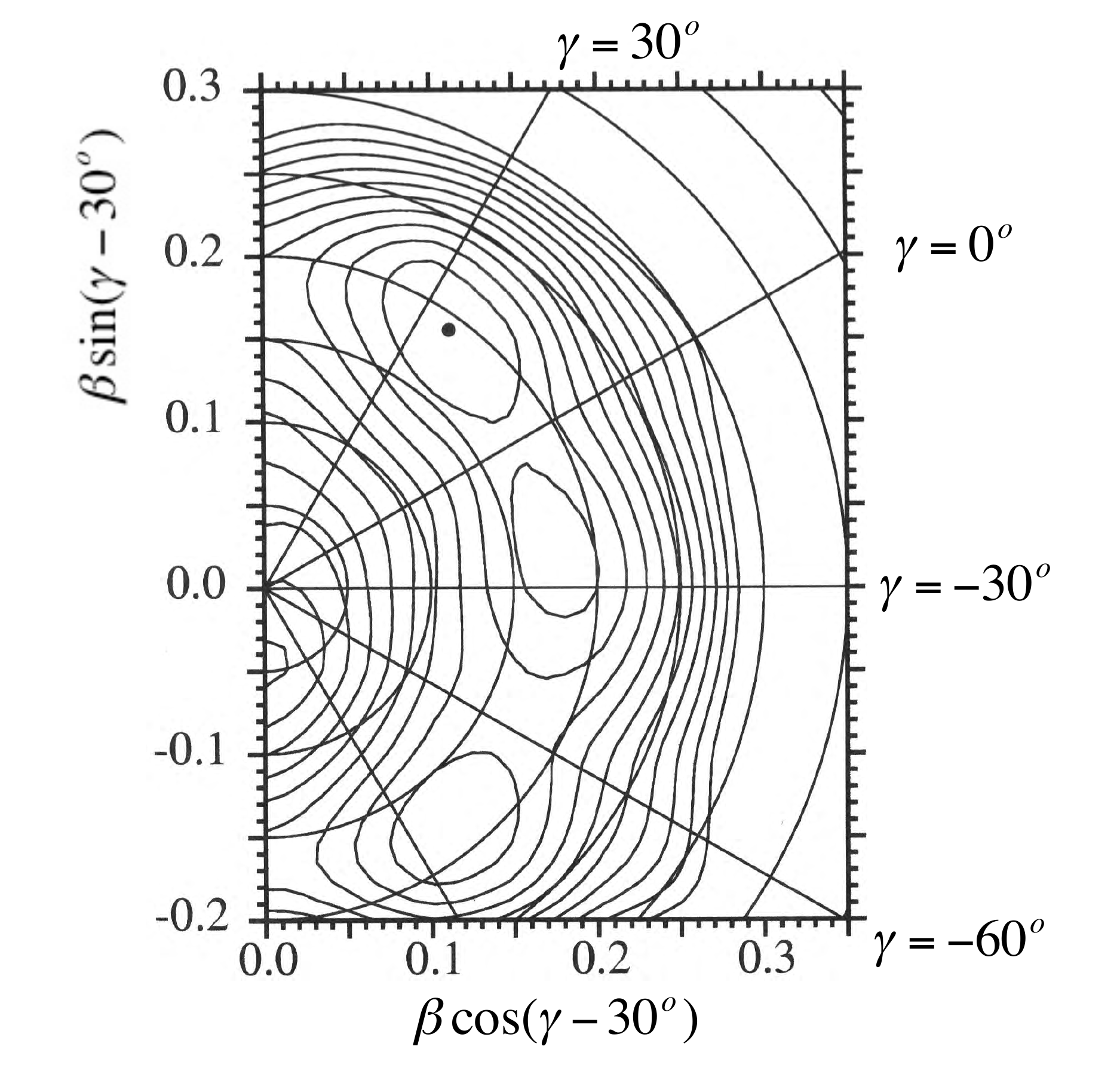}  
  \caption{\label{f:SDTRS} Left: Minima of the total routhian surface for $^{150}$Gd calculated by means of the cranked shell correction approach
  based on the Nilsson potential without pairing. Reproduced from  \cite{Neergard76}. 
    Right:  Total routhian surface for $^{134}$Pr calculated by means of the TRS code \cite{trs} based on the Woods-Saxon potential
     including pairing. The configuration 
  is one h$_{11/2}$ \qpr and one h$_{11/2}$ \qnd. The rotational frequency is $\hbar \om$=0.35 MeV. The contour lines are spaced by 0.2 MeV. 
   }
  \end{center}
 \end{figure}

Calculating both $E'(\beta,\ga,\om)$ and
$J(\beta,\ga,\om)$, the energy $E(\beta,\ga,J)$ can be generated  by interpolation.
Extended  collections of such total energy  surfaces based on the Woods-Saxon potential and parametrized pairing
have been published in Ref. \cite{pes}. For more detailed structure analysis one
 constructs diabatic quasiparticle configurations by various interpolation schemes, which will not be detailed here.  
 A popular  method of this type is the Cranked Nilsson-Strutinsky (CNS) approach reviewed by Afanasjev {\it et al.} \cite{tbrev}, 
 It represents an efficient tool for exploring the high-spin structure of nuclei, because it does not include pair correlations. 
 A public domain computer code, the Ultimate Cranker,  is based on the same techniques but includes pairing. It is available online \cite{uc}.
 In their contribution to this Focus Issue, P. M. Walker and F. R. Xu \cite{NCWalker} present a further approach along these lines.
 The shell correction approach applied to a Woods-Saxon potential is combined with a particle number concerving
 treatment of the pair correlations, which allows them to study the quenching of pairing 
 and the development of the nuclear shape  subsequent excitation of \qpsd.   They advanced the approach by 
 taking projection onto good \am into account, which removes problems of the rotating mean field  approximation 
 at band crossings (section \ref{sec:crdiabatic}) and the transition between different symmetries (section \ref{sec:ChangeSym}).

It has to be stressed here that in addition to the discussed 
cranked shell correction method the various  versions of the self consistent  cranking model which are based on energy density functionals 
have been  very successfully applied to study the structure of rapidly rotating nuclei.
This area beyond the Unified Model will not be covered in my contribution, which focuses on the aspects of symmetries and coherence of the rotating mean field.
The contributions to this Focus Issue by L. Egido \cite{NCEgido} and P. G. Reinhard \cite{NCReinhard} and the review articles by M. Bender {\it et al.}
\cite{Bender03} and D. Vretenar {\it et al.} \cite{RMFreview} are good starting points for exploring these developments.

\subsection{Rotation about a tilted axis}\label{sec:TAC}
\subsubsection{The spinning clockwork picture}\label{sec:clockwork}
Stable uniform rotation appears for the  axis with the largest  \momid.
As discussed in Sect. \ref{sec:UM}, the Unified Model assumes that the nucleus rotates like an irrotational quantum liquid, which  
  implies the following. Rotation about a symmetry axis is impossible. 
   The shape dependence of the \momis is given 
  by Eq. (\ref{eq:momiIF}), which assigns  the largest \momi to  the medium axis of the   triaxial nucleus.
  It needs to be emphasized  that this is in contrast to a rigid ellipsoidal body,  for which 
  the maximal \momi belongs to the  short axis. The irrotational-liquid behavior 
  emerges because the nucleus is composed of two kinds of indistinguishable 
  fermions. Accordingly,  the ground state band  of an even-even triaxial nucleus corresponds to
rotation about the medium axis with the maximal moment of inertia, which becomes the one of the axis perpendicular to 
the symmetry axis in the special case of axial symmetry. 

This traditional view misses essential aspects of nuclear rotation which the study of the rotating mean field  has revealed.
 Unlike the particles in a macroscopic liquid the nucleons have a mean free path that is larger than the size of the nucleus. 
 As a consequence their motion is restricted by quantization. The quantization of the \am of the nucleonic orbits  determines  the 
 way nuclei rotate. In order to illustrate the essential consequences it is useful  
viewing the nucleus as a clockwork of small gyroscopes, which carry a fixed \amd.
 The essence of the analogy is the approximately constant   angular momentum of the high-j nucleonic orbitals, which is fixed by quantization.
 \setcounter{footnote}{0}
 \footnote{We consider nuclei with moderate deformation. For superdeformed nuclei the constancy of \am caused by quantization becomes eroded.}
The gyroscopes represent the nucleonic orbitals, which are coupled  to the deformed potential by their non-isotropic density distribution.
 The left panel of Fig. \ref{f:qpcouplings}  depicts a gyroscope representing a high-j orbital. (The density distribution of low-j orbitals 
 is less torodial but still deformed, except s-orbitals.) 
 As discussed in section \ref{sec:orbits} and in more  detail in the \ref{sec:semi}, the orientation   of the gyroscopes with respect to the
 density distribution results from the balance between the inertial forces and the torque exerted by the
 deformed potential. The resulting arrangement in the clockwork generates nucleon density distribution which 
  generates the average field in a selfconsistent way.  The interplay is governed by the principle that the short-range attractive 
 interaction favors large overlap of the densities of the orbitals and the Pauli Principle.

   \begin{figure}
  \begin{center}
 \includegraphics[width=\linewidth]{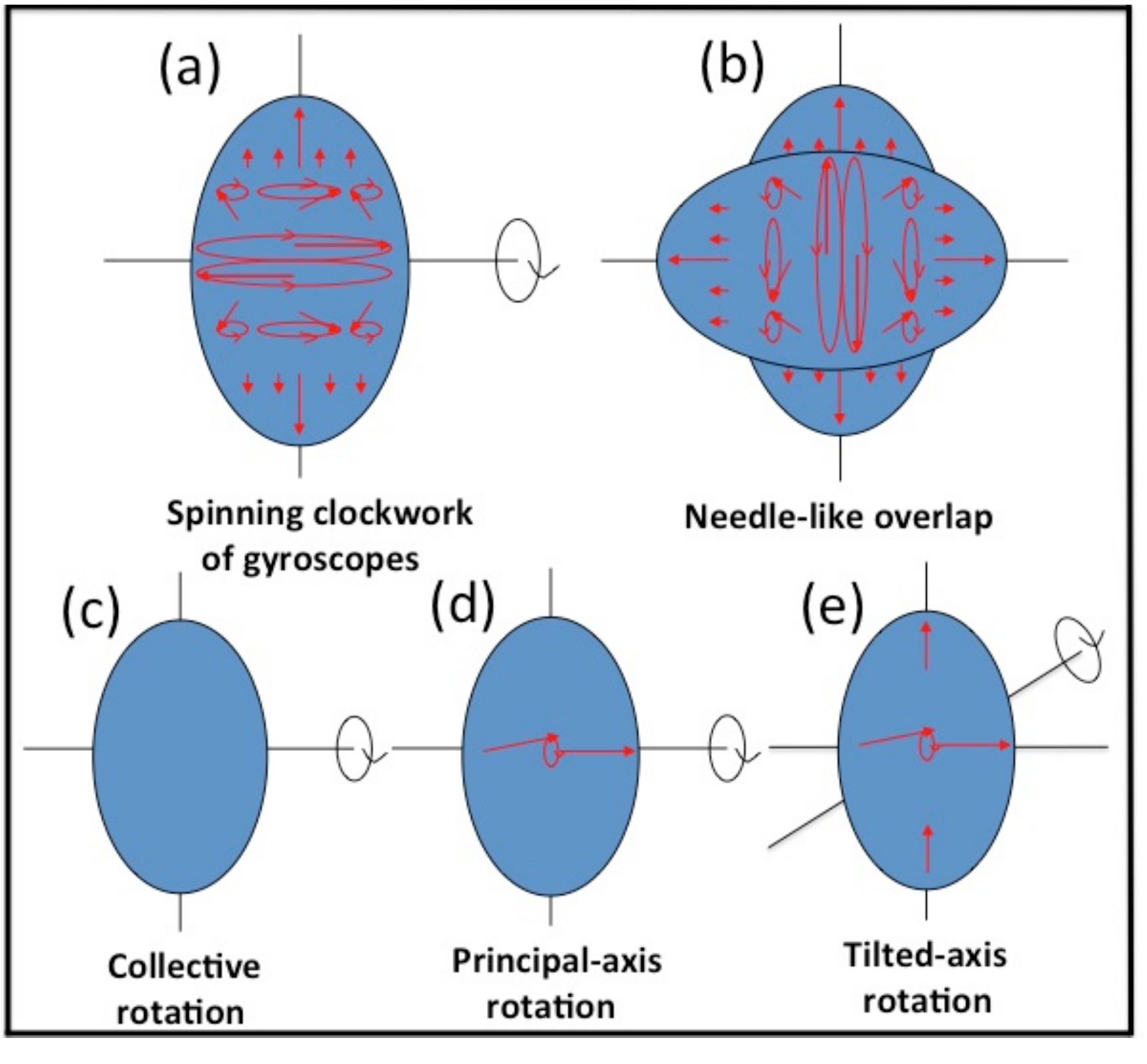} 
  \caption{\label{f:ClockCRPRTR} The nuclear clockwork of gyroscopes. The figure shows in a schematic way the
  orbits of the \am vectors of the occupied orbitals. The vectors are distributed over the volume for graphical reasons. They should be 
  thought originating from the center.  The lower row shows only the rearrangement compared to (a), that is (c) is 
  simplified (a).
   }
  \end{center}
 \end{figure}

 Fig. \ref{f:ClockCRPRTR} (a) illustrates the arrangement of the gyroscopes for the ground state of an even-even axial nucleus. 
 The  figure depicts the precessional cones of the \am vectors of the gyroscopes.
 \footnote{ The vectors are distributed over the volume for graphical reasons. They should be 
  thought originating from the center. }
   The orbitals  arrange  back-to-back, which gives good overlap between their density distributions
  and is also favored by the pair correlations. For a partially filled shell this results in a prolate density distribution  and a prolate potential. 
 (For the ground state of a triaxial  nucleus the  arrangement of the gyroscopes consists of one set of back-to-back orbital pairs aligned with the long axis and and another set 
  of pairs aligned with the short axis, which generates a triaxial potential (see \ref{sec:semi})). 
  
  Turning the clockwork around the symmetry axis does not change the system. Fig. \ref{f:ClockCRPRTR} (b) shows that turning the clockwork
  around the short axis makes a large difference which mainly comes from the reorientation of the many \am vectors. The displacement of the surface
  bump of the density distribution is much less important, keeping in mind that the axis ratio is largely exaggerated in the figure.
  This observation is at variance with the Unified Model,
  which considers the deformation of the nuclear surface as the origin of the rotational degrees of freedom.  
    In fact, the overlap of the product of the quantal wave functions of the orbitals in
  Fig. \ref{f:ClockCRPRTR} (b) falls off rapidly with the rotation angle  (see section \ref{sec:coherence}). As a consequence, one may generate 
  a collective wave functions with a large number of nodes by superposing the different orientation of the needles, which means an 
  extended  rotational band.

  The \am of the ground band is generated by gradually aligning the gyroscopes with the rotational axis. 
  The needed energy  determines  the \momid.         
  Section  \ref{sec:coherence} will discuss the emergence of the rotational degree of freedom in a more quantitative way.   
  Generating the  angular momentum of a rotational band by
gradually aligning two long  vectors composed of nucleonic angular momenta
was first discussed by Danos and Gillet \cite{stretch}.  Fig.  \ref{f:ClockCRPRTR} (a) illustrates their "stretch" scheme.
 One of the  two vectors is thought to be constructed
by stretched coupling of the
valence nucleons  shown in the upper half  and the other vector by stretched coupling the valence nucleons
shown in the lower half. In the ground state the two vectors have opposite direction. They gradually align along the band
increasing the \amd.  The capability of the stretch scheme 
to account for the energetics of  realistic rotational bands  was never demonstrated for the 
ground state bands of even-even nuclei, for which it was conceived. 
However in a modified version, it  very well describes Magnetic Rotation, which will be discussed in section \ref{sec:MR}. 

 From Fig.  \ref{f:ClockCRPRTR} (b) it is clear that a rotation by 180$^\circ$ about the short axis brings the clockwork back to its original state, which implies that 
  the ground band contain only even spins. As a consequence the ground state band contains only even $I$.

If the rotational motion is slow enough, the alignment of the individual orbitals with the rotational axis is small and proportional to the rotational frequency $\om$,
and the total \am is proportional to $\om$ as well.
The rotational motion is characterized by  three principal moments
of inertia ${\cal J}_\mu$ and  the well known linear relation
between the angular momentum $\vec R$  and the angular velocity $\vec \om$,   
\beq\label{eq:jomclass} 
R_\mu={\cal J}_\mu \om_\mu.
\eeq 
Only uniform rotation about  principal axes is possible. 

The rotational response of clockwork (a) is the one of the collective rotor of the Unified Model. To simplify the following discussion 
In the lower row of Fig. \ref{f:ClockCRPRTR} the many orbitals that generate the collective \am are left away in order to simplify the  
discussion, i. e. (c) represents (a).
 Part (d) illustrates case when one or more quasiparticles align with the short  axis, which is the rotational axis. 
 Turning   the arrangement of the gyroscopes  by  180$^\circ$ about the principal short axis leaves it invariant.
 Therefore the signature remains a good quantum number  that labels the $\De I=2$ rotational band. 
 More generally, one of the principal axes of a triaxial nucleus remains  the rotational axis when one or more \qps align with it. The
 arrangement appears as a $\De I=2$ band corresponding to the signature of the arrangement.   
  
\subsubsection{Appearance of tilted rotation}\label{sec:AppTAC}

  Fig.  \ref{f:ClockCRPRTR} (e) illustrates a case when
  the gyroscopes  are asymmetrically oriented, 
 such that their angular momenta add up to a finite 
component $J_3$ and along the  symmetry axis and a perpendicular component  $i_1$,
which with the \am of remaining nucleons ${\cal J}_1 \om_1$  adds to $J_1=i_1+{\cal J}_1 \om_1$. The
axis of uniform rotation will be tilted away from the principal axes. 
 Uniform rotation appears when the  rotational axis $\vec \om$ has
 the direction of   total \am $\vec J$, which requires that 
 $\tan\vth=J_3/\left(i_1+{\cal J}_1 \om \sin\vth \right)$.  This condition 
 can be met by the appropriate value of $\vth$.
 
 The example demonstrates that a nucleus may uniformly rotate about an axis 
 that differs from one of the principal axes of its mass density distribution. 
 The tilted rotation of nuclei is due to 
the quantized angular momentum of the nucleonic orbitals, which 
carry a fixed amount of \am large enough to tilt the rotational axis.
Such a mode does not appear
 for a classical rotating rigid body, which obeys the relations  (\ref{eq:jomclass}) well known from
 textbooks on  mechanics. The reason is that its constituents do not carry \am of their own, because
 they are considered as point masse, not as gyroscopes.
  A rotating drop of ideal liquid obeys Eqs. (\ref{eq:jomclass}) as well if the flow is irrotational (see Ref. \cite{BMII} section 6A-5).
   However, a classical liquid may also  uniformly rotate about
an axis different from the principal axes of its density distribution. This possibility
 was already pointed out by Riemann \cite{Riemann} for an ellipsoidal self-gravitating fluid.
 The origin of the tilt of the rotational axis are deviations from irrotational flow 
 which generate vorticity. Thus, the tilt of the  rotational axis is due to  an intrinsic
vorticity of the system. This applies to nuclei as well, because the high-j orbitals appear
as pronounced vortices in the microscopic current pattern of a rotating nucleus (see discussion in Ref. \cite{Frauendorf05}).

\subsubsection{Tilted axis cranking  solutions of axial nuclei}\label{sec:axialTAC}
The \am geometry has to be planar.
As discussed in section \ref{sec:planarTAC}, planar tilted axis cranking  solutions show up 
as $\De I =1$ bands, i.e. the two signatures are degenerate, merged  into one sequence. 
  The  $E2$ transitions between states with $I$ and
 $I \pm 1$ are strong, because the 
 charge distribution is asymmetric with respect to the tilted 
axis of rotation. 
  The stretched $M1$-transitions are  enhanced,
because the transverse magnetic moments of the different quasiparticles 
add up. The $B(M1)$ values need
not be large in all cases, because the transverse magnetic moments of the
contributing quasiparticles can have different signs.
For principal axis cranking  solutions,   both types of transitions connect two \confs of
opposite signature, which  are non-degenerate in general. The transition
can only be of the order of single-particle transitions, because they    
connect  different quasiparticle configurations. Hamamoto and Sagawa \cite{HaSa79} studied the 
electromagnetic transition matrix elements between different Cranked Shell Model  \confsd.

Let us continue to use  $^{163}_{66}$Er$_{97}$ as an illustrative example. 
Fig. \ref{f:spagthEr163} shows the \qp levels as functions of the tilt angle
$\vth$. Such  diagrams permit a first guess of the  
equilibrium  angle  by adding the \qp routhians
$e'_i(\om, \vth)$ to the vacuum routhian 
and looking
for the minimum of the sum with respect to $\vth$.
 The vacuum behaves like a collective rotor (see section \ref{sec:clockwork}, Eq. (\ref{eq:jomclass}). If
 we consider an axial nucleus, it has only the 
component $R_1 \approx \om {\cal J} \sin \vth $ perpendicular 
to the symmetry axis and its routhian is 
 $E'_0 \approx -(\om \sin \vth)^2 {\cal J}/2$. 
 
 The quasineutron levels E and F emanating from the Nilsson state
$[523]5/2^-$ are strongly coupled to the
deformed potential. They have $j_3\approx \pm 5/2$ and $j_1 \approx 0$.
This is reflected by the $\vth$ dependence, which is close to 
$\mp 5/2~\om \cos\vth$.  Occupying E
results in an equilibrium angle $\vth < 90^o$. 
As expected for a tilted axis cranking  solution, the band [E]
appears as a $\De I =1$ sequence in Fig. \ref{f:er163e}. 
Further examples of strong coupling to the
deformed potential are 
quasineutron level $[505]11/2^-$ and  
the quasiproton levels $[404]7/2^+$ and $[523]7/2^-$, 
which are all seen as $\De I=1$ bands.

  \begin{figure}[t]
  \begin{center}
 \includegraphics[width=\linewidth]{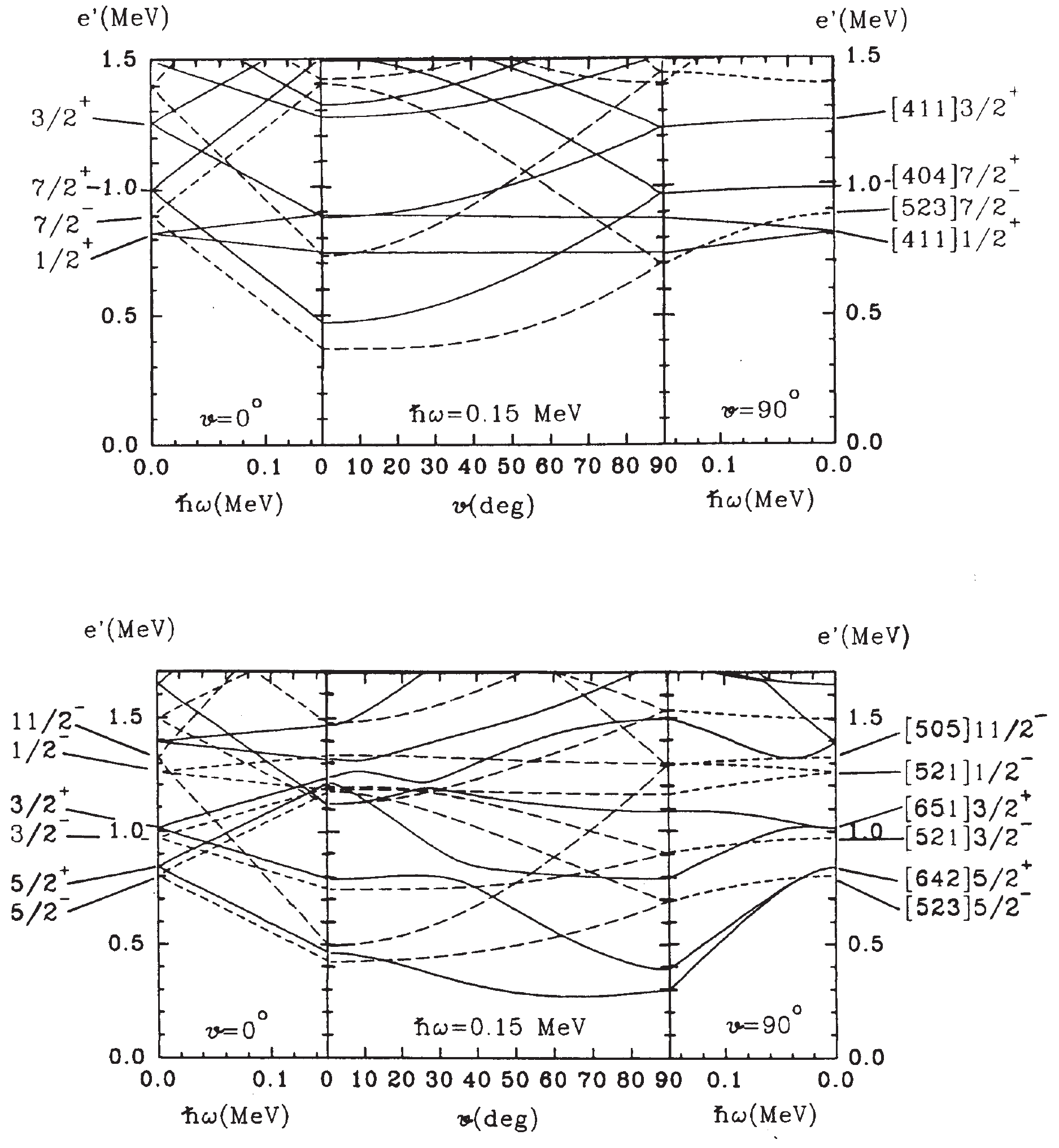} 
  \caption{\label{f:spagthEr163}  Quasiproton routhians for $Z\approx 66$ (upper  panel) and quasineutron routhians 
for $N\approx 97$ (lower  panel) 
 as functions of the tilt angle $\vth$ at the frequency
$\om=0.15$ MeV$/\hbar$ (middle parts). The variation with
$\om$ is shown in the left-hand parts for $\vth=0^o$ and 
in the right-hand parts for $\vth=90^o$. 
Full lines: positive parity. Dashed-dotted lines: negative parity.
 The parameters are $\eps=0.252,~\ga=0^o,
~\eps_4=-0.004, \De_p=0.87~\mathrm{MeV},~\De_n=0.80~\mathrm{MeV}$. From Ref.
\cite{Brockstedt94}.     }
  \end{center}
 \end{figure}

\begin{figure}[t]
  \begin{center}
 \includegraphics[width=\linewidth]{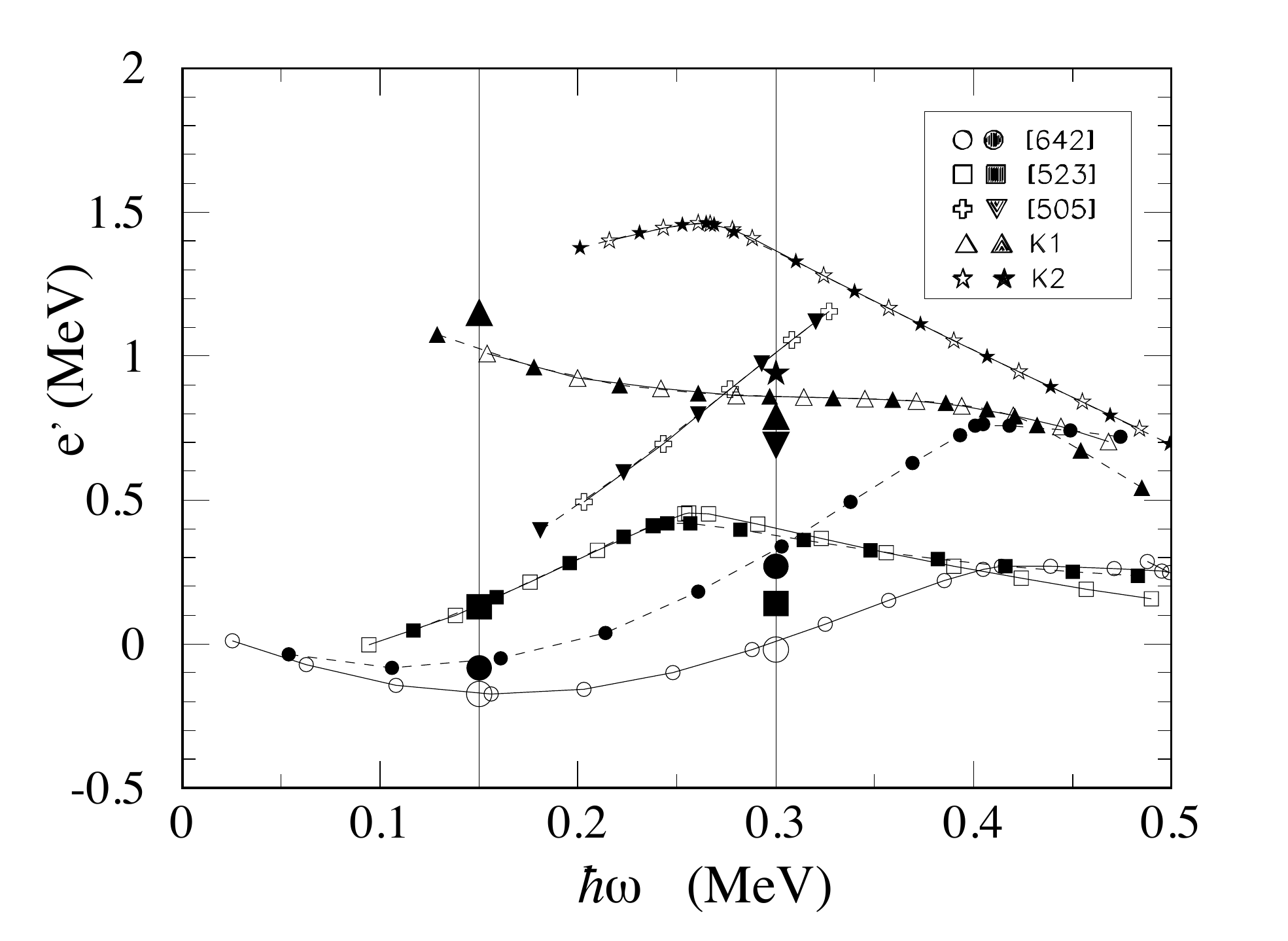} 
  \caption{\label{f:Er163TAC} \label{f:er163e}
Experimental  routhians  of the rotational bands in   $^{163}$Er.
The bands  are labeled by the Nilsson quantum numbers.
In Fig. \ref{f:Er163spec}, [642] is labeled by A and B and
[523] by E and F.
  The tilted axis cranking  routhians are shown as the large symbols on the two vertical lines
where the calculations have been carried out. The same symbols are used for 
the experiment and calculations.  A rigid rotor
reference routhian $-\om^2\times 31.7 \hbar^2\mathrm{MeV}^{-1}$ is subtracted.
From Ref. \cite{berk94} after Ref. \cite{Brockstedt94}.  
 }
  \end{center}
 \end{figure} 
 \begin{figure}[t]
  \begin{center}
 \includegraphics[width=0.9\linewidth]{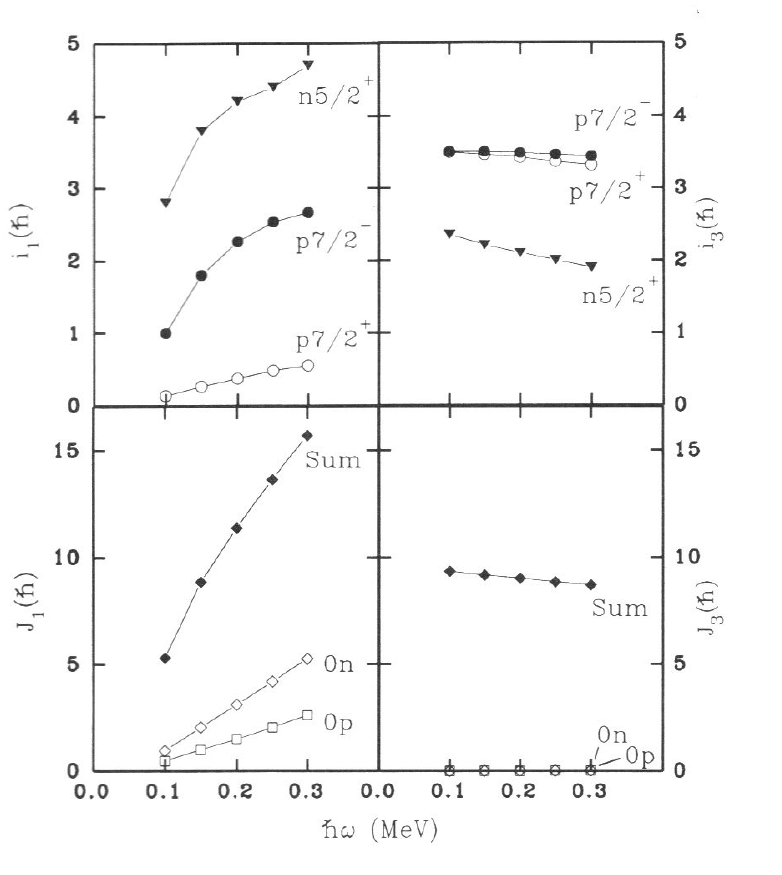} 
  \caption{\label{f:Er163TACJ1J3} 
 The angular momentum composition of the three-quasiparticle configuration K1 in $^{163}$Er
 calculated in the tilted axis cranking  approach.
  The \am expectation values for the short 1-axis and the long 3-axis are displayed.
 The total \am $J_{1,3}$ is the sum of the \qp vacuum parts denoted by $[p0]$ and $[n0]$ 
 and the \qp contributions $i_{1,3}$, where the short hand notation is used:  $p7/2^+ =\pi[404]7/2^+,~
 p7/2^-=\pi[523]7/2^-,~n5/2^+=\nu[642]5/2^+$.
From  Ref. \cite{Brockstedt94}.  }
   \end{center}
 \end{figure} 
 \begin{figure}[t]
  \begin{center}
 \includegraphics[width=0.9\linewidth]{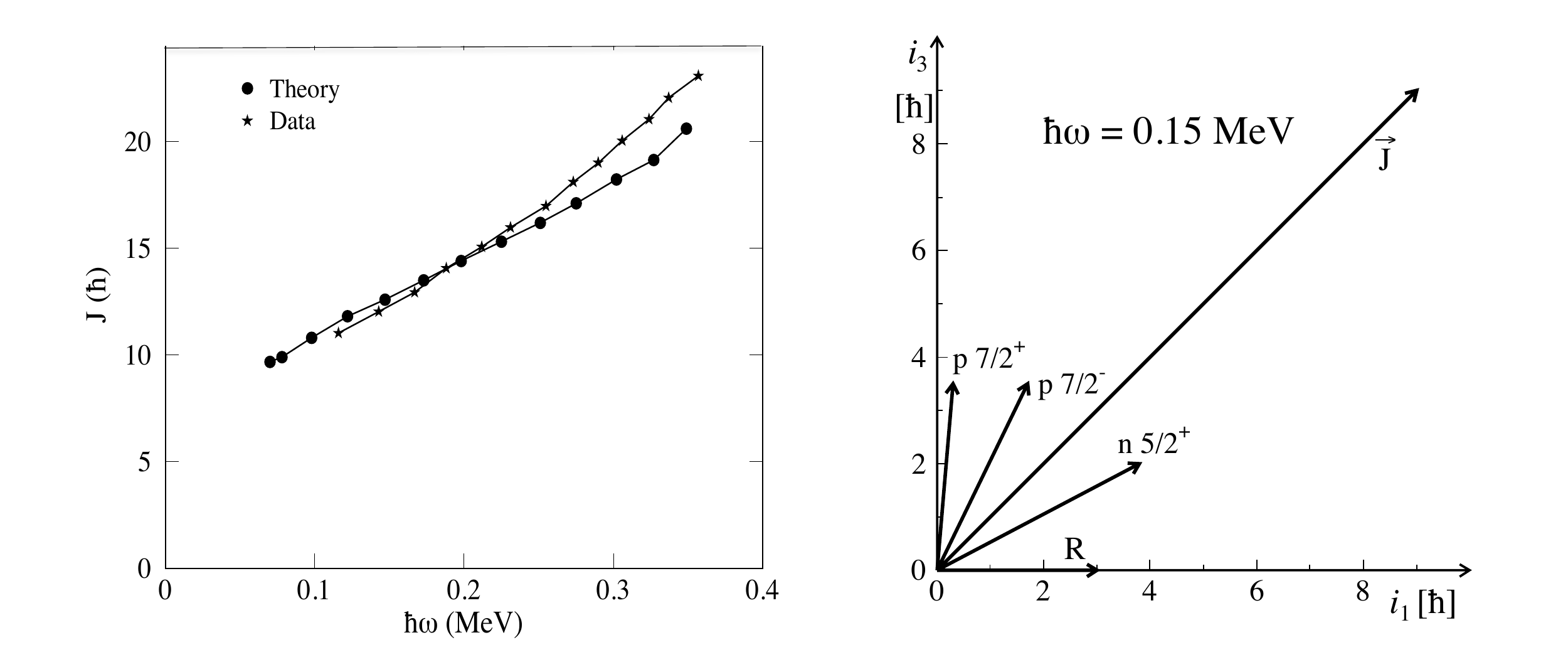}
  \caption{\label{f:Er163TACJarrows}
Left panel: Comparison of the total \am  calculated by means of tilted axis cranking 
with experimental values of the three-quasiparticle configuration K1 in $^{163}$Er.
Right panel: Vector diagram of
the angular momentum composition at $\om=0.15~\mathrm{MeV}/\hbar$ obtained by 
the calculations shown in the left panel and Fig. \ref{f:Er163TACJ1J3}.  
 The collective angular momentum of the \qp vacuum is denoted by $R$. Taken from  Ref. \cite{Brockstedt94}.  }
   \end{center}
 \end{figure} 
 \begin{figure}[t]
  \begin{center}
\includegraphics[width=0.6\linewidth]{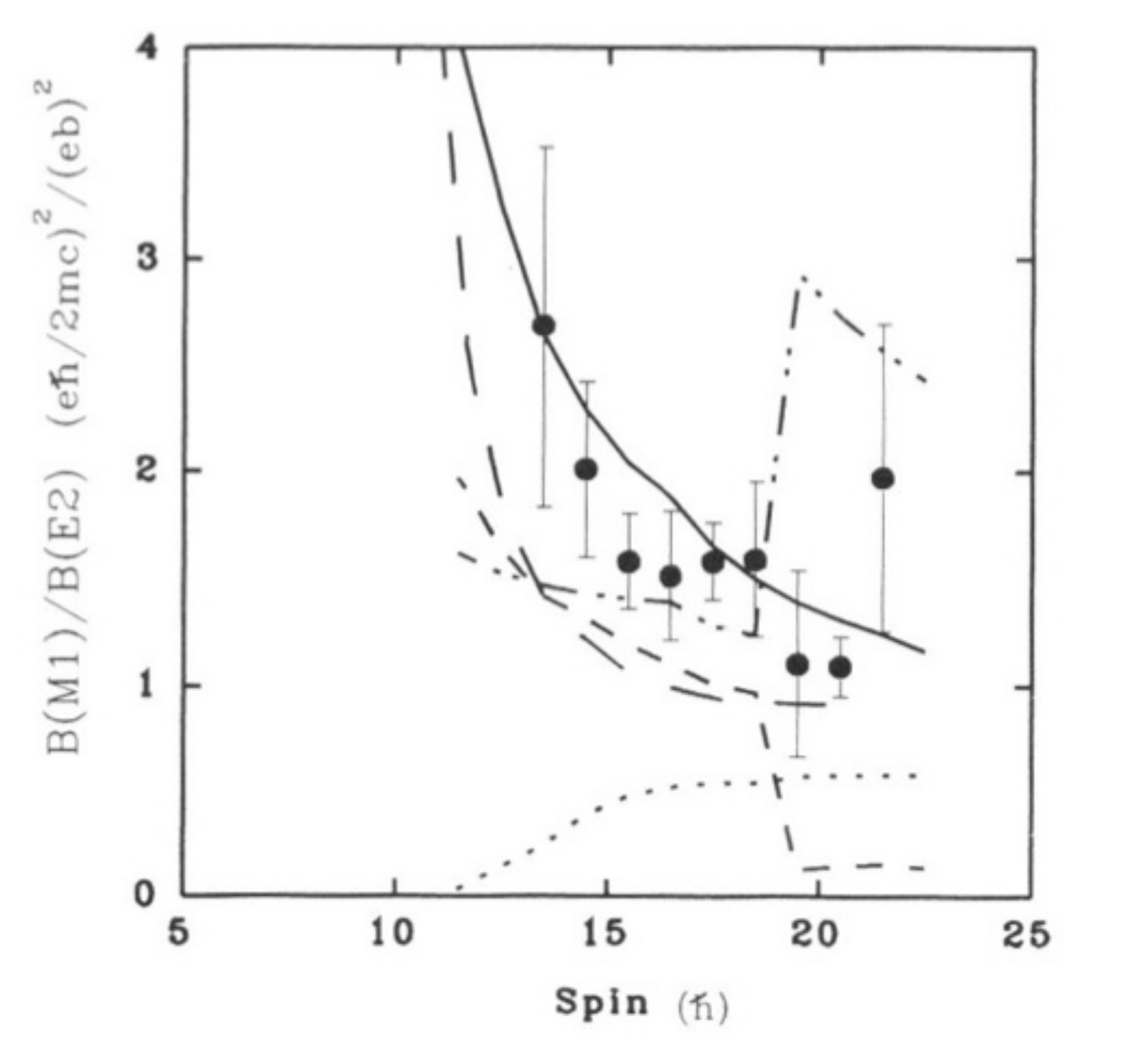}   
  \caption{\label{f:Er163BM1BE2} 
  Experimental ratios of \mbox{$B(M1,I\rightarrow I-1)/B(E2,\rightarrow I-2)$} for the K1 band in $^{163}$Er. They are compared with the 
  tilted axis cranking  calculations  (long dash) and the D\"onau-Frauendorf vector model Eqs. (\ref{eq:DF}).
    Different \confs are  shown:
  $[p7/2^+p7/2^-n5/2^+]$ (full), $[n11/2^-n3/2^-n5/2^-]$ (medium dash),  $[n11/2^-n3/2^-n5/2^+]$ (dot),
    $[p7/2^+p7/2^-n5/2^-]$. The short hand notation is the same as in Fig. \ref{f:spagthEr163} and 
    $n11/2^-=\nu[505]11/2^-,~n3/2^-=\nu[521]3/2^-,~n5/2^-=\nu[523]5/2^-$. The experimental $i_1(\om_1)$
    is derived from the $\alpha=+1/2$ branch of $n5/2^+$. The transition matrix element $Q_t=7eb$ (defined by footnote to Eq. (\ref{eq:Qt})) is used in 
    the calculation of the $B(E2)$ values.
    Taken from Ref. \cite{Brockstedt94}.}
     \end{center}
 \end{figure}

 The \qn routhians A and B emanating from $[642]5/2^+$   show a more complex
behavior. The lower routhian  A  has a very shallow minimum at $\vth \approx 60^o$. 
Fig. \ref{f:MinGamTh} shows the location of  the minimum of the  lowest \qp state 
with respect to $\vth$ and $\ga$ for different shell filling. 
For $\vth=60^\circ$ and $\ga=0^\circ$  the shell filling parameter is close to -0.2, which
is about the location of the neutron chemical potential $\la_n$ in the i$_{13/2}$ shell.   
The lowest \qn is only weakly coupled to the deformed potential, because its 
quadrupole moment is strongly reduced by the pair correlations.
(It is half particle and half hole.)   For this reason,  the minimum is shallow.
 The upper routhian B has its minimum at $\vth=90^o$, 
because it changes (through a quasicrossing with C) into the $\Omega=-5/2$ state at $\vth=0^\circ$. 
Combined with the vacuum routhian, both \confs 
  [A] and [B] are principal axis cranking  solutions with $\vth=90^o$, which have been
discussed in section \ref{sec:CSM}. 
 
The  pair of parallel 
trajectories emanating from the Nilsson state  $[521]1/2^-$ is an example for a 
pseudospin doublet.
Their routhians  do not change with $\vth$ and their distance  equals 
$\hbar\om$. They behave in this way, because their angular momentum  
is approximately equal to the pseudo 
spin, which is decoupled from the deformed field. It takes 
  the direction of $\vec \om$ or opposite to it, corresponding to a 
routhian $e'(\vec \om)\approx e(\om=0)\pm\hbar \om /2$.
  The concept of pseudo spin and the reason for the decoupling are  explained in Ref.  \cite{RMP}. 
  The vacuum routhian keeps
$\vec J$ parallel to the 1-axis and the solution is of the principal axis cranking  type. 
 The levels of the $I+1=-1/2+2n$ sequence have the same energy as the levels 
 with $I=1/2+2n$. The ground state of $^{171}$Yb is $[521]1/2^-$.  The rotational band consist of 
 the expected doublets. 
 The  routhians of the \qpr level $[411]1/2^+$    
do not change with $\vth$ too, because they combine 
 to pseudo spin singlet state.
Doublet $\De I=2$ sequences that correspond to the
signatures  $\al=\pm 1/2$  of the orbital $[411]1/2^+$ 
are observed in the odd-$Z$ Tb isotopes. 
 
 The band K1 in Fig. \ref{f:Er163spec} is assigned to the three-\qp  configuration \mbox{$[\pi
7/2^+,\pi 7/2^-,\nu 5/2^+]$}, where each of the \qps occupies the 
lower of the two branches emanating at $\vth=0^\circ, 90^\circ$ and $\om=0$ in Fig. \ref{f:spagthEr163}.
The   composition of the total angular momentum is displayed in Fig.  \ref{f:Er163TACJ1J3} and 
the right panel of Fig. \ref{f:Er163TACJarrows}. 
The  tilt angle $\vth (\om)$  
grows as a consequence of
the increasing collective angular momentum $R$, but remains
below $90^\circ$. Accordingly, band K1 is  observed as a $\De I=1$ sequence.
The left panel of Fig. \ref{f:Er163TACJarrows}  compares the experimental function $J(\om)$ 
of band K1 with the tilted axis cranking  calculation. Fig. \ref{f:Er163TAC} includes 
the routhians calculated by means of tilted axis cranking
 for two values of the frequency $\om$.  
The agreement between theory and experiment is typical for 
the other tilted axis cranking  calculations in well deformed nuclei.     
 Tilted axis cranking calculations account for the transition intra band $M1$ and $E2$
 rates as well.  As an example, Fig. \ref{f:Er163BM1BE2} shows the 
 ratios $B(M1)/B(E2)$ for the K1 band.  The calculation well follows the experiment.
It is somewhat smaller than the experiment, which can be  
traced back to general inaccuracies of 
the mean field  $g$-factor values of the involved quasi-particles. 

The example $^{163}$Er is typical for the accuracy of tilted axis cranking  calculation in reproducing the experiment. 
Further phenomena that have been studied  in the framework of the tilted axis cranking  approach can be found in Ref. \cite{RMP}, as the interplay
of high-K and low-K bands in the mass 180 region or the decay of high-K isomers.  P. M. Walker and F. R. Xu \cite{NCWalker} 
extensively discuss the high-K isomers in their contribution to this Focus Issue, 

\subsubsection{Approximate tilted axis cranking  solutions - relation to principal axis cranking}\label{sec:approxTAC}

 In the  strong coupling limit, tilted axis cranking approximates   the Unified Model semiclassically. 
The total angular momentum is composed only of the collective part $J_1=\hbar R$ and $J_3=\hbar K$,
which is constant. The energy
\beq\label{eq:Escl}
E=\frac{R^2}{2{\cal J}}=\frac{I(I+1)-K^2}{2{\cal J}}
\eeq
is the same as in the Unified Model. 
The tilt angle is given by
\beq\label{eq:scl}
\cos \vth=\frac{K}{I+1/2}.
\eeq 
For this value of $\vth$, the tilted axis cranking  expressions 
(\ref{eq:be22}-\ref{eq:bm1}) agree with 
the transitions probabilities in the 
Unified Model when the Clebsh-Gordan coefficients are approximated  by
their high-spin asymptotic values. 
 This approximation is very accurate except near $I=K$. 
 
The Cranked Shell Model  uses a less stringent approximation. 
The assumption $J_3=\hbar K$ is kept but 
$J_1$ is calculated by means of  principal axis cranking  at 
the frequency is $\om_1=\om \sin \vth$, where
$\vth$ given by (\ref{eq:scl}).  It
amounts to neglecting the term $-\om \hat J_3 \cos \vth $ in
the mean field routhian.
If $J_3$ is not too large, the band starts at a relatively low frequency
 and  $\vth$ changes rapidly from $0^o$ to $90^o$.  
 The neglected term is not too important if the coupling of
the  quasi-particle orbital 
to the deformed potential is much stronger than
$-\om  \hat J_3\cos \vth$.  Then it is a good
approximation to take only  $-\om \hat J_1\sin \vth $
into account. This is not the case
for   weakly coupled orbitals or small deformation. 
Of course, the Cranked Shell Model approximation
becomes better and better when $\vth$ approaches $90^o$. 

D\"onau and Frauendorf  \cite{vector,vectors} worked out   the transition
probabilities for the Cranked Shell Model approximation. They are given by  Eqs.
(\ref{eq:bm1}) -  (\ref{eq:delta}) 
using  the strong coupling limit (\ref{eq:scl}) for the tilt angle $\vth$..
The intrinsic components of the magnetic moments
of the  excited quasi-particles $i$  
are calculated  by means of the relations         
\bea\label{eq:DF}
\mu_\nu=\sum\limits_i g_{i} j_{\nu,i},~~
\mu^c_\nu=g_R(J_\nu-\sum\limits_i j_{\nu,i}),~~\nu=1,~3,
\eea
where $\mu^c_1$ denotes the collective magnetic moment of the  \qp vacuum.  
The global estimate $g_R\approx Z/A$ is usually good enough.
The gyromagnetic ratios $g_{i}$  
are either taken from the experiment or calculated from the mean field
solutions at $\om_1=0$. The components $j_{3,i}$ are set equal to
the angular momentum projection $\hbar\Omega_i$ at $\om_1=0$. The components $j_{1,i}$ are either
calculated as $\langle j_1\rangle$ by means of the Cranked Shell Model or extracted as the  aligned
angular momenta $i_i$ from the differences
between the experimental functions $J_1(\om_1)$ with and 
without the quasi-particle $i$ present.
This calculation scheme for transition probabilities is  referred to as the "semiclassical vector
model" or "D\"onau-Frauendorf model".
 The possibility of extracting the aligned angular momenta and the 
g-factors from experiment substantially improves the accuracy of the 
calculation.

 Fig. \ref{f:Er163TACJ1J3}  shows that  $j_{i,3}=\hbar\Omega_i$ is a good approximation
 for the well deformed nucleus $^{163}$Er.   
Fig. \ref{f:Er163BM1BE2}  includes calculations of the \mbox{$B(M1,I\rightarrow I-1)/B(E2,\rightarrow I-2)$} 
ratios which exploit the flexibility of the vector model.
The \qp alignments $i$ are calculated from taking the  differences between  appropriate
\qp \confs in $^{163}$Er and $^{163}$Tm. 
The calculations use  experimental gyromagnetic ratios $g_{K,i}$,
 which  are average values derived from $B(M1)$ values of the pertinent one-\qp bands for the rare earth region. 
 Since information of the same kind from related \confs is used, it is 
 not surprising that the agreement with the experiment is better than for the microscopic tilted axis cranking  calculation.

The Cranked Shell Model is quite commonly used because of its simplicity. 
The spectrum of rotational bands can be constructed  by occupying the 
\qp levels in a single \qp diagram like Fig.  \ref{f:spagN96}. This allows 
a quick semi-quantitative classification of the multiple band spectra
provided by  $\ga$ spectroscopy.  A certain price of this
simplicity is the fact that high-K \confsd, which gain much energy by the tilt, appear as relatively
highly excited \qp configurations in a Cranked Shell Model  \qp routhian, whereas they are close to the yrast line and
strongly  populated in experiment.

The  Cranked Shell Model can be seen as an approximation to  tilted axis cranking, which generates the best possible mean field solution.  
The Cranked Shell Model amounts to  extrapolating  the tilted axis cranking  \qp routhians
from the values calculated at $\vth=90^o$ by 
means of the expression
\beq\label{eq:pactac}
e'_i(\om,\vth)=e'_i(\om,90^o)-\om\left(j_{1,i}(\sin\vth-1)+ \hbar\Omega_i\cos \vth\right).
\eeq 
Fig.  \ref{f:spagthEr163} demonstrates that this is a quite decent approximation for a
number of quasi-particles. But it does not fully account  
 for the complex behavior of the $i_{13/2}$ \qns as functions of $\vth$.
 Within the approximation (\ref{eq:pactac}), the Cranked Shell Model transformation (\ref{eq:om1EpCSM})
 of the experimental energies $E(I)$ to experimental routhians $E'(\om_1)$
  accommodates the energy gain due to tilt.
    The possible \confs for high-K bands can be singled out of the low-K \confs
  by the criterion that  they must contain signature-degenerate   \qp routhians.   

In their contribution to this Focus Issue, P. M. Walker and F. R. Xu \cite{NCWalker} discuss the phenomenon of high-K
isomerism. Their theory combines  cranking about the principal 1-axis with the shell-correction method and a particle number conserving treatment
of pairing to account for the attenuation of the pair correlations by \qp excitations. As discussed above for the Cranked Shell Model,  
 the \am component along the symmetry axis is approximated by fixing it to $J_3=\hbar K$, which is a good approximation for 
 well deformed nuclei and large values of $K$.       
The fixed-$K$ approximation becomes bad for the weakly deformed nuclei, because the
responses to $-\om j_1$ and $-\om j_3$ are comparable. It is also problematic for
triaxial nuclei because the principal axis cranking  calculations provide only the 1-component of the \amd. An estimate 
of the $J_3$ component becomes rough if not impossible.  

\subsubsection{Change of symmetry}\label{sec:ChangeSym}

The discussed  symmetry types are reflected by the multiplicity of the bands. 
If the rotational axis coincides with a principal axis one observes a  $\Delta I=2$ sequence.
If the rotational axis lies in  a  principal plane  one observes  a  $\Delta I=1$ sequence.
 If the rotational axis points out of a principal plane   one observes  two degenerate $\Delta I=1$ sequences. 
With increasing rotational frequency, the direction of the rotational axis may change from one to another symmetry type, which implies  
changing the interpretation of the mean field solution.  In an infinite system sudden symmetry changes appear as singularities at a phase transition.
In a finite system, as the nucleus, the change is gradual. When changing from  lower to higher symmetry the band with the higher multiplicity splits into 
two bands of lower multiplicity, which are associated with different mean field configurations. Vice versa,   when changing from  higher to lower symmetry 
two bands of lower multiplicity associated with different mean field configurations merge into one band of higher multiplicity. Both cases are observed.
In the following we  discuss symmetry changes by analyzing energies and transition probabilities obtained by diagonalizing  the quasiparticle triaxial rotor Hamiltonian (\ref{eq:HQR}).
Since the  quasiparticle triaxial rotor model conserves \amd,  it describes the smooth transition from one to the other symmetry type. In this respect we consider 
the quasiparticle triaxial rotor model results  as "exact" and compare them with the  cranking approximation to the same model case.

For now we consider the case of axial nuclei, we will return to the symmetry changes in section \ref{sec:TriaxTAC}}.
 A typical scenario is that some quasiparticles generate angular momentum along the 3-axis ($\vth=0^\circ$) and that there is   
 collective \am $\vec R$ pointing along the 1-axis ($\vth=90^\circ$), which results in a tilt angle  $0^\circ<\vth<90^\circ$.
   As $R$ increases  along the band the tilt angle $\vth$ increases, which 
 for certain  \qp \confs  may approach 90$^\circ$, where  the ${\cal R}_z(\pi)$ is restored. Correspondingly, the $\Delta I=1$ sequence
 changes into two $\Delta I=2$ sequences. 
\begin{figure}[t]
\begin{center}
\includegraphics[width=0.4\linewidth]{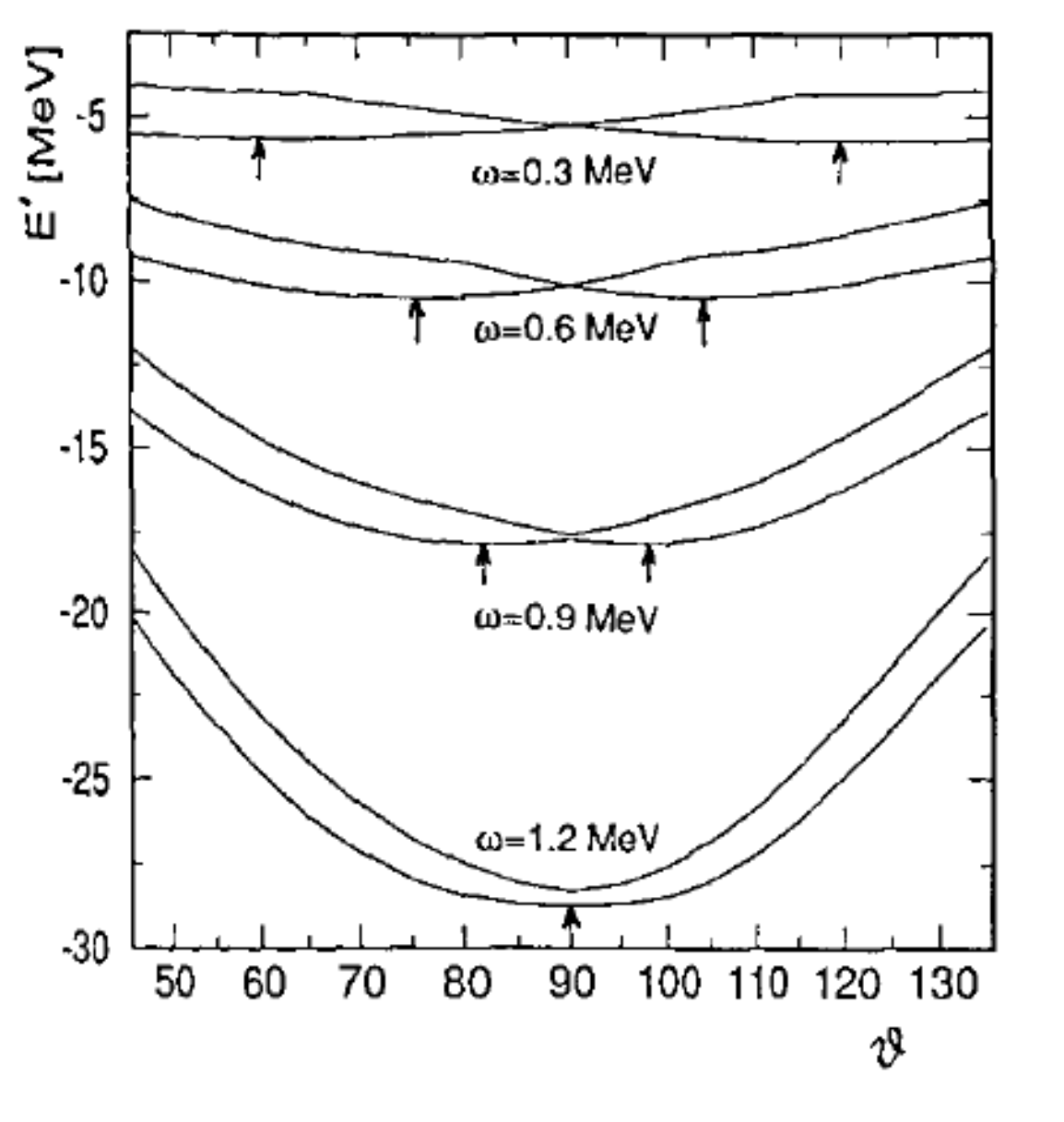} 
\includegraphics[width=0.58\linewidth,angle=0.6]{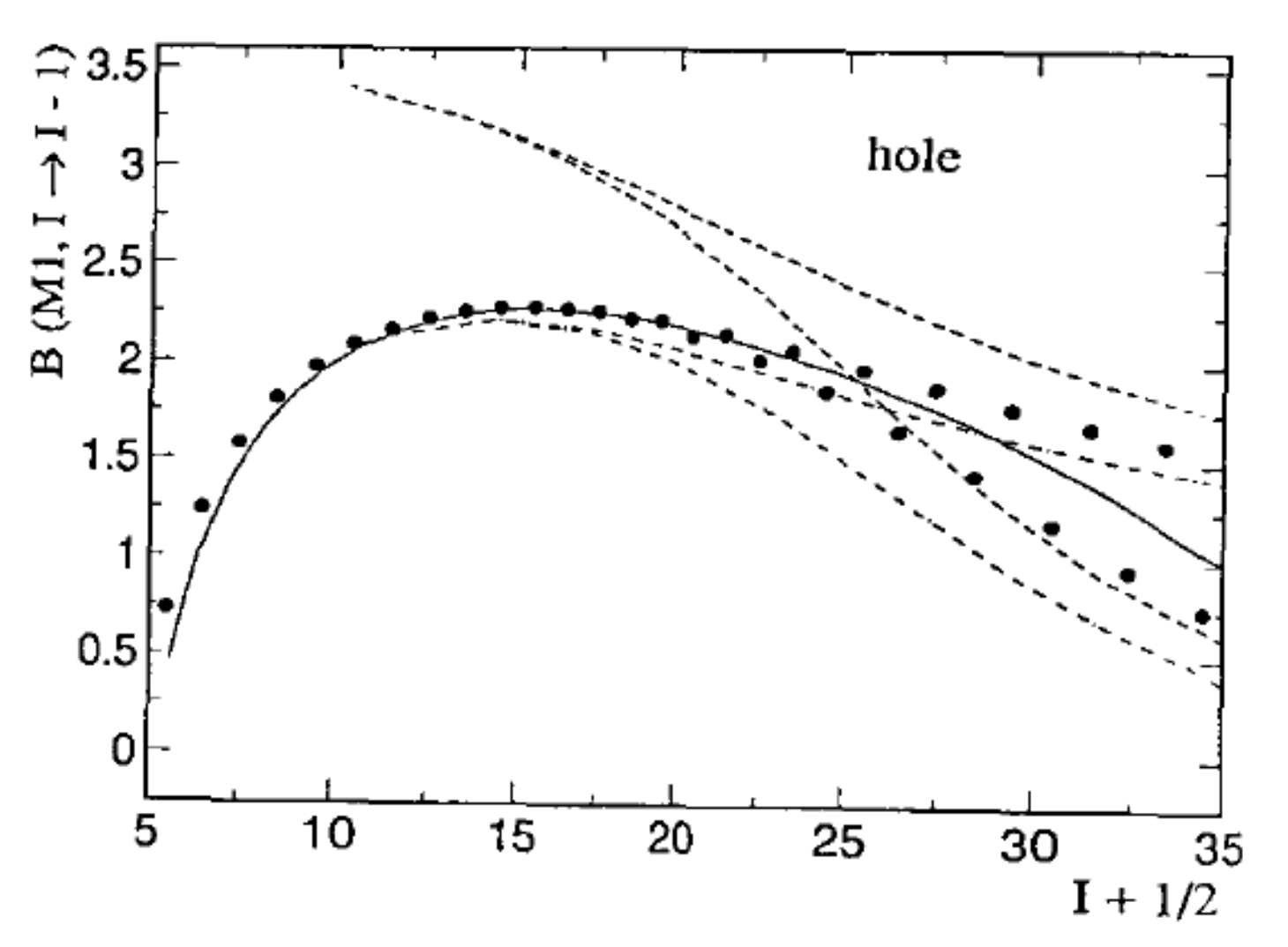} 
\caption{\label{f:SymmetryChange} Left panel: Total routhian of an h$_{11/2}$ hole coupled to an axial rotor as function 
of the tilt angle $\vth$ for several rotational frequencies $\hbar\om$ ($\hbar=1$). The arrows indicate the location of the minima.
 The configuration changes at the kinks in the upper curves.
Right panel: Reduced transition probabilities  $B(M1, I\rightarrow I-1)$ of an h$_{11/2}$ hole coupled to an axial rotor as a function 
of the spin $I$. The unit is $(g_{qp}-g_R)\mu^2_N$. The dots are the results of the exact results obtained by diagonalizing the hole-rotor model.
The full drawn curve shows the tilted axis cranking values. The lower pair of dashed curves show the values obtained by means of the 
D\"onau-Frauendorf vector model  Eqs. (\ref{eq:DF}). The upper pair of dashed curves shows the values obtained 
by means of Eq. (\ref{eq:bm1out}). 
 From Ref. \protect\cite{prtac}.}
\end{center}
\end{figure}

 Frauendorf and Meng \cite{prtac} investigated the transition region in the framework of the quasiparticle triaxial rotor model 
 presented in section \ref{sec:QTR}. 
     For comparison
 the total cranking routhian  is constructed for the model,
 \beq\label{eq:eppr}
 E'(\om,\vth)=e'(\om,\vth)-\om^2\sin ^2\vth \frac{{\cal J}_1}{2}.
 \eeq
The quasiparticle routhian  $e'(\om,\vth)$ is eigenvalue of the cranking Hamilonian (\ref{eq:hpbf}) restricted to an h$_{11/2}$
subshell. The rotor \momi ${\cal J}_1$ describes the routhian of all other nucleons. It is set equal to the \momis
${\cal J}_1={\cal J}_2$ used in the   
quasiparticle triaxial rotor (\ref{eq:HQR}) (${\cal J}_3=0$). The deformation $\beta=0.3$ chosen. 
From the examples studied in Ref. \cite{prtac}, we discuss the case of a pure   h$_{11/2}$ hole coupled to the axial rotor. 
The left panel of Fig. \ref{f:SymmetryChange} shows the total routhian   (\ref{eq:eppr}) as a function of the tilt angle $\vth$.
The high-j hole prefers $\vth=0$ because its \am aligns with the 3-axis. The collective term prefers $\vth=90^\circ$.
The equilibrium angle lies at the minimum of the total rothian, where the two tendencies are balanced. 
With increasing $\om$   the minimum moves toward $\vth=90^\circ$, which it reaches for $\hbar \om\approx 1$MeV.
Below this frequency the minimum $\vth<90^\circ$ is associated with the two degenerate  signatures of the $\Delta I=1$ band.
Above,  e. g. at   $\hbar \om= 1.2$ MeV, the lower of the two minima represents the $\De I=2$ band with $\al =-1/2$ and 
the upper minimum the   $\De I=2$ signature partner band $\al =1/2$. 

In order to avoid over counting the number of configurations that represent rotational bands, it is helpful to refer to a 
symmetric graph like the left panel of Fig. \ref{f:SymmetryChange}. Below  $\hbar\om = 1$MeV there are always two degenerate 
minima symmetric to 90$^\circ$, which is the manifestation of breaking the ${\cal R}_1(\pi)$ symmetry. These two
minima generate the $\De I=1$  band by combining their pertinent \confs  to even and odd superpositions, which are the two 
signatures comprising the band.  As seen in the figure, the two configurations represent the continuation of  
the \confs with the hole on the higher level (inside the kink), which are already taken into account.  Therefore they must be discarded
as configurations assigned to bands. For other one-\qp configurations 
the spurious ones can be eliminated by the same reasoning. Elimination becomes more involved for a larger number of
excited \qpsd, where rules are formulated in  Refs. \cite{tacdic} and \cite{prtac}.  

The tilted axis cranking solution very accurately approximates the energy of the exact hole-rotor solution up to $\hbar \om=0.7$ MeV, where the 
two signatures begin to separate. For $\hbar \om> 1.2$ MeV the two principal axis cranking solutions at $\vth=90^\circ$ very well reproduce the energy
of the two $\De I=2$ bands, in particular  the small signature splitting. In the transition region the tilted axis cranking energy  lies half-way
between the energies of the two signatures branches. 

\begin{figure}[h]
\begin{center}
\includegraphics[width=0.78\linewidth]{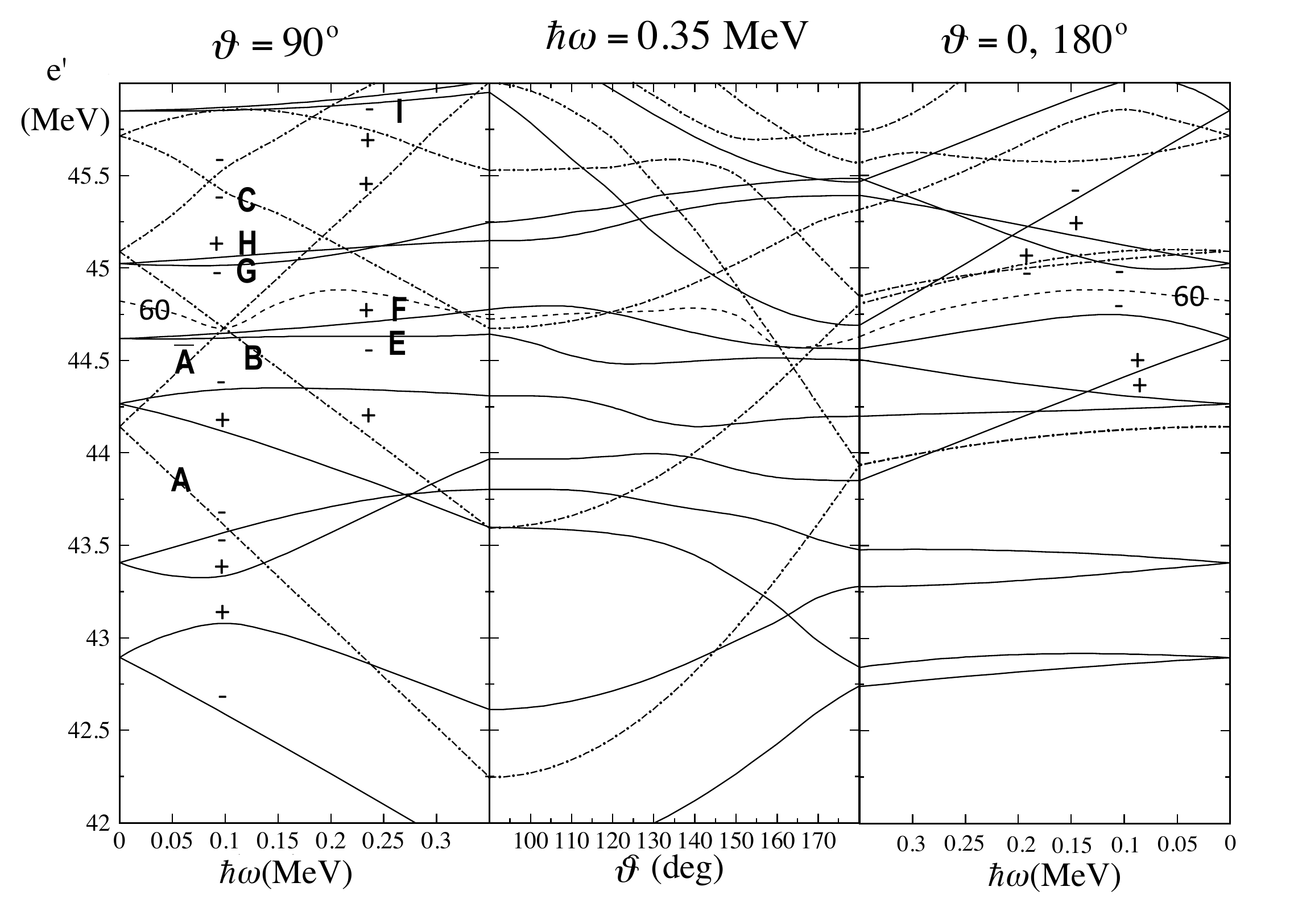} 
\includegraphics[width=0.78\linewidth]{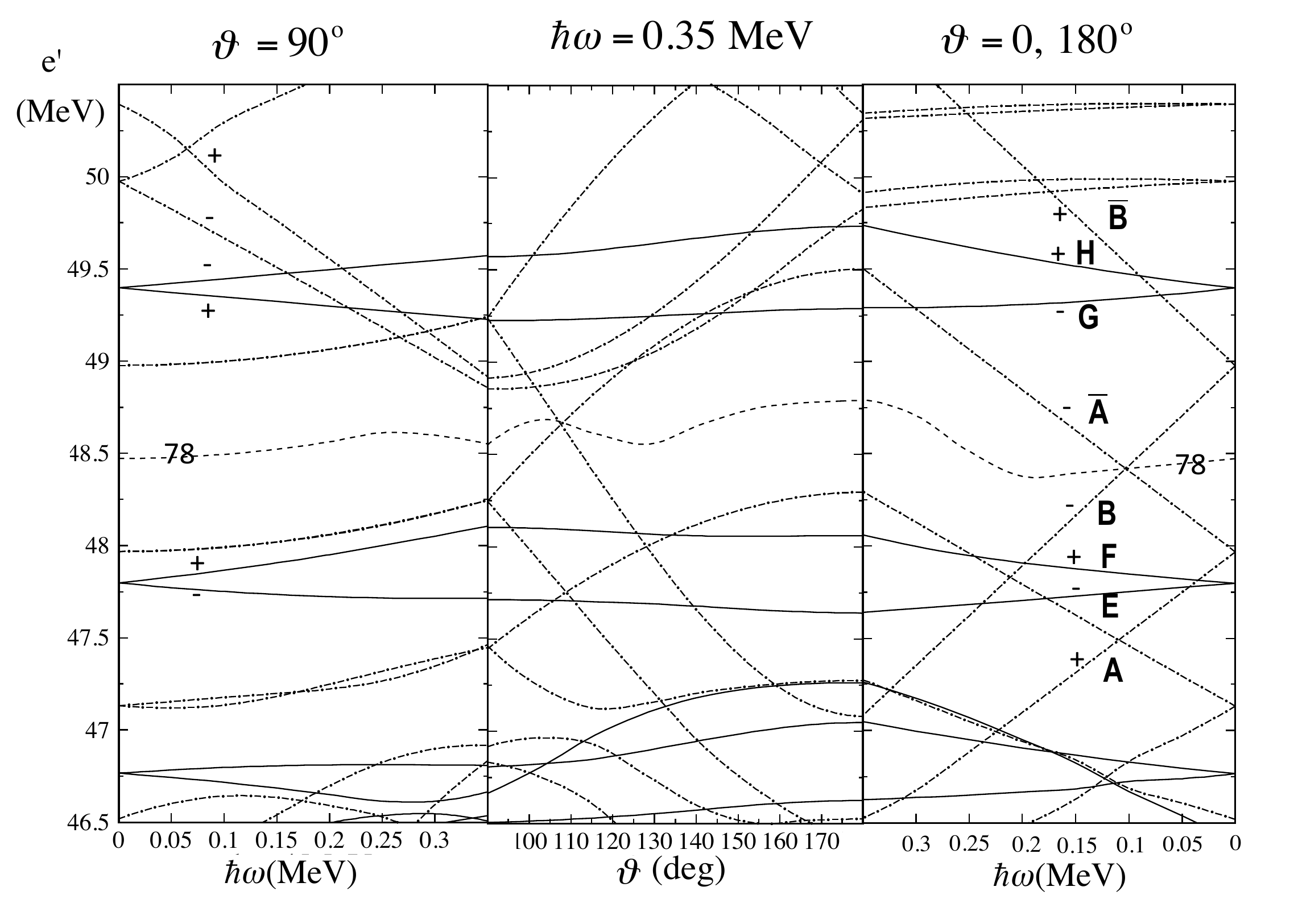} 
\caption{\label{f:spagNd138}
Single-particle  routhians  (protons uper panel , neutrons lower panel)   around the Fermi surface of $^{138}$Nd calculated for the deformation parameters $\varepsilon_2=0.17$, 
$\gamma=30^\circ$. Full line $\pi$=+, dash-dot line $\pi=-$. In case of rotation about the long axis, $\vth=0$,
and about the short axis, $\vth=90^\circ$, the signatures $\alpha=\pm 1/2$ are indicated by $\pm$, respectively.
The middle panel connects the two axis at the indicated frequency. The angle in the middle part is either $\vth$ or $180^\circ-\vth$, because 
the pertinent  rotations are related by inverting the orientation of the l-axis, which leaves the potential invariant.
The h$_{11/2}$ orbitals are labelled by A, B, C and the $\pi=+$ orbitals by E, F, ...\ . Taken from  Ref. \protect\cite{Nd138}.}
\end{center}
\end{figure}

The right panel of Fig. \ref{f:SymmetryChange} illustrate the transition from broken to conserved
${\cal R}_1(\pi)$ symmetry for the reduced M1 transition probability. Up to $I\approx 20$ the tilted axis cranking values calculated by means of 
Eq. (\ref{eq:bm1}) agree very well with the exact hole-rotor values.  Above, they are about half-way between  the two
types of  transitions, the larger  ones $\al=-1/2\rightarrow 1/2$ and the smaller ones $\al=1/2\rightarrow -1/2$. The signature dependence
 becomes earlier apparent in the transition rate than in the energies ($I=25$ corresponds to $\hbar\om=0.7$ MeV). In the principal acid cranking regime 
 the two $\De I =2$ bands of opposite signature are different \confsd.   Hamamoto and Sagawa \cite{HaSa79} suggested 
 calculating the transition probabilities between two  \confs 1 and 2   by means of the expression  
 \bea\label{eq:bm1out} 
B(M1,\{I,1\}\rightarrow \{I-1,2\})=<{1\vert\cal M}_{-1}(M1)\vert 2>^2\nonumber\\
=\left[\sqrt{\frac{3}{8\pi }}<1\vert\hat\mu_3\vert 2>\right]^2.
\eea
As seen in Fig. \ref{f:SymmetryChange}, the $B(M1)$ values calculated this way very well reproduce the signature dependence. They 
approach the exact hole-rotor values above $I=30$, which corresponds  to $\hbar \om=0.9$ MeV. At low spin the expression 
overestimates the exact values because the tilt of the rotational axis is not taken into account.     
 
The transition region is well approximated by means of the semiclassical vector model discussed 
in section \ref{sec:approxTAC} with the signature correction suggested in Ref. \cite{vectors}.
The model  uses the Cranked Shell Model concept for   
the 1-axis component  $\om_1$ of the angular velocity, which is based on 
an estimate the \am projection on the 3-axis. Using, as common,  the $K$ value of the band head may not be 
good enough, because of the response  of the hole to the inertial forces. In Fig. \ref{f:SymmetryChange} this is reflected by the 
somewhat  low values as compared to the exact hole-rotor results  at high spin.
  
As demonstrated by the example, a proper description of the transition region  requires
 going beyond the rotating mean field  approximation to an \am conserving approach. The multi-quasiparticle triaxial - rotor model
(see section \ref{sec:QTR} and Refs. \cite{BMII,NR,Ringbook}) is one well proven approach. 
More microscopic models are based on \am projected deformed mean field approaches. The contributions to this Focus Issue
by  J. L. Egido \cite{NCEgido}, Y. Sun \cite{NCSun} and J. A. Sheikh {\it et al. } \cite{NCSheikh} are examples.

\begin{figure}[t]
\begin{center}
\includegraphics[width=0.8\linewidth,angle=-90]{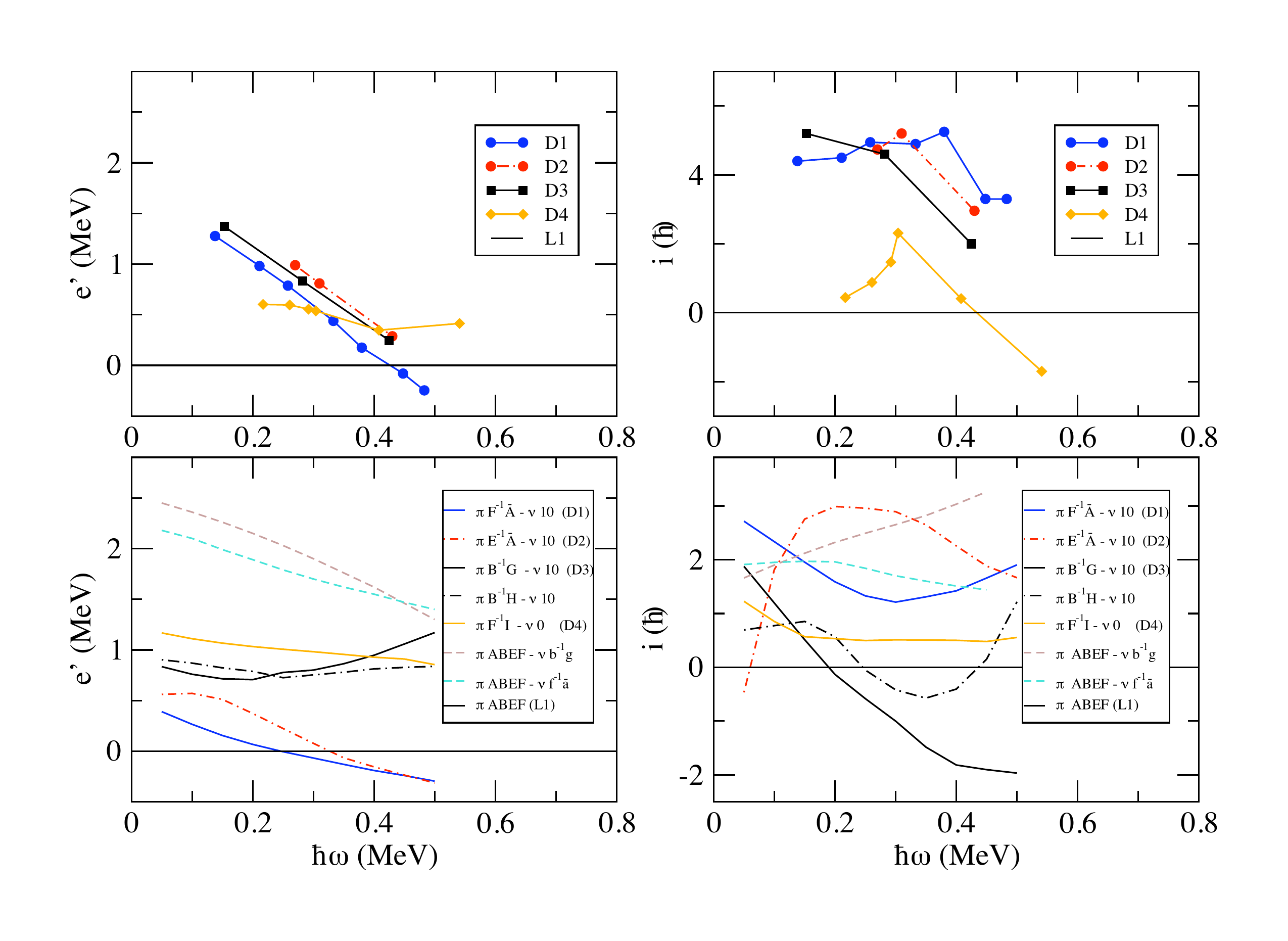} 
\caption{\label{f:Nd138Dipole}Experimental and calculated routhians and single-particle aligned angular momenta relative to band
L1 for the $\Delta I=1$ bands D1-D4 in $^{138}$Nd. The line type indicates the parity: full $\pi=+$, dash-dot $\pi=-$. 
The label $\nu 0$ means no neutron excitation and $\nu 10$ is short-hand for the neutron \conf  B$\bar \mathrm{A}$.  
The calculated routhians, except $\pi  ABEI\nu 0$, are shifted by
  \mbox{ $E_{exp, L6}(10^+ )-(E_{calc,\nu 10}(\omega=0)-E_{calc,\nu 0}(\omega=0) )$}  in order to account
   for the energy needed  breaking a neutron pair.
Taken from Ref. \protect\cite{Nd138}.}
\end{center}
\end{figure}

\subsubsection{Tilted axis cranking  solutions of triaxial nuclei}\label{sec:TriaxTAC}

Triaxial deformation at moderate spin is predicted for the regions around $Z=44,~N=64$, $Z=64,~N=76$ and  $Z=78,~N=116$ \cite{Moller06}. 
 When the  ${\cal R}_z(\pi)$ symmetry
is broken, triaxial  tilted axis cranking  solutions are observed as $\De I =1$ bands, as it is the case for axial solutions. 
Fig. \ref{f:spagNd138} shows the single particle  routhians   around $Z=60,~N=78$. The proton h$_{11/2}$  orbitals  from the bottom 
of the shell drive toward $\vth=90^\circ$ because they have best overlap when aligned with the s-axis, while the neutron h$_{11/2}$  orbitals  from the top
of the shell drive toward $\vth=0^\circ$ because they have best overlap when aligned with the l-axis (see discussion in \ref{sec:semi}). Combining them 
results in  a tilt of the rotational axis into the s-l-plane. These \confs appear as $\De I=1$ dipole bands.  Fig. \ref{f:Nd138Dipole} shows the 
dipole bands observed in $^{138}_{60}\mathrm{Nd}_{78}$ by Petrache {\it et al.} \cite{Nd138}. The routhians and alignments are displayed relative to the band L1, which
is the proton \conf  at $\vth=90^\circ$ with the routhians ABEF and below  occupied  and the neutron  \conf at $\vth=0^\circ$ with
the routhians  A$\bar\mathrm{A}$EF and below occupied.
This \conf has $\vth=90^\circ$ and appears as an even-$I$ band. The \confs of the dipole bands are labelled by the p-h excitations relative to L1, where $\nu 0$
means no neutron excitation and $\nu 10$ is short-hand for the neutron \conf  B$\bar \mathrm{A}$. The relative routhians and aligned angular momenta are the sums of the excited
particles and holes. Such simple scheme roughly accounts for the experiment and provides a \conf assignment. 
Better agreement with the experiment cannot be expected for a fixed deformation and zero pairing.

\paragraph{Wobbling}

One important difference to axial nuclei is that for trixial  nuclei
the moments of inertia of all three principal axes are finite,
which represent an additional degree of freedom: the orientation of the
core \am $\vec R$ with respect to the principal axes.  It appears  as  low-energy collective quadrupole 
states which are generated by exciting  rotational quanta about the axes with the smaller \momisd.
They form the spectrum of the triaxial rotor Hamiltonian (\ref{eq:HROT}), which for molecules as water is 
observed up to a large number of quanta along the unfavored axes. In the case of nuclei, only
the lowest excitations are realized, which have been called the wobbling mode \cite{BMII}. 
 Its presence is a clear indication of triaxiality.  
  We discuss two  phenomena, transverse wobbling and chiral vibrations, which 
appear as a consequence of the strong coupling of the wobbling mode with the \qpsd.

For the discussion it is important to note that
the \momi of the collective core is largest for the m-axis. This is the case  for the irrotational flow \momisd.
 Rotating mean field  calculations confirm this order, which is also implied  by the observation that the 
collective \momi increases with the deviation  from symmetry with respect to rotational axis. The deviation is largest for the m-axis, because the
perpendicular  l- and s-axes have largest length difference. In the following discussions we
  assign the s-axis to 1,  the m-axis to 2 and the l-axis to 3.  

The harmonic limit of wobbling has been discussed by Bohr and Mottelson \cite{BMII}, which we sketch.
The yrast states with no wobbling excitation correspond to rotation about the 2-axis with the largest \momid. 
The rotor Hamiltonian  (\ref{eq:HROT}) is rearranged as
\beq\label{eq:HWOB}
H=\frac{\hat R^2}{2{\cal J}_2}+H_W,
~~~H_W= R\left( \frac{1}{2{\cal J}_3}-\frac{1}{2{\cal J}_2}\right)\hat r_3^2+
R\left( \frac{1}{2{\cal J}_1}-\frac{1}{2{\cal J}_2}\right)\hat r_1^2.
\eeq
with \mbox{$\hat r_1=\hat R_1/\sqrt{R}$} and \mbox{$\hat r_3=\hat R_3/\sqrt{R}$}. The 
 rotor energy is minimal for uniform rotation about the 2-axis, which is given by the first term as $R(R+1)/(2{\cal J}_2)$. It is assumed 
that the deviation of $\hat R_2$ from its eigenvalue $R$ is small such that $[\hat R_3,\hat R_1]=-i\hat R_2\approx -iR$. Within this 
harmonic approximation,
$\hat r_1$ and $\hat r_3$ are canonical operators and $H_W$ represents a harmonic oscillator with the spectrum 
\beq\label{eq:EWOB}
E_W=\hbar \om_W(n+1/2),~~ \om_W=\frac{R}{{\cal J}_2}\sqrt{\left(\frac{{\cal J}_2}{{\cal J}_3}-1\right)\left(\frac{{\cal J}_2}{{\cal J}_1}-1\right) },
\eeq
 where $R$ has to be taken as $R+1/2$ in order to account for semiclassical corrections (cf. \ref{sec:semi}).
Wobbling means that the \am vector precesses on an elliptic cone about the 2-axis. In the laboratory frame it corresponds to the same 
precession of the m-axis about the \am vector $\vec R$.  The wobbling phonon carries  the signature $\alpha=1$ 
because $\hat r_1$ and $\hat r_3$
are odd under the rotation ${\cal R}_2(\pi)$. That is,  the zero-phonon state has $\alpha=0$,  the one-phonon state $\alpha=1$,
and  the two-phonon state $\alpha=0$.
The expressions for the quadruple transitions matrix elements are given in Refs. \cite{BMII} (section 4.5e) and \cite{FD14}. 

\begin{figure}[t]
\begin{center}
\includegraphics[width=0.48\linewidth]{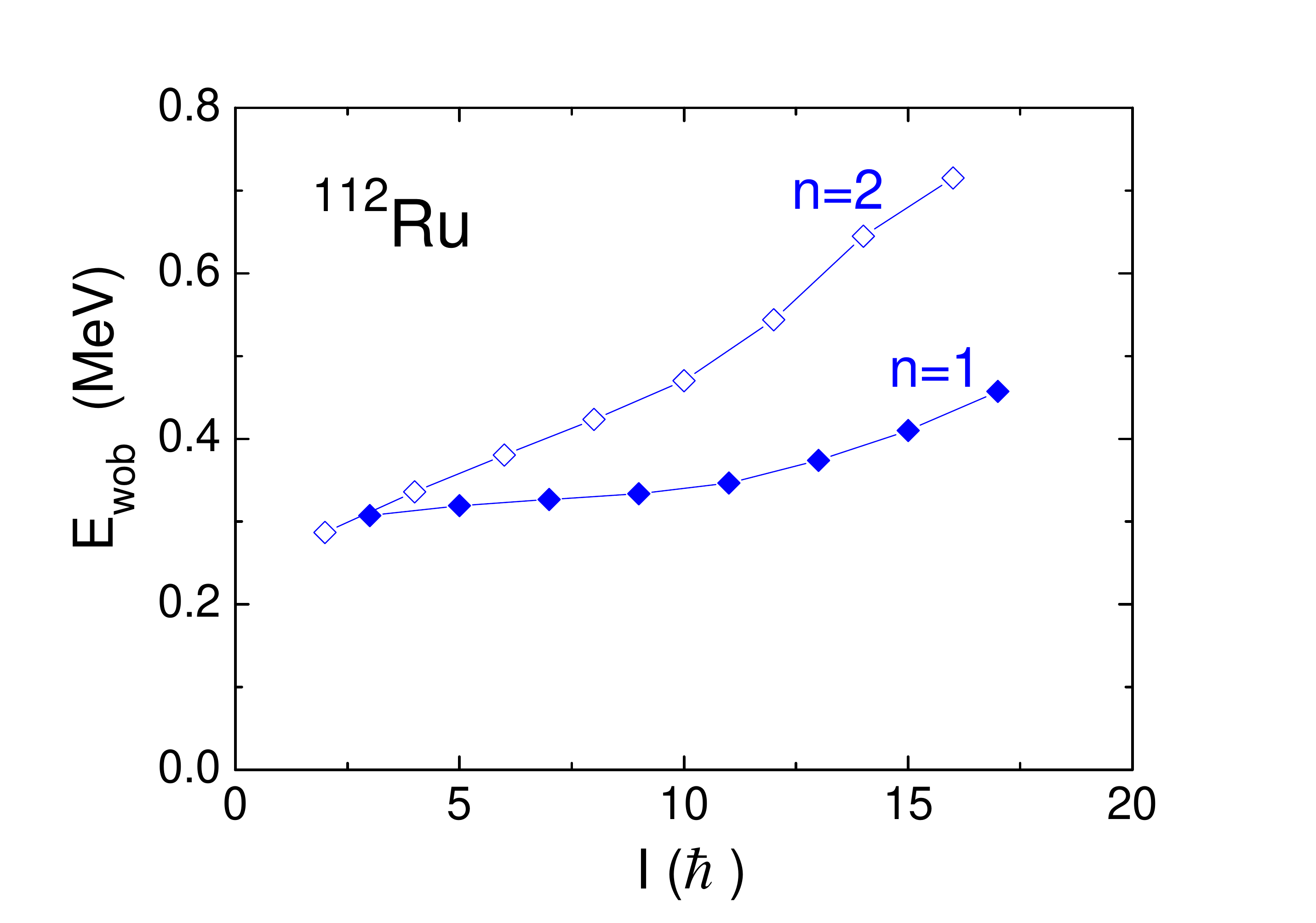} 
\includegraphics[width=0.48\linewidth]{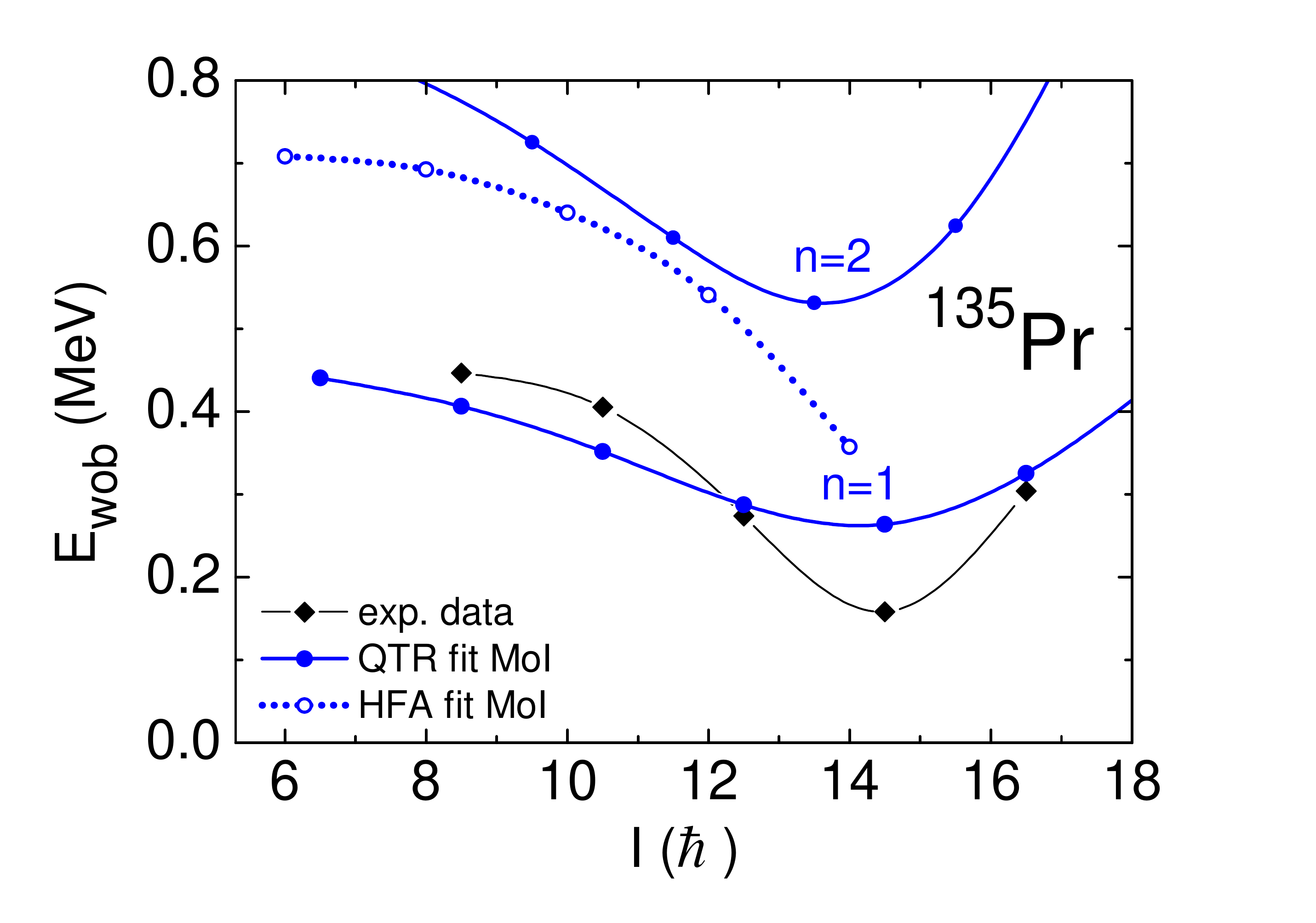}
\caption{\label{f:wobbler}
Left: Experimental energies  of the two lowest wobbling bands $n = 1$ (odd $I$) and  $n = 2$ (even $I$)  relative to the  $n = 0$ yrast 
sequence in $^{112}$Ru as an example for a simple wobbler.\\Right: Excitation energies of the $n=1~(\alpha=1/2)$ and $n=2
~(\alpha=-1/2)$ of the $\pi h_{11/2}$ transverse 
wobbling bands relative to the  $n = 0~(\alpha=-1/2)$ yrast sequence in $^{135}$Pr. 
Solid blue lines and full dots: \qp triaxial rotor calculation (QTR);  dotted blue lines and open dots: HFA calculation (\ref{eq:om2wob});
black line and full diamonds: experimental data.
From  \protect\cite{FD14}.}
\end{center}
\end{figure}

The left panel of Fig. \ref{f:wobbler} shows the lowest bands in $^{112}_{44}$Ru$_{68}$ as the best example for this kind of "simple" wobbling observed so far. 
 The  expected sequence even $I$, odd $I$, even $I$ for the respective zero-,  one-,  and two-phonon states is observed.  However, only above
$I=10$ one sees the expected increase of $\om_W$ with $I$. The equidistance between the zero-, one- and two-phonon states, 
 characteristic for the harmonic,  limit is strongly perturbed.    

\paragraph{Transverse wobbling}
In axial nuclei the tilt of the rotational axis reflects the combination of \qp \am along the symmetry axis and
collective angular momentum perpendicular to it, which increases along the band. As a consequence, the tilt 
 angle $\vth$ increases with $\om$.  As discussed in section \ref{sec:ChangeSym}, this  may lead 
to a transition from broken to restored ${\cal R}_z(\pi)$ symmetry 
(i. e. signature splitting) when $\vth$ approaches $90^o$. The analog can happen in triaxial nuclei
when a \qp aligns with the 1- or 3- axis causing a tilt of the rotational axis into the 1-2- or 1-3- planes, respectively.
 However,  there is also the possibility
of a transition from conserved to broken ${\cal R}_z(\pi)$ symmetry 
with {\em increasing} frequency. This occurs when a high-j  particle
couples to the triaxial core. As discussed in \ref{sec:semi}, it aligns its \am with the s-axis, 
because this orientation maximizes the overlap with the triaxial potential.
When  $\vec R$ aligns with the s-axis the energy 
gain due to the cranking term $-\vec \om \cdot \vec J$ is maximal.
With this orientation the ${\cal R}_z(\pi)$  symmetry  is good, 
and the band is a $\De I=2$ sequence of good signature.
When  $\vec R$ has the direction of the
m-axis the energy of the collective rotation is minimal. 
With this orientation the ${\cal R}_z(\pi)$  symmetry is broken,
and the band is a $\De I=1$ sequence. The balance of these two energies decides the symmetry.
Since the cranking term is linear and
the collective rotational energy quadratic, the former 
dominates at low and the latter at high $\om$. 

The essence of the interplay is seen when one   assumes  that the \am $j$ of the \qp is 
rigidly aligned with the s-axis, which is the frozen alignment (FA) approximation studied in Ref. \cite{FD14}. 
 The classical routhian augmented by the Lagrangian multiplier $\lambda \om^2/2$ is
\beq\label{eq:EptransWob}
 E'(\vec\om)+\lambda \om^2/2=-\frac{1}{2}\sum\limits_i\om_i^2{\cal J}_i-\om_1j+\frac{1}{2}\lambda\sum\limits_i\om_i^2.
\eeq
The Lagrangian multiplier is added to find the stationary points under the constraint of fixed $\om$,
i. e. with respect to the orientation of the rotational axis. Taking the derivatives with respect to $\om_i$
the stationary points are determined by the equations
\bea\label{eq:Jc}
({\la-\cal J}_1)\om_1=j_1,~~({\la-\cal J}_2)\om_2=0,~~({\la-\cal J}_3)\om_3=0,
\eea
which have to be fulfilled simultaneously.
The first solution 
\beq\label{eq:omc1}
\om=\om_1, ~~\om_2=0, ~~\om_3=0, ~~\lambda={\cal J}_1+j/\om
\eeq
  represents rotation about the 1-axis with
  \beq
  J={\cal J}_1\om+j,~~E=\frac{(J-j)^2}{2{\cal J}_1}.
  \eeq
The second solution 
\beq\label{eq:omc2}
\om_1=\frac{j}{{\cal J}_2-{\cal J}_1}\equiv \om_c,~~\om_2\not=0,~~\om_3=0,~~\lambda={\cal J}_2.
\eeq
represents rotation about an axis in the 1-2-plane tilted by the angle $\vth=\arctan (\om_1/\om_2)$ with respect to the 1-axis.  
Energy and \am are given by
\beq\label{eq:JE2nd} 
J_1=\om_c{\cal J}_1+j=\frac{j{\cal J}_2}{{\cal J}_2-{\cal J}_1}\equiv J_c,~~J_2={\cal J}_2\om_2,~~~E=\frac{J_c^2}{2{\cal J}_1}+\frac{J^2-J_c^2}{2{\cal J}_2}.
\eeq
  For $J<J_c$
 the first solution has lower energy and is a minimum. It has ${\cal R}_1(\pi)$ symmetry
  and represents a $\De I=2$ band with the signature of the aligned \qpd.
  
 For $J>J_c$ the second solution has the lowest energy and is a minimum where the  first becomes  a saddle point.
  The second solution  represents  rotation about the tilted axis. It breaks the ${\cal R}_1(\pi)$ symmetry and is assigned to a
  $\De I =1$ band. 
 Thus $J_c$ marks the transition from a $\De I=2$ band of given signature to a  $\De I=1$ band. 
 
 The energy (\ref{eq:JE2nd})
 has a similar form as Eq. (\ref{eq:ErotI}) for a \qp rigidly aligned with the symmetry axis of an axial nucleus. In this case  
 ${\cal J}_1=0$ and $j=\hbar K$. The first solution (\ref{eq:omc1}) does not represent a band. The second solution (\ref{eq:omc2}) represents 
 the $\De I=1$ sequence built on the band head state with $I=K=J_c/\hbar$ (cf. Eqs. (\ref{eq:ErotI},\ref{eq:Escl})). With increasing \am
 the  \qp may decouple from the 1-axis and reorient its $\vec j$ toward the 2-axis. This corresponds to a transition to  
${\cal R}_2(\pi)$ symmetry, which is analog to the transition discussed in section \ref{sec:ChangeSym} before.

The transition from solution 1 assigned to a $\De I=2$ band to solution 2 assigned to a $\De I=1$ band  has to be gradual.  That is, 
 a second  $\De I=2$ sequence of opposite signature must merge with the first. The second band is the 
 transverse wobbling excitation, which has been introduced by Frauendorf and D\"onau \cite{FD14}.  
 Let us discuss the transition in the framework of the harmonic frozen alignment approximation (HFA) \cite{FD14},
 which considers a Hamiltonian that corresponds to the classical routhian (\ref{eq:EptransWob}),
 \beq\label{eq:HFA}
H_{FA}=\sum\limits_i\frac{\hat R_i^2}{2{\cal J}_i}=\frac{(\hat J_1-j)^2}{2{\cal J}_3}
 +\frac{\hat J_2^2}{2{\cal J}_2}+\frac{\hat J_1^2}{2{\cal J}_3},
 \eeq 
where the aligned \am $j$ is assumed to be fixed ("frozen"). 
For $J<J_c$ rotation  about the 1-axis is preferred. Assuming that the components $\hat J_2$ and $\hat J_3$ are small, one 
approximates $\hat J_1\approx J- \hat J_2^2/2J- \hat J_3^2/2J$. This leads to a harmonic wobbler Hamiltonian that looks like 
$H_W$ for the simple wobbler given by Eq. (\ref{eq:HWOB}) when one changes the axes as 
$2\rightarrow 1,~ 3\rightarrow 2,~1\rightarrow 3$ and  replaces  
\beq
R\rightarrow J ~~{\mathrm{and}} ~~{\cal J}_1 \rightarrow \bar {\cal J}_1= \frac{{\cal J}_1}{1-j/J}.
\eeq
The resulting modified expression (\ref{eq:EWOB}) can be rewritten as
\beq\label{eq:om2wob}
\omega_w =
\frac{j}{{\cal J}_1} \left[ \left( 1 + \frac{J}{j}\left( \frac{{\cal J}_1}{{\cal J}_2} - 1\right) \right) 
\left( 1 + \frac{J}{j}\left( \frac{{\cal J}_1}{{\cal J}_3} - 1 \right) \right) \right]^{1/2}.
\eeq
For the considered case of transverse wobbling  the \qpd's \am is aligned with the 1-axis, which means  that  ${\cal J}_1<{\cal J}_2$ and 
${\cal J}_1>{\cal J}_3$. Then,  the factor $1+J({\cal J}_1/{\cal J}_2-1)/j$ in  Eq.~(\ref{eq:om2wob}) decreases with $J$ and
 the wobbling energy will also decrease  for sufficiently large $J$. It reaches 
zero at $J_c$  the  critical \am (\ref{eq:Jc}).  

Frauendorf and D\"onau \cite{FD14} classified the wobbling modes as transverse when the \qpd's \am is perpendicular to  the m-axis and as longitudinal when it 
is aligned with the m-axis.  A hole coupled to the triaxial core will align its \am with the 3-axis (l-)  
which also corresponds to transverse wobbling. 
For the typical order ${\cal J}_2>{\cal J}_1>{\cal J}_3$ the decrease of $\om_W$ with $J$ is more 
rapid than for the high-j particle as can be seen
from Eq. (\ref{eq:om2wob}) after exchanging the axes $1\leftrightarrow 3$.   
 A mid-shell \qp tends to align with the 2-axis (m-), which corresponds to longitudinal wobbling. 
 The wobbling frequency increases with $J$,  as can be seen from Eq. (\ref{eq:om2wob}) after exchanging the axes $1\leftrightarrow2$, 
 because both factors increase with $J$.

 The  right panel of Fig. \ref{f:wobbler}
shows $^{135}_{59}$Pr$_{76}$ as an example for 
transverse wobbling. The yrast \conf  is $[\pi(\mathrm{AEF})\nu( \bar\mathrm{A}\mathrm{AEF})]$ in the single routhian diagram Fig. \ref{f:spagNd138}.
The calculations use \mbox{${\cal J}_m=21~\hbar^2\mathrm{ MeV}^{-1},~{\cal J}_s=13~\hbar^2\mathrm{ MeV}^{-1},~{\cal J}_l=4~\hbar^2\mathrm{ MeV}^{-1}$}, 
which were adjusted to 
optimize the agreement with the experimental energies. For the diagonalization of the full quasiparticle triaxial rotor Hamiltonian
(\ref{eq:HQR}) a pure h$_{11/2}$ proton is assumed and a deformation of $\beta=0.17,\ \ga=26^\circ$.
 
  The orbital  A is  aligned with the s-axis which is transverse to the m-axis.  Accordingly, the wobbling frequency 
decreases with $I$. The HFA expression (\ref{eq:om2wob}) gives $\om_W=0$ at the critical spin which  is $J_c=14.4\hbar$ 
by Eq. (\ref{eq:Jc}).
In addition the full \qp + triaxial rotor (QTR) calculation (see section \ref{sec:QTR}) is shown, to which  HFA is 
an approximation that  
becomes invalid near and above $J_c$. In contrast to the HFA, the exact eigenvalues of  $H_{QTR}$ (\ref{eq:HQR}) 
 smoothly pass 
 $J_c$. The h$_{11/2}$ \qpr is not rigidly aligned with the s-axis. The quasiparticle triaxial rotor model takes into account its response 
   to the inertial forces by gradually 
  tilting $\vec j$ toward the m-axis. For the considered case  this realignment is substantial, such that for $J>J_c$   longitudinal coupling 
 becomes prevalent, which is seen as the increase of the wobbling frequency. 
 
 Wobbling is a motion of the triaxial charge density relative to the 
 space fixed \am vector, which generates enhanced E2 radiation. Frauendorf and D\"onau  \cite{FD14}  
 predicted  large $B(E2, I\rightarrow I-1)$  values for the transitions from the 
 wobbling to the yrast band, which were confirmed by Matta {\it et al.} \cite{Matta15}. 
 These strong E2 transitions discriminate the wobbling band against the signature 
 partner band with  the \conf  $[\pi(\mathrm{BEF})\nu( \bar\mathrm{A}\mathrm{AEF})]$ in Fig. \ref{f:spagNd138}, 
 which has the same signature $\alpha=1/2$ as the wobbling band.
  The pertinent band was also identified in addition to the wobbling and found to have the expected weak E2 transitions to the yrast band.
  
 Frauendorf and D\"onau  \cite{FD14} calculated  microscopic \momis 
  \mbox{ ${\cal J}_m=17~\hbar^2\mathrm{ MeV}^{-1},~{\cal J}_s=7~\hbar^2\mathrm{ MeV}^{-1},~{\cal J}_l=3\hbar^2~\mathrm{ MeV}^{-1}$ }
  by means of cranking with  small $\om$ about the three principal axes. For these values  Eq. (\ref{eq:Jc}) gives the critical spin  $J_c=9.3\hbar$, which is 
   lower than 14.4$\hbar$ obtained with the fitted \momisd. A direct tilted axis cranking  calculation using the same parameters
   gives $\vth=90^\circ$ and $\f=0$ for low $\om$, i. e. rotation about the s-axis.   
   Above $\hbar \om_c=0.28$ MeV the tilt angle $\f$ moves rapidly toward $90^\circ$, i.e. the rotational axis swipes through the s-m-plane.  
   The critical frequency is smaller than  $\hbar \om_c=0.55$ MeV obtained by means of the HFA expression  
   (\ref{eq:omc2}) for the cranking \momisd.
   Possible reasons for the discrepancy may be that the \qpr is not rigidly  aligned with the s-axis and that the \momis are not
   constant up to $\om_c$ as assumed in the estimate (\ref{eq:omc2}).    
    Transverse wobbling appears to be  rather sensitive to the value of $\ga$ 
    and of the pair gaps $\De$, which strongly influence the ratios between the three \momisd.
    The sensitivity to such details  makes it difficult  obtaining precise results by microscopic calculations.
   However, for  transverse wobbling of the triaxial strongly deformed $^{163,165,157}$Lu isotopes, the microscopic calculations
   that combined principal axis cranking with quasiparticle random phase approximation (see section \ref{sec:UMmicro}) provided quite 
   promising results \cite{Shimizu09,FD15}.
    In their contribution to this Focus Issue, Sheikh {\it et al.}  \cite{NCSheikh}   present the microscopic study of transverse wobbling
     in $^{135}$Pr in the framework of the Triaxial Projected Shell Model, which
     describes  the wobbling energy as  well as the quasiparticle triaxial rotor model 
     calculation in shown in Fig. \ref{f:wobbler} right. 
  I. Hamamoto discusses wobbling in her contribution to this Focus Issue as well \cite{NCHamamoto}.
  
   \paragraph{Chirality}
   When the angular momentum does not not lie in one of the three principal planes of a
triaxial nucleus the combination of the three different principal axes 
with $\vec J$
becomes chiral (see section \ref{sec:symmetry}). There are   
the left-handed $|l\rangle$ and right-handed $|r\rangle$
mean field solutions, which a related to each other  by 
\beq
|l\rangle={\cal T R}_y(\pi) |r\rangle.
\eeq
In the ideal case of strongly broken  ${\cal T R}_y(\pi)$ symmetry, 
the matrix elements $\langle l|Q_{22}|r\rangle,~\langle l|Q_{21}|r\rangle$ and 
$\langle l|\mu_{11}|r\rangle$ are small because the electromagnetic
field does not provide enough angular momentum to turn the long vector $\vec J$ from   
the left- to the right-handed position. 
In order to calculate the transition matrix elements for this ideal case one
must take into account that the exact states of the two bands have good 
${\cal T R}_y(\pi)$ symmetry.
  One may choose the phases  such that 
\beq\label{eq:tr}
{\cal T R}_y(\pi)|IM\pm\rangle=|IM\pm\rangle
\eeq
 when  acting on the  states $|IM\pm\rangle$ of good angular momentum, 
 which describe the two degenerate  bands
(see Ref. \cite{BMI}). The linear combinations
\beq\label{eq:chpm}
|+\rangle=\frac{1}{\sqrt{2}}(|r\rangle+|l\rangle),~~
|-\rangle=\frac{i}{\sqrt{2}}(|r\rangle-|l\rangle)    
\eeq
fulfill the relation (\ref{eq:tr}), i. e. they restore the broken symmetry. 
Thus for $E2$-transitions one has instead of (\ref{eq:be22}) 
\bea\label{eq:be22pm}
B(E2, I~\pm\rightarrow I-2~\pm)= \frac{5}{8\pi}|\langle r| Q_{22}| r\rangle 
+ \langle l| Q_{22}| l\rangle |^2,\\
B(E2, I~\pm\rightarrow I-2~\mp)= \frac{5}{8\pi}|\langle r| Q_{22}| r\rangle     
- \langle l| Q_{22}| l\rangle |^2
\eea
Eqs. (\ref{eq:be21}) and (\ref{eq:bm1}) are modified in the same way
\cite{pr134tac}.
  
 \begin{figure}[t]
\begin{center}
\includegraphics[width=0.5\linewidth]{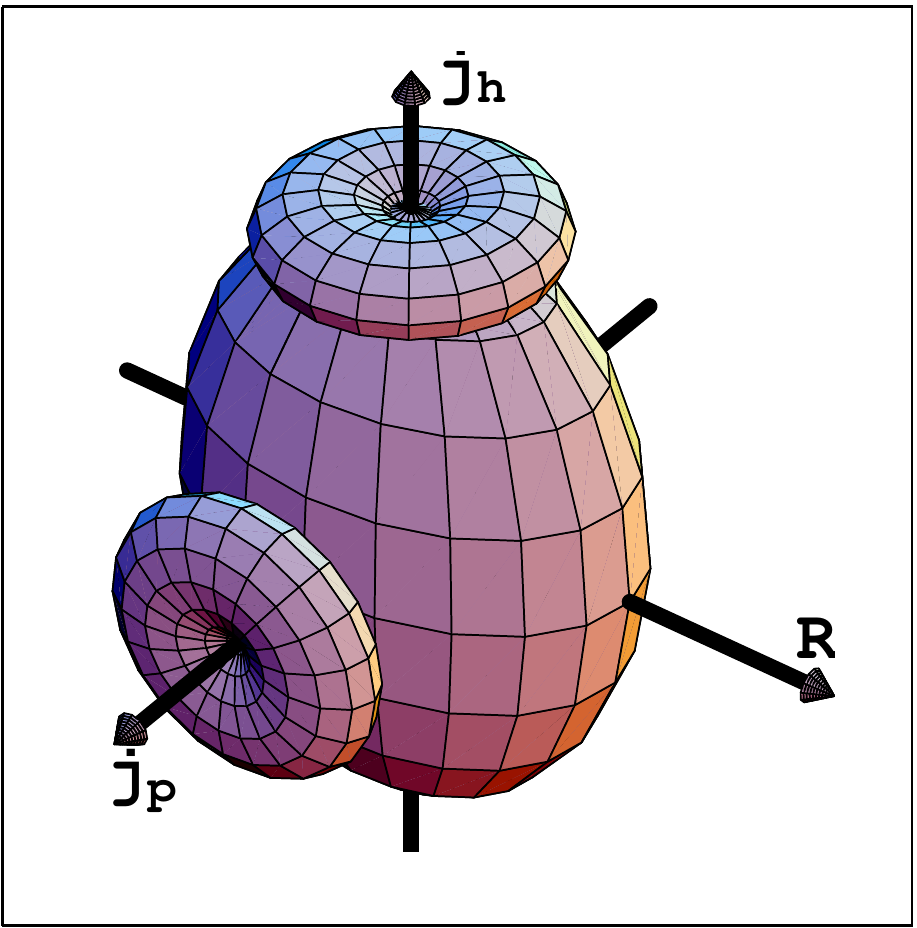}
\caption{\label{f:orbitschiral} Orbitals of a high-$j$ proton and a high-$j$ neutron hole coupled to the
 triaxial density distribution. The collective \am $\vec R$ of
the rotor prefers the medium axis, which has the 
largest moment of inertia. The total \am $\vec J=\vec R + \vec j_p + \vec j_h$ 
 points out of the plane of the 
drawing. The short, medium and long axes form a 
right-handed screw with respect to $\vec J$.  }  
\end{center} 
\end{figure}  
 \begin{figure}[t]
\begin{center}
\includegraphics[width=0.8\linewidth]{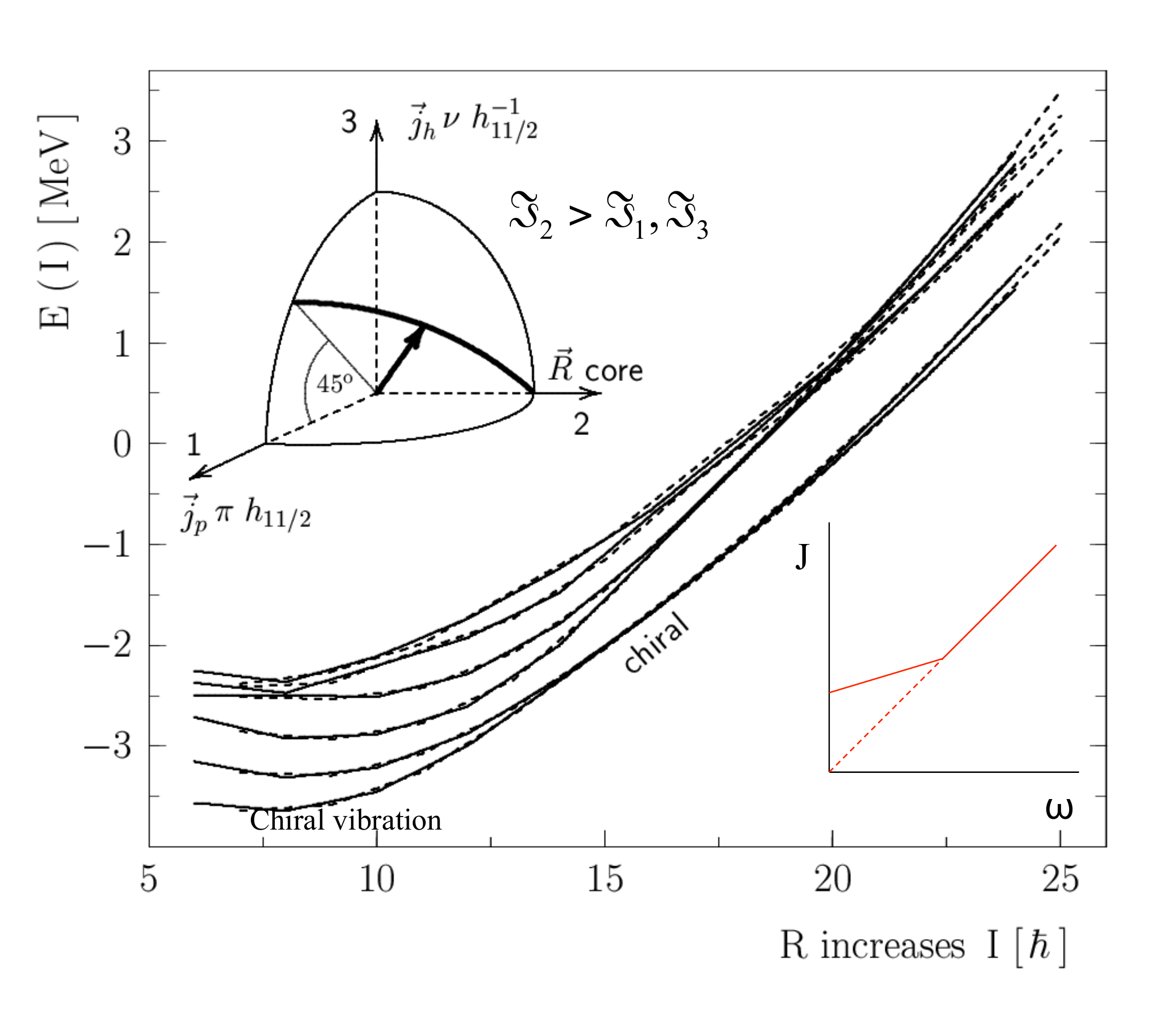}
\caption{\label{f:chiralpr} Rotational levels of a $\pi h_{11/2}$ particle and  a $\nu h_{11/2}$ hole coupled to
a triaxial rotor with $\ga=30^o$. 
 Full lines: $\al=0$ (even $I$). 
Dashed lines to $\al=1$ (odd $I$). The left inset shows the orientation of the
angular momentum with respect to  the triaxial potential, where 1, 2 and 3 correspond
to the s-, m- and l-  axes, respectively.
The angular momentum vector moves along the heavy arc. The  position displayed 
corresponds to the spin interval $13 <I<18$, where the two lowest bands
are nearly degenerate.  The right-handed position is shown.
The left-handed is obtained by reflection through the 1-3 plane.
The calculation uses $\eps=0.24$, $\ga=30^\circ$ and the irrotational-flow ratio ${\cal J}_{2}/{\cal J}_{1,3}=4$.
The right inset shows the function $J(\om)$ for the frozen alignment  approximation.
Adapted from Ref. \protect\cite{chiralpr}.  }
\end{center} 
\end{figure}  

The  ${\cal T R}_y(\pi)$ symmetry is broken when the angular momentum
has components on all three principal axes. Fig. \ref{f:orbitschiral} illustrates 
the typical  \am geometry.
 High-j particles of one kind of nucleons
and  high-j  holes of the other kind couple with the triaxial deformed
potential.    
The particles tend to align their angular momentum with the short
axis and 
the  holes with the long axis, because these orientations
maximizes the overlap of their  density distributions with the
potential (see \ref{sec:semi}).  
 The collective angular momentum $\vec R$  provides the third component
 along the medium axis.  
In analogy to transverse wobbling, the  collective \am  $\vec R$ increases along the band.
  At the band head, 
only the particle and the hole generate the angular momentum, which lies in the 1-3
plane.  At low frequency, $\vec R$ stays within this s-l- plane 
aligning with the angular momentum of the particle and hole, because this orientation 
minimizes the cranking term $-\vec \om \cdot \vec J$.  
For larger values of $\om$ it is energetically favorable
that $\vec R$ increases into the direction of the m-axis, because the moment of inertia is
larger. The vector $\vec J$ moves out of the 1-3 plane towards the 2-axis. 
Accordingly the symmetry changes from ${\cal R}_z\not =1,~{\cal T R}_y=1$, corresponding to one 
$\De I=1$ band, to ${\cal R}_z\not =1,~{\cal T R}_y\not=1$, corresponding to two degenerate $\De I=1$ bands.
As for transverse wobbling, there is a critical frequency for which the planar tilted axis cranking  solution becomes chiral. 

To qualitatively discuss the transition we use again the frozen alignment approximation, i. e. we assume that the particle \am $j_1$
is rigidly aligned with the 1-axis (s-) and the hole \am $j_3$ with the 3-axis (l-). 
 To find the stationary points with respect to the orientation of  $\vec \om$  the classical routhian is augmented by 
the  Lagrangian multiplier $\frac{1}{2}\lambda\om^2$ 
in order to keep  $\om^2$  constant, 
\beq
E'(\om_1,\om_2,\om_3)+\frac{1}{2}\lambda\om^2=-\frac{1}{2}\sum\limits_i{\cal J}_i\om^2_i-\om_1j_1-\om_3j_3 +\frac{1}{2}\lambda\sum\limits_i\om_i^2.
 \eeq
 Taking the derivatives with respect to $\om_i$,
 the stationary points are given by the equations
 \beq 
({\la-\cal J}_1)\om_1=j_1,~~({\la-\cal J}_2)\om_2=0,~~({\la-\cal J}_3)\om_3=j_3,
 \eeq
which have to be fulfilled simultaneously \cite{tacdft1}.  As for transverse wobbling, there are two solutions,
which are determined by  second equation. The solution is planar when $\om_2=0$, or chiral when
$\la={\cal J}_2$. For the chiral solution $\om_2$ increases while the other two components stay constant 
$ \om_1=j_1/({\cal J}_2-{\cal J}_1)$ and $ \om_3=j_3/({\cal J}_2-{\cal J}_3)$. At the critical frequency $\om_c$  both $\om_2=0$ and  
$\la={\cal J}_2$, which means
\beq\label{eq:Jcchiral}\om_c=\sqrt{\left(\frac{j_1}{({\cal J}_2-{\cal J}_1)}\right)^2+\left(\frac{j_3}{({\cal J}_2-{\cal J}_3)}\right)^2},~~J_c={\cal J}_2 \om_c.
\eeq

As in the case of transverse wobbling, the chiral doubling develops in a gradual way.
 Below $J_c$  the second $\De I=1$ band appears as the  one-phonon chiral vibration excitation.  
Frauendorf and D\"onau  \cite{FD16} worked out the small amplitude solution of the frozen alignment approximation. 
The expressions are  too complex to be included here. 
Frauendorf and Meng \cite{chiralpr} carried out  quasiparticle triaxial rotor 
 model calculations for the special case $j_1=j_2=11/2$  and $\ga=30^\circ$, which are  shown in Fig. \ref{f:chiralpr}. 
 The irrotational-flow ratios of the \momis were assumed, which are ${\cal J}_2/{\cal J}_1={\cal J}_2/{\cal J}_3=4/1$.
  According to the frozen alignment estimate (\ref{eq:Jcchiral}), the one-phonon band merges the zero-phonon band at  
 $J_c=\sqrt{2}j_1{\cal J}_2/({\cal J}_2-{\cal J}_1)=\sqrt{2}\times11/2\hbar\times4/3=10.3\hbar$. In the full \qp triaxial rotor calculation,
 the two bands meet somewhat above at $J=14\hbar$.
 Based on their quasiparticle triaxial rotor model, the authors realized for the first time that the doublets manifest the left-handed and right- handed
 geometry of the three types of \am illustrated in Fig. \ref{f:orbitschiral}.  They suggested that the two $\De I=1$ sequences 
 observed in $^{134}_{59}$Pr$_{75}$ represent chiral partner bands.

 \begin{figure}[t]
\begin{center}
\includegraphics[width=0.8\linewidth]{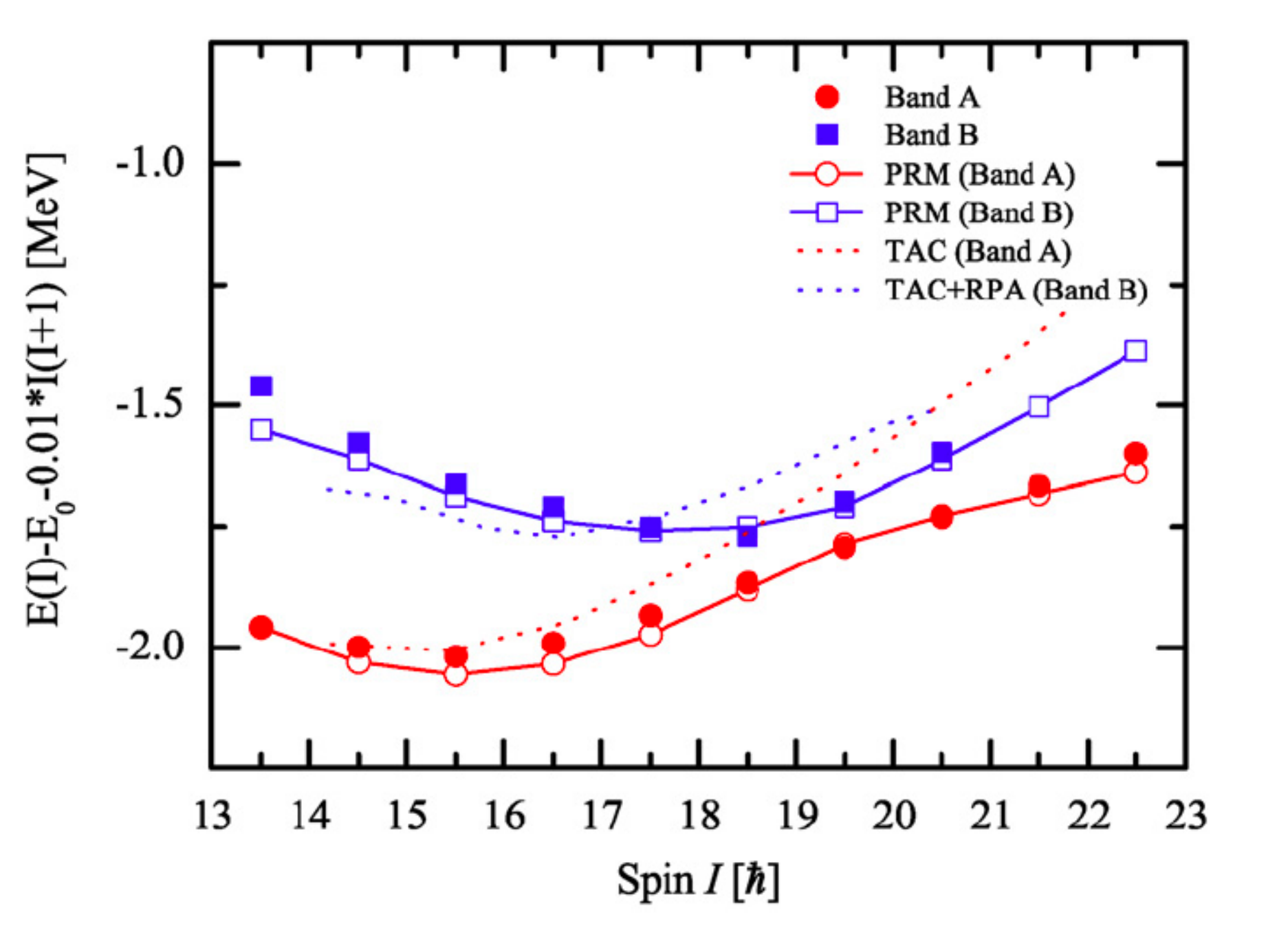}
\caption{\label{f:Nd135E} The excitation energies $E(I)$ for the chiral sister bands in $^{135}$Nd
with the \conf $[\pi h_{11/2}^2\nu h_{11/2}^{-1}]$. 
Filled symbols: experiment \cite{Zhu03,Mukhopadhyay07};
open symbols (PTR):  two protons+one neutron hole+triaxial rotor ($\ga=30^\circ$) calculations;
dotted lines: TAC +QRPA calculations \cite{Zhu03,Mukhopadhyay07}.
Taken  from Ref. \cite{Qi09}.}
\end{center} 
\end{figure}  
 \begin{figure}[h]
\begin{center}
\includegraphics[width=0.8\linewidth]{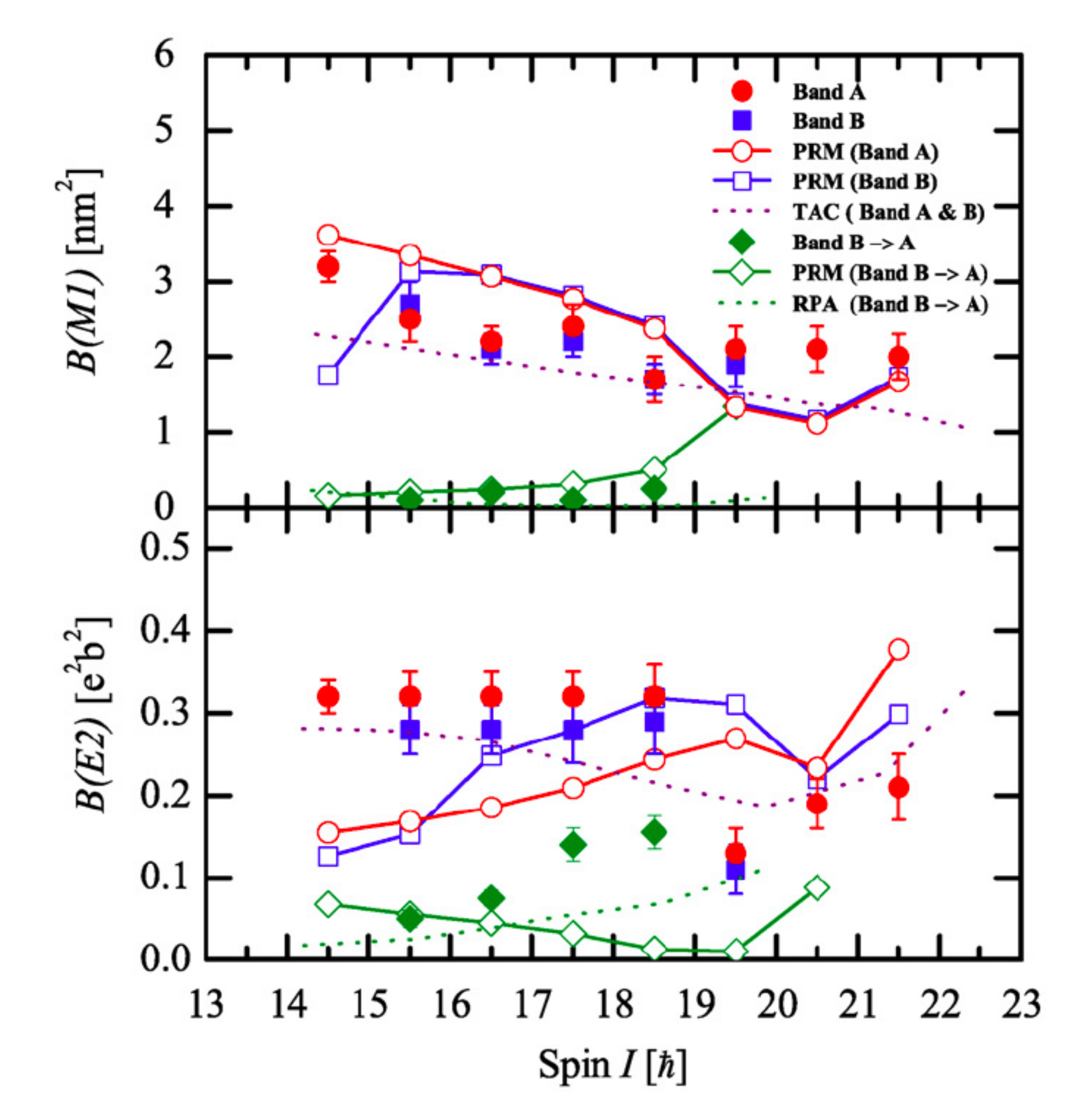}
\caption{\label{f:Nd135M1E2}
Reduced transition probabilities of the chiral sister bands in $^{135}$Nd
with the \conf $[\pi h_{11/2}^2\nu h_{11/2}^{-1}]$. 
Filled symbols: experiment \cite{Zhu03,Mukhopadhyay07};
open symbols (PRT):  two protons+one neutron hole+triaxial rotor ($\ga=30^\circ$) calculations; 
dotted lines (TAC): inband transitions calculated by means of TAC;
dotted lines (RPA): interband transitions calculated by means QRPA based on the TAC mean field \cite{Zhu03,Mukhopadhyay07}.
Taken from Ref. \cite{Qi09}.}
\end{center} 
\end{figure}

  Dimitrov {\it et al.}  \cite{pr134tac} found the first chiral rotating mean field solution in  the framework of shell correction tilted axis cranking approach 
  (SCTAC)  for $^{134}_{59}$Pr$_{75}$.
  The \conf is composed of a particle-like $h_{11/2}$ \qpr and a hole-like $h_{11/2}$ \qnd. 
  Without pairing the \conf is $[\pi AEF\nu A]$ in Fig. \ref{f:spagNd138}, which has a finite tilt  $\vth$ in the s-l-plane. 
 The collective \am $\vec R$ drives the rotational axis out of the 
  s-l-plane at the critical spin $J_c =10.7\hbar$ corresponding to \mbox{$\hbar\om_c=0.30$ MeV} ($\vth=60^\circ$ and $\beta=0.19, ~\ga=28^\circ$). 
  The result is consistent with the observed encounter of the  two $\De I=1$ sequences  at $I=14$. The encounter is actually
  a crossing of the two sequences, which depart from each other for higher spin.  
   Almehed {\it et al.} \cite{ADF11} described the
  excited $\De I=1$ bands in odd-odd nuclei around $A=134$ 
  as the one-phonon chiral vibrational excitations in the framework of the QRPA  (see section \ref{sec:UMmicro}) based on the 
  planar rotating mean field solution found by SCTAC (TAC+QRPA). 
  The method  provides the vibrational energy $\hbar\om_V$ without any new parameter. In case of $^{134}$Pr,
the value of $\hbar\om_V=0.30$ MeV at the rotational frequency $\hbar \om$=0.30 MeV is somewhat lower than the experimental value of
0.37 MeV. The QRPA values stay below the experimental ones, reaching zero at $J_c$. 
  
  Figs. \ref{f:Nd135E} and \ref{f:Nd135M1E2} shows $^{135}_{60}$Nd$_{75}$ as another example for the transition from the 
  planar to a chiral mean field solution. The \conf is composed of two particle-like $h_{11/2}$ \qprs and a hole-like $h_{11/2}$ \qnd.
   Without pairing the \conf is $[\pi ABEF\nu A]$ in Fig. \ref{f:spagNd138}, which has a finite tilt  $\vth$ in the s-l-plane.
 In the SCTAC calculations carried out by Zhu {\it et al.}  \cite{Zhu03}, the collective \am $\vec R$ drives rotational axis out of the 
  s-l-plane at the critical spin $J_c =20.5\hbar$ corresponding to $\hbar\om_c=0.47$ MeV ($\vth=68^\circ$ and $\eps\approx 0.20, ~\ga\approx30^\circ$).
  In the experiment the two partner bands come closest at $J=(I+1/2)\hbar=20\hbar$. 
  In order to describe the approach of the one-phonon band Mukhopadhyay {\it et al.} \cite{Mukhopadhyay07} carried out  TAC+QRPA calculations along the lines 
  described in Ref. \cite{ADF11}.  As seen in   Fig. \ref{f:Nd135E},
  the calculation well reproduces the experiment without any parameter adjustment up to $J_c$ where the QRPA stops working.  
    Fig. \ref{f:Nd135M1E2} shows that the TAC+QRPA calculations reproduce the $B(M1)$ and $B(E2)$ values of the two bands fairly well. 
    Qi {\it et al.}  \cite{Qi09} studied the transition from chiral vibration to chiral rotation in the framework of the quasiparticle particle 
     triaxial rotor model (cf. section \ref{sec:QTR}).
    Two $h_{11/2}$ protons and one $h_{11/2}$ neutron hole were coupled to a triaxial rotor with irrotational-flow ratios between the \momisd. 
    The model parameters, the average value of the \momisd, the  triaxiality parameter  $\ga$ and
    the coupling strength to the deformed potential, were adjusted to reproduce the experimental energies.
     The results  in Figs.   \ref{f:Nd135E} and \ref{f:Nd135M1E2}
   rather well account for the  experiment.   The analysis of the \am geometry showed that around $J=20$ the two bands are to good approximation
   represented by  the two linear combinations (\ref{eq:chpm}). Above, the two bands move apart again because the two protons are tilted toward the m-axis like
   the one proton generating transverse wobbling in the neighbor $^{135}_{59}$Pr$_{76}$ (see discussion  above). 
   
Alc\'antara-N\'u\~nez {\it et al.} \cite{Rh105low} and Tim\'ar {\it et al.} \cite{Rh105high} found two chiral tilted 
axis cranking  solutions in the same nucleus
 $^{105}_{45}$Rh$_{60}$, which belongs to the second region of triaxiality  around $Z =44, ~N =64$.
The tilt into the s-l-plane is generated by the combination of the hole-like $g_{9/2}$ \qpr with particle-like $h_{11/2}$ \qnsd. 
For the \conf $[\pi g_{9/2}\nu h_{11/2}^2$] the tilted axis cranking  solutions becomes chiral  at $J_c=21\hbar$ ($\vth=75^\circ,~\beta=0.23,~\ga=29^\circ$, SCTAC). The 
pertinent experimental bands approach each other like for  the discussed $A=134,$ 135 cases. At $I=45/2$ the distance reaches its minimum of 100 keV.
For the second \conf   [$\pi g_{9/2}\nu h_{11/2}(g_{7/2}d_{5/2})$] the tilted axis cranking  solution becomes 
chiral at $J_c=9\hbar$ ($\vth=57^\circ, \beta=0.23,~\ga=29^\circ$).
The pertinent bands are nearly degenerate for $I=17/2 $ and 19/2 ($\De E<$50 keV). 
For larger spin the partners depart from each other as in the case of 
$^{134}_{59}$Pr$_{75}$. Ayangeakaa {\it et al.}  \cite{Ayangeakaa13} found two chiral bands in $^{133}$Ce with the \confs
$[\pi\mathrm{g}^{-1}_{7/2}\pi\mathrm{h}^1_{11/2}\nu\mathrm{h}^{-1}_{11/2}]$ and 
$[\pi\mathrm{h}^2_{11/2}\nu\mathrm{h}^{-1}_{11/2}]$.


Zhu  {\it et al.} \cite{Mochiral} and Luo {\it et al.} \cite{Ruchiral} interpreted a pair of bands   in the even - $N$ Mo and Ru isotopes 
as chiral partners with a structure that differs from the one discussed so far. The suggested \conf is 
$[\nu\mathrm{h}_{11/2}^1 (\mathrm{d}_{5/2}\mathrm{g}_{7/2})^{-1}]$. 
The tilted axis cranking  calculations showed that the \am of the normal parity $(d_{5/2}g_{7/2})$ hole is strongly aligned
with the l-axis and the \am of the $h_{11/2}$ with the s- axis.
The total angular momentum moves from the l-s-plane through the aplanar region to the s-m-plane, 
which is caused by the collective \am driving toward the m-axis.  
For $Z\approx 44$, $N\approx 66$ 
  the two-\qn \conf $[h_{11/2}(d_{5/2}g_{7/2})]$ with $\De_n=0.7$ MeV was assumed \cite{Ruchiral}. 
 The rotating mean field  solution stayed in the s-l-plane being very soft in the m-direction. 
  The TAC+QRPA calculations 
 gave a vibrational frequency of about $300$ keV. 
  The   experimental bands in  $^{106}_{42}$Mo$_{64}$ and $^{110,112}_{44}$Ru$_{66,68}$ 
are separated by about 200 keV or less and have similar decay 
properties, which   qualifies them as chiral partners.

The dynamics of chirality beyond the harmonic approximation have mainly been studied in the 
framework quasiparticle triaxial rotor model model. The details of the approach relevant to chirality are presented in the contributions by 
K. Starosta and T. Koike \cite{NCStarosta} and I. Hamamoto \cite{NCHamamoto} to this Focused Issue, which also   contain a comprehensive  review 
of the experimental evidence for chirality. Meng {\it et al.}  \cite{ChiralReviewMeng} reviewed work on chirality from  a more general perspective.
The studies of doublet bands in odd-odd $^{104,106}$Rh,$^{104,106}$Ag and $^{98,100}$Tc in the 
framework of the Triaxial Projected Shell Model \cite{Dar15,NCSheikh}
 represent a promising perspective for a microscopic large-amplitude description of chiral doubling.

\section{Emergence and disappearance of collective degrees of freedom}\label{sec:emergence}
 The Unified Model bases on the dichotomy of the  "collective" degrees of freedom, which
describe the shape of the nucleus and the  "intrinsic" degrees of freedom, which are particle-hole configurations or quasiparticle configurations.
Any number of collective excitations can be superimposed on the fermionic intrinsic states.  Clearly this is an idealization. 
Nuclei are composed of a relative small number of nucleons 
compared to other many-body systems. The "granular structure" of the collective degrees of freedom appears  after 
the excitation of a finite number of quanta, which results in a progressive decoherence of the collective modes.
This section is devoted to phenomena that require a refinement of the concept of the collective degrees of freedom.
The focus will be on the rotational degrees supplemented by some thoughts about the $\beta$ degree of freedom.

As a condition for  the appearance of rotational bands in nuclei, Bohr and Mottelson
\cite{BMII} state: ''A common feature of systems that have rotational
 spectra is the existence of a ``deformation'', by which is implied
a feature of anisotropy that makes it possible to specify an orientation
of the system as a whole. In a molecule, as in a solid body, 
the deformation reflects the highly anisotropic mass distribution,
 as viewed from the intrinsic coordinate frame defined by 
the positions of the nuclei. In the nucleus, the rotational degrees 
of freedom are associated with the deformations in the nuclear 
equilibrium shape that result from the shell structure.''
The first statement is very general, providing for all kinds of 
mechanisms that break the symmetry.
The low spin data seemed to indicate that the only possibility to meet this
condition is a substantial deformation of the nuclear density distribution,
which is measured by the electric quadrupole moment.
One suggestive evidence is the correlation between  the moment of inertia 
and the nuclear deformation pointed out by Grodzin \cite{grodzin}.  Raman {\it et al.} \cite{Raman}
found that ratio of the \momi and the reduced probability for the $0^+\rightarrow 2^+$  transition 
can be well approximated by the expression 
\bea\label{eq:RJB}
RJB(2^+)= \frac{\cal J}{B(E2,0^+\rightarrow 2^+)}=\frac{3}{E(2^+)B(E2,0^+\rightarrow 2^+)}\nonumber\\
\approx 1150 Z^{-2}A^{2/3} \hbar^2\mathrm{MeV}^{-1}(e\mathrm{b})^{-2}, 
\eea
where $E(2^+)=3\hbar^2/{\cal J}$ is used. 
The ratio is nearly constant
because both the numerator and the denominator are proportional to the square
of the deformation.  In the following we discuss phenomena 
that can only be understood by invoking the granular structure of nuclear rotation reflecting its  
microscopic underpinning.

  \subsection{Band termination}\label{sec:termination}

  Band termination appears naturally within the clockwork picture of the rotating nucleus. 
Consider a  nucleus that is not too strongly deformed,
such that the grouping of the
single-particle levels into spherical shells still exists.
The angular momentum of the high-j orbitals remains 
close to $j$, their value in the spherical potential.
The total angular momentum  $J$ is generated by gradually
aligning the  angular momentum vectors  $\vec j$ of the particles and holes
in the incompletely filled shells.
 Eventually all vectors $\vec j$ are aligned in accordance with the Pauli
principle. The mean field state $|\rangle$ is 
${\cal R}_z(\psi)$-symmetric for this stretched coupling. That is, 
the band is terminated. 
In order to increase the angular momentum further, new \conf must be  generated
by exciting particles from the core.
The   shape  may change appreciably before attaining
 ${\cal R}_z(\psi)$ symmetry at termination.
  
  Experimentally, band termination is seen as a
break of the smooth relation $J(\om)$ between angular momentum and rotational frequency.
Unlike after a band crossing, where another smooth sequence $J(\om)$ starts, 
after a termination
the level distances become irregular and there are competing parallel decay 
paths. 

Termination may occur also for nuclei with a substantial deformation.
Nucleons in a deformed harmonic oscillator potential, which rotates about   
a principal axis, is an illustrative example.  
 Band termination within this model was comprehensively discussed by Afanasjev {\it et al.}
 \cite{tbrev}, who also gave the pertaining references to the numerous earlier
 studies. The rotating harmonic oscillator separates into independent
oscillations along the three principal axes, as the non-rotating
oscillator does. A band is defined by
a fixed number of oscillator quanta along each of the principal axes. 
Consider rotation about the 1-axis. The maximal angular momentum of a given \conf is 
\beq
J_{1,max}=|\Sigma_2-\Sigma_3|, ~~
\Sigma_{2}=\sum\limits_{k,occ}\left(n_{2,k}+\frac{1}{2}\right),~~
\Sigma_{3}=\sum\limits_{k,occ}\left(n_{3,k}+\frac{1}{2}\right),
\eeq
where $n_{2,k}$ and $n_{3,k}$ are the numbers of quanta on the respective 2-, 3- axes of the single particle state
$k$. When this value is reached the density becomes symmetric with respect
to the 1-axis and the band terminates. Again, termination is due to
of the restriction of the accessible angular momentum by quantization.   

Termination does not have to take place. For near spherical nuclei, the
deformation may change  in  such a way that the spherical shell structure
dissolves. Then core angular momentum becomes accessible in addition to 
the angular momentum of the valence particles. For the harmonic oscillator 
\cite{tbrev} Afanasjev {\it et al.} delineated the shapes which terminate and which not.

The two  mechanisms idealize  different features of real nuclei.
The  high-j intruder orbitals behave like spherical ones
whereas the low-spin orbitals are better accounted for by
the deformed harmonic oscillator. Only realistic self consistent cranking calculations 
are capable of describing band termination  quantitatively.

Bohr and Mottelson discussed band termination for light nuclei in the
$sd$ shell \cite{BMII}.  
R\"opke and Endt \cite{roepke1,roepke2} 
used mean field configurations for classifying many
bands, which all terminate when the amount of angular momentum available
for the particles in the $sd$ shell is exhausted. 
The oscillator model accounts fairly well for
the  termination of bands in these  light nuclei.

 Nilsson and Ragnarsson \cite{NR} and Afanasjev {\it et al.} \cite{tbrev} reviewed 
 the theoretical and experimental work on band termination in heavier nuclei.
 It commonly appears in nuclides with several valence particle or holes outside closed shells.
 A large number of terminating bands have
  been observed in the regions $Z\geq 64,~N\geq 82$ and $Z\approx 50,~N\approx50$. 
Extensive theoretical analyses have been carried out in the frame of the
Cranked Nilsson-Strutinsky (CNS) model (see section \ref{sec:SCPAC}) 
as well as by means of configuration-confined selfconsistent approaches that are based on 
energy density functionals.  

\begin{figure}[t]
  \begin{center}
 \includegraphics[width=0.8\linewidth]{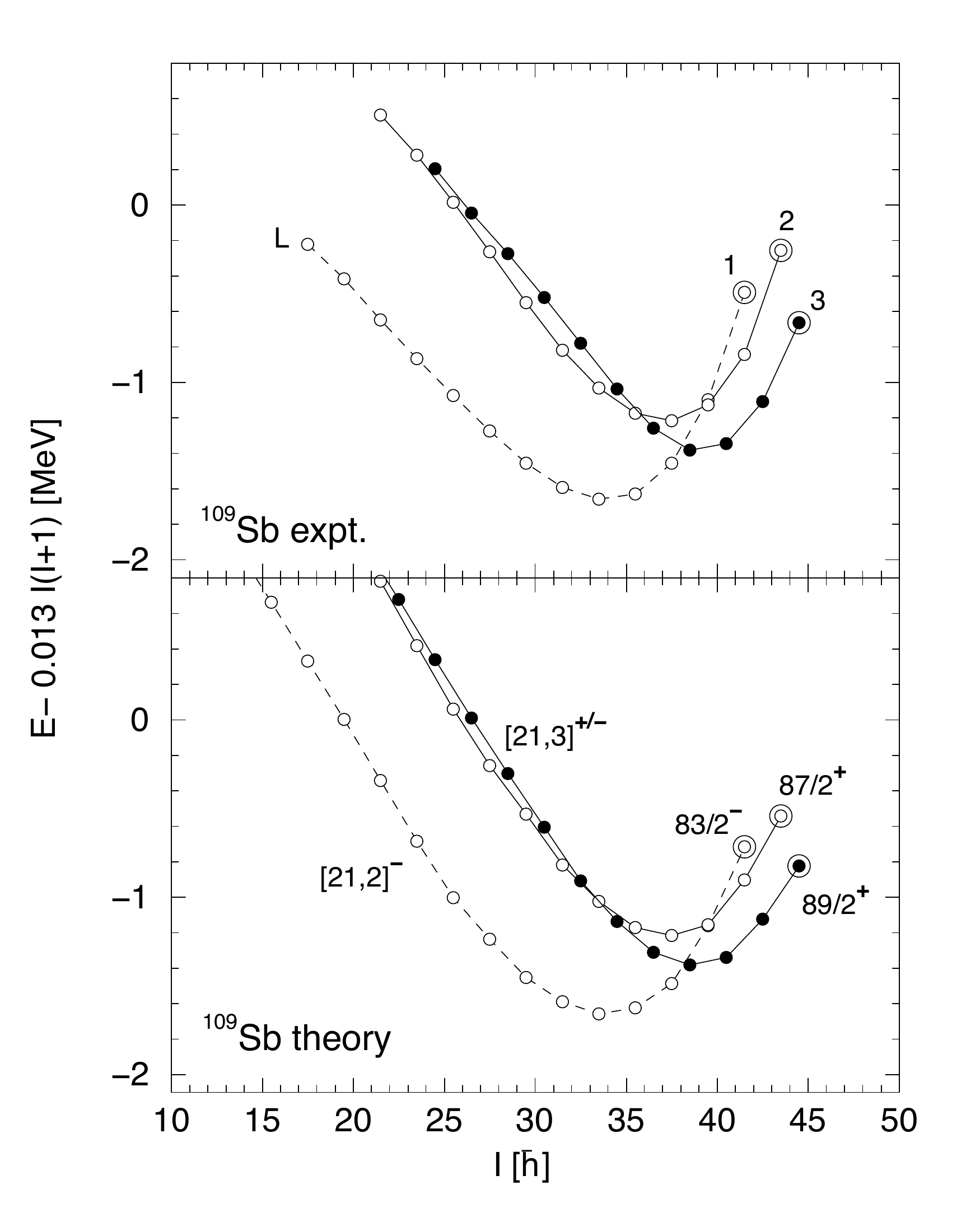} 
\caption{\label{f:sb109e}
The energies of the bands 1, 2,  3 in $^{109}_{51}$Sb$_{58}$
relative to the energy of a rigid rotor with a moment of inertia
of ${\cal J}=38\hbar^2\mathrm{MeV}^{-1}$. The terminating states are shown as large circles 
labeled 
by the spin and parity $I^\pi$.  Band 2 and 3 
are not linked to the known part of the spectrum. The unknown
  energy  of the lowest level is chosen
  to give agreement with the Cranked Nilsson-Strutinsky 
calculations in the lower panel.
The configuration assignment $[lm,n]^\pm$ in the lower panel is as follows.
l: number of $g_{9/2}$ proton holes, m: number of $h_{11/2}$ proton
particles, n: number of $h_{11/2}$ neutron particles and $\pm$ gives the signature
$\al=\pm 1/2$.
From Ref. \protect\cite{tbrev}}
\end{center}
\end{figure}

\begin{figure}[t]
\begin{center}
\includegraphics[width=0.8\linewidth]{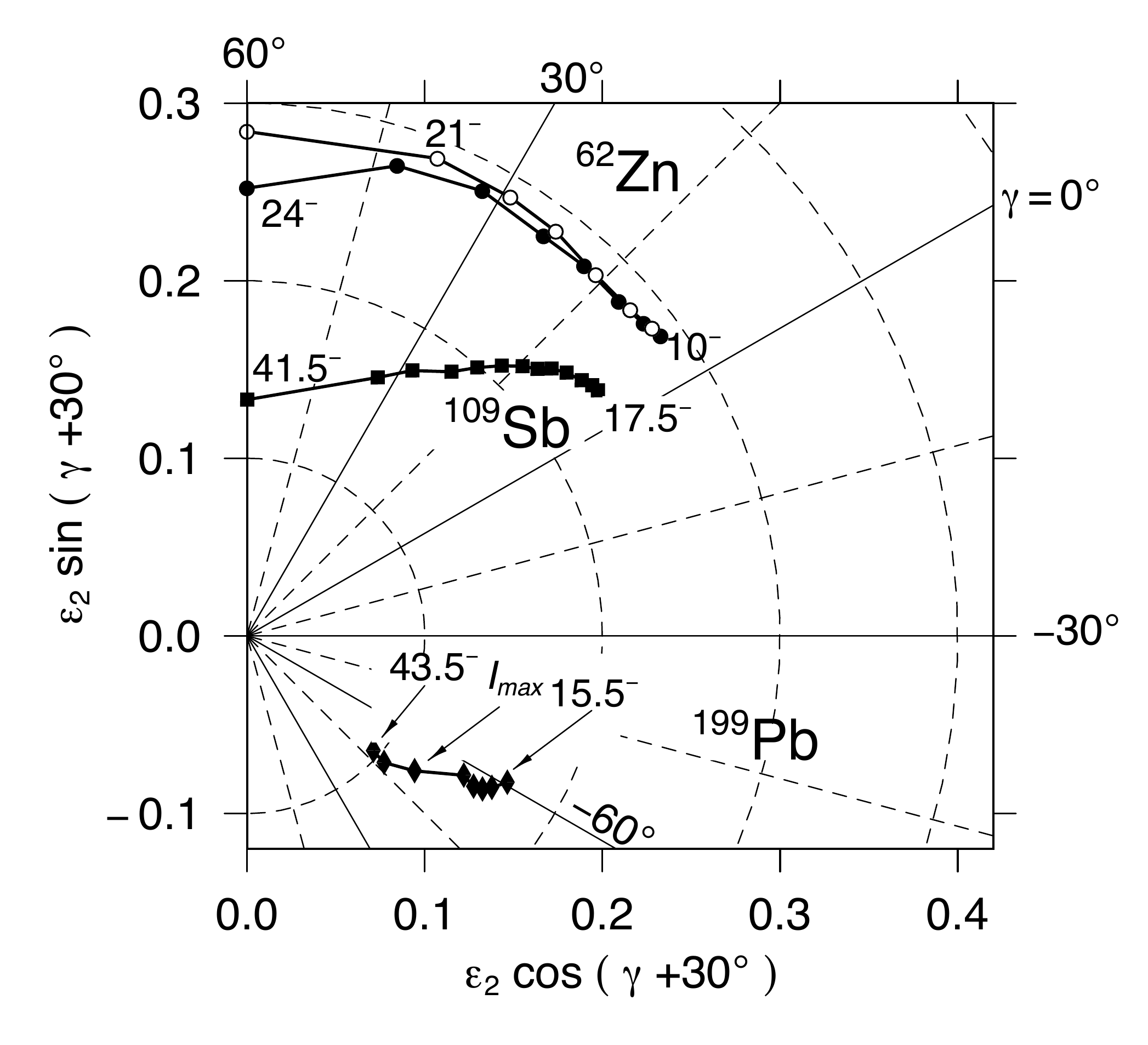} 
\caption{\label{f:tbdef}
The calculated deformation parameters $\eps(=\eps_2)$ and $\ga$ of bands in 
$^{199}_{82}$Pb$_{117}$, $^{109}_{51}$Sb$_{58}$ and $^{62}_{30}$Zn$_{32}$
as functions of the
angular momentum. Each spin value is represented by a symbol, where $\De I =2$.
From Ref. \protect\cite{tbmr}.}
\end{center}
\end{figure}

\begin{figure}[t]
\begin{center}
\includegraphics[width=0.8\linewidth]{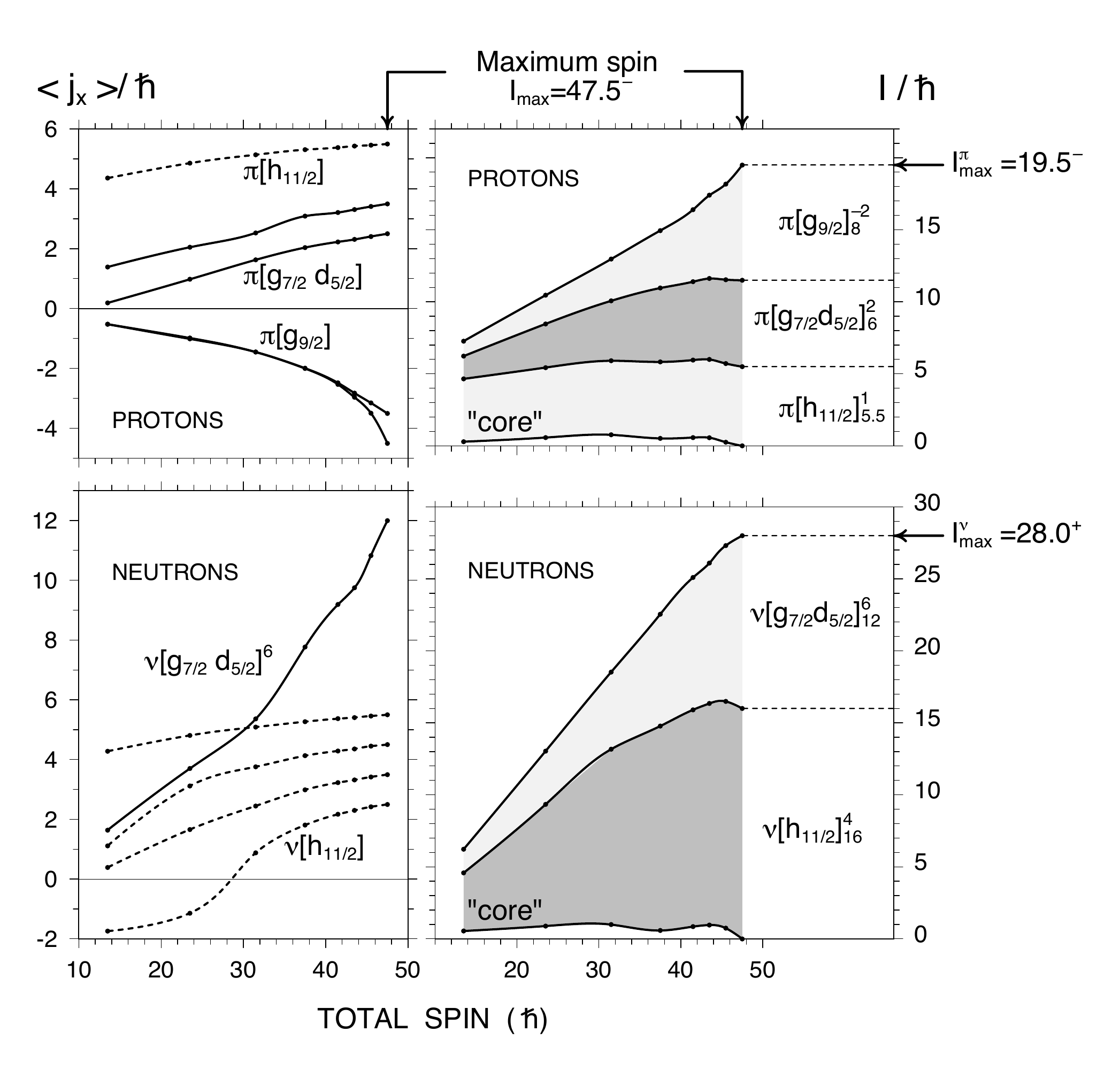} 
\caption{\label{f:sb111comp}
Contributions of the valence particles and holes to
the total angular momentum of a terminating band in $^{111}_{51}$Sb$_{60}$.
The calculations follow the deformation path relevant for the specific
configuration, which is  shown in Fig. \ref{f:tbdef}.
From Ref. \protect\cite{tbrev}}
\end{center}
\end{figure}

As an example, 
Fig. \ref{f:sb109e} shows three terminating bands in
$^{109}_{51}$Sb$_{58}$. Band 1 with $(\pi,\al)=(-,-1/2)$ has the proton
configuration $[(g_{9/2})^{-2}_8(d_{5/2}g_{7/2})^2_6(h_{11/2})^1_{11/2}]$ and the
neutron configuration $[(h_{11/2})^2_{10}(d_{5/2}g_{7/2})_{12}^6]$ relative to the $Z=N=50$ closed shell configuration.
 Stretched coupling of
all the angular momenta gives $I^\pi=83/2^-$, as observed.  The bands 2 and 3
with $(\pi,\al)=(+,-1/2)$ and $(+,1/2)$
 have the same proton configuration  combined with   neutron configurations 
$[(h_{11/2})^3_{27/2}(d_{5/2}g_{7/2})_{21/2}^5]$ and
$[(h_{11/2})^3_{27/2}(d_{5/2}g_{7/2})_{23/2}^5]$, respectively.
   As expected,  they terminate at $I^\pi=87/2^+$ and $89/2^+$.
Fig. \ref{f:tbdef} shows that  band 1 starts with the substantial 
prolate deformation of $\eps \approx 0.25$. Generating angular momentum by gradual
alignment of the valence particle angular momenta,
the shape becomes triaxial and then oblate and symmetric with
respect to $\vec J$ at  termination.
 The corresponding decrease of the $B(E2)$ values has been confirmed
experimentally and agrees well with the Cranked Nilsson-Strutinsky (CNS) calculations \cite{sb109l,tbrev}.
 The dispersion $\De J$, which measures the symmetry breaking (see Sec.
\ref{sec:coherence}) and the
moment of inertia $\J2$ decrease along this path.
 The path in the $\eps - \ga $ plane is similar for bands 2 and 3. 

Wadsworth {\it et al.} \cite{sb109l} and Afanasjev {\it et al.} \cite{tbrev} compare the CNS energies  to the ones of a  rotor
with the rigid body moment of inertia, which has become the common way to 
present the data and calculations.  They  
call the characteristic U-shape of  the function  $E(I) - E(I)_{rotor}$ 
  in Fig. \ref{f:sb109e} a "smooth unfavored
termination". "Smooth" is just one of the  criteria for a 
sequence of levels to qualify for a rotational band. The possibility
to observe the bands  up to termination is due to
the low level density, which is caused by the gaps in the single-particle 
spectrum at
$Z=50$ and $N=58$. If such gaps appear the smoothness condition
is well satisfied (cf. Eq. (\ref{alpha}) in section \ref{sec:coherence}). "Unfavored" means that the rotational
energy is larger than that of the reference rotor.
 This feature, which is seen in  Fig. \ref{f:sb109e} as the
sharp  increase of the relative energy at the highest spins,  
reflects the decrease of the \momi $\J2$. 

Fig. \ref{f:sb109e} is an exampled for the general  behavior of  the terminating bands
in the mass 110 region and lighter nuclei. Afanasjev {\it et al.}  \cite{tbrev} characterize it 
 as  starting with  "collective rotation" and then "gradually losing the collectivity". In this way they
 describe the gradual restoration of the
${\cal R}_z(\psi)$ symmetry, which is   substantially broken at the bottom
of the band. This terminology is quite commonly used in connection with the appearance or 
disappearance of regular rotational sequences.   
It is discussed in detail in section \ref{sec:coherence} that
the substantial symmetry breaking does not necessarily imply
"high collectivity" in the sense of a large number of nucleons being involved. 
 Fig. \ref{f:sb111comp} demonstrates that the rotation is 
not very collective. The angular momentum is generated by a 
relatively small number of particles. In the case of shears bands discussed in section \ref{sec:MR} 
the number of active nucleons is 4-6.   
A more general and appropriate terminology 
is to characterize termination as a gradual loss of coherence of the rotational motion.

 \subsection{Magnetic rotation}\label{sec:MR}
 As already discussed in section \ref{sec:AppTAC}  the orientation of the
 nuclear mean field is to be attributed to the  arrangement of the nucleonic orbitals, in particular the strongly un-isotropic high-j  orbitals.
 The discovery of Magnetic Rotation is a striking example for this picture of a spinning clockwork of gyroscopes.
   In 1991 Baldsiefen {\it et al.}  \cite{mrda} and Fant {\it et al.}  \cite{mrb1} reported the
observation of very regular pattern of $\ga$ rays in  $^{197-200}$Pb,
which can be arranged into rotational bands according to the accepted
criteria. At low spin irregular level
spacings are observed that are typical for these semi magic nuclei
with a near spherical shape.
 The regular band structures appear for $I>10$.
 The moments of inertia are $\J2 \sim 20 \hbar^2MeV^{-1}$, 
about one half of the moments of inertia
of the $2^+$ states of the well deformed nuclei of the rare earth region.
However, the transitions  were shown to be 
of stretched dipole type and  
arguments in favor of a magnetic character were  presented.
The   weak or missing stretched $E2$-transitions  
pointed to  the very small deformation expected for these nuclei. The observation,
completely unexpected in the context of the Unified Model, was explained by Frauendorf \cite{tac} by means of TAC
calculations.
Reviews on Magnetic Rotation are given in Refs. \cite{RMP,magrevCM,magrevH,revMeng} and in the contribution by   
Meng and Zhao to this Focus Issue \cite{NCMengZhao}.

The experimental results are in 
contradiction with  the  traditional Unified Model view,
according to which nuclear rotation is a collective phenomenon which only
occurs in well deformed nuclei.
 The ratio ${\cal J}^{(2)}/B(E2,0^+\rightarrow 2^+)$, which 
is $\sim 10 (eb)^{-2} \hbar^2\mathrm{MeV}^{-1}$ for   well
deformed heavy nuclei and $\sim 5 (eb)^{-2} \hbar^2\mathrm{MeV}^{-1}$ for superdeformed nuclei,
exceeds $ 100 (eb)^{-2} \hbar^2\mathrm{MeV}^{-1}$ for the dipole bands. 
 There must be something rotating that
carries a long transverse magnetic dipole  moment but almost
no charge quadrupole moment.

\begin{figure}[t]
  \begin{center}
\includegraphics[width=0.6\linewidth]{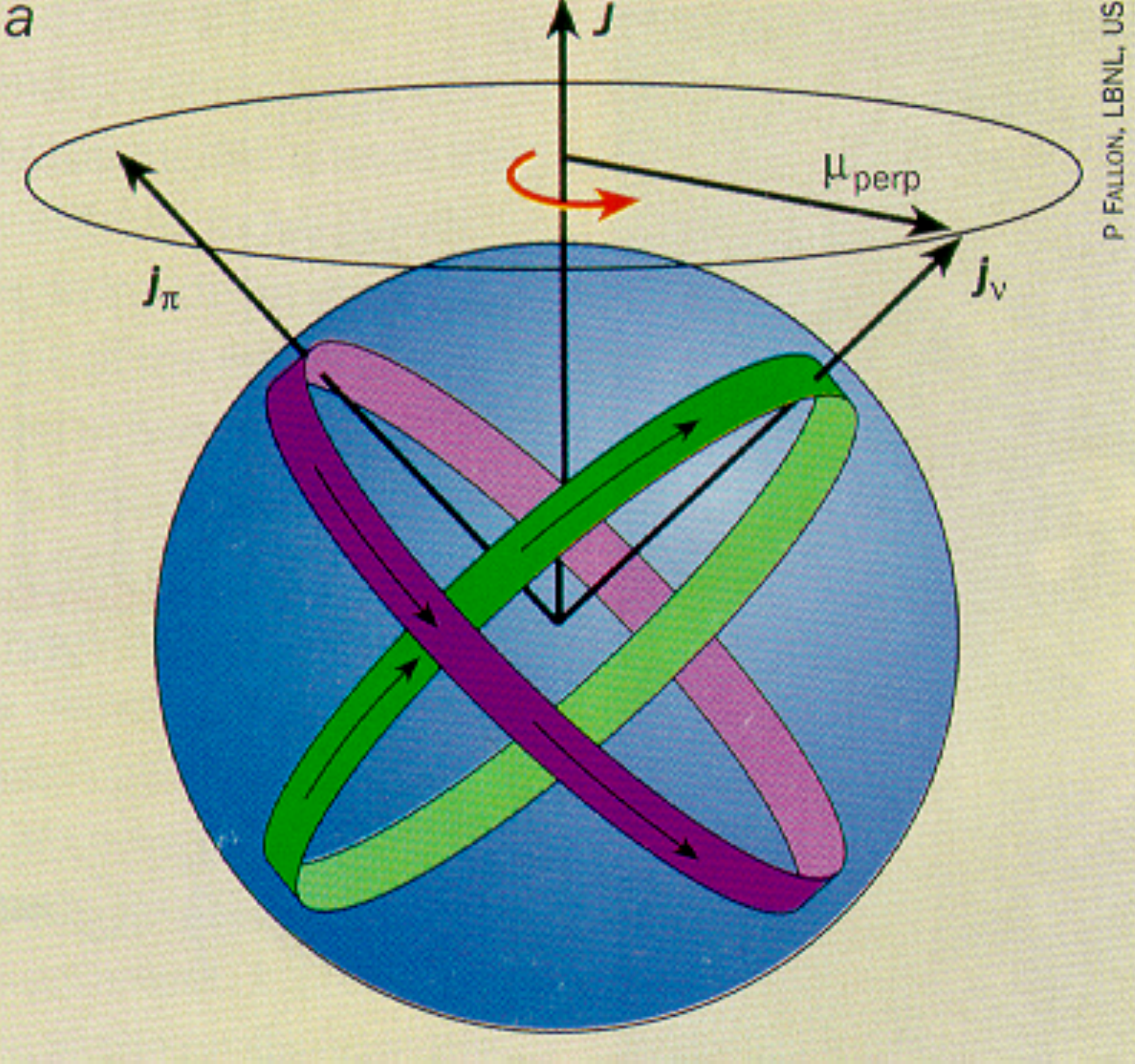}   
  \caption{\label{f:currents} 
 Magnetic rotation. The
high-$j$ proton particles ($\pi$)  and neutron holes ($\nu$) form current loops embedded in
the near spherical mass distribution of the nucleus. These current loops
as well as the associated transverse magnetic moment $\mu_{perp}$
allow one to specify the  angle $\psi$ of a rotation  around the
 axis  $\vec J$. The total angular momentum $J$ increases by the
gradual alignment of the  particle and hole
 angular momenta $\vec j_\pi$ and  $\vec j_\nu$.
 This is called the
Shears Mechanism. The interaction due to shape polarization  tries to
keep $\vec j_\pi$ and  $\vec j_\nu$ at a right angle, like the
 spring of a pair of shears.
 Figure from \cite{PhysWorld} }
     \end{center}
 \end{figure}

\begin{figure}[t]
  \begin{center}
\includegraphics[width=\linewidth]{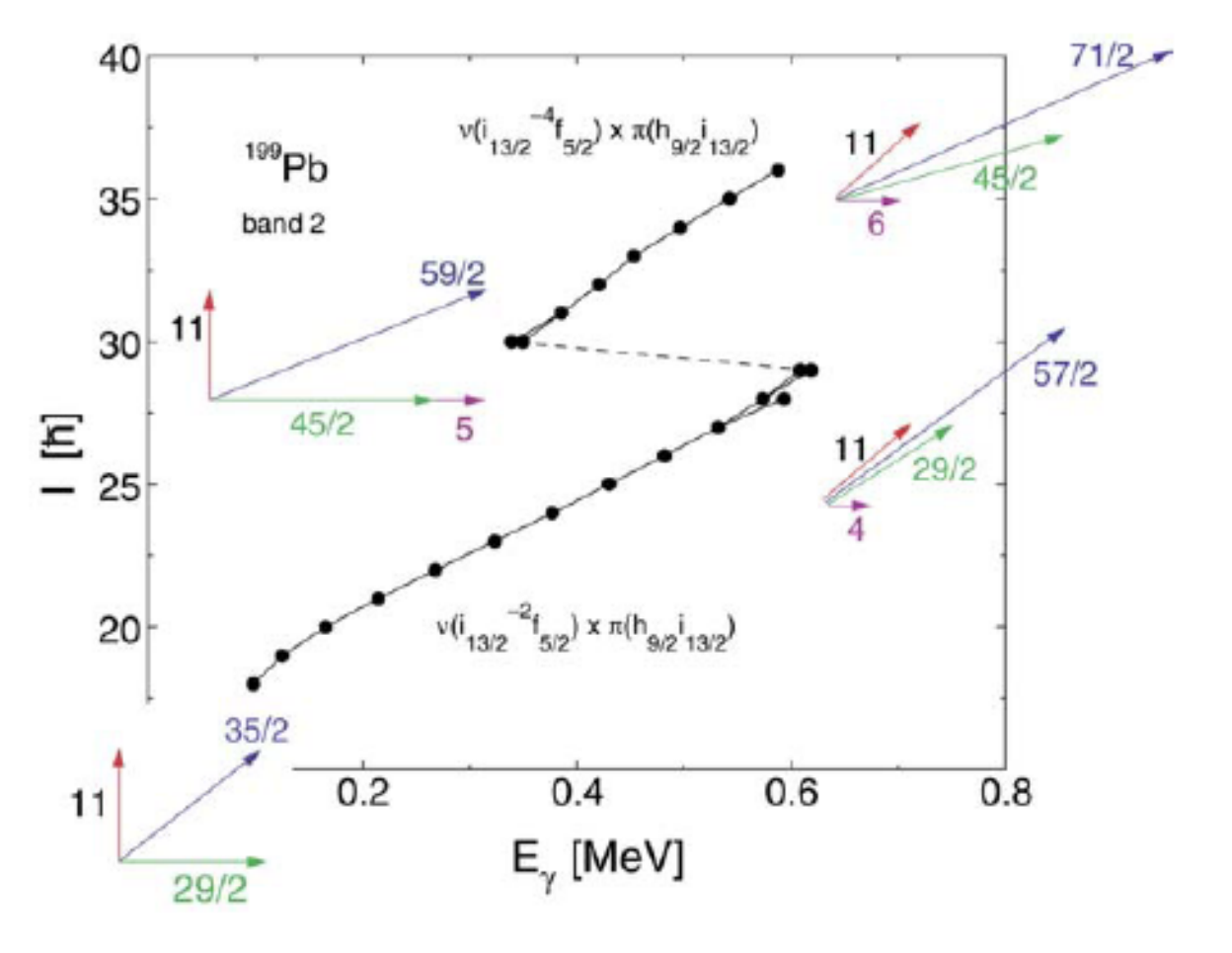} 
      \caption{\label{f:MRvectors} Angular momentum as a function of the $\De I=1$  transition energies
      $E_\ga=\hbar \om$ for band 2 in $^{199}$Pb. The arrows display the following \confsd:
       11 red $\left[\pi( h_{9/2}i_{13/2})_{11^-}\right]$, 29/2 green $\left[\nu( i_{13/2}^{-2})_{12^+}f_{5/2})\right]$,
45/2 green $\left[\nu( i_{13/2}^{-4})_{12^+}f_{5/2})\right]$.  No particle-hole character
is indicated for the $f_{7/2}$ \qnd. The total \am $\vec J$ is displayed by  black arrows and the
core \am $\vec R$ by purple arrows. 
             From Ref. \cite{magrevH}.}
     \end{center}
 \end{figure}

 Figs. \ref{f:currents} illustrates the structure  this novel rotor,
 which is an example that few high-j orbitals establish sufficient
 orientation such that quantal rotation emerges. 
 As a typical case let discuss the magnetic dipole band 2 in $^{199}_{82}$Pb$_{117}$, which is illustrated in
  Fig.  \ref{f:MRvectors} in more detail.
A pair of protons is excited across the $Z=82$ shell gap
into the \conf $[\pi\left( (h_{9/2}i_{13/2})_{11^-}(s_{1/2})^{-2}\right)]$,
where the subscript on the parenthesis quotes the angular momentum for stretched coupling of
the particles within the sub-shell. We simplify the notion for the \conf to $[\pi(h_{9/2}i_{13/2})_{11^-}]$ because 
$s_{1/2}$ holes do not contribute to the \amd. 
The proton \conf is combined with the neutron hole  \conf $[\nu\left( (i_{13/2}^{-2})_{12^+}(pf)^{-7}\right)]$.  
With seven  neutron holes in the $N=126$ shell the pair correlation must be taken into account. The  pertinent \qn \conf 
 includes the two $i_{13/2}$  and one $f_{5/2}$  excited  \qpsd. The remaining $pf$ holes form the core,
which is the \qp vacuum. In Fig. \ref{f:MRvectors}, we display the core \am 
as $\vec R$. In the following we use the  notation 
$[\nu (i_{13/2}^{-2})_{12^+}f_{5/2}]$, which indicates the hole character of the  $i_{13/2}$  \qnsd.

 The angular momenta of the high-j 
 protons  and of the high-j
 neutron holes, which  are separately lined up,  are
represented by the two arrows $\vec j_\pi$ and $\vec j_\nu$, respectively.
Since this arrangement breaks the ${\cal R}_z(\psi)$ symmetry,
a rotational band is the consequence, as observed. 
There is no ${\cal R}_z(\pi)$
symmetry in accordance with the observed $\De I =1$ sequences.
The separate configurations
$[\pi(h_{9/2}i_{13/2})_{11^-}]$ and
$\left[\nu( i_{13/2}^{-2})_{12^+}f_{5/2})\right]$, have ${\cal R}_z(\psi)$ symmetry.
They are well known isomeric states in the even-$N$ Pb isotopes, which are collected in Table \ref{t:PbmuQs}. 
No rotational levels are found on top of these states. The
 dipole bands appear only when the two structures are  combined including a substantial angle.   

The high-$j$ orbitals have  toroidal
density distributions, which are illustrated by the two loops.
The interaction between the particles and the holes is repulsive and
favors an angle of $90^o$, at which the two loops are as far from each other
as possible. Along the band,
the total angular momentum is increased by gradually aligning
$\vec j_\pi$ and $\vec j_\nu$. This process has been dubbed the "shears mechanism" \cite{mrb6}
because the motion resembles the closing of a pair of sheep-shears, which have a 
spring to keep them open. For this reason 
the magnetic dipole bands are also referred to as "shears bands" \cite{mrb6}.
Closing the blades of the shears  increases the energy  because
the loops are aligning. If the two blades are long it takes many
steps (increments of $J$ by 1$\hbar$) until the shears are closed. The energy
increases gradually, resulting in the observed smooth increase of the frequency
$\om$ ($=dE/dJ$) with $J$. The function $\om(J)$ and its derivative,
the inverse of the moment of inertia ${\cal J}^{(2)}$, are determined the
interaction between the high-$j$ orbitals, which will be analyzed below.

\begin{figure}[t]
  \begin{center}
 \includegraphics[width=0.8\linewidth]{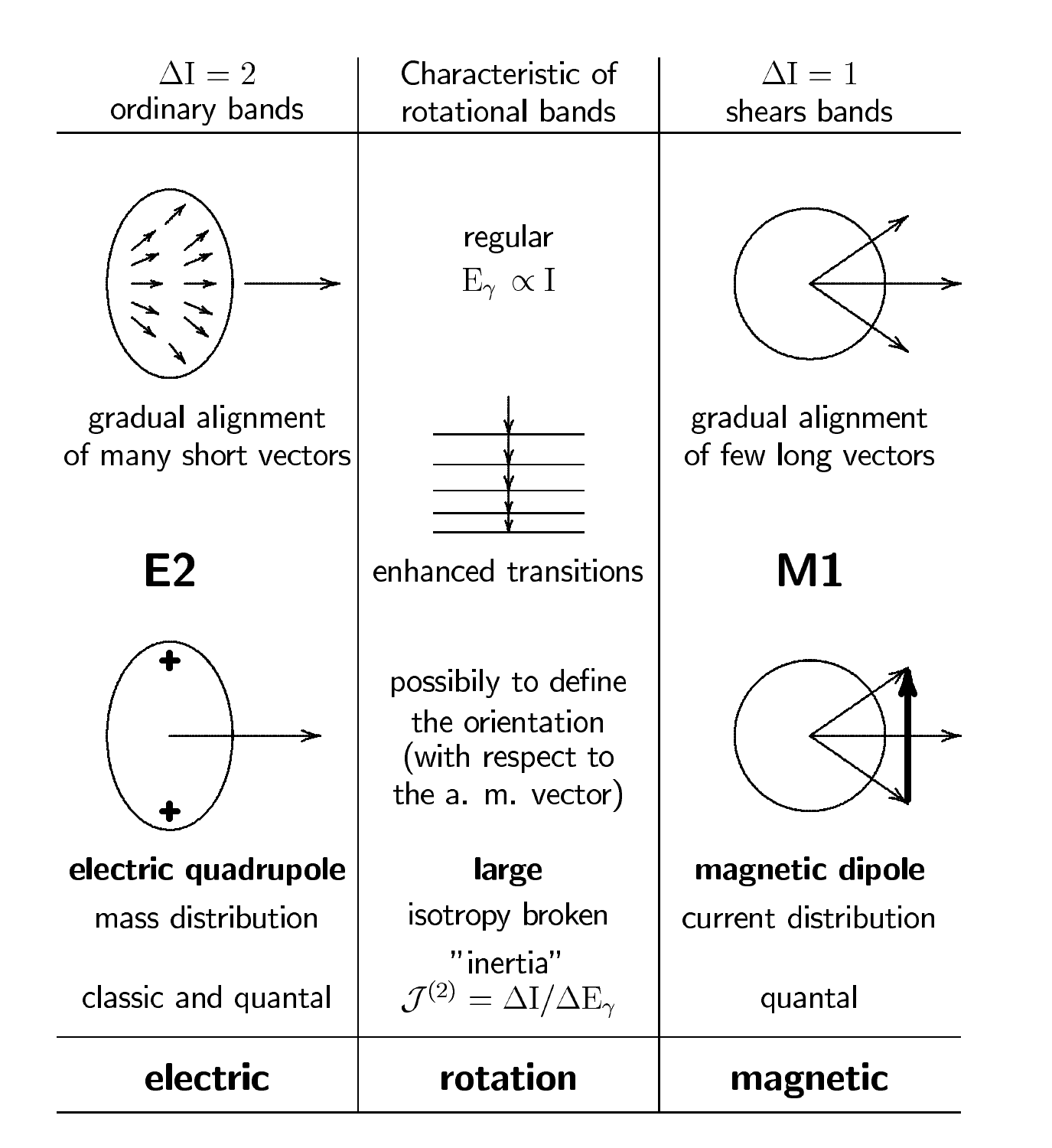}
   \caption{\label{f:MRER} The relation between electric and magnetic rotation. From Ref. \cite{Berkeley94}. }
 \end{center}
 \end{figure}

The shears arrangement of the high-$j$ orbitals gives
rise to a large transverse
magnetic dipole moment $\mu_\perp$.
The protons contribute an  orbital part and a spin
part ($g_p>0$) to the magnetic moment $\vec \mu$,
which align to $\vec j_\pi$. The spin part of the  neutrons is anti-parallel to  $\vec
j_\nu$ ($g_\nu < 0$). Thus all transverse components add up. 	It is
this long transverse dipole that rotates and  generates the strong magnetic
radiation. As seen in Fig. \ref{f:currents}, the rotating  transverse
magnetic moment decreases as the blades close, which means that the shears mechanism can be directly 
seen as a decrease of the $B(M1)$ values. 
 For this type of rotation,  the transverse magnetic dipole moment   
is a measurable quantity that  specifies the orientation angle. The rotating magnetic vector 
generates strong $M1$ transitions between  the adjacent members of the rotational band, which allows the experimentalist to 
arrange the transitions into a rotational sequence. In this sense it is an order parameter that indicates the coherence of 
the nucleonic motion. In case of ordinary rotation it is the large  
electric quadrupole moment that plays this role in generating the strong $E2$ transitions, which are use to identify a rotational band.
The analogy suggests adopting the  respective terminology "magnetic" and "electric" to distinguish between the two types of rotation \cite{Berkeley94},
 the relation of which is  illustrated by Fig. \ref{f:MRER}.

\begin{figure}[t]
  \begin{center}
 \includegraphics[width=0.405\linewidth]{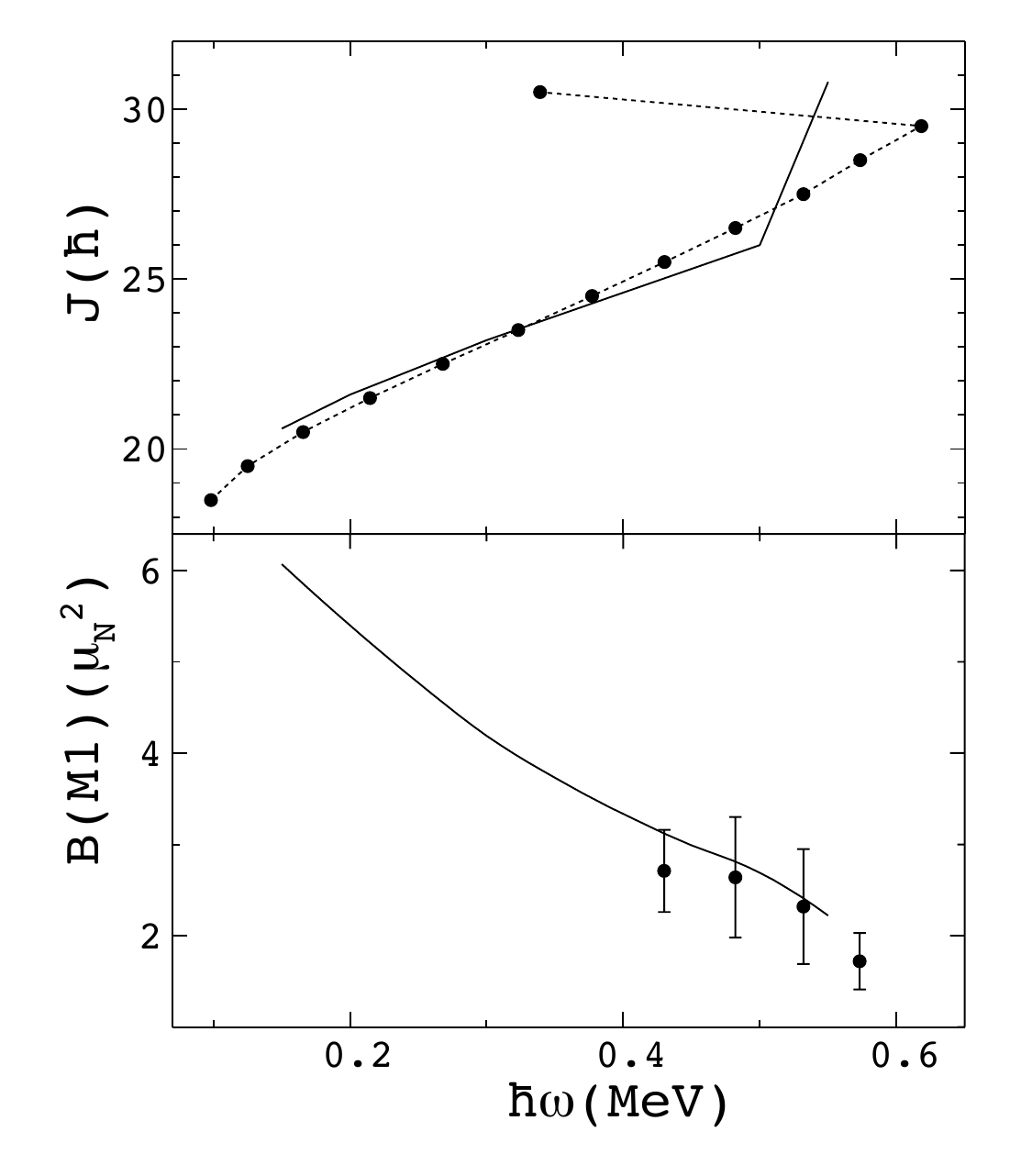} 
 \includegraphics[width=0.585\linewidth]{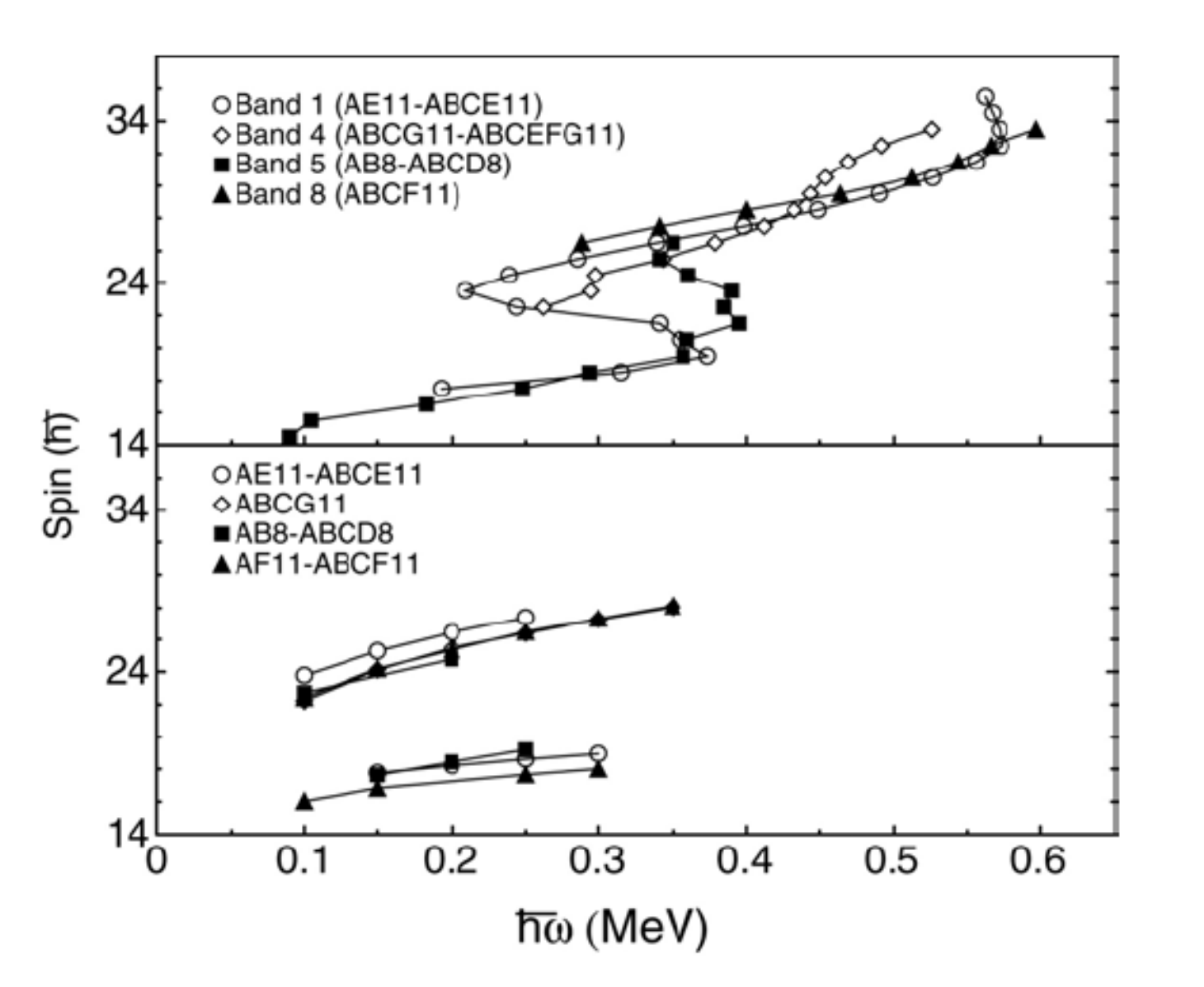}  
   \caption{\label{f:MRJBM1} 
Left panel: Angular momentum and $BM1$ values as functions of the rotational
frequency for the high-j \qp \conf
$[(\pi h_{9/2}i_{13/2})_{11^-})\times(\nu i_{13/2}^{-2})_{12^+}f_{5/2}]$
 in $^{199}$Pb (band 2). There is an additional $f_{5/2}$ \qn excited, which is not explicitly indicated.  Full 
lines are the tilted axis cranking  calculation.  Points are the experimental data.
The details of the tilted axis cranking  calculations are given in Refs. \cite{tac,mrb6}. From Ref. \cite{RMP}.\\  
Right panel: Experimental (upper panel) and calculated (lower panel) \am as a function of the angular frequency
of the positive parity bands in $^{196}$Pb. The Cranked Shell Model letter code is used, which denotes the $i_{13/2}$ \qns by A, B, C, D
and the $\pi=-$ \qns by E, F, G. The proton \confs $[(\pi h_{9/2}i_{13/2})_{11^-}]$ and $[(\pi h_{9/2}^2)_{8^+}]$  are denoted by 11 and 8. 
The tilted axis cranking  calculations are described in Ref. \cite{chmel07}.
 From Ref. \cite{magrevH}. }
     \end{center}
 \end{figure} 
 
  \begin{figure}[t]
  \begin{center}
\includegraphics[width=\linewidth]{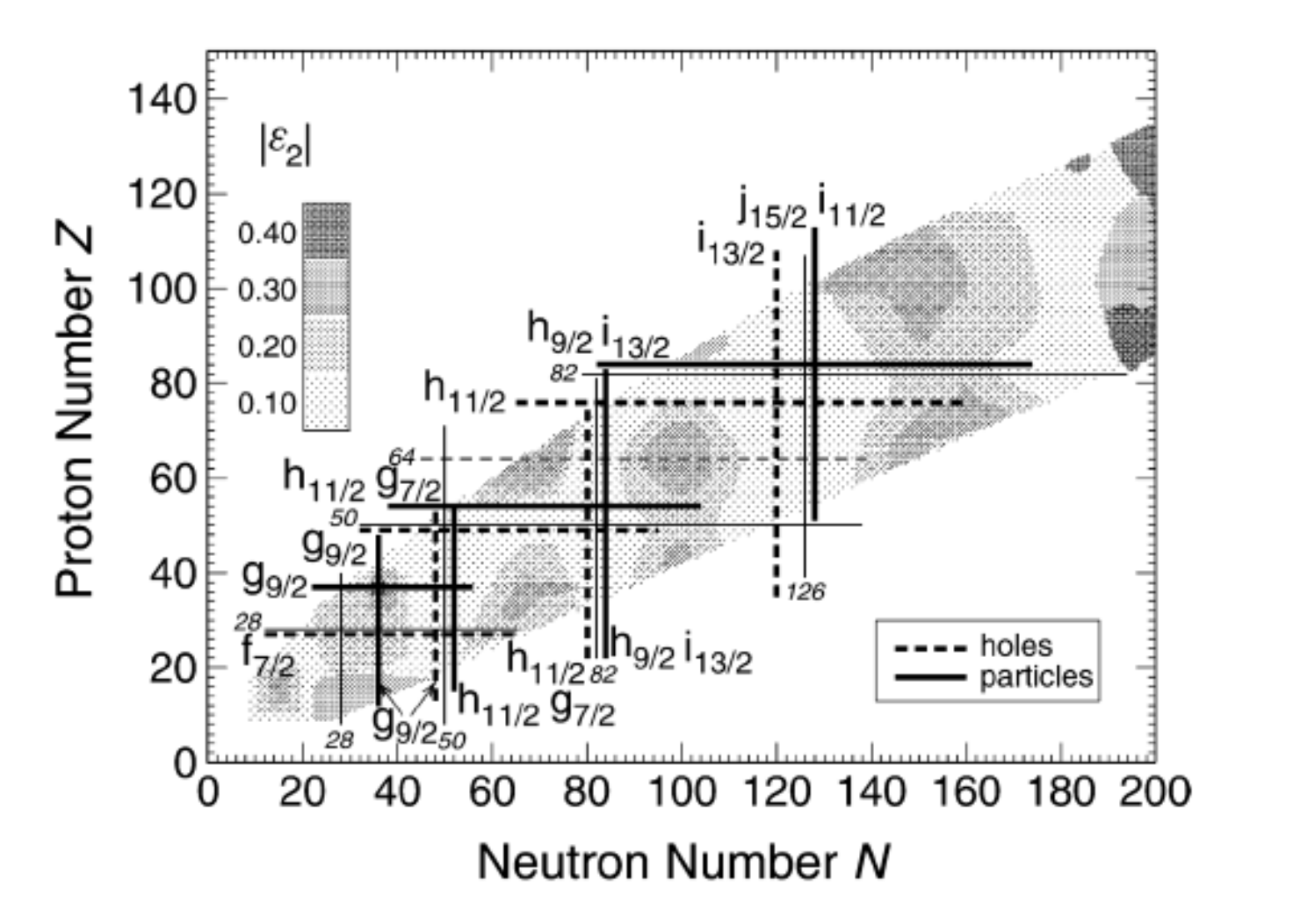} 
      \caption{\label{f:MRchart} Schematic diagram of regions of expected magnetic rotation. Full lines indicate the location of
high-j particles and dashed lines those of the high- j holes. The grey scale gives the deformation. The regions
where full and dashed lines cross are particularly favored for magnetic rotation
 From Ref. \cite{Berkeley94}.}
     \end{center}
 \end{figure}

Tilted axis cranking calculations  quantitatively
reproduce  the salient features of the dipole bands:
the nearly linear relation  between frequency and angular momentum, 
the large values of $B(M1)$, which decrease with \amd,  and the small values of $B(E2)$. As an example,
Fig. \ref{f:MRJBM1} left shows the tilted axis cranking  results  \cite{tac,mrb6,BM1199Pb} for the above discussed
high-j \conf
$[(\pi h_{9/2}i_{13/2})_{11^-}\times(\nu i_{13/2}^{-2})_{12^+}f_{5/2}]$
in $^{199}_{82}$Pb$_{117}$ (band 2).    In the frame of the  tilted axis cranking  approach, the  interaction between 
the high-j orbitals is due to
the slightly oblate  average potential, which is induced by them.
It also ensures the alignment
of the  protons to one blade and of the  neutron holes to the other.
As discussed in the  \ref{sec:semi},   
the high-j
holes tend to align their angular momentum with the symmetry axis,
whereas the high-j particles prefer  the perpendicular  orientation. 
The angle between the two blades $\vec j_\pi$ and $\vec j_\nu$
results from the balance between the inertial forces generated by
$-\vec \om \cdot \vec J$, which try to align the
two vectors with the axis of rotation $\vec J$ and the restoring force
of the slightly deformed potential, which
tries to keep the two loops at $90^o$. The opening angle
decreases with increasing $\om$, because the inertial forces get stronger.
The angular momentum vector $\vec J$ keeps an angle of 
about $45^o$ with the principal axes of the
density distribution.
It should be stressed at this point that the
substantial moments of inertia of the shears bands cannot be attributed to
the slight deformation. According to the
tilted axis cranking calculations the shears mechanism contributes two thirds
to the moment of inertia $\J2$, whereas only  one third
is due to the low-j particles, which generate the slightly deformed  
mean field.

The right  panel of Fig. \ref{f:MRJBM1} summarizes the functions $J(\om)$ of magnetic dipole bands observed in
$^{196}$Pb. The backbends appear when the shears  are  closed (or almost closed).    Then a  pair  
of neutron holes is excited which extend the length of the neutron blade. The shears open again releasing some of the 
stress exerted by the force. Fig. \ref{f:MRvectors} illustrates   this in-out-in  motion of the blades for band 2 in $^{199}$Pb.
 The red arrows show the
proton \conf $\left[(\pi h_{9/2}i_{13/2})_{11^-}\right]$. The green arrows  the  show the neutron \conf $\left[\nu( i_{13/2}^{-2})_{12^+}f_{5/2})\right]$
below the backbend and $\left[\nu( i_{13/2}^{-4})_{20^+}f_{5/2})\right]$ above.  
The left panel of Fig. \ref{f:MRJBM1} shows that the tilted axis cranking  calculations describe the  energies and $B(M1)$ values of band 2 
rather well.

Magnetic rotation is expected in other regions of the nuclear chart, where \mbox{high-j} proton particles combine with \mbox{high-j} neutron holes,
or vise-versa,  and the nuclear deformation is small. In Fig. \ref{f:MRchart} we show our early predictions of its appearance in
Ref. \cite{Berkeley94}.  By now \mr has been found in all predicted regions that are experimentally accessible (see reviews 
\cite{RMP,Amita00,magrevCM,magrevH,revMeng}). 
The  agreement between experiment and theory  in Fig. \ref{f:MRJBM1} is typical  
for tilted axis cranking  calculations in other regions, for which Table \ref{t:MRexamples} provides examples.
Meng and Zhao discuss tilted axis cranking  calculations
based on the relativistic mean field (RMF) approach in their contribution to this Focus Issue \cite{NCMengZhao}).

 The experimental indicators for magnetic rotation can be summarized as
follows.
\begin{enumerate}
\item A $\De I =1 $ sequence of strong magnetic dipole transitions, corresponding to $B(M1) >1 \mu_N^2$,
\item A smooth increase  of the $\gamma$ transition energy with  angular momentum,
\item A substantial moment of inertia of $\J2>0.2{\cal J}_{rig}$,
\item Small deformation of $\beta <0.15$,
\item Weak or absent quadrupole transitions, corresponding to  $\beta_t <0.10$, 
\item A  large ratio of $B(M1)/B(E2) > 10 (\mu_N/\mathrm{eb})^2$,
\item A large ratio \mbox{$RJB=\J2/B(E2) > 100\hbar^2 \mathrm{MeV}^{-1} (\mathrm{eb})^{-2}$} or $RJB/RJB(2^+)>10$.
\end{enumerate}
Here $RJB/RJB(2^+)$ is given by Eq. ({\ref{eq:RJB}).

 Table \ref{t:MRexamples}  exemplifies the essential  features of magnetic rotation from different mass regions.
The transition deformations $\beta_t$ expresses the quadrupole transition matrix element by the liquid drop deformation parameters (\ref{eq:LDR}-\ref{eqn-bohr-q})
   \beq\label{eq:betat}
  { \cal M}_{-2}(E2)=\frac{3}{4\pi}Ze(1.2\mathrm{fm})^2A^{2/3}\beta_t.
      \eeq
The $B(E2)$ values are proportional to the square of the transition deformation parameter $\beta_t$.
The values  listed in Table  \ref{t:MRexamples} are consistent with the equilibrium deformations  $\beta_{TAC}$ 
from tilted axis cranking  calculations. The reduction reflects the tilt angle $\beta_t=\beta_{TAC}\sqrt{3/8}\sin^2\vth$. 
To emphasize  the difference to the collective rotational mode of the Unified Model the ratio of the $RJB/RJB(2^+)$ is added to the table
(see discussion at the beginning of this section before Eq. (\ref{eq:RJB})). 
 
    \begin{table}[t]
\begin{center}
\begin{tabular}{|cccccccccccc|}
\hline
$A$&$Z$&$\#$&$\J2$&$\frac{\J2}{\Jr}$&$RJB$&$\frac{RJB}{RJB(2^+)}$&$B(M1)$&$B(E2)$&$\beta_t$&$\frac{B(M1)}{B(E2)}$&$\beta_{TAC}$\\
\hline
199&82&B&22&0.23&280&48&2.0&0.078&0.029&25&-0.10$^a$\\
139&62&G&12&0.23&43&5.3&1.5&0.29&0.094&5.2&0.10$^b$\\
142&64&J&22&0.29&220&28&1.1&0.10&0.052&11&-0.095$^c$\\
110&48&Q&13&0.37&130&11&3.8&0.10&0.083&38&0.12$^d$\\
108&50&B&32&0.94&800&78&1.0&0.041&0.051&24&-0.09$^e$\\
84&37&G&8.8&0.39&150&9.3&0.58&0.052&0.093&11&0.13$^f$\\
\hline
 \end{tabular}
 \end{center}
 \caption{\label{t:MRexamples} Examples of magnetic rotational bands in different mass regions. The column $\#$ quotes the label of the band in the ENSDF compilation
 \cite{ENSDF}. The ratio  ${\cal J}/{B(E2)}$ is denoted by $RJB$, its value for the 2$^+_1$ state is taken from the global expression (\ref{eq:RJB}).
  Units: ${\cal J}(\hbar^2$ MeV$^{-1}$), $RJB(\hbar^2$MeV$^{-1}$ (eb)$^{-2})$, $B(M1)(\mu^2_N)$, $B(E2)$((eb)$^{2})$. 
  The transition deformation is given by Eq. (\ref{eq:betat}).
  Data and tilted axis cranking  calculations are taken from Refs. $^a$\cite{BM1199Pb,chmel07},  $^b$ \cite{BM1139Sm}, $^c$\cite{BM1142Gd}, 
 $^d$\cite{BM1110Cd,RMP}, $^e$ \cite{BM1108Sn},   $^f$ \cite{BM184Rb}. }

\end{table}

Naturally there is a gradual transition between magnetic
and electric (collective) rotation and the limits are to some extent  arbitrary.
One may consider which are the important transitions that 
constitute the rotational sequence.  These are the $\De I=2$ electric quadrupole transitions for the electric (collective) rotation,
and the $\De I=1$ magnetic dipole transition for magnetic rotation. The branching ratio of a decay from a given level favors
 M1 if 
 \beq\label{eq:branching}
 \frac{B(M1, I\rightarrow I-1)}{B(E2, I\rightarrow I-2)}>2.8 (\hbar \om)^2,
 \eeq 
where the units MeV, $\mu_N^2$ and (eb)$^2$ are used.  With a frequency of $\sim 0.4 \mathrm{MeV}/\hbar$, which is typical for 
 the middle of the bands,
the ratio should be larger than 0.5 $(\mu_N/\mathrm{eb})^2$.  The ratios in Table  \ref{t:MRexamples} are much larger, 
which correspond to the strong suppression of the 
E2-transitions. The reason is  the small deviation of the charge density from symmetry with respect to the rotational axis. 
 The very small values of $\beta_t$ listed in Table  \ref{t:MRexamples} 
can be compared with the transition deformation of well deformed axial nuclei. For low-K ($<$2) bands $\vth\approx90^\circ$
$\beta_t=\sqrt{3/8}\beta=0.61\times0.3=0.18$, which corresponds to
$B(E2)=1.4 (\mathrm{eb})^2$. The gyromagnetic ratios $\vert g_j-g_R\vert$ are typically below 0.5$\mu_N$,
which  gives $\mu_3\sim  \vert g_j-g_R\vert K <1\mu_N$ and $B(M1)<0.12\mu_N^2$. Putting these estimates into Eq. (\ref{eq:branching}),
 one finds that the E2 transitions dominate for $\hbar \om>0.170$ MeV. Accordingly,
the bands have electric character, only few M1 transitions  are seen near the band head, if any. The signature partners  (A, B) and
(E, F) in Fig. \ref{f:Er163spec}  are examples.  High-K bands have a mixed character. The magnetic moment $\mu_3$ is larger and so the $B(M1)$ 
values. For example, the band K1 in $^{163}$Er has a ratio $B(M1)/B(E2)\approx1.5(\mu_N/\mathrm{eb})^2$ (see Fig. \ref{f:Er163BM1BE2}).
The M1- transitions are seen up to $I=55/2$ for the experiment shown in Fig. \ref{f:Er163spec}.
   
\subsection{Shears geometry}\label{sec:SG}

The simple geometry shown in Fig. \ref{f:MRvectors} and the left panel of Fig.  \ref{f:Eshears}  opens a more phenomenological
perspective taken by Macchiavelli {\it et al.} \cite{mrbe11,mrbe8,mrbe9} (see also Refs. \cite{RMP,magrevCM,magrevH}). We 
discuss it, slightly generalized, for the  shears band 1 in $^{199}$Pb, to which the \conf $[(\pi h_{9/2}i_{13/2})_{11^-}\times,(\nu i_{13/2}^{-3})_{33/2^+}]$
is assigned. The spinning clockwork is very  simple. Only two gyroscopes called the shears blades generate most of  the \am by gradually aligning.  
The two blades are composed of high-j orbitals in stretched coupling, $[(\pi h_{9/2}i_{13/2})_{11^-}]$ and $[(\nu i_{13/2}^{-3})_{33/2^+}]$. 
Experimentally, the band starts at $I_h=39/2$ and terminates at $I_t=59/2$.   It is assumed that the band terminates
when the two blades align.  The \am of the aligned blades  is $I=55/2$. The difference is attributed to the core of the remaining 6 $pf$ neutron holes.
 To keep things simple and guided by the tilted axis cranking  calculations it is assumed that core \am $\vec R$ is aligned with neutron blade and proportional to $\om$,
 \bea\label{eq:shearscore} 
 R(I)=\Theta_R\om(I),~~E_R(I)=\om(I)R(I)/2,~~\Theta_R=R_t/\om_t,\nonumber \\
 J=I+1/2,~~J_\nu=j_\nu+R(I)+1/2,~~J_\pi=j_\pi+1/2.
 \eea  
   For $\hbar\om(I)$ the experimental transition energies are taken, and $\hbar\om_t=0.554$ MeV is the energy of the last transition before the termination of the band.
  The terms 1/2 represent the common quantal
  correction to the classical expressions.  The geometry of the arrows in Fig. \ref{f:Eshears} (left panel) gives 
  \beq\label{eq:shearsangles}
  \cos \vth=\frac{J^2-J_\pi^2-J_\nu^2}{2J_\pi J_\nu},~~\sin\vth_\pi=\sin\vth\frac{J_\nu}{J},~~\sin\vth_\nu=\sin\vth\frac{J_\pi}{J},
  \eeq
  where $\vth$ is the angle between the blades and $\vth_\pi$ and $\vth_\nu$ the respective angles of the blades with $\vec J$. The value $R_t$ is taken such that 
  $\vth=0$ at the terminating spin. For the studied \conf $j_\pi=11\hbar$, $j_\nu=16.5\hbar$ and $R_t=2\hbar$.  
   The semiclassic expressions (\ref{eq:m1}-\ref{eq:delta}) 
  can be directly applied to the shears bands. The expressions become more illustrative  when re-expressing them in terms of the angles $\vth_\pi$ and $\vth_\nu$. 
  The two separate blades are observed as isomer states in the Pb-isotopes \cite{ENSDF}. Their static magnetic moments \mbox{$\mu_s=\langle II\vert\mu_z\vert II\rangle$}
  and quadrupole moments \mbox{$Q_s=\langle II\vert\hat Q_s\vert II\rangle$}  are measured, which are 
  collected in  Table \ref{t:PbmuQs}. 
     According to Eqs. (\ref{eq:mu},\ref{eq:Qstat}) ($\vth=0$) static moments are related to intrinsic moments  by $\mu_s=\mu/(1+1/2I)$ and $Q_s=Q/(1+3/2I)$.
        Expressed in terms of $\vth_\pi$ and $\vth_\nu$, Eq. (\ref{eq:bm1}) becomes
   \beq\label{eq:bm1SB}
   B(M1,\,I\rightarrow I-1)=\frac{3}{8\pi} \left(\sin\vth_\pi\mu_\pi-\sin\vth_\nu\mu_\nu\right)^2.
    \eeq
The left panel of Fig. \ref{f:199PbBM1BE2} shows that Eq. (\ref{eq:bm1SB}) reproduces well the 
experimental $B(M1)$ values with $\mu_{\pi}=10.5\mu_N$ and $\mu_\nu=-2.0\mu_N$
which correspond to $\mu_s=10.0\mu_N$ and $-1.8\mu_N$, respectively. 
  Although somewhat on the low side, the adopted values are consistent with the static magnetic moments of the isomers in Table \ref{t:PbmuQs}.

   \begin{figure}[t]
  \begin{center}
  \vspace*{1cm}
   \includegraphics[width=0.48\linewidth]{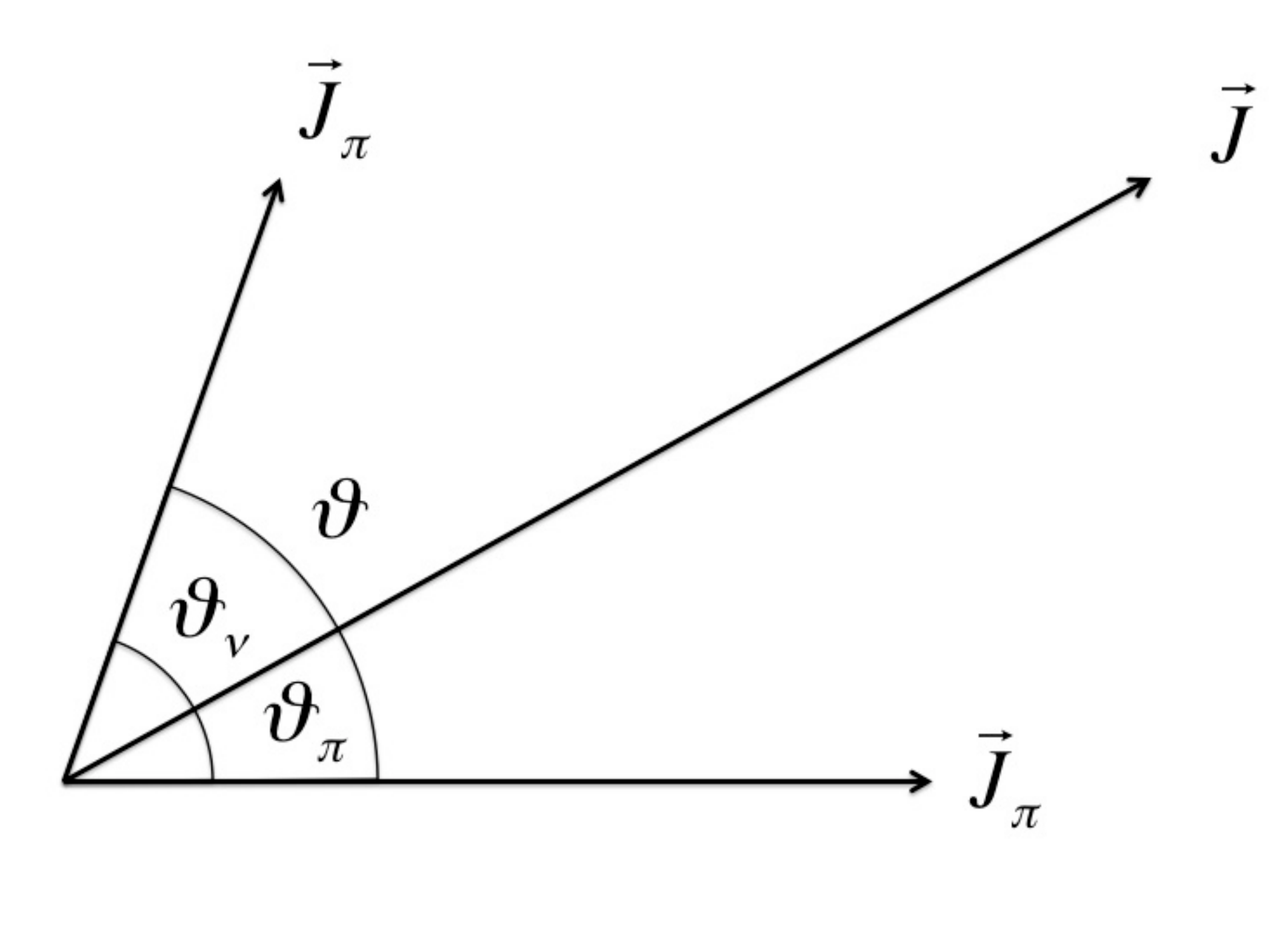} \includegraphics[width=0.48\linewidth]{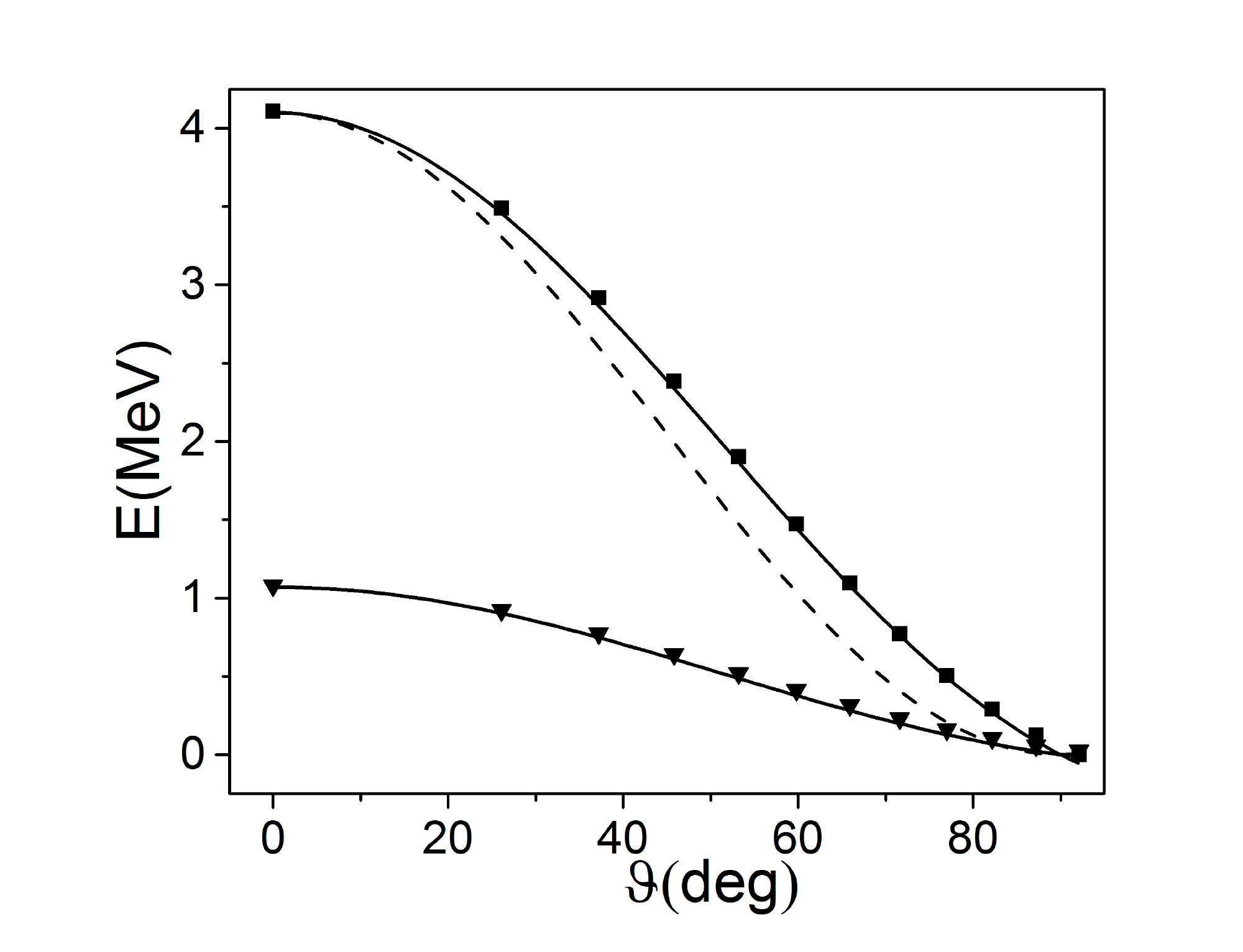}   
  \caption{\label{f:Eshears} 
Left panel: Geometry of the angular momenta in a shears band. 
Right panel: Experimental rotational energies (squares)  of the high-$j$ \qp \conf
$[(\pi h_{9/2}i_{13/2})_{11^-})\times(\nu i_{13/2}^{-2})_{12^+}f_{5/2}]$
 in $^{199}$Pb  (band 2) as a function of the shears angle $\vth$ calculated by means of Eqs. (\ref{eq:shearscore}) and (\ref{eq:shearsangles}).
   The squares correspond to
  the rotational rotational sequence in the left panel of Fig. \ref{f:MRJBM1}. The triangles show the core rotational energy. 
  The full curves show  fits
  of the expression $A\cos^2\vth+B\cos\vth$  to the points. The dashed curve shows a fit of the expression   $A\cos^2\vth$.
  
  }
     \end{center}
 \end{figure}

  \begin{figure}[t]
  \begin{center}
\includegraphics[width=0.48\linewidth]{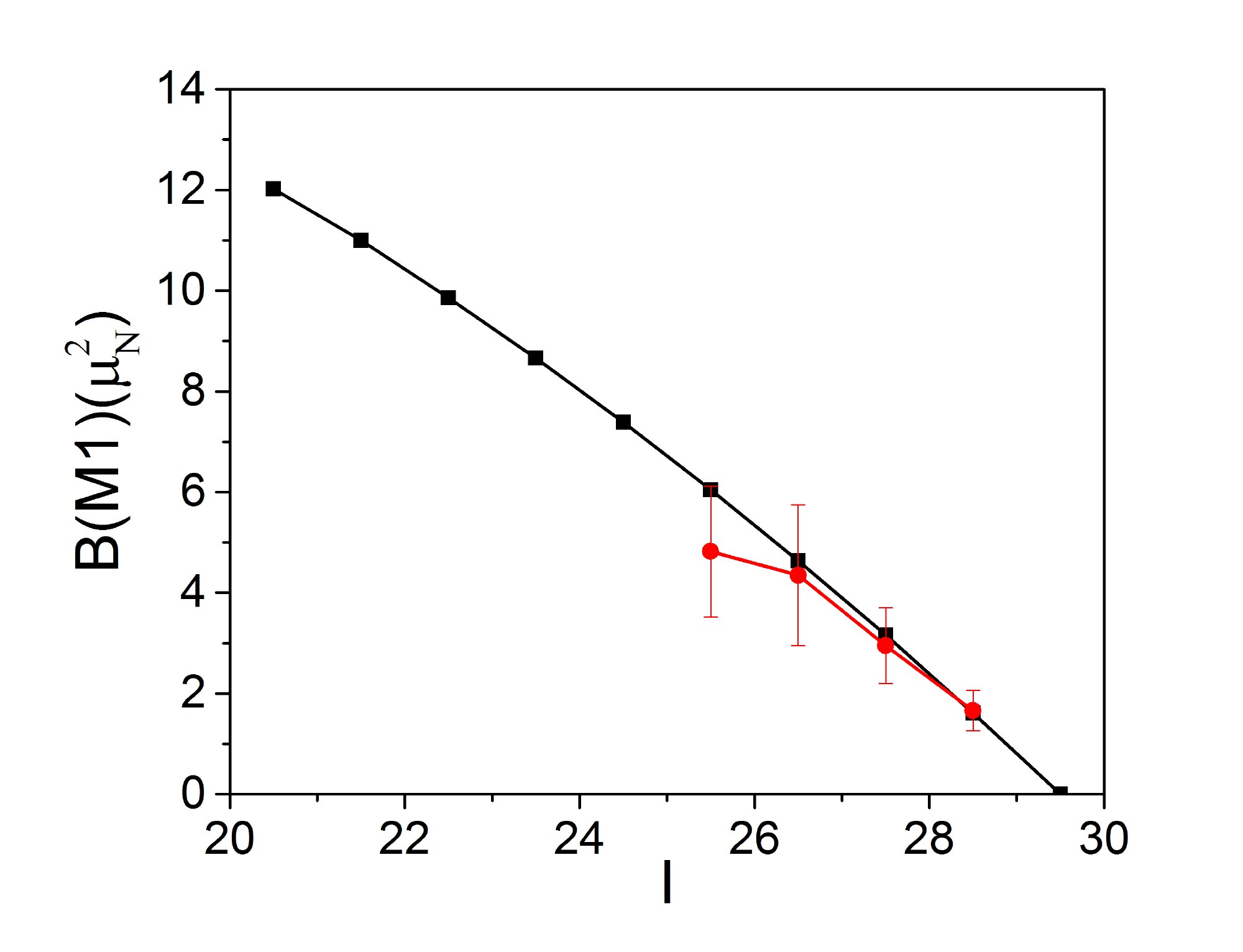} 
  \includegraphics[width=0.48\linewidth]{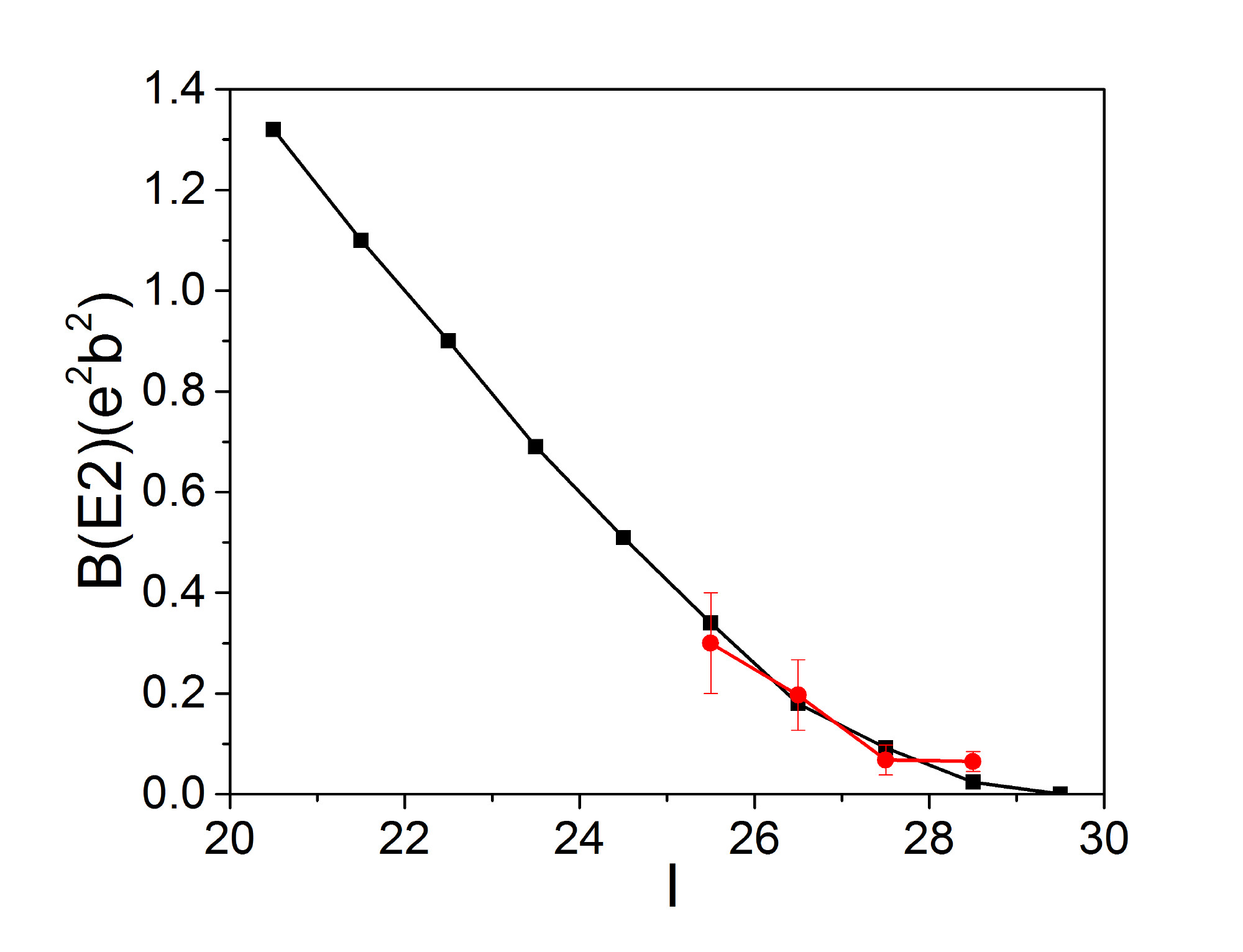}
    \caption{\label{f:199PbBM1BE2} 
Reduced transition probabilities of the high-$j$ \qp \conf
\conf $[(\pi h_{9/2}i_{13/2})_{11^-}\times,(\nu i_{13/2}^{-3})_{33/2^+}]$
 in $^{199}$Pb (shears band 1)  calculated by means of Eqs. (\ref{eq:bm1SB}) and (\ref{eq:be2SB}) compared with the data  
 from Ref. \cite{BM1199Pb}. }
     \end{center}
 \end{figure} 
 
    \begin{table}[t]
\begin{center}
\begin{tabular}{|ccccccc|}
\hline
&&$\left[\pi( h_{9/2}i_{13/2})_{11^-}\right]$&&&$\left[\nu( i_{13/2}^{-2})_{12^+}\right]$& \\
A&$E_x$(MeV)&$\mu_s(\mu_N)$&$Q_s$(eb)&$E_x$(MeV)&$\mu_s(\mu_N)$&$Q_s$(eb)\\
\hline
192&2.743&&-2.9(3)&2.624&-2.08(2)&0.32(4)\\
193&2.584+x$^a$&9.9(4)$^a$&-2.8(3)$^a$&$x\approx0.165^b$&-1.16(1)$^b$&0.19(1)$^b$\\
193&&&&2.612+x$^d$&-2.82(15)$^d$&0.45(4)$^d$\\
194&2.933&11.3(2)&-4.5(9)&2.628&-2.07(12)&0.48(3)\\
196&3.192&11.0(3)&-3.4(7)&2.694&-1.90(6)&0.65(5)\\
198&3.489&&&2.821&-1.86(2)&0.75(5)\\
199&&&&2.559$^c$&-1.076(3)$^c$&\\
199&&&&3.491$^d$&-2.39(15)$^d$&\\
200&3.182&&&3.005&-1.84(1)&0.79(3)\\
\hline
 \end{tabular}
 \end{center}
 \caption{\label{t:PbmuQs}  Energies, static magnetic and electric quadrupole moments of Pb isotopes.
The superscripts indicate \confs that deviate from  the heading:
\mbox{ a: $\left[\pi( h_{9/2}i_{13/2})_{11^-}\times \nu i_{13/2}^{-1}\right]$}, \mbox{ b: $\left[ \nu i_{13/2}^{-1}\right]$},
\mbox{c: $\left[\nu( i_{13/2}^{-2})_{12^+}f_{5/2})\right]$}, \mbox{d: $\left[\nu i_{13/2}^{-3})_{33/2^+}\right]$}.
Data from \cite{ENSDF}. If several measurements are quoted the weighted average is given. }

\end{table}

 The $B(E2)$ values in the right panel of Fig.  \ref{f:199PbBM1BE2} are calculated assuming axial  shape for the two blades and that
 the quadrupole moments of the blades add, i.e. $Q_{0}=Q_{0\pi}+Q_{0\nu}$ and $Q_{2}=Q_{2\pi}+Q_{2\nu}$. 
 Eq. (\ref{eq:be22}) can be re-written as
 \beq\label{eq:be2SB}
 B(E2,\,I\rightarrow I-2)=\frac{15}{32\pi}\left[Q_{0\pi}\sin^2\vth_\pi+Q_{0\nu}\sin^2\vth_\nu\right]^2.
 \eeq
 The right panel of Fig. \ref{f:199PbBM1BE2} shows that Eq. (\ref{eq:be2SB}) reproduces well the experimental $B(E2)$ values 
 with $Q_{0\pi}=-5.0$ eb and $Q_{0\nu}=0.9$ eb, which correspond to $Q_{s\pi}=$-4.4 eb and $Q_{s\nu}$=0.8 eb. 
 The proton values are consistent with the experimental static quadrupole 
  moments in Table \ref{t:PbmuQs} ($Q_{s\pi}=$-4.5 (9) eb ($A=$194)   and 
  $Q_{s\pi}=$-3.4 (7) eb  ($A=$196)).   The static quadrupole moment of the $33/2^+$    neutron \conf is not measured
  but it is expected to be similar to $Q_{s\nu}\approx0.8$ eb for the $12^+$ isomers in the even -$N$ neighbors (compare $A=$193 with 194).

The static magnetic and quadrupole moments of the band head $29/2^-$ of the shears band in $^{193}$ Pb are measured.
For the assigned \conf $\left[\pi( h_{9/2}i_{13/2})_{11^-}\times \nu i_{13/2}^{-1}\right]$   
Eq. (\ref{eq:shearsangles}) gives $\vth_\pi=26^\circ$ and  $\vth_\nu=47^\circ$ for the blade angles ($\Theta_R=0$ no estimate possible). 
Re-expressed in $\vth_\pi$ and $\vth_\nu$ Eq. (\ref{eq:mu}) reads 
\beq
\mu_s=\frac{I}{I+1/2}[\mu_\pi\cos\vth_\pi+\mu_\nu\cos\,\vth_\nu].
\eeq
 Using the  values $\mu_\pi=12.4\mu_N$ and $\mu_\nu=-1.2\mu_N$ derived from the static magnetic moments of the 
 blades in Table \ref{t:PbmuQs} ($A=194$ and 193), respectively) gives $\mu_s=9.9\mu_N$
 for the static magnetic moment of the isomeric band head, which agrees  with the experimental value $\mu_s=9.9(4)\mu_N$.
Assuming that the quadrupole moments of the blades  add up, Eq. (\ref{eq:Qstat}) becomes   
 \bea
 Q_s=\frac{I}{I+2/3}\Bigl[<Q_{0\pi}>\left((\cos\vartheta_\pi)^2-{1\over2}(\sin\,\vartheta_\pi)^2\right)+\nonumber\\
<Q_{0\nu}>\left((\cos\vartheta_\nu)^2-{1\over2}(\sin\,\vartheta_\nu)^2\right) \Bigr]. 
  \eea
 Combining the  values $Q_{0\nu}$=0.23 eb ($A=$193) with $Q_{0\pi}=$-5.1 eb ($A=$194) 
 and  -3.8 eb ($A=$196) derived from the static quadrupole moments of the 
 blades in Table \ref{t:PbmuQs}  gives the respective  the static quadrupole moments 
 -3.1 eb and -2.4 eb for the isomeric band head, which agree with with the experimental 
 value $Q_s=-2.8(3)$ eb within the experimental uncertainty.

The right panel of Fig. \ref{f:Eshears}  shows the experimental rotational 
energy of band 2 in $^{199}$Pb as a function of  the shears angle $\vth$. In this case the 
 blades  $[(\pi h_{9/2}i_{13/2})_{11^-}]$ and $\left(\nu i_{13/2}^{-2})_{12^+}f_{5/2})_{29/2}\right)]$
have the lengths $j_\pi=11\hbar$ and $j_{\nu}=14.5\hbar$.  In order to have $\vth=0$ at  terminating spin 59/2  the core \am is chosen $R_t=3.5\hbar$, 
which with the energy of the last transition  gives $\Theta_R=3.5\hbar$/(0.618 MeV/$\hbar$). 
The major part of the rotational energy is generated by the repulsive interaction between the particles in one blade and the holes in the other.
Macchiavelli {\it et al.}  \cite{mrbe11, mrbe9,mrbe8} suggested  that this  "bladon interaction"  
represents the interaction between the quadrupole moment of one blade and the 
quadrupole moment induced by the other blade. Such effective interaction mediated by the exchange of a quadrupole phonon depends on blade angle 
$\vth$ as 
\beq\label{eq:Vbladon1}
V(\vth)=V_2D^2_{00}(\vth)=(3\cos^2\vth-1)/2.
\eeq 
 The dashed line in the figure shows that the rotational energy changes roughly as $\cos^2\vth$. The strength $V_2$ is consistent within a factor of 2 with 
 experimental coupling strength of \qps with quadrupole vibrations in odd-A nuclei \cite{BMII,magrevCM}.  
 The same mechanism works in tilted axis cranking  description of the shears bands.  The slight oblate equilibrium deformation is
 induced by the high-j orbitals, and the rotational energy  appears as a consequence of the interaction of the high-j orbitals with the deformed 
 potential (see the discussion in \ref{sec:semi}).  A detailed discussion can be found in Ref. \cite{RMP}.
 
 As shown by the full line in Fig. \ref{f:Eshears}, the experimental rotational energy   
is  reproduced by the function \mbox{$E(\vth)=4.1(0.6 \cos^2\vth+0.4\cos\vth)$ MeV}. 
One quarter of the total is the core rotational energy, which is not far from the ratio 1/3  found by tilted axis cranking  calculations.  
The remaining part, the bladon interaction, is  described by 
  \beq \label{eq:Vbladon2} 
   V(\vth)=E(\vth)-E_R(\vth)=3.0(0.6 \cos^2\vth+0.4\cos\vth)\mathrm{ MeV}.
  \eeq
The curve $J(\om)$ from the tilted axis cranking  calculation for the band 2 shown in Fig. \ref{f:MRJBM1} can be well 
reproduced  by  \mbox{$V(\vth)=2.3(0.43 \cos^2\vth+0.57\cos\vth)$ MeV}, which suggests that the  
deviations of the energy from the pure $\cos^2\vth$ shape  can be attributed to additional mechanisms that are excluded from the simple phenomenological 
approach but taken into account in  the  tilted axis cranking  calculations. Candidates are  the interaction blades with the core, the possibility that the high-j constituents
of the blades are not rigidly coupled and pairing correlations.   
 Van Isacker and Macchiavelli \cite{IM13}  suggested another possibility. They  derived  a bladon interaction
 of the form $V(\vth)=(a+b\cos\vth)/\sin\vth$ by calculating the 
expectation value of a typical shears configuration composed of two high-j protons and two high-j neutron holes with short range interactions.    
The first term  has a minimum at $\vth=90^\circ$  as expected 
from the overlap of the wave functions of the blades. The second term shifts the minimum to a higher angle, as the  $\cos \vth$ term in Eq. (\ref{eq:Vbladon2}).

\setcounter{footnote}{0}
\subsection{Emergence of rotational bands}\label{sec:coherence}
 \footnote{  In order to simply notation, the \am $J$ is assumed to be measured in units of $\hbar$ in the present section \ref{sec:coherence}.  }
The nucleus increases its angular momentum in two different ways. One is coherent rotation of the nucleus, which results in regular rotational bands.
 The other is exciting  quasiparticles that align their individual angular momenta in an irregular way. Figs. \ref{f:Er164Ehigh} and \ref{f:Er163Ehigh} are 
 examples for the competition of the two modes.  As discussed in section \ref{sec:RMF}, the rotating mean field  accounts for both on equal footing. 
 In the present section we focus on the emergence of regular rotational sequences as a consequence of spontaneous symmetry breaking, which was already
 preliminary addressed in the preceding sections. To start, we realize that the quasiparticle routhian (\ref{eq:h'sp})
 derives from an effective two-body routhian that is invariant with respect to rotation about the $\vec \omega$ axis. 
 Rotational invariance implies that there is a family 
 mean field  solutions that break the rotational symmetry. They  are related by rotation about $\vec \omega$ by the angle $\psi$ and have the same energy. 
 In the space fixed coordinate system each of these mean field states 
 rotates  uniformly about the $\vec \omega$ axis. 
 This motion is 
 associated with the angle $\psi=\om t$ which specifies the orientation of the degenerate mean field solutions.  
 Hence, the orientation angle $\psi$ of the mean field  represents the microscopic realization of the angle variable of the collective rotor 
 wave function of the Unified Model. When the mean field is rotated   all the nucleonic orbitals (gyroscopes) are rotated by the same angle, i. e.
  their quantal states  change in a coherent way.  
Such  orientation angle $\psi$ and the collective wave function that lives on it exist only if the mean-field 
solution breaks rotational symmetry with respect to the   $\vec \omega$ axis.

 \begin{figure}[t]
  \begin{center}
\includegraphics[width=0.8\linewidth]{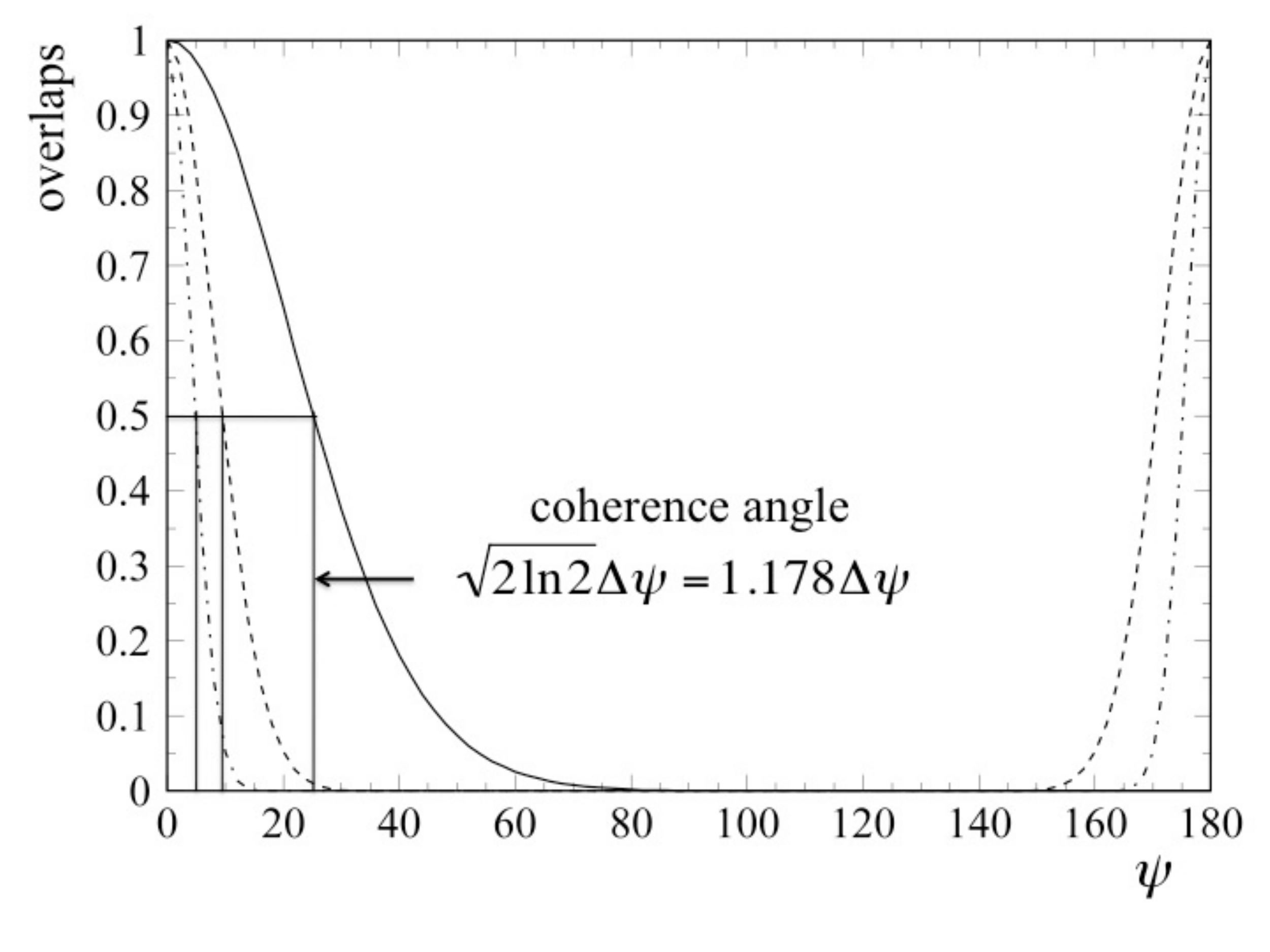} 
\end{center}
\caption{\label{f:overl}
The overlap $\vert{\cal R}_z(\psi)\vert$ for a shears band (full), a normal deformed band (dash) and a superdeformed band (dash-dot).
The details of the calculation  are given in Tab. \ref{t:crmr}.}
\end{figure}

  \subsubsection{Coherence length}\label{sec:cohl}
 
  To quantify the degree of symmetry breaking it is instructive to introduce the notation of a "coherence length",
  which is used in other fields of many-body physics. It is the minimal length that a collective wave function can resolve. In the case of superconductivity the coherence length
  $\xi=\hbar v_F/(\pi\De)$ is the size of a Cooper pair 
 \setcounter{footnote}{0}
  \footnote{
  For a more profound  discussion of the coherence length  see e. g. the textbook {\em Solid State Physics} 
  by Ashcroft and Mermin \cite{Ashcroft}. The Fermi velocity  $v_F=\sqrt{2e_F/m}$ is the velocity of the electrons at the Fermi surface $e_F$,
  which is the energy of highest occupied electron states.}.
  The wave function of the pair condensate cannot change more rapidly than $\xi$.  When the pair condensate flows through a wire,
  its wave function acquires the phase $ipx/\hbar$. The phase cannot change more rapidly than $\xi$, that is $p<p_{max}=\hbar/\xi$. When the  current through the wire
  is increased, superconductivity breaks down at the critical current density \mbox{ $j_{max}=e\rho p_{max}/2m=10^3-10^4 A/m^2$}.

In analogy, there exists a coherence angle $\Delta \psi$ that limits the  resolution 
  of the  wave functions of collective rotation, which I
   call "coherent" because the nucleons are correlated such that a common orderly motion results. The coherence angle can be determined from the overlap 
  between the different mean field solutions $\vert \psi\rangle$  that specify the angle $\psi$. 
As illustrated by Fig. \ref{f:overl},   
the overlap between two mean field solutions  rotated with respect to each
other by the angle $\psi$ about $\vec J$
can be well approximated by
a periodic Gaussian \cite{pga,Ringbook},
\beq \label{eq:overlf}
\vert\langle{\cal R}_z(\psi)\rangle\vert \approx \exp\left[-\frac{\sin^2(\psi)}{2\De \psi ^2}\right] ~
or~ \approx \exp\left[-\frac{2}{\De \psi^2}\sin^2
\left(\frac{\psi}{2}\right)\right],
\eeq
where the first expression holds when there is  no ${\cal R}_z(\pi)$  symmetry and the second when  ${\cal R}_z(\pi)\vert\rangle=\vert\rangle$.
Fig. \ref{f:overl} displays three examples: a superdeformed band, a normal deformed band similar to the ground band of $^{176}$Hf and a
shears band   in $^{199}$Pb (see sections \ref{sec:reg} and \ref{sec:3examples}).

The mean field state $\vert\rangle$ can be thought as a wave packet composed of the
angular momentum eigenstates $\vert I,M=I\rangle$ forming the band,
\beq\label{eq:decomp}
\vert \rangle=\sum c_I \vert I,M=I\rangle,   
\eeq
where $I=\alpha+2n$  when ${\cal R}_z(\pi)$ is good.
The periodic Gaussian form (\ref{eq:overlf}) implies a near Gaussian
distribution of the coefficients
\beq\label{eq:overli}
\vert c_I\vert \propto  \exp\left[-\frac{(I-J+1/2)^2}{2\De J^2}\right],
\eeq
which are the Fourier coefficients of the overlap function $\langle{\cal R}_z(\psi)\rangle$ (one-dimensional \am projection).
The width of the overlap $\De \psi $ is the inverse of 
the angular momentum width  $\De J$, because
\beq
\De \psi \De J\approx 1
\eeq
according to the uncertainty principle for angle and \amd. 
 In this way,  $|\rangle$
 represents a coherent state, which is  
a wave packet for which the product $\De p \De x $ of the momentum and
coordinate is as small as possible. It behaves as much as a classical object as 
permitted by the laws of quantum mechanics (see  Blaizot and Ripka \cite{blaizotbook}).

 The coherence angle sets the limit of how much phase $\psi I$   the rotational wave function $\exp(i\psi I)$ can  acquire. That is,  
  it restricts the number of states in the rotational band. The condition for emergence of a collective
  rotational wave function  can be written as
\beq \label{orientation}
\De\psi   < \pi~~~\mathrm{if}~~~{\cal R}_z(\pi)\vert\rangle\not=\vert\rangle~~~ \mathrm{or}~~~ \De\psi   < \pi/2~~~\mathrm{if}~~~{\cal R}_z(\pi)\vert\rangle=\vert\rangle.
\eeq
The coherence angle must be small enough to resolve a phase increment of a half   or a quarter  turn, respectively, 
to generate two different rotational wave functions.
 The complementary \am width $\De J\approx\De \psi^{-1}$ measures how much angular momentum can be generated from the
mean field state $\vert\rangle$. The \am width is also given by the dispersion 
\beq \label{disp}
\De J^2=\langle  \hat J_z^2 \rangle - \langle  \hat J_z \rangle^2 =
 \sum\limits_{ph} \langle ph|\hat J_z|0 \rangle^2.
\eeq
The  sum runs over the particle-hole or two-\qp excitations,
depending on the version of the mean field theory. 

In order to  elucidate some points, we consider a simplified model 
for the shears bands.  We assume that the rotating mean field  is composed of  the $I=10$ proton state $\vert 10,10, p\rangle$  
and  the $I=10$ neutron-hole state $\vert 10,10, n\rangle$,
which include the shears angle of $\vth$,
\bea
\vert \psi=0\rangle= {\cal R}_y(\vth/2)\vert 10,10, p\rangle
\times{\cal R}_z(\pi){\cal R}_y(\vth/2)\vert 10,10, n\rangle  \nonumber\\
=\sum\limits_m D_{10,m}^{10}(0,\vth/2,0)\vert 10,m, p\rangle \sum\limits_{m'} D_{10,m'}^{10}(\pi,\vth/2,0)\vert 10,m', n\rangle.
\eea
At the band head $\vth=\pi/2$. The overlap and the dispersion are given by 
\bea
\langle 0\vert\psi\rangle= \left [\sum\limits_m D_{10,m}^{*10}(0,\pi/4,0)D_{10,m}^{10}(\psi,\pi/4,0)\right ]^2,\label{eq:overlvm}\\
J=\langle J_z\rangle=2\sum\limits_m D_{10,m}^{10}(0,\pi/4,0)^2m,\label{eq:Jvm}\\
\langle J_z^2\rangle=2\sum\limits_m D_{10,m}^{10}(0,\pi/4,0)^2m^2.\label{eq:dispvm}
\eea 
Numerical evaluation of Eqs. (\ref{eq:Jvm},\ref{eq:dispvm}) gives $J=14.14\approx\sqrt{2} \times10$, as expected from geometry and   $\De J=2.23$ for the dispersion.
 The overlap (\ref{eq:overlvm}) 
is near-Gaussian with a Half Width at Half Maximum (HWHM) of 30.3$^\circ$, which corresponds to $\De \psi=0.53 $ and $\De J=1/\De \psi=2.23$ assuming 
genuine  Gaussians. The Fourier coefficients (projection on fixed $J=m+m'$) have a close-normal distribution 
centered around  $J=14.5$ with a right-side HWHM of 2.5, which corresponds to  $\De \psi=2.12$. 

The band terminates at $\vth=0$, when the mean field 
state becomes the $J=20$ state. For the simple model, termination appears after an increase of \am by  $\De I  \approx \pi\De J$. This may represent 
a general  estimate of the band length, provided the configuration does not change such that additional \am becomes available. 
The collective wave functions that describes the band above the band head $I=14$ oscillate on the angle interval $\lambda=2\pi/(I-14)$.
 At termination the mean field cannot any longer resolve the oscillations. This appears for $2 \pi/(\pi\De J)\sim2\De \psi$. That is, 
 collective rotational states emerge only up to $\lambda\sim2\De \psi$, which is about the FWHM of the Gaussian.   In case the mean field
 conserves signature (${\cal R}_z(\pi)$ symmetric) the resolution is reduced by a factor of 2, i. e. to the HWHM. The discussion above estimates the
 \am available at the band head to built the rotational sequence. Closing the shears along the band reduces the dispersion $\De J$, which is 
 zero for the terminating state $I=20$. Of course it has to be like this, because the difference $I-20$ shrinks.

\subsubsection{Regularity and collectivity}\label{sec:reg}
In the framework of the Unified Model the members of a rotational band have the same intrinsic state.
A less stringent condition for a rotational band is that its intrinsic structure changes
in an adiabatic way with the \amd. That is, the \qp states
gradually change with $\om$, and the mean field stays within the same $\om$-dependent configuration.   
 Let us formulate the condition  for similar intrinsic structure of adjacent levels
in a  quantitative way. Since we are
interested in small changes of the wave function
we may use  perturbation theory for comparing the adjacent states of the band.
When   the frequency is incremented
by $\De \om$, the state $|\om\rangle$ changes as
\beq \label{pert}
|\om +\De \om \rangle =  |\om \rangle+ \De \om \sum\limits_{ph} |ph \rangle
\frac{\langle ph|\hat J_z|0 \rangle }{e_p+e_h},
\eeq
and the total angular momentum increases as
\beq\label{DeltaJ}
J(\om+\De \om)=J(\om)+{\cal J}^{(2)}\De \om.
\eeq
The dynamical moment of inertia 
\beq \label{momi}
 {\cal J}^{(2)}=dJ/d\om=2\sum\limits_{ph}\frac{\langle ph|\hat J_z|0 \rangle^2 }{e_p+e_h}
\eeq
measures the local increment of the angular momentum with the frequency $\om$.
The  state $|\om (I+1)  \rangle$ has a structure  similar to
$|\om (I) \rangle$ if it differs only by particle-hole excitations
with small amplitudes
\beq \label{alpha}
\alpha_{ph}=\frac{|\langle ph|\hat J_z|0 \rangle| }{{\cal J}^{(2)}(e_p+e_h)} \ll 1.
\eeq
If this relation is fulfilled, $ {\cal J}^{(2)}(\om)$ will  change little from $I$
to $I+1$, and the spacing $\om$ between the levels $I$ and $I+1$ will be a smooth function of $I$.

If   the more stringent
 condition  $\De I\alpha_{ph}\ll 1$ holds for an interval $\De I \gg 1$,
the band is not only regular. 
  The relation 
between the spin and the level spacing will be nearly 
linear within the interval $\De I$,
because the nonlinear terms of a perturbation expansion of $J(\om)$
 are of higher order in $\alpha_{ph}$. The $I(I+1)$ rule of the Unified Model
follows from the assumption that the Hamiltonian is quadratic in the
angular momentum, which means a linear relationship between angular momentum and
frequency.

An overall measure of the structural similarity is the overlap
\beq\label{eq:overlo}
\vert\langle \om(I) |\om(I+1)\rangle\vert^2=1-D, ~~ 
D=\sum\limits_{ph}\alpha_{ph}^2.
\eeq
Its deviation from one , the overlap defect $D$, should be small as compared with 1.
The overlap defect is also invoked in the contribution by Nakatsukasa {\it et al.} to this Focus Issue \cite{NCNakatsukasa}.

One may quantify the degree
of collectivity by  counting the number of particle-hole excitations
in the sums (\ref{disp}) and (\ref{momi}), which is a measure how many single particle orbitals 
become active in generating one or two units of \amd.  There is the problem that the
sums contain  many tiny terms. In order to come up with a definite number one has to
set a lower limit for the matrix element $\langle ph|\hat J_z|0 \rangle^2$.
Table \ref{t:crmr} shows two cases. If the limit is set to 0.1,
 the truncated sums
 exhaust almost the full value.

\begin{table}[h]
\begin{center}
\begin{tabular}{|crrr|}
\hline
deformation&super &normal &weak\\
\hline
${\cal J}^{(2)}$&97&56&14\\
$\De J$&14&7.1&2.9\\
$HWFM $&$5^o$&9$^o$&25$^o$\\
$\De \psi $&0.087&0.13&0.37\\
$1/\De \psi $&11.4&7.5&2.7\\
$\alpha_{max}$&0.003&0.01&0.15\\
$D$&0.005&0.03&0.05\\
$Q_t$&5.2&2.6&0.7\\
$\mu_t$&0&0&3.5\\
\hline
$\langle ph|\hat J_z|0 \rangle^2 >0.1$&&&\\
${\cal J}^{(2)}$&96&52&11\\
$\De J$&14&6.6&2.0\\
$n_{ph}$&96&76&14\\
\hline
$\langle ph|\hat J_z|0 \rangle^2 >0.5$&&&\\
${\cal J}^{(2)}$&92&44&7\\
$\De J$&14&5.6&1.5\\
$n_{ph}$&58&22&3\\
\hline
$Z$&64&72&82\\
$N$&88&104&117\\
$\eps$&0.6&0.3&0.1\\
$\De_p$&0&0.75&0\\
$\De_n$&0&0.70&0.75\\
\hline
\end{tabular}
\caption{\label{t:crmr}
Upper panel: Character of the  different types of nuclear  rotational bands.
The listed quantities  are calculated by means of tilted axis cranking  at $\hbar\omega=0.3$ MeV with the  parameters in the lowest panel. 
The coherence length $\De \psi$
is derived from the numerical calculations of the overlap functions shown in Fig. \ref{f:overl}. 
The overlap defect $D$ is calculated for a $\De I=2$ transition in the case
of super and normal deformation and for a $\De I=1$ transition in the case
of weak deformation. 
The amplitude (\protect\ref{alpha}) of the strongest p-h transition
is given in the line $\alpha_{max}$.
Only the terms with $\langle ph|\hat J_z|0 \rangle^2$ larger
than  indicated are included in the sums  (\protect\ref{disp}) and 
(\protect\ref{momi}). The number of terms is given by $n_{ph}$.   
The moments of inertia are in  units $\hbar^2~\mathrm{MeV}^{-1}$,
the quadrupole moments in $eb$, the magnetic moments in $\mu_N$ and
the pair gaps in MeV. The transition quadrupole moment $Q_t$ is defined by Eq. (\ref{eq:Qt}).}
\end{center}
\end{table}

\subsubsection{Three examples}\label{sec:3examples}

Fig. \ref{f:overl} and Tab. \ref{t:crmr} compare the  indicators of rotational behavior for
 nuclei with super,
normal and weak deformation. As respective   examples we take  tilted axis cranking  calculations for
the yrast bands of $^{152}$Dy and $^{174}$Hf and  the  shears band 
in $^{199}$Pb discussed in section  \ref{sec:MR}.

\paragraph{Superdeformation}
The nuclei are  very well oriented. 
As qualitatively discussed in section \ref{sec:clockwork},
 they are much sharper oriented than one expects from
the anisotropy  of the density distribution.
The   superdeformed nucleus has an axis ratio of 2:1. 
Two density distributions with this axis ratio  still 
have an appreciable
overlap at a  relative angle of 90$^o$, whereas  the overlap of the mean
field states becomes already very small  at an angle of 10$^o$.   
This needle-like  behavior can be attributed
 to the nodal structure of the incompletely filled spherical
states which represents a strong element of anisotropy.
The number of nodes  of the wave functions determines
 the momentum of the particles. Hence one may say that also in the
case of well and superdeformed nuclei the symmetry breaking is primarily
due to the anisotropy caused by the momentum distribution of the particles at
the Fermi surface.  The picture of a spinning clockwork of gyroscopes (see section \ref{sec:clockwork}) visualizes 
the anisotropic  momentum distribution in a schematic way.

Alternatively, one may invoke the 
stretch picture \cite{stretch}, which  
separates the  particles into two groups, one with $j_1>0$ and the other with   
$j_1<0$.  
Each group generates a strong current pattern in the 2-3 plane, which represents the
element of anisotropy. The fact that the net current in the 2-3 plane
is zero is not relevant for the orientation. 

 The dispersion $\De J=14$ indicates that the superdeformed mean field supports a rotational sequence of $\De I \sim 40$. Very regular 
 superdeformed bands that stretch over this interval are quite commonly observed in the $A=150$ region.  
 The overlap defect $D=0.005$ is very small.  Superdeformed nuclei come   closest to the 
 assumption  $D=0$ of the Unified Model.
The small value of $\alpha_{max}=0.003$ ensures that the terms in expression  (\ref{pert}) remain small enough, such  that
the linear relation  $J(\om)={\cal  J}\om$ holds 
over an extended $\om$ range. Indeed, the \momi ${\cal J}^{(2)}$ of  the lowest rotational band in superdeformed  $^{152}$Dy
changes only by 10\% over the observed \am range of $\De J=38$ (see e. g. picked fence spectrum in Fig. 5 in the contribution to this Focus Issue
by M. A. Riley, J. Simpson and E. S. Paul \cite{NCRiley}).       
These features are in line with the large number $n_{ph}=96$ of terms in the sums (\ref{disp}) and (\ref{momi}).
In order to generate one or two units of \amd, very many orbitals align their individual \am with the rotational axis  by  tiny amounts.
This is genuine collective rotation.

The quadrupole moment of the charge distribution with respect
to the axis of rotation $Q_t$ is  large.
The rotation has  electric character because 
it is the asymmetric charge distribution that 
rotates and generates the strong $E2$-radiation connecting
the members of the band.

\paragraph{Normal deformation}
 The larger overlap defect $D=0.03$ and 
the larger value of $\alpha_{max}$ imply a shorter   \am interval within which function  $J(\om)$ is approximately linear. 
Indeed, the  experimental \momi ${\cal J}^{(2)}$ of  the ground band of $^{164}$Er increases by 10\% 
between $I=0$ and $I=10$ (see Fig. \ref{f:Er164Elow}).
The number $n_{ph}=22$ indicates a good degree of collectivity.
 According to the dispersion $\De J=7$ the
 mean field supports a rotational sequence of $\De I \sim 20$.  
  At $I=12$, well  before the termination estimate, the ground band in even-even nuclei is crossed 
 by the s-band, which contains two aligned high-j quasiparticles.   One may consider
 such a band crossing as a first response  to  stress toward termination. 
 Termination is observed in nuclei with $N$ around 88  
 at $I_t\approx 40$ (see Fig. 2 in the contribution to this Focus Issue by M. A. Riley, J. Simpson and E. S. Paul \cite{NCRiley}). 
 Before termination the rotational alignments  of two i$_{13/2}$ \qns and two h$_{11/2}$ \qprs  generate $I_{qp}\approx20$ units of \amd. 
 The difference of $I_t-I_{qp}\approx20$ is generated by collective rotation,   which well agrees with estimate of the amount of \am 
 that can be carried by the deformed mean field.

\paragraph{Weak deformation}
The example is the magnetic rotational  band 2
in $^{199}_{82}$Pb$_{117}$ (band 2) with the \conf  $[(\pi h_{9/2}i_{13/2})_{11^-}\times(\nu i_{13/2}^{-2})_{12^+}f_{5/2}]$.
 Although  the value of $\Delta \psi$ is two and a half times
 larger than for a normal deformed nucleus,  the
 nucleus is still sufficiently well oriented to develop quantal rotation.
 The $i_{13/2}$    and $h_{9/2}$ protons and two  $i_{13/2}$ 
quasi-neutrons with hole character contribute 5.1 to  $\De J^2$,
 the remaining 3.3 come from the
 low-$j$  neutrons in the $fp$ orbitals. Thus  most of what
is going round are the four high-$j$ orbitals, which form the current loops
in Fig. \ref{f:currents}.
The  value of $Q_t$ reflects the  almost symmetric distribution of   charge
 with respect to the axis $\vec J$. The rotation
has magnetic character, because it is the magnetic dipole that goes
round  generating the observed strong $M1$ transitions which connect the
members of the band.

The dispersion  $\De J\sim 3$ suggests  that mean field supports a rotational
sequence of $\De I \sim9$. The observed shears  band 2 in $^{199}$Pb  extends from 35/2 to 59/2, where it terminates.
The difference of $\De I=12$ somewhat exceeds the estimate. 

The value $\alpha_{max}=$0.15 ensures the regularity of the
band, but it is too large to imply a linear relation
$J(\omega)$ over many transitions.
The largest amplitudes $\alpha_{ph}$ belong to the four
high-$j$ quasi-particles, which contribute 10$\mathrm{MeV}^{-1}$ to the total moment
of inertia ${\cal J}^{(2)}$.
This part of the
angular momentum is generated by the shears mechanism,
the energetics  of which 
 are discussed in  section \ref{sec:MR}.
The low-$j$ neutrons  in the $fp$ orbitals
contribute $4\mathrm{MeV}^{-1}$ to  ${\cal J}^{(2)}$. Because of their small particle-hole  amplitudes $\alpha_{ph}<0.02$,
they add a linear contribution to $J(\om)$.

With $n_{ph}=14$, magnetic rotation is much less collective than the rotation of the normal and 
superdeformed nuclei. Nevertheless, it fulfills the criteria for
 rotational bands and shows up as such.
If the limit on $\langle ph|\hat J_z|0 \rangle^2$ is increased to 0.5, there are only  3 terms,
which come from the high-$j$ particles and holes. They account for 
half of ${\cal J}^{(2)}$ and  $\De J^2$. This part is not very collective
indeed. The other half comes from nine $fp$ neutron terms and two high-$j$
terms, which are below the limit. These numbers illustrate 
qualitative statement made in section
\ref{sec:MR} that \mr
consists of the rotation of few high-$j$ current loops accompanied
by some collective rotation of the core.       

{\em The examples  indicate that substantial coherence does not necessarily imply strong collectivity in the above sense.}
Traditionally the two notations are used synonymous. The decisive criterion is a sufficiently short coherence length $\De \psi$  which supports the
phase change of the rotational wave function.   The complementary \am width $\De J$ limits the number rotational states  $\De I\approx 3\De J$
 that can be generated from the 
mean field state before the band terminates.
As already pointed out, termination is only observed under favorable circumstances. The band may be crossed by 
other bands. 
Along the band, the  pairing correlations may decrease and   nuclear shape may change  such that  $\De J$ increases, which postpones termination.

  \begin{figure}[t]
\centerline{\includegraphics[width=0.5\linewidth]{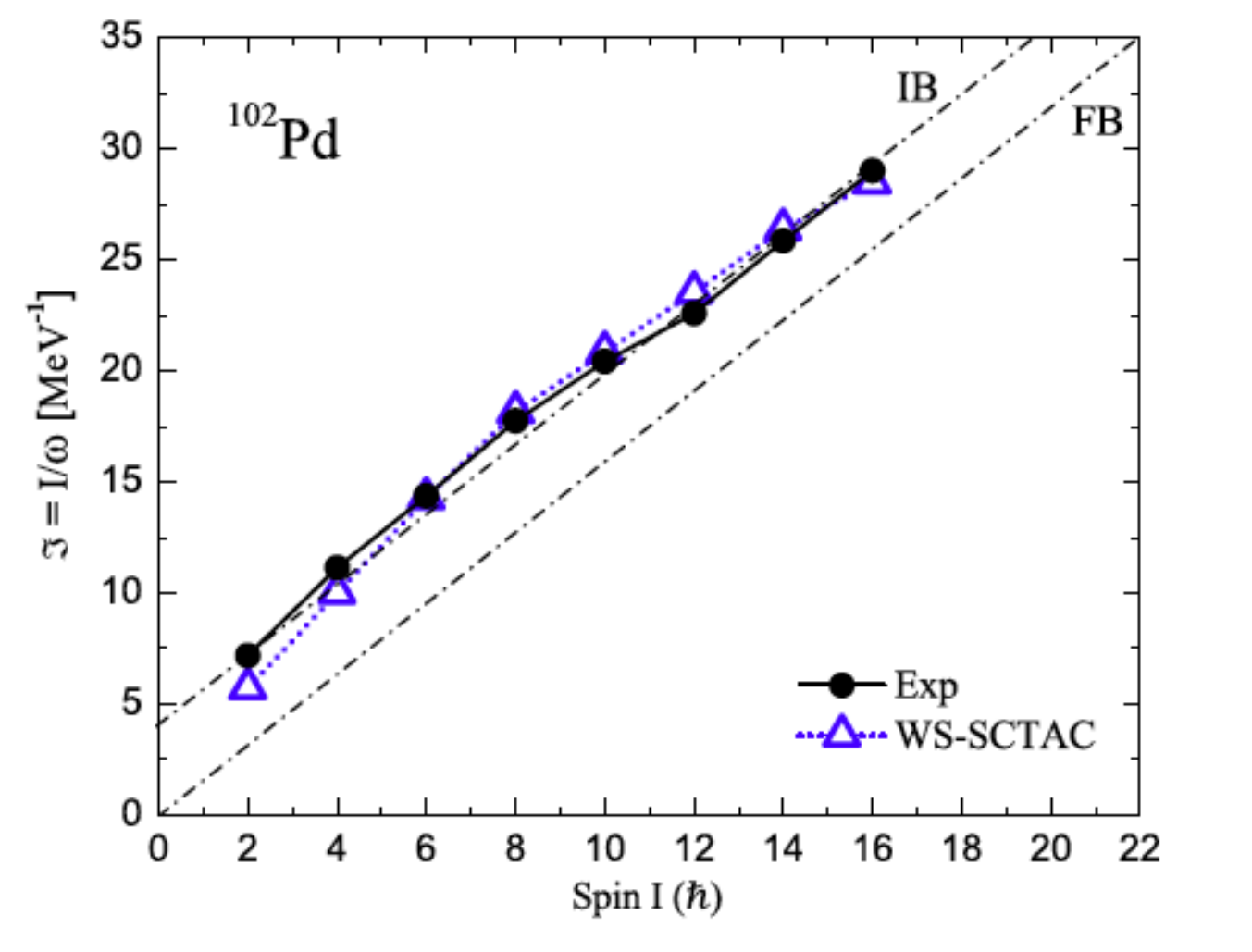}\includegraphics[width=0.5\linewidth]{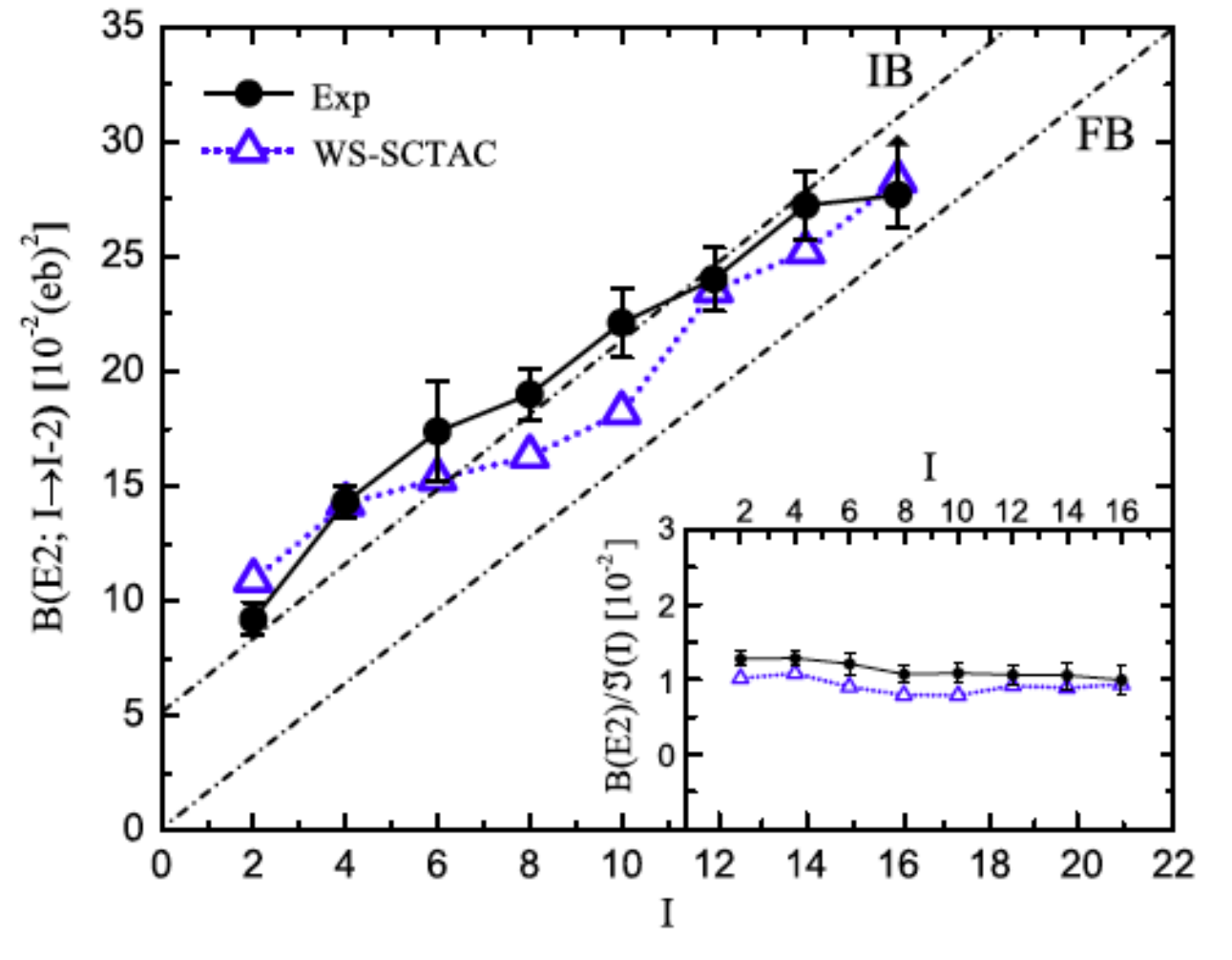}}
\caption{\label{fig:102PdPRL} 
Left panel: Experimental moments of inertia of the ground band 
of $^{102}$Pd, where \mbox{$\omega=(E(I)-E(I-2))/2$} and ${\cal J}^{(1)}=I/\om$.  Right panel:  Experimental $B(E2, I\rightarrow I-2)$ values of the
ground band of $^{102}$Pd.
The blue lines WS-SCTAC show the calculations  \cite{FGS11} by means of the cranked shell correction method using a Woods Saxon potential
 (see section \ref{sec:SCPAC}). The acronym FB stands for free bosons. It 
denotes the limit of harmonic quadrupole oscillations described by Eqs. (\ref{eq:TidalHV}). The acronym IB stands for interacting bosons. 
It denotes a  deviation from the harmonic limit that can be ascribed to a quadratic interaction between the bosons. The inset shows the ratio
$B(E2,I\rightarrow I-2)/{\cal J}^{(1)}$ in units $\hbar^{-2}\mathrm{MeV(eb)}^2$.
 Taken from Ref. \cite{Frauendorf15}.  } 
\end{figure}

 \subsection{Tidal Waves}\label{sec:tidal}
 
 The Unified Model treats vibrations and rotation on equal footing assuming that they are well decoupled from the intrinsic degrees of freedom. 
 However as illustrated by Fig.  \ref{f:HV-qp}, the collective states near 
 the yrast line are much less coupled to the \qp background than the states of small \am at the same energy. In other words, the yrast states of a vibrational multiplet
 are much less damped than the low-spin members. The Unified Model misses this aspect because it disregards damping. The weak damping of the vibrational yrast states 
 allows one to describe them in the framework of the rotating mean field,  which is  
 Tidal Wave concept introduced by Frauendorf {\it et al.}  \cite{FGS11} (and Ref. \cite{FGS10} with  complimentary material). 
 
 Consider the phenomenological  Bohr Hamiltonian    (\ref{eq:BH}-\ref{eq:momiIF}) with inertial parameters  $B_{\beta\beta}=B_{\gamma\gamma}=B_i=\sqrt{5}/2B$.  
 Uniform rotation about the  axis with the maximal moment of inertia has the lowest energy for a given angular momentum, i. e. it corresponds to the yrast states when quantized. 
In the co-rotating frame, the deformation parameters $\beta$ and $\gamma$ do not depend on time. 
 Their values are given by minimizing the energy 
 \begin{equation}\label{Ebeta}
E(\beta,\gamma)=\frac{J^2}{2{\cal J}(\beta,\gamma)}+V(\beta,\gamma),~~
{\cal J}=4B\beta^2\sin^2\gamma
\end{equation}
at given \am $J=I\hbar$.
In the case of a harmonic vibrator  $V=\frac{C}{2}\beta^2$. Minimizing the energy one finds
\bea\label{eq:TidalHV}
\gamma_e=\frac{\pi}{2},~~~ \beta^2_e=\frac{J}{2\sqrt{BC}},~~~ {\cal J}=4B\beta^2_e=2J\sqrt{\frac{B}{C}},\nonumber \\
\omega=\frac{J}{{\cal J}}=\frac{1}{2}\sqrt{\frac{C}{B}}=\frac{1}{2}\om_V,~~~E=\omega J=\om_V\frac{J}{2}=C\beta^2_e.
\eea
The wave travels with the angular velocity $\omega$ being one half of the oscillator frequency
$\om_V$. The angular momentum is generated by increasing the deformation $\beta^2$, while the angular velocity stays constant.
These are the yrast states  of the vibrator multiplets described in a semiclassical way. 
Frauendorf {\it et al.} called the mode  "Tidal Wave"  \cite{FGS11}, because it has wave character: the energy and angular momentum increase with the wave amplitude while
the frequency stays constant. Using a suitable  potential   one can easily incorporate 
unharmonicities and cover  the transition to stable rotation. Bohr and Mottelson discussed the preceding
 in more detail in Appendix 6B-3 (Yrast Region of Harmonic Vibrations) of their Monograph \cite{BMII}. 

 \begin{figure}[h]
\centerline{\includegraphics[width=0.48\linewidth]{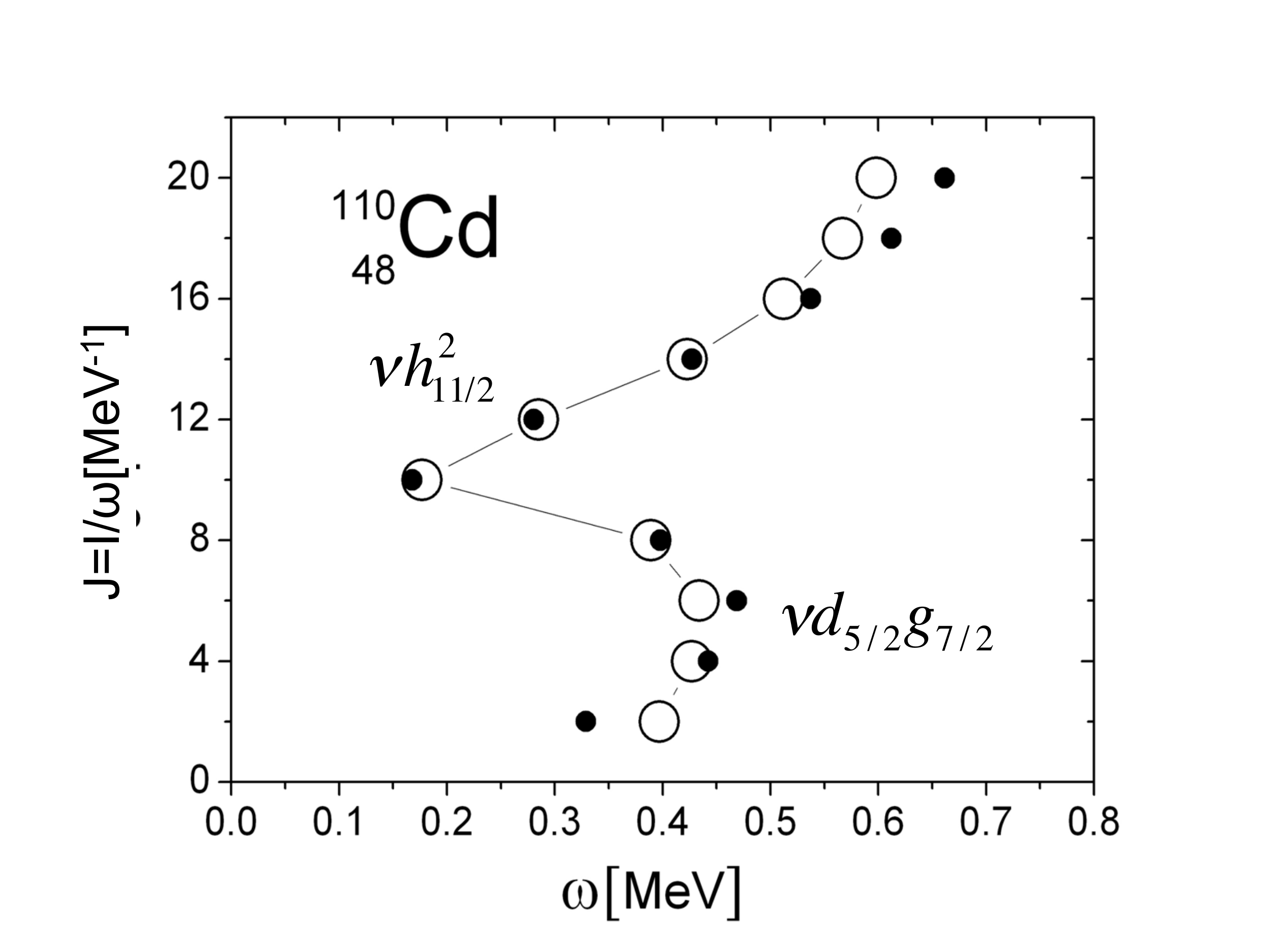}\includegraphics[width=0.48\linewidth]{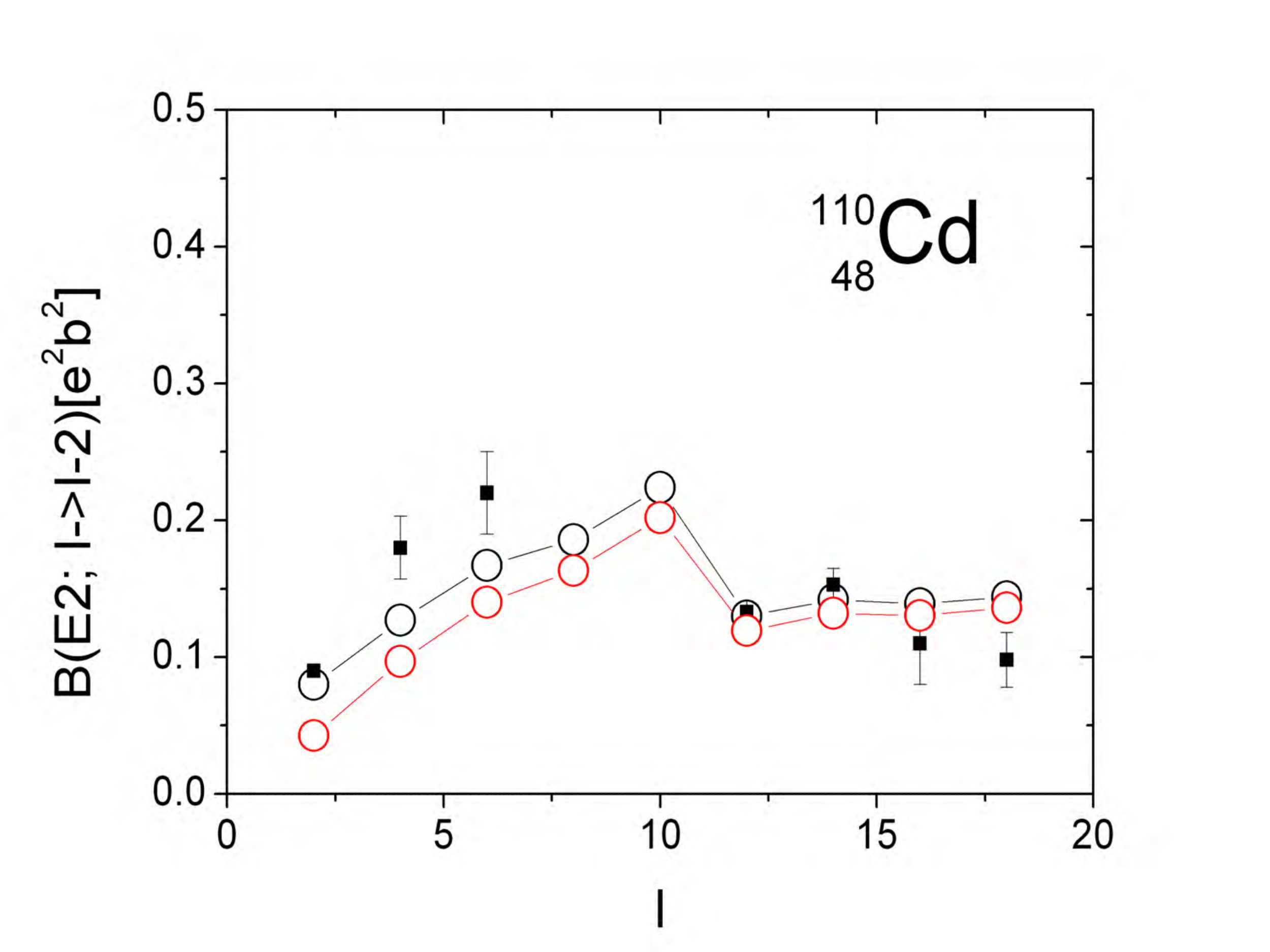}}
\caption{\label{fig:110CdTAC} 
Left panel: Experimental moments of inertia of the ground band  (black dots)
of $^{110}$Cd, where \mbox{$\omega=(E(I)-E(I-2))/2$} and $J=I$ compared with the  calculations by means of the Cranking model \cite{FGS11} (black circles).
 Right panel:  Experimental $B(E2, I\rightarrow I-2)$ values of $^{110}$Cd compared with the  
 calculations (black circles without red circles with quantal correction, see Ref.
 \cite{FGS11}). From Ref. \cite{Frauendorf15}.  } 
\end{figure}    

The yrast states of $^{102}$Pd shown in Fig. \ref{fig:102PdPRL} are a beautiful example of a slightly anharmonic \twd. 
Ayangeakaa {\it et al.}  \cite{102PdTidalPRL}  interpreted the states as a condensate of $d$ bosons. 
Accordingly, up to seven bosons are observed, which align their angular momenta. 
If the bosons were free, the function 
${\cal J}^{(1)}(I)$ would be a straight line starting at the coordinate origin (FB in Fig. \ref{fig:102PdPRL}). 
The experimental \momi can be very well approximated by ${\cal J}^{(1)}=\Theta_0+\Theta_1I$ (IB in Fig. \ref{fig:102PdPRL}).
The displacement by $\Theta_0$  
was attributed  to an interaction between the bosons that is quadratic in the boson number \cite{MAFGC14}.

The inset in Fig. \ref{fig:102PdPRL} demonstrates the wave character. The \am increases solely due to the increase of the \momid,
which is proportional to the square of the wave amplitude $\beta_e$.  As the $B(E2, I\rightarrow I-2)$ values are also proportional 
to $\beta_e$ the ratio is constant. The global estimate Eq. (\ref{eq:RJB}) gives 1.7$\times10^{-2}\hbar^{-2}\mathrm{MeV(eb)}^2$,
which is not far from  the ratio 1.3$\times10^{-2}\hbar^{-2}\mathrm{MeV(eb)}^2$ in Fig. \ref{fig:102PdPRL}.

The fact that the \tw is represented by a static deformation 
in the co-rotating frame of reference allows one to microscopically calculate its properties by means of the rotating mean field  approaches. 
Frauendorf {\it et al.}  \cite{FGS10,FGS11} calculated the energies of the yrast states and the $B(E2)$ of the intra band transitions up 
to spin $I=16$ for the nuclides with $Z=44-48,~N=65-66$.  
Figs. \ref{fig:102PdPRL} and \ref{fig:110CdTAC} exemplify the accuracy of the parameter-free calculations. In particular the change of the yrast states from the 
purely collective tidal wave (g-band) to the  configuration  with two rotational aligned h$_{11/2}$ \qps  (s-band) is reproduced in detail. 
In the case of $^{102}$Pd (Fig. \ref{fig:102PdPRL}) the collective g-band can be followed up to $I=14$, where it is at higher energy than the s-band. 
This is a consequence of almost no mixing between the two configurations.
In the case of $^{110}$Cd (Fig. \ref{fig:110CdTAC})  the collective g-band is crossed by the s-band earlier and the two bands interact stronger. The
two aligned h$_{11/2}$ quasiparticles  in the s-band reduce the deformation but stabilize it such that the sequence becomes more rotational. The example
suggests that the method should apply to odd-$A$ and odd-odd  nuclei without any further sophistication. 

  \begin{figure}[t]
\centerline{\includegraphics[width=0.8\linewidth]{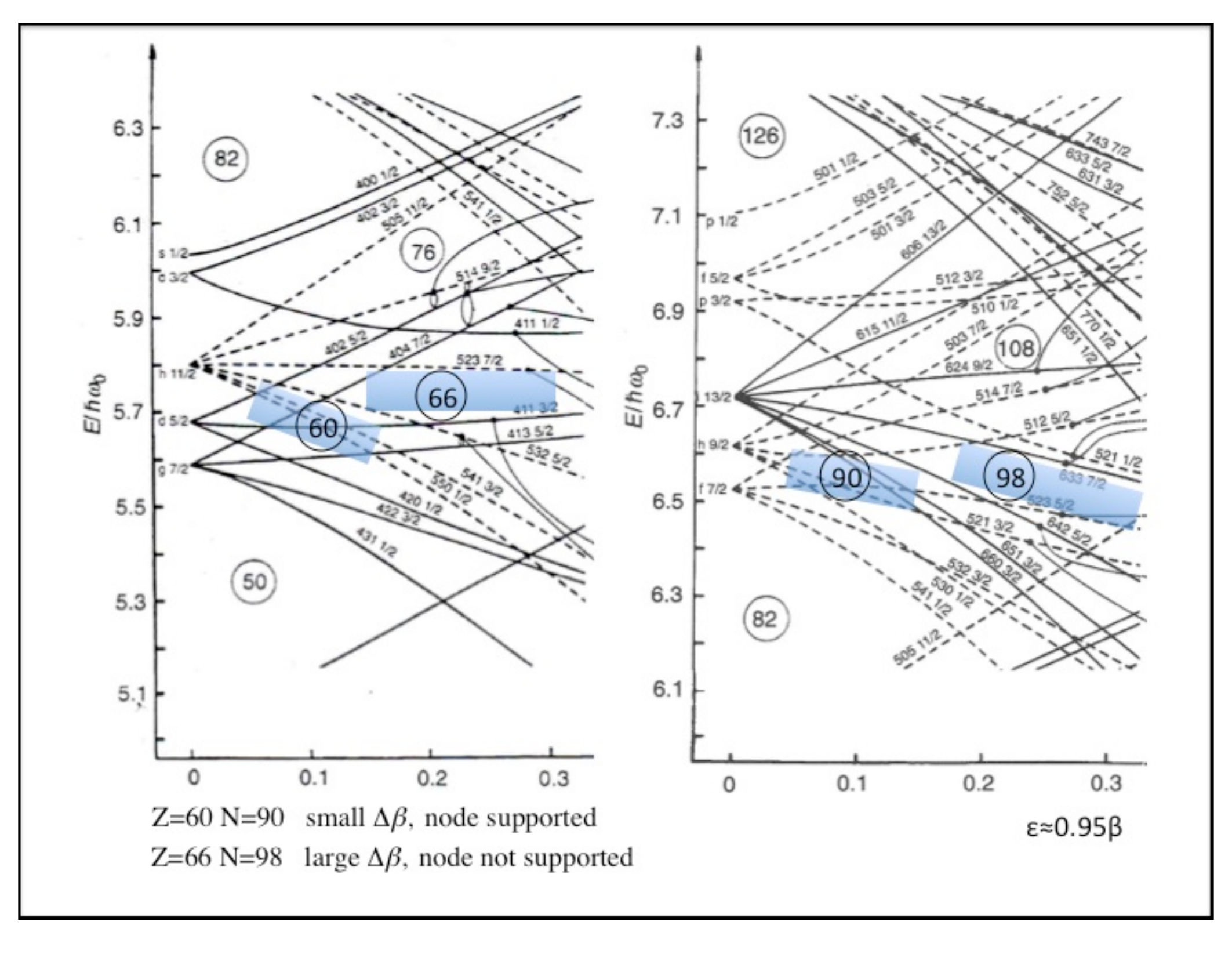}}
\caption{\label{fig:CohBet}  Energies of the modified oscillator as function of the deformation parameter $\eps$. 
The regions of large and small coherence length are highlighted.
Taken from Ref. \cite{NR}.} 
\end{figure}    

 \begin{figure}[t]
\centerline{\includegraphics[width=0.8\linewidth]{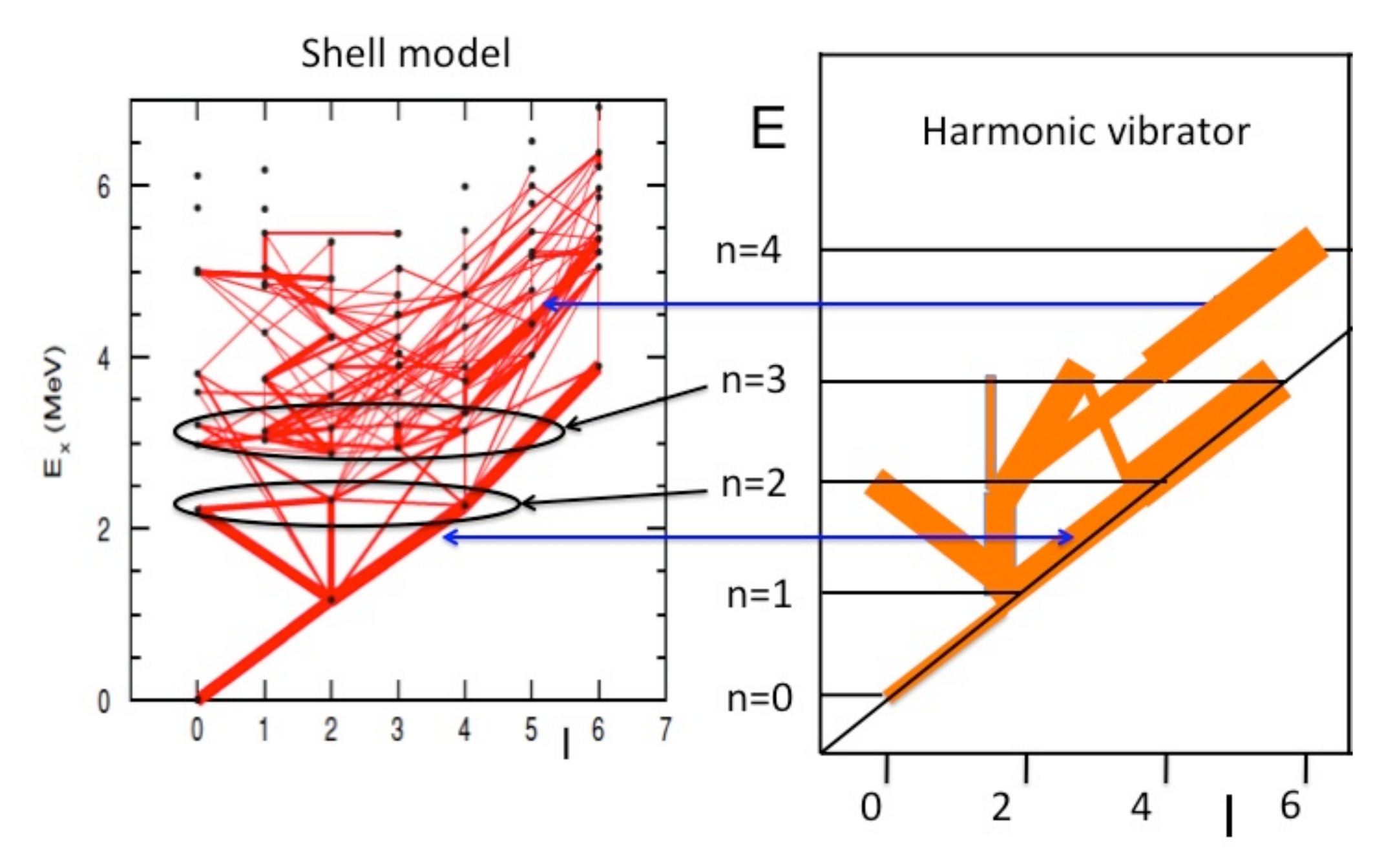}}
\caption{\label{fig:decoherenceVIB}Left panel:  Spectrum of $^{62}$Ni calculated by means of the 
Shell Model \cite{Chakraborty11}. The states are displayed as dots. The E2 transitions between the states are shown as the bars
that connect the dots.  The width of the bars is 
proportional to the $B(E2)$ value of the connecting transitions.\\
Right panel: Spectrum of the harmonic vibrator limit of the Bohr Hamltonian. The horizontal lines represent the multiplets
of states of increasing boson number $n$. The bars indicate the E2 transition between the states, which are not explicitly marked.  
The width of the bars is 
proportional to the $B(E2)$ value of the connecting transitions.\\
The recognizable enhanced transitions of the Shell Model calculation are associated by means of arrows with the transitions of 
collective quadrupole vibrator. The ellipses enclose the regions where the two- and three-phonon multiplets are expected.} 
\end{figure}

\subsection{Coherence  of the deformation degrees of freedom}\label{sec:CL}

The resolution of the collective wave function of deformation degree of freedom $\beta$  is also limited by a coherence length. It appears in the overlap of two 
mean-field solutions with different deformation \mbox{$\vert \langle \beta\vert \beta'\rangle\vert^2\approx\exp \left[-(\beta-\beta')^2/\Delta \beta^2\right]$}.
The overlap plays a central role in describing the shape dynamics by means of the Generator Coordinate Method. Here we address it only in a qualitative way.   

Fig. \ref{fig:CohBet} shows the single particle levels as functions of the deformation variable $\beta$. The overlap falls off the stronger the more the occupation of
the states near the Fermi surface changes over an interval of $\beta$. For $Z\approx 60$ and $N\approx 90$  many up-sloping levels cross many down-sloping. This results
in a considerable re-occupation over the highlighted  $\beta$ interval. A relatively small coherence length $\De \beta$  is expected.  For $Z\approx 66$ and $N\approx 98$ 
there is no re-occupation within the highlighted deformation interval. 
The overlap will still be smaller than one, because the single particle wave functions change with $\beta$. A 
relatively large value $\De \beta$ will result. The different coherence lengths acount for the following. In the transitional nuclei around $N=90$ there is
a low-lying  0$_2^+$ state with the properties of the collective one-phonon $\beta$ vibration, but no evidence for the two-phonon state \cite{CZ01}. 
The coherence length $\De \beta$
is small enough to resolve one node of the vibration but too large to resolve two nodes. For the well deformed nuclei around $N=98$ there is no evidence for a collective 
$\beta$ vibration. The coherence length $\De \beta$ is too large to even resolve one node.

The left panel of Fig. \ref{fig:decoherenceVIB} shows the E2 transitions obtained in a Shell Model calculation for spherical nucleus  $^{62}$Ni. 
They are compared with the transition strengths of the harmonic vibration limit of the Bohr Hamiltonian in the right panel. 
As discussed in section \ref{sec:UMsph}
(c. f. Fig. \ref{f:HV-qp}) the quadrupole vibrations become increasingly de-coherent when moving away from the yrast line. The Shell Model
gives a collective 2$^+$ state  interpreted as the one-phonon state and at twice the energy the states 0$^+$, 2$^+$, 4$^+$ interpreted as the two-phonon triplet.
  As expected for the  harmonic vibrator,  the transition strengths to the one-phonon state are enhanced and there are no transitions to the zero-phonon state. 
 At variance with the harmonic oscillator, the reduced transition probability  $B(E2,2^+_2\rightarrow2^+_1)$  is much smaller than the $B(E2)$ values for
 the   $4^+_1\rightarrow2^+_1$  and $0^+_2\rightarrow2^+_1$ transitions,
 which are not twice as large as the value for the $2^+_1\rightarrow0^+_1$ transition. The transition between the yrast states of the Shell Model are strong and 
 may be accounted for by an anharmonic   tidal wave. The Shell Model calculation moreover shows enhanced transitions
  parallel to the yrast sequence, which can be assisted with the 
 yrare members of the vibrational multiplets (see Bohr and Mottelson \cite{BMII}, Appendix 6B-3). 
 The remaining part of the Shell Model
 transition pattern looks chaotic. The coherence of the vibrational motion is lost. 
    
\section{Outlook}
The Unified Model with its esthetic dichotomy of collective and intrinsic degrees of freedom represents  the long wavelength 
limit of the collective modes compared to the granular structure of the underpinning nucleonic microstructure. The granularity becomes visible 
and has to be taken into account already  after exciting few collective quanta. The resulting entanglement of intrinsic and rotational degrees 
is accounted for by      
the rotating mean field, which provides the possibly simplest interpretation of  the multi-band structure of the yrast region. 
New phenomena arising from the strong coupling between the rotational and nucleonic degrees of freedom have been 
discovered  within this framework, some of which have been discussed in
this contribution.  The price for the simplicity is the restriction to uniform rotation and a semiclassical description that violates \am conservation.
Theoretical tools  beyond the mean field and the small-amplitude approximations, which combine 
the Generator Coordinate Method with configuration mixing (see e.g. the contributions to this Focus Issue 
by Egido \cite{NCEgido}, Sun \cite{NCSun}, Walker and Xu \cite{NCWalker}, Sheikh {\it et al.} \cite{NCSheikh} and the review by Bender {\it et al.} \cite{Bender03}),
remove these restrictions at the price of increasing the complexity of the description and drastically  the computational effort. This type of calculations may  
describe  the experimental data and allow for reliable predictions. In my view it seems important developing methods that reveal the underlying simple 
structures, which may be the ones discussed in this contribution or others that have not been covered (e. g. dynamical symmetries) or not anticipated yet.  
Particularly 
challenging is the question how to address the regions where the coherence of the nucleonic motion sets in or disappears. One example is the damping of   
 rotation in "warm nuclei" few MeV above yrast, which is discussed in the contribution to this Focus Issue by S. Leoni and A. Lopez-Martin \cite{NCLeoni}.   
 Conceptually, one has to give up on describing  the properties of individual quantum states.  
 The rapid increase of the level density makes such predictions  impossible because the results become exceedingly sensitive to the details of the 
 Hamiltonian and to the  solution scheme  of the many-body problem. Part of the theoretical results must be considered as "random". The challenge is to 
 decide what should be classified as random fluctuations and what as average properties described by the theory. On the other hand, allowing for a 
 degree of randomness favors alternative many-body approaches based on sampling techniques, which open new avenues.  
\newline
\newline
\noindent      
{\bf Acknowledgments}
I was lucky to come to the Niels Bohr Institute early in my scientific career. These were the hay-days of highspin physics. Copenhagen/Lund
was a powerhouse of this new exciting direction, which has fascinated me until now. I became acquainted with a new style of doing theoretical work:
Discuss with the experimentalists their latest results and expose your ideas to their scrutiny. These were the NBI Mondays, when the Ris\o  \ group came 
to meet with  Aage and Ben and us other theoreticians. For me these meetings were  a great source of inspiration and the origin of lasting friendships. 
Aage and Ben had an ingenious way seeing the simple principles behind complex observations or computational results. 
Searching for them has become the goal of my own work.
Working with these two great minds and the other outstanding scientists I met at the NBI was such a privilege.   THANK YOU!    \\
Support by the DoE grant DE-FG02-95-ER40934 is acknowledged.

 \newpage
 \appendix

\setcounter{footnote}{0}
 
 \section{Semiclassical analysis of high-j quasiparticle states in a triaxial quadrupole potential rotating about an axis in one of its principal planes }\label{sec:semi}
  The appendix extends the discussion  of the geometry and rotational response of high-j  \qp orbitals in section \ref{sec:orbits}. It is based on the
  work published in Refs.  \cite{MotANL,FrauDrexel,FR81,HamMot83,FraSob86}, which is supplemented by unpublished material.  
 
      The coupling of the high-j intruder state with the remaining states of the major shell to which it belongs is disregarded. This makes the quantum system one dimensional:
The number of radial knots is zero, i. e. the particle moves on an orbit of fixed radius and
     the particle motion is restricted to the sphere of constant \amd\footnote{We use $\hbar=1$ in this appendix.},     
     \beq\label{eq:jsphere}
     j^2=j_1^2+j_2^2+j_3^2=j(j+1).
     \eeq
 Working out the semiclassical theory it is useful  to introduce the canonical operators momentum $p$ and position $q$  (canonical variables in the classical mechanics),
 which obey the standard commutation relation    $[p,q]=-i$.     
One may take the \am projection $\hat j_3$
as the momentum  operator $q$. Then the angle operator $\hat \phi$ which fixes the orientation of the \am vector $\vec j$ with respect to the 3-axis becomes the conjugate
position operator $q$ and
\beq\label{eq:can}
[p,q]=[\hat j_3,\hat\phi]=-i.
\eeq 
As the momentum takes only the $2j+1$ discrete values $k=-j$, ...,$j$, the angle can also take only take $2j+1$ discrete values on the unit circle.
These are the eigenvalues of $\hat \phi$, which can be chosen as 
\beq
\phi_n=\frac{2\pi}{2j+1}n,~~~n=-j,~,-j+1,~....,~j-1,~j.
\eeq  
It is common to use the momentum eigenstates $\vert k\rangle$ as a basis, which are related to  the angle eigenstates $\vert n\rangle$ by
the transformation
\beq \label{eq:phi-k}
\vert k\rangle=\frac{1}{\sqrt{2j+1}}\sum\limits_{n=-j}^{j}e^{ik\frac{2\pi}{2j+1}n }\vert n\rangle,
~~~\vert n\rangle=\frac{1}{\sqrt{2j+1}}\sum\limits_{n=-j}^{j}e^{-i\frac{2\pi}{2j+1}n k}\vert k\rangle.
\eeq 
The amplitude 
\beq
\langle n\vert k\rangle=\frac{1}{\sqrt{2j+1}}e^{ik\frac{2\pi}{(2j+1)}n}=\frac{1}{\sqrt{2j+1}}e^{ik\phi_n}
\eeq
is the discrete version of the well known expression $1/\sqrt{2\pi}\exp{[i k \f]}$ for the $\hat j_3$ eigenfunctions in the full
orientation space.  
The discreteness of $j_3$ and $\phi$ should be kept in mind in the semiclassical analysis below.

In discrete space it is convenient to work with the operator $\exp[i\hat \phi]$ instead of $\hat \phi$ itself. Its matrix elements in the $k$ basis are
\beq
\langle k\vert e^{i\hat \phi}\vert k'\rangle=\delta_{k,k+1},
\eeq 
which is seen using Eqs. (\ref{eq:phi-k}) and noticing their orthonormality. Then the matrix of the operator 
\beq
\langle k\vert\sqrt{j-\hat j_3} e^{i\hat \phi}\sqrt{j+\hat j_3}\vert k'\rangle=\sqrt{(j-k)(j+k+1)}\delta_{k,k+1}
\eeq
is recognized as the matrix of the standard $\hat j_+$ operator. Thus the operators $\hat j_\pm$ can be exposed as
\beq\label{eq:j+j-}
\hat j_+= \sqrt{j-\hat j_3} e^{i\hat \phi}\sqrt{j+\hat j_3}, ~~~\hat j_-=(\hat j_+)^\dagger=\sqrt{j+\hat j_3} e^{-i\hat \phi}\sqrt{j-\hat j_3}.
\eeq
The standard commotion relations $[\hat j_3,\hat j_\pm]=\pm \hat j_\pm$ and $[\hat j_+,\hat j_-]=2\hat j_3$ are fulfilled because the matrices agree with 
the ones of the three operators within a set of states $\vert j,k\rangle$. They can also be directly verified 
using $\left[\hat j_3,\exp[\pm i\hat \phi]\right]=\pm \exp[\pm i\hat \phi] $.

           \begin{figure}[t]
 \begin{center}
 \includegraphics[width=0.48\linewidth]{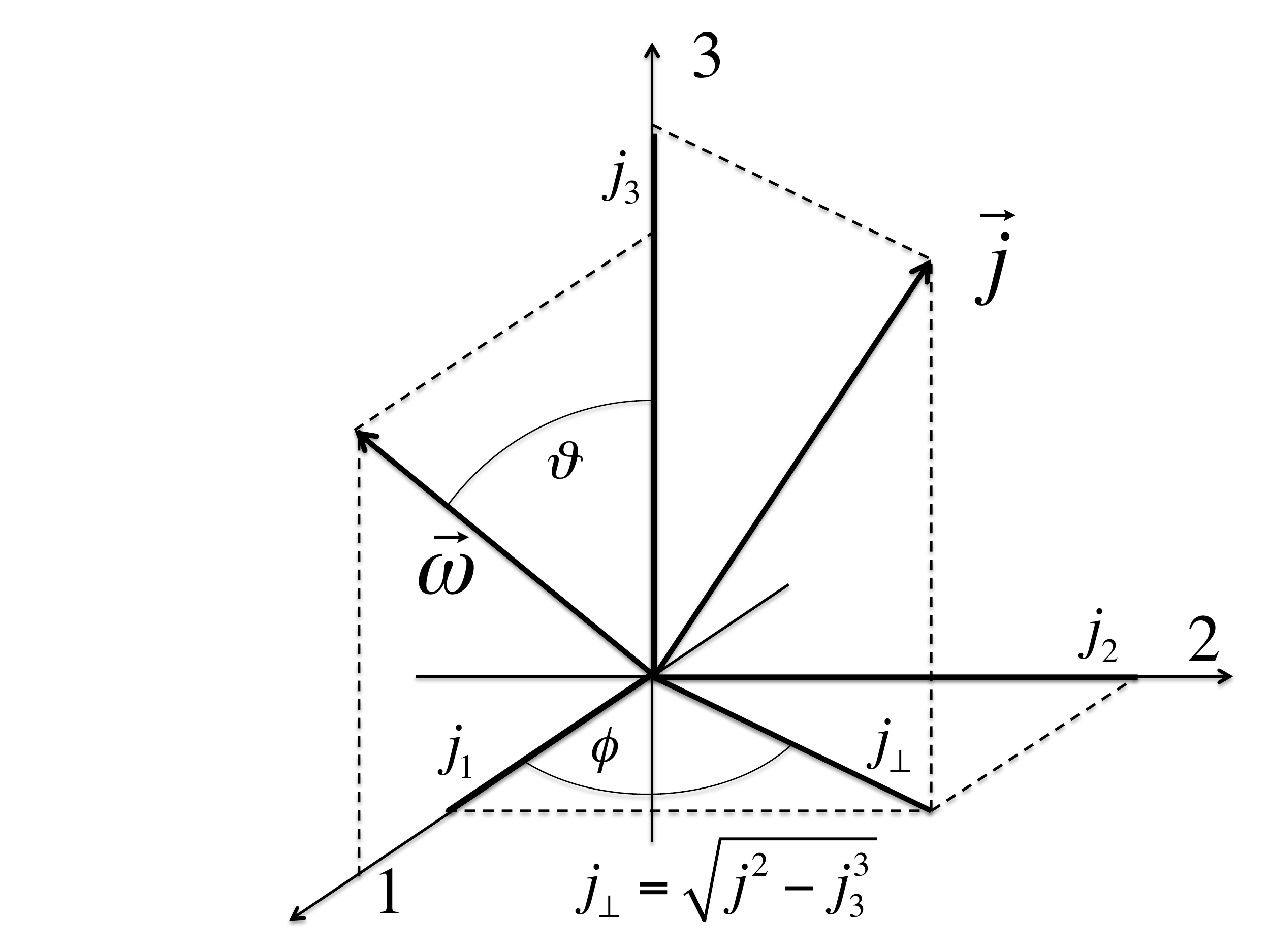} 
  \includegraphics[width=0.48\linewidth]{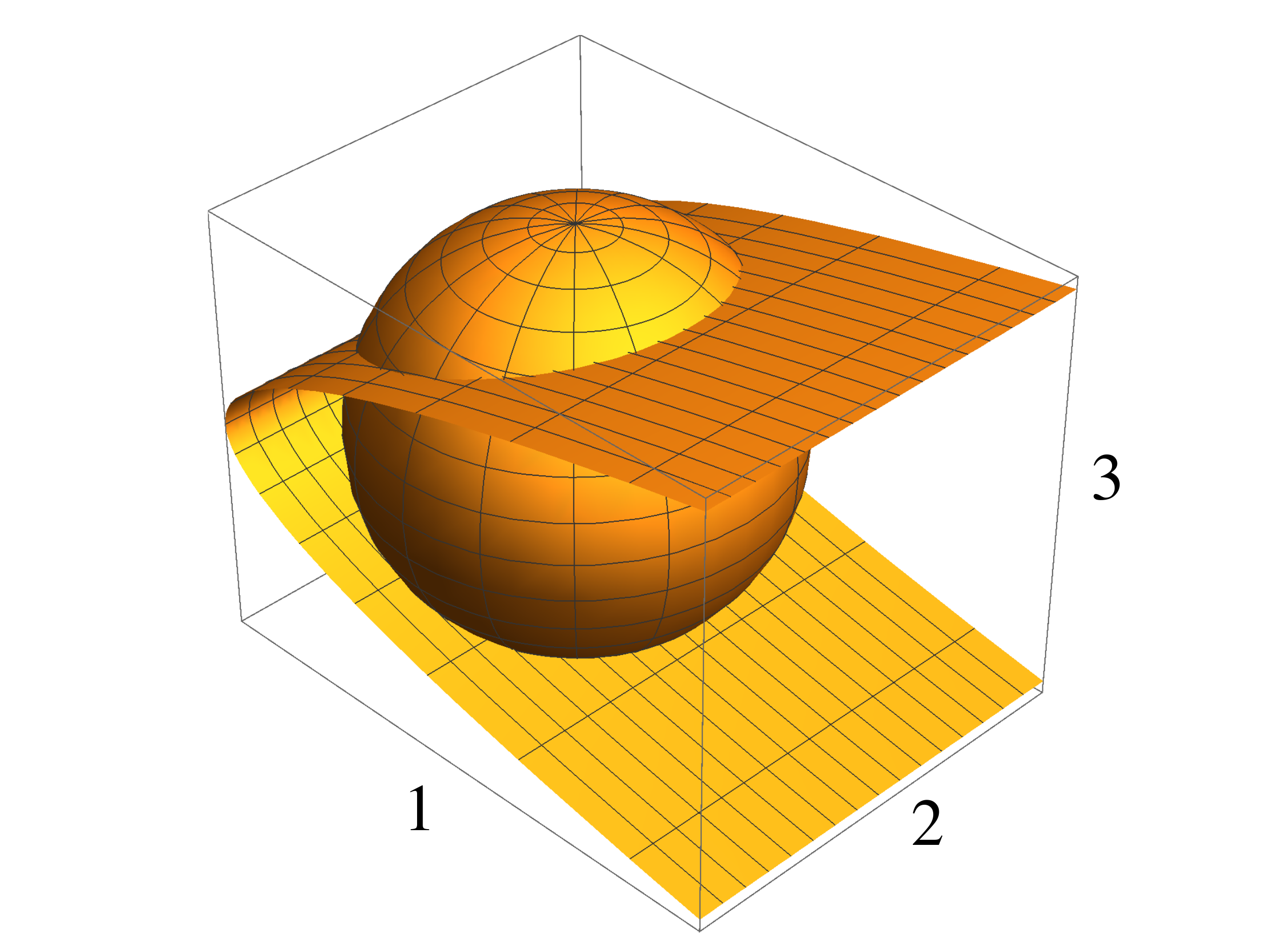}
    \caption{\label{f:j3phi}  
 Left: Geometrical relation between the three Cartesian \am components $j_1,~j_2,~j_3$ of the  i$_{13/2}$ particle and the
 angle $\phi$, which is taken as the momentum conjugate to $j_3$. In addition the orientation of 
 angular velocity vector $\vec \om$ is shown, which is chosen to lie in the 1-3 -plane. \\
 Right: Classical orbits of the \am vector $\vec j$ of an i$_{13/2}$ particle in an axial potential that rotates about the 1-axis. The orbits are the line of 
 intersection between the sphere of constant length $j$ and the parabolic cylinder of constant  routhian (energy in the rotating frame).
 They represent two of the orbits in the lower left panel of Fig. \ref{f:orbitsax}, which shows the projection of the three-dimensional orbits onto the 1-3-plane.}
   \end{center}
 \end{figure}

In semiclassical approximation $\hat {\vec j}$ is considered as a classical vector. The commutators are replaced by the Poisson brackets. 
The classically corresponding momentum $j_3$ and position  $\phi$   are   canonical variables, which commute.   The classically corresponding
expressions for $j_+$ and $j_-$, given by Eqs. (\ref{eq:j+j-}), can be rearranged to 
\beq
j_1=j_\bot\cos\phi,~~ j_2=j_\bot\sin\phi,~~j_\bot=\sqrt{j^2-j_3^2}.
\eeq
A first correction to the classical expressions is obtained replacing $j^2$ by $j(j+1)$.
{\bf We introduce  the following convention.
Instead of $\sqrt{j(j+1)}\approx j+1/2$ we simply use $j$. It is understood that evaluating the expressions one has to use $j+1/2$
instead of $j$, e. g.  $j=7$ for an i$_{13/2}$ particle.  }

Quantization is realized by the Bohr-Sommerfeld rules
\bea
\oint pdq=\oint \phi\left(j_3\right)dj_3=S=2\pi n,\\
\oint qdp=\oint j_3(\phi)d\phi=S=2\pi n,\\
\int\int dpdq=\int\int d\phi dj_3=S=2\pi n.
\eea
Geometrically this means the following. The phase space is the projection of the surface of the \am sphere on 
the cylinder with the radius $j$ and $-j\leq j_3 \leq j$.  The area between two quantized orbits is $2\pi$.

   The  energy in the deformed potential is determined by the orientation of $\vec j$ with respect to the principal axes, 
     which can be expressed in terms of the components of the \amd. The expression 
     can be derived from Eqs. (\ref{eq:hMO},\ref{eq:deflund}) by the consideration that the quadrupole operator
     is constructed by coupling the two vectors $\vec x$ to a spherical tensor operator of rank 2.
     Coupling the two \am vectors in the same way $\vec j$ results in a spherical tensor operator of rank 2, which transforms under rotation like the quadrupole tensor. 
     As the energy of the high-j orbital depends only on its orientation
     with respect to the potential, it must depend on the components of $\vec j$ in the same way as given by Eqs. (\ref{eq:hMO},\ref{eq:deflund}). 
     Using the explicit expressions for the quadrupole moments in terms of the components of $\vec x$, one obtains 
       \beq\label{eq:ellipsoid}
     e'=e-\om(\sin \vth j_1+\cos\vth j_3),~~e=\kappa\left[ \left(3j_3^2-j^2\right)\cos\ga-\sqrt3\left(j_1^2-j_2^2\right)\sin\ga\right].
     \eeq
     For the following examples the parameter $\kappa$ is determined from the distance between the i$_{13/2}$ Nilsson levels at the deformation $\varepsilon=0.26$, 
     which gives $\ka=0.0086\om_0$. The energies  
    of the  orbits  are taken as $e'=e'(\om)$  from the quantal routhians shown in the upper panels of the figures to be discussed.  
As discussed in context of Table \ref{t:axes}, one can make any two of the three principal axes to be the 1- and  3-axes
     by choosing the appropriate $\ga$-sector.   In the following  we consider only a planar tilt of the rotational axis into the 1-3- principal plane. 
     Quantization is achieved by diagonalizing the Hamiltonian  $h'=e'$ given by Eq. (\ref{eq:ellipsoid}) within the basis of \am eigenfunctions $\vert j,m\rangle$.  
     The eigenvalues are called  "quantal routhians" in the following.
          

The orbits are determined by \am  and energy conservation.
Without pairing, they are given by the intersection of the \am sphere (\ref{eq:jsphere}) and the 
energy ellipsoid  (\ref{eq:ellipsoid}). The time derivatives of the   momentum and coordinate are given by
\bea
\dot p=v_3=-\frac{\partial e'\left(j_3,\phi\right)}{\partial\phi}=4\ka\sqrt{3}\sin\ga j_2\left(j_1-j_{1c}\right)\label{eq:qdot},\\
\dot q=\dot \phi=\frac{\partial e'\left(j_3,\phi\right)}{\partial j_3}\nonumber\\
=4\ka\sqrt{3}\left[\cos(\ga+30^\circ)(j_3-j_{3c})+
\sin\ga\frac{j_3j_1\left(j_1-j_{1c}\right)}{ j^2-j_3^2}\right]\label{eq:pdot},\\
j_{1c}=-\frac{\om\sin\vth}{4\ka\sqrt{3}\sin\ga},~~j_{3c}=\frac{\om\cos\vth}{4\ka\sqrt{3}\cos(\ga+30^\circ)}\label{eq:center}.
\eea
The stationary points of the classical motion are located at $\dot q=\dot p=0$. Minima and maxima are singular points of no motion.
Any saddle point S is localized on a special orbit called separatrix, which will be labelled by S.  
 A particle moving on a separatrix will reach the saddle point only after infinite time because velocity and acceleration go to zero. 
 As a consequence the expectation value of $\vec j$ is given by the value  at the saddle point.
 
If the stationary point is located outside the 1-3 plane, i.e.  $j_2\not=0$, Eqs. (\ref{eq:qdot},\ref{eq:pdot}) imply that 
its coordinates are $j_1=j_{1c},~j_2=\pm\sqrt{ j^2-j_{1c}^2-j_{3c}^2},~j_3=j_{3c}$. If it is located in the 1-3- plane 
its coordinates are $j_1=j_1(\al_0),~j_2=0,~j_3=j_{3}(\al_0)$, where $\al_0$ is the solution of $\dot p(\al)=0$ with
$j_1(\al)=j\sin\al,~j_3(\al)=j\cos\al$ in Eq.(\ref{eq:pdot}). Except for certain simple cases, it is best finding the solution numerically.

The classical probability for the particle to be within the interval $(j_3,~j_3+dj_3)$ is
\beq\label{eq:Pj3}
dP(j_3)=\frac{dt}{T}=\frac{1}{\vert v_3T\vert}dj_3=\frac{1}{\vert4\ka\sqrt{3}\sin\ga j_2\left(j_1-j_{1c}\right)T\vert}dj_3,
\eeq
where $T$ is the period of the orbit. The major contributions to expectation values of physical quantities come from regions where $v_3$ is small.  
The classical probability for the particle to be within the interval $(j_1,~j_1+dj_1)$ is
 \bea\label{eq:Pj1}
dP(j_1)=\frac{1}{\vert v_1T\vert}dj_1=\frac{1}{\vert v_3T\vert}\left|\frac{dj_3}{dj_1}\right|_{e'}dj_1\nonumber\\
=\frac{1}{\vert4\ka\sqrt{3}\cos(\ga+30^\circ) j_2\left(j_3-j_{3c}\right)T\vert}dj_1.
 \eea

In the following  the projections of the orbits onto the 1-3 plane will be shown. They are given by eliminating $j_2$ from $e'$ in Eq. (\ref{eq:ellipsoid}) by means of
     Eq. (\ref{eq:jsphere}). The resulting expression for $\ga\not=0,\pi$ is
     \bea\label{eq:ellipsoid13}
     e'=2\kappa\sqrt{3}\left[ \cos(\ga+30^\circ)(j_3-j_{3c})^2-\sin\ga\left(j_1-j_{1c}\right)^2\right]\nonumber\\
     -2\ka\sin(\ga-30^\circ) j^2+\frac{\om^2\sin^2\vth}{4\ka\sqrt{3}\sin\ga}+\frac{\om^2\cos^2\vth}{4\ka\sqrt{3}\cos(\ga+30^\circ)}.
     \eea
     The curves of constant energy are ellipses or hyperbolas centered at $j_{1c},~j_{3c}$.
     The axial case is treated by taking the limit $\ga \rightarrow0$, which gives parabolas centered at the $j_{3c}$- axis. 
       
           \begin{figure}[t]
  \begin{center}
 \includegraphics[width=0.48\linewidth]{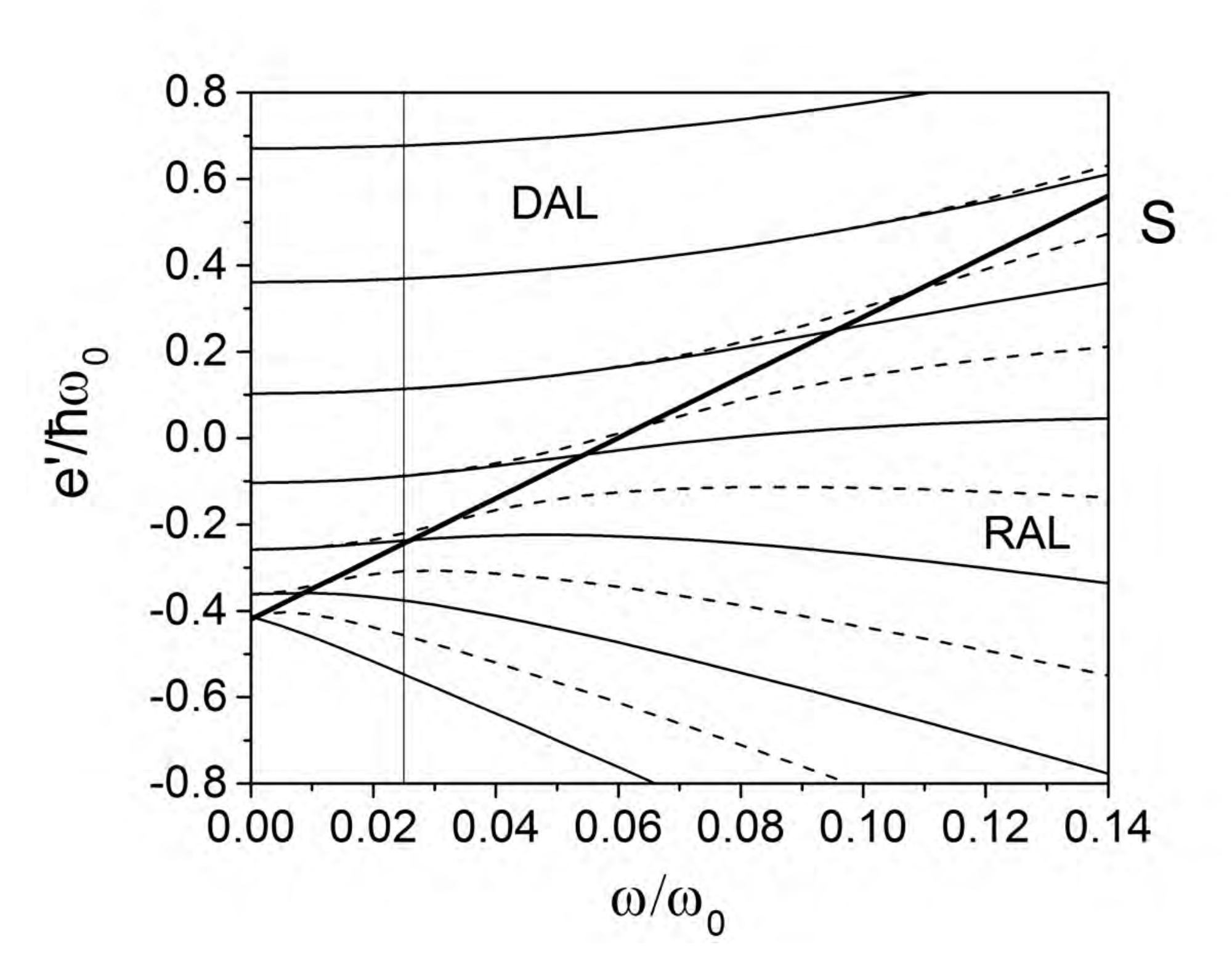} 
 \includegraphics[width=0.48\linewidth]{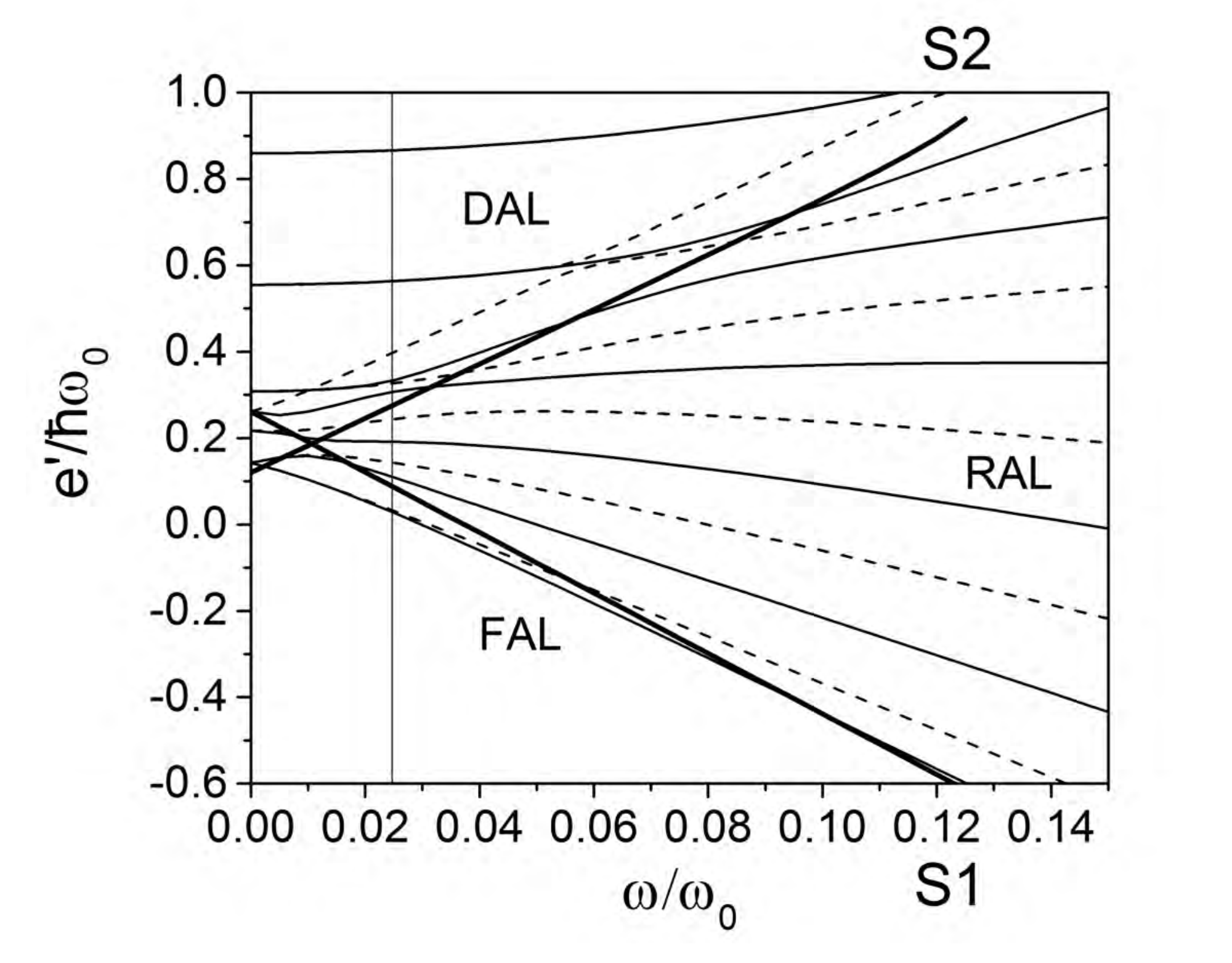}  
 \includegraphics[width=0.43\linewidth]{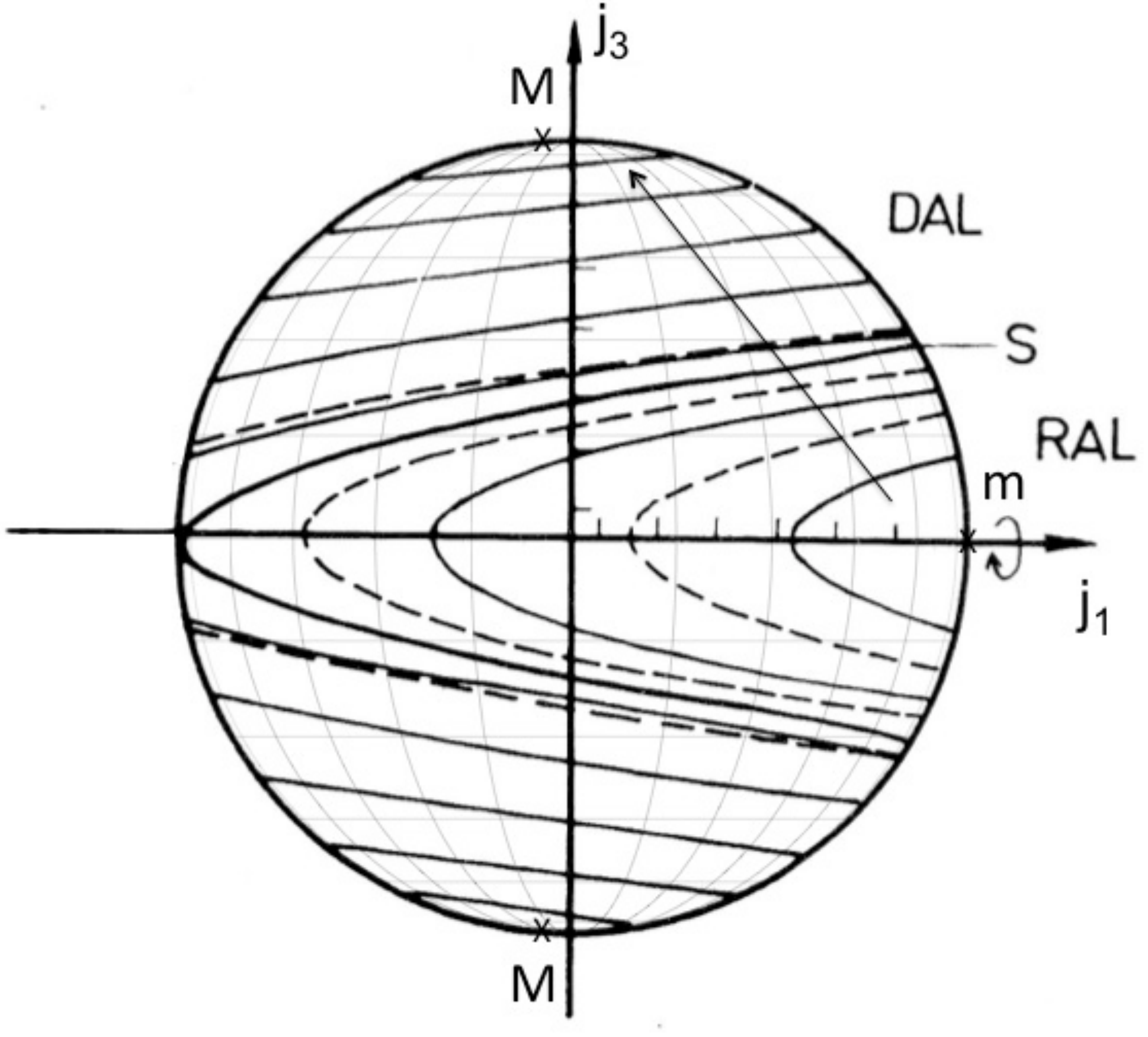} 
 \includegraphics[width=0.54\linewidth]{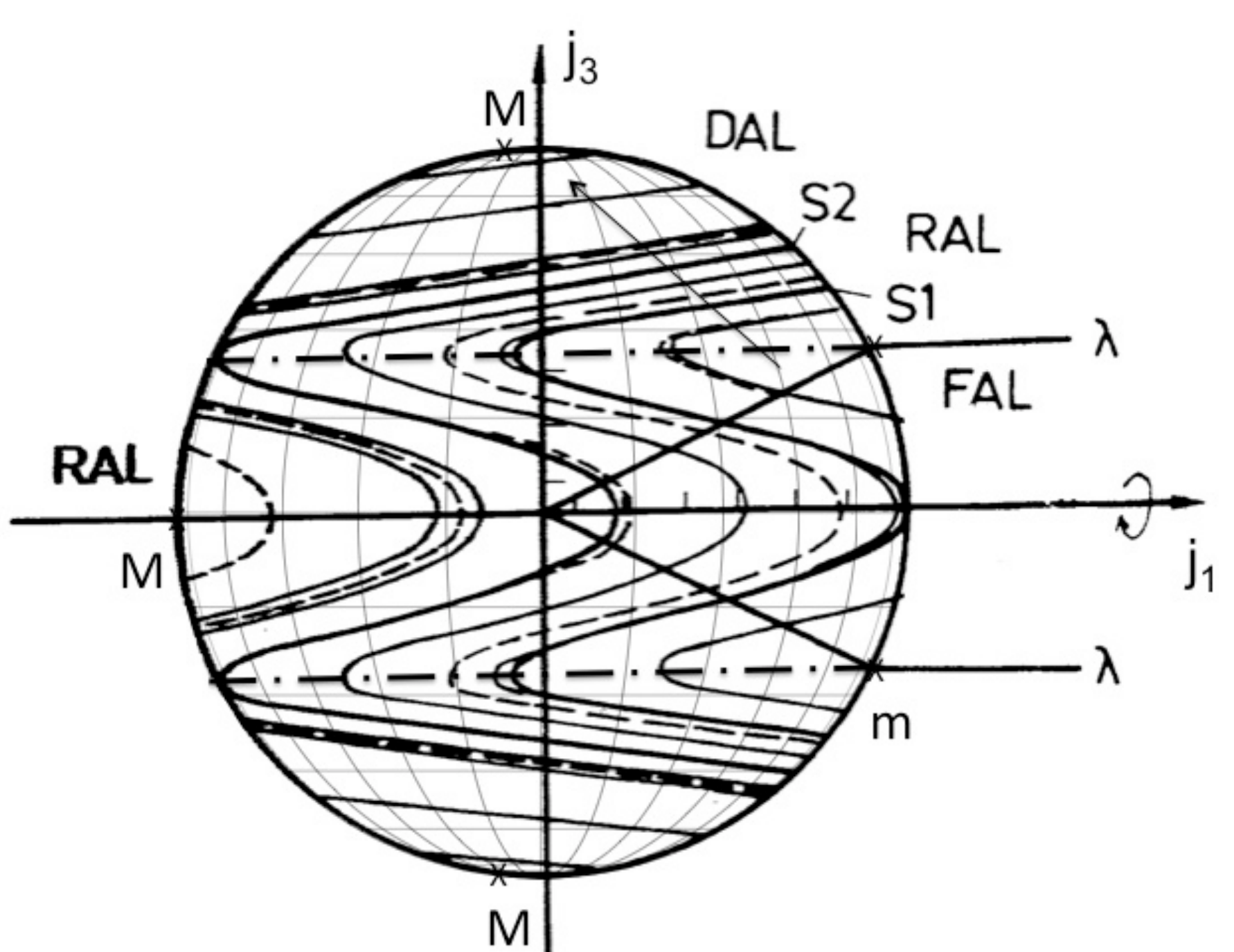}  
    \caption{\label{f:orbitsax}  Routhians and classical orbits of an i$_{13/2}$ \qp
          in the potential with $\varepsilon=0.26, \gamma=0$, rotating with the frequency $\om$ about the 
   1-axis  perpendicular to the symmetry 3-axis.\\ 
   Upper left panel: Quantal routhians for zero pairing. The thick curve shows the energy of the separatrix S. At $\om=0$,  the  levels
     with increasing energy have the  \am projection on the symmetry axis 
    \mbox{$\vert j_3\vert=\Omega=$1/2, 3/2, 5/2, .... .} The full and dashed lines correspond to signature $\alpha=$1/2 and -1/2, respectively.\\
     Upper right panel: Quantal routhians for $\Delta=0.12~\hbar\om_0$ and $\lambda$ between the 3/2 and 5/2 levels.
    The thick  lines show the energies of the two separatrices S1 and S1  in the low-right panel. \\
    Lower left panel:  Classical orbits  in \am space for zero pairing. The figure shows the 1-3-projection of the orbits, which  lie on the surface of the 
    sphere of fixed \amd. The frequency $\om=0.025\om_0$ which is indicated by the thin line in the upper panels. The energy of the orbits
    is the energy of the levels in the upper panels at the thin line.
     The energy of the orbits increases in direction of the arrows. The points of lowest possible energy are labelled with m and the point of highest
     possible energy with M. 
     In order to emphasize the location of the orbits on the  \am sphere the
     lines of constant altitude and latitude are shown. \\
    Lower right panel: Analogous to the lower left panel for finite pairing. The dash-dotted line is the orbit with $e'(0)=\lambda$.}
  \end{center}
 \end{figure}

       \subsection{Axial shape, rotation about the short axis}
       
        The left panels of Fig.~\ref{f:orbitsax} illustrate  an i$_{13/2}$ particle in the prolate potential, which rotates
       about the 1-axis perpendicular to the symmetry axis 3 ($\vth=90^\circ,~j_{3c}=0$). For $\gamma=0$ the equi-energy ellipsoids become 
       parabolic cylinders, which  move to the left with increasing energy. 
       The orbits lie at the intersection curves between the energy cylinders and the \am sphere (see Fig. \ref{f:j3phi}).
        The  orbits displayed  in the lower left panel have the energies of the quantal levels at $\om=0.025\om_0$ 
       in the upper left panel. 
       
       As seen, there are two topologically different kinds of orbits, which are separated by the  separatrix S.  
       The separatrix  lies at the intersection of the \am sphere with the parabolic cylinder whose vertex line is tangent to the sphere. The touching point 
       lies at $j_1=-j,~j_2=0,~j_3=0$. It specifies the  saddle point S, as seen by putting it into Eqs. (\ref{eq:qdot},\ref{eq:pdot}).
       Since it contains the saddle point, the expectation value $\langle j_1 \rangle=-j$ for the separatrix S.   
       The curvature of the parabolas is $6\ka/\om$. For $\om\rightarrow0$ the parabolas become horizontal lines and all lie outside S (which becomes  the $j_3=0$ line).
       The orbits revolve the symmetry axis 3. This motion is caused by the torque excerted by the deformed potential and is of the same type as the precession 
       of a top in the gravitational field. Since the precession cone is aligned with the deformation axis  this type of motion is called Deformation ALigned (DAL). 
       For each DAL state $j_3>0$ there is a degenerate one with $j_3<0$. Combining them to even and odd superpositions
     results in the characteristic signature doublets.
    For $\om\rightarrow\infty$ the parabolas become vertical lines, which corresponds to negligible torque of the deformed potential  compared to torque 
       by the inertial forces. The vector $\vec j$ revolves the    
        the 1-axis, which is the rotation axis. Accordingly, this type of motion is called Rotational ALigned (RAL).  
        As in any symmetric potential, where even and odd states alternate, consecutive RAL states 
     correspond to opposite signature. For intermediate values of $\om$ the topology of the orbits is either  RAL or DAL.
          
      The upper left panel of Fig. \ref{f:orbitsax} illustrates how the separatrix divides the spectrum into the DAL region above S 
     and the RAL region below S. The separatrix is a straight line with the slope -$\langle j_1 \rangle=j$. 
     The  fingerprints of the two coupling schemes  are easily recognized.
      More details of the routhian diagram can be related to the classical 
     orbits. The axis of the DAL precession cones is somewhat tilted toward  $j_1<0$.  It pierces the \am sphere at the point M, which is maximum 
     of  the energy. M lies at $j_1=-\om/(6\ka),~j_2=0,~j_3=\pm\sqrt{ j^2-\left(\om/6\ka\right)^2}$   (c. f. Eqs. (\ref{eq:qdot},\ref{eq:pdot}).) 
     As the consequence, the average of the $j_1$ value of the DAL orbits is slightly negative, which is
       reflected by the positive slope  of the quantal routhians, which increases with $\om$.  
          For the RAL orbits the average value of $j_1$ can be estimated  by means of expression (\ref{eq:Pj3}). The right turning points contribute most
          because $j_2\rightarrow0$. For the lowest RAL orbits the  left turning point of $j_1$ at the vertex contributes less because $j_2$ is finite,
          and  the average of $j_1$  is somewhat less than the values at the right turning points.    
    When approaching S for the higher orbits,  the left turning point becomes more important  because $j_2$ decreases.
     That is, the average $j_1$ becomes more negative.  For S it reaches $-j$ because the turning point  becomes a stationary  point. 
     Near the separatrix quantum effects  become important, which smoothly
     connect the DAL and RAL  routhians through  S. They can be interpreted as tunneling through the classically forbidden regions. For example,
      some signature splitting between the DAL doublets emerges, when the upper and lower orbit come close at the left turning point.

    With increasing 
     frequency the parabolas become flatter and flatter, their curvature being $6\ka/\om$. This means that the RAL part of the 
     phase space increases at the expense of the DAL part. This is illustrated by
    the left panel of Fig. \ref{f:sepax}, which show how the separatrix changes with $\om$. The separatrix disappears for $\om>6\ka/ j$ when the curvature of the parabola 
    exceeds the one of the circle. Only RAL orbits remain. The separatrix $\om=0.3\om_0$ comes close to this limit.
            
   Fig. \ref{f:qpcouplings} sketches  the two coupling schemes: 
     The fingerprint for RAL coupling is  a well-split sequence of states 
     of alternating signature with approximately constant aligned \am $i=\langle j_1\rangle=-de'/d\om$, which
     decreases with energy.  The fingerprint of DAL coupling are signature doublets and small negative aligned \amd.    
  
   The geometry of the orbits in the presence of pairing is obtained by replacing the energy of the particle in the deformed field by its \qp energy,
     \beq\label{eq:orbitsp}
     j_1^2+j_2^2+j_3^2= j^2, ~~~e'=\sqrt{(e-\lambda)^2+\Delta^2}-\om j_1.
     \eeq 
      This corresponds to introducing non-rotating \qps exposed to the cranking term. Loosely speaking, it amounts to leave away the
     negative-energy states in Fig. \ref{f:spagN96} and their interactions with the positive-energy states. 
     The Bogoljubov transformation is carried out for the non-rotation frame, and the equations of motion (\ref{eq:eom}) are solved in the subspace of the positive-energy \qpsd. 
     The quantal routhians in the upper right panel of Fig \ref{f:orbitsax}   are obtained using this approximation.The same holds for the following examples that take pairing into account.
           \begin{figure}[t]
 \begin{center}
  \includegraphics[width=\linewidth]{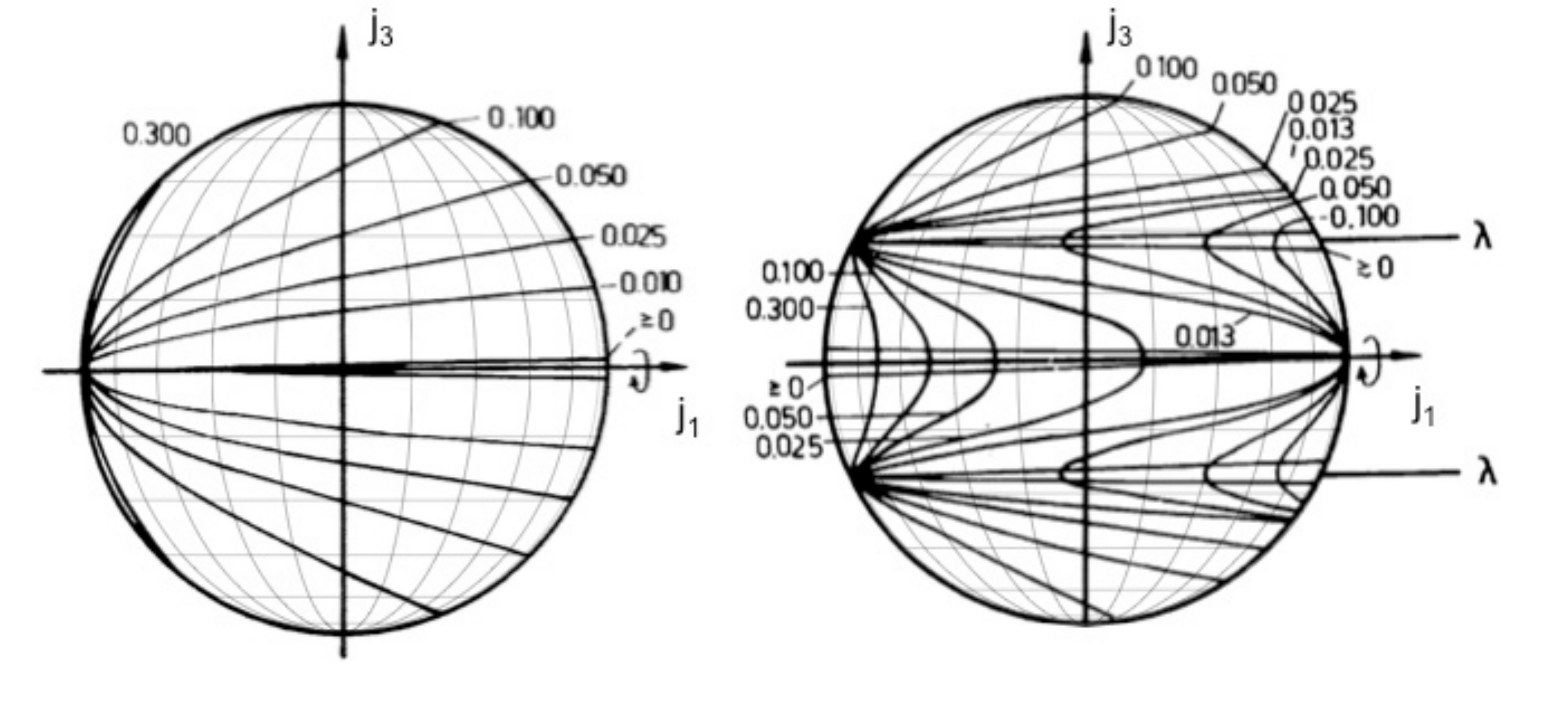}
    \caption{\label{f:sepax}  Sequence of separatices in Fig. \ref{f:orbitsax}. The separatrices are labeled by $\om/\om_0$.}
  \end{center}
 \end{figure}
      
     The right panels of Fig \ref{f:orbitsax} illustrate the case of an i$_{13/2}$ \qp in the axial potential. The gap parameter is $\Delta = 0.12  \om_0$, and $\lambda$ lies between
     the $\Omega=3/2$ and 5/2 levels.   The dash-dotted line shows the projection of the orbit $e=e'(\om=0)=\lambda$. 
     Well above it, the quasiparticle orbit becomes the particle orbit 
     and well below  the hole orbit.   The latter is obtained by reflecting the particle orbit on
      the 2-3 plane, $j_1\rightarrow -j_1$, which can be seen by taking the limit $\Delta\rightarrow 0$ 
     of Eq. (\ref{eq:orbitsp}). The particle-like branch is smoothly connected with the hole-like branch for finite $\Delta$. The resulting energy surfaces are  fourth order cylinders with three vertices, 
     which move from the right to the left with increasing $e'$. The  orbits are the intersection curves of the cylinders with the \am sphere. 
     
      The resulting topology is more complex. A new type of orbits appears, which revolve the  axis $\vec j_\lambda$ marked by $\lambda$. 
      Frauendorf and Sobeslavsky \cite{FrauDrexel,FR81,FraSob86} called it Fermi axis, because  its direction is determined 
      by $\lambda$, which is close to  the Fermi level of the unpaired system.
      They called the  orbits that revolve the Fermi axis Fermi ALigned (FAL). The FAL orbits have a finite
      \am projection on both the 3- and 1-axes.
     Like for the DAL orbits there two distinct degenerate FAL orbits with $j_3\approx\pm j_{3\lambda}$.
     In contrast to the DAL orbits they have a positive value of $j_1\approx  j_{1\la}$, because of the tilt of the Fermi axis. 
     When the quasiparticle revolves  the Fermi axis it changes from a particle into a hole. The quasiparticle partially decouples from the
     deformed potential, because the particle and hole quadrupole moments have opposite sign and tend to compensate each other.
     In summary, the  FAL orbits have finite \am components along the rotation axis 1 and the deformation (symmetry) axis 3. They appear in
      signature doublets, which may be slightly split  by tunneling.

     The FAL orbits are separated by the separatrix S1 from the RAL region, which is separated by the  separatrix S2 from the DAL regions. 
     The separatrix S2 divides the RAL region into the left part, which contains hole-like \qps that anti-align with the 
     rotational axis, and the right part, which  contains particle-like \qps that align with the 
     rotational axis. As seen in the right panel of Fig. \ref{f:sepax}, the FAL region  only exists for  low $\om$ and $\lambda$ about in the middle of the shell.
     The two separatrices divide the spectrum into the three regions, which are easily recognized in the upper right panel of Fig. \ref{f:orbitsax}  by their fingerprints. 
      The crossing of S1 and S2 corresponds to the single separatrix $\om=0.013\om_0$ in the right panel of Fig. \ref{f:sepax}.
       For smaller frequency  $\om$ a different  topology appears, which is
       indicated by the separatrices  labeled by $\om > 0$. S1 divides the FAL region from particle-like DAL above and hole like DAL below. S2 divides the hole-like RAL interior region 
       from the hole-like DAL regions above and below.  The right panel of Fig. \ref{f:sepax}  shows how the separatrices change with $\om$, delineating the phase spaces of the three coupling schemes.

     \begin{figure}[t]    
    \begin{center}
  \includegraphics[width=0.51\linewidth]{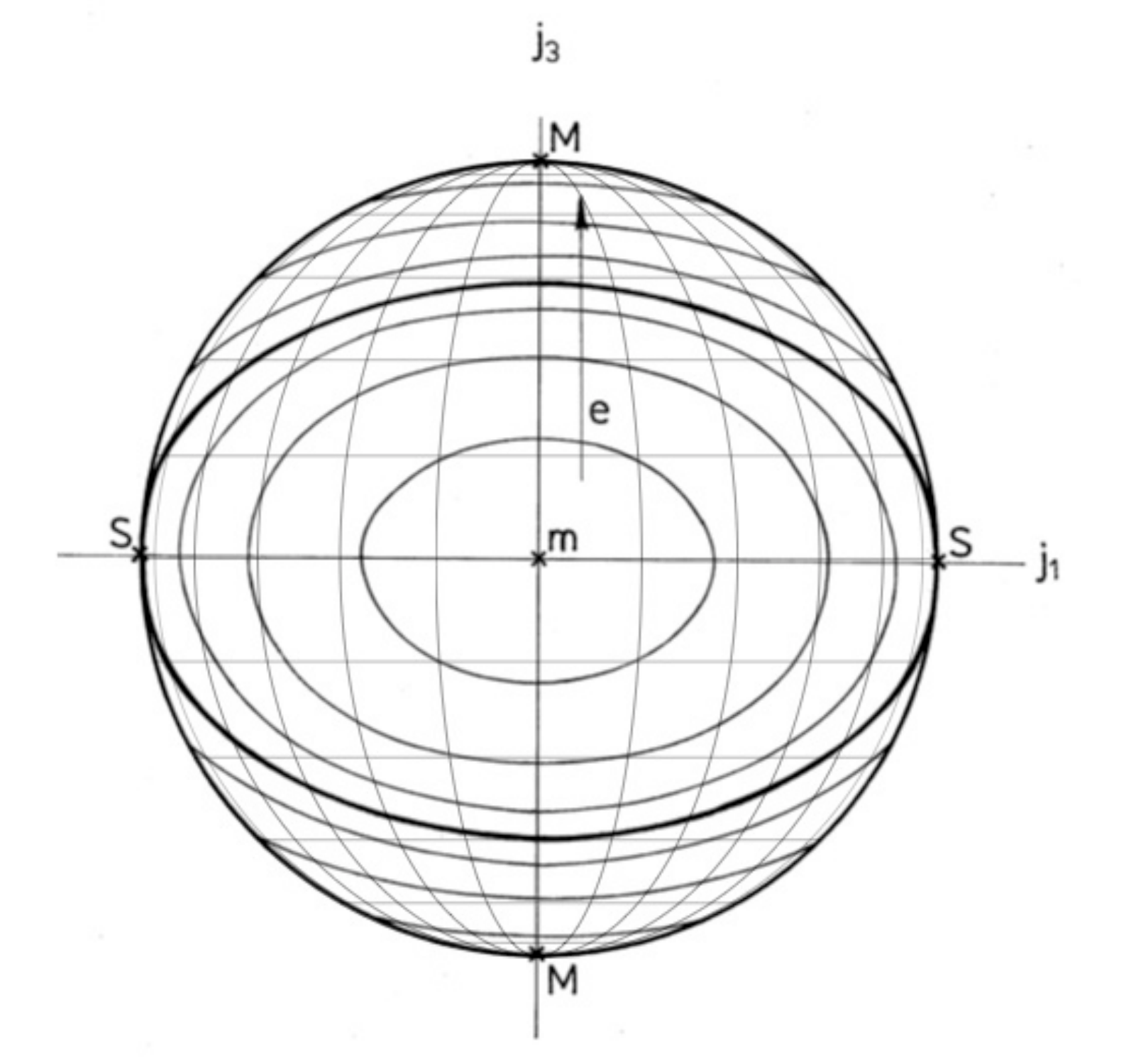}
   \includegraphics[width=0.48\linewidth]{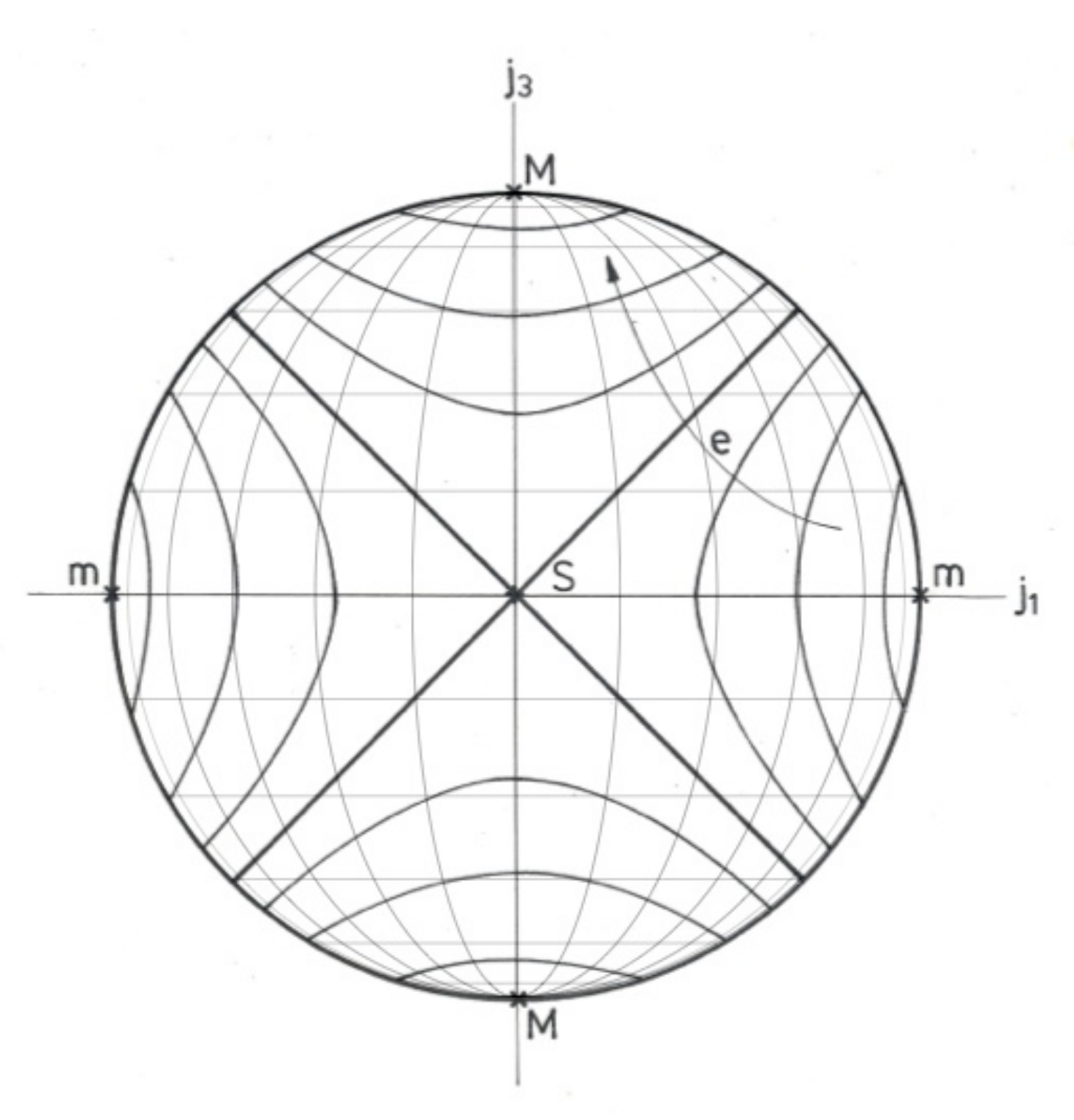}
  \caption{\label{f:orbitstrax}
   Classical orbits  in \am space for zero pairing and $\om=0$. Left: Medium-long projection of the orbits ($\ga=-30^o$), which  lie on the surface of the 
    sphere of fixed \amd.    
    Right: Short-long  projection of the orbits ($\ga=30^o$). The minima and maxima of the energy are labeled by m and M, respectively.
    The energy of the orbits increases in direction of the arrow. 
  }
   \end{center}
 \end{figure}
 \begin{figure}[t]
  \begin{center}
  \includegraphics[width=0.48\linewidth]{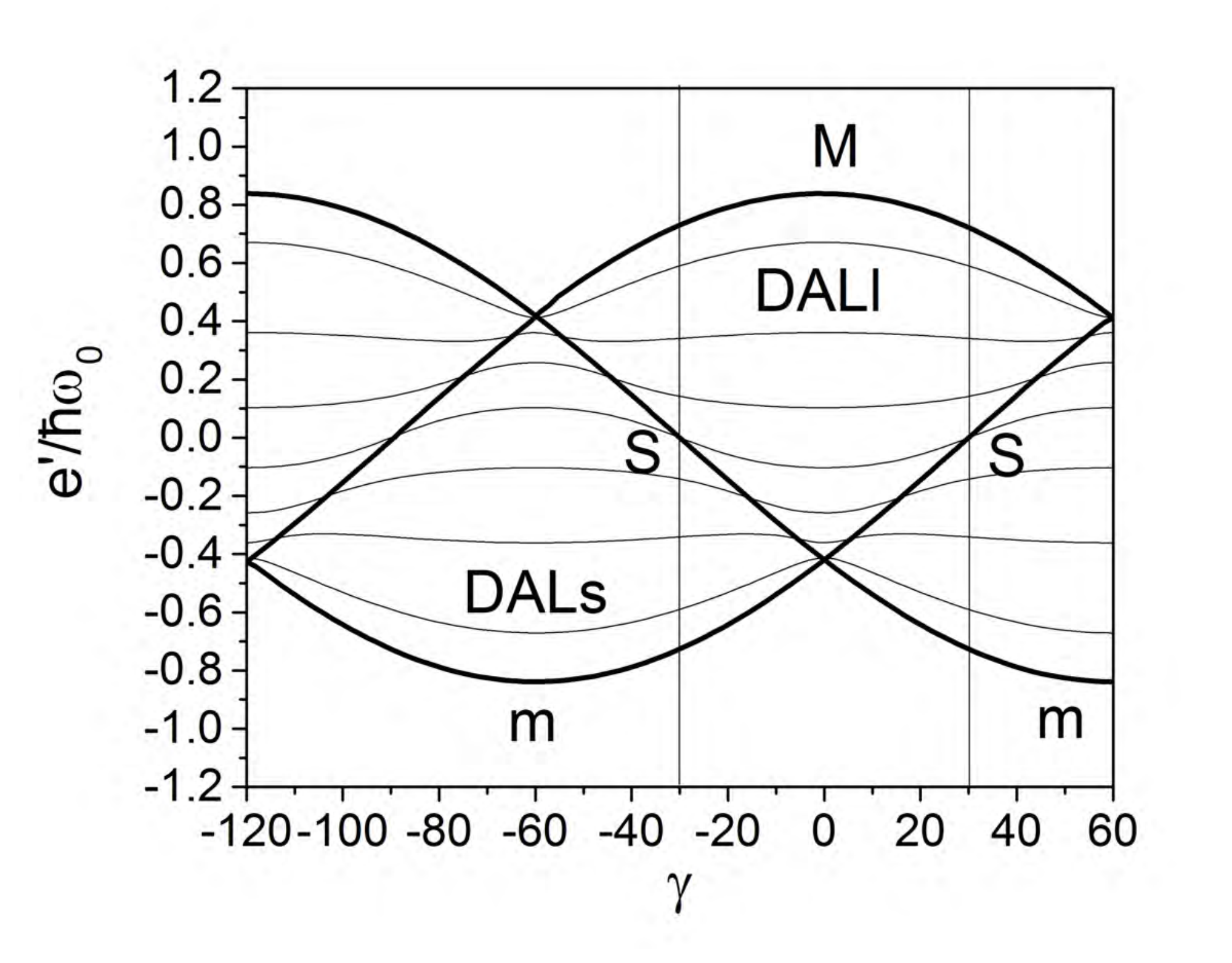} 
  \includegraphics[width=0.50\linewidth]{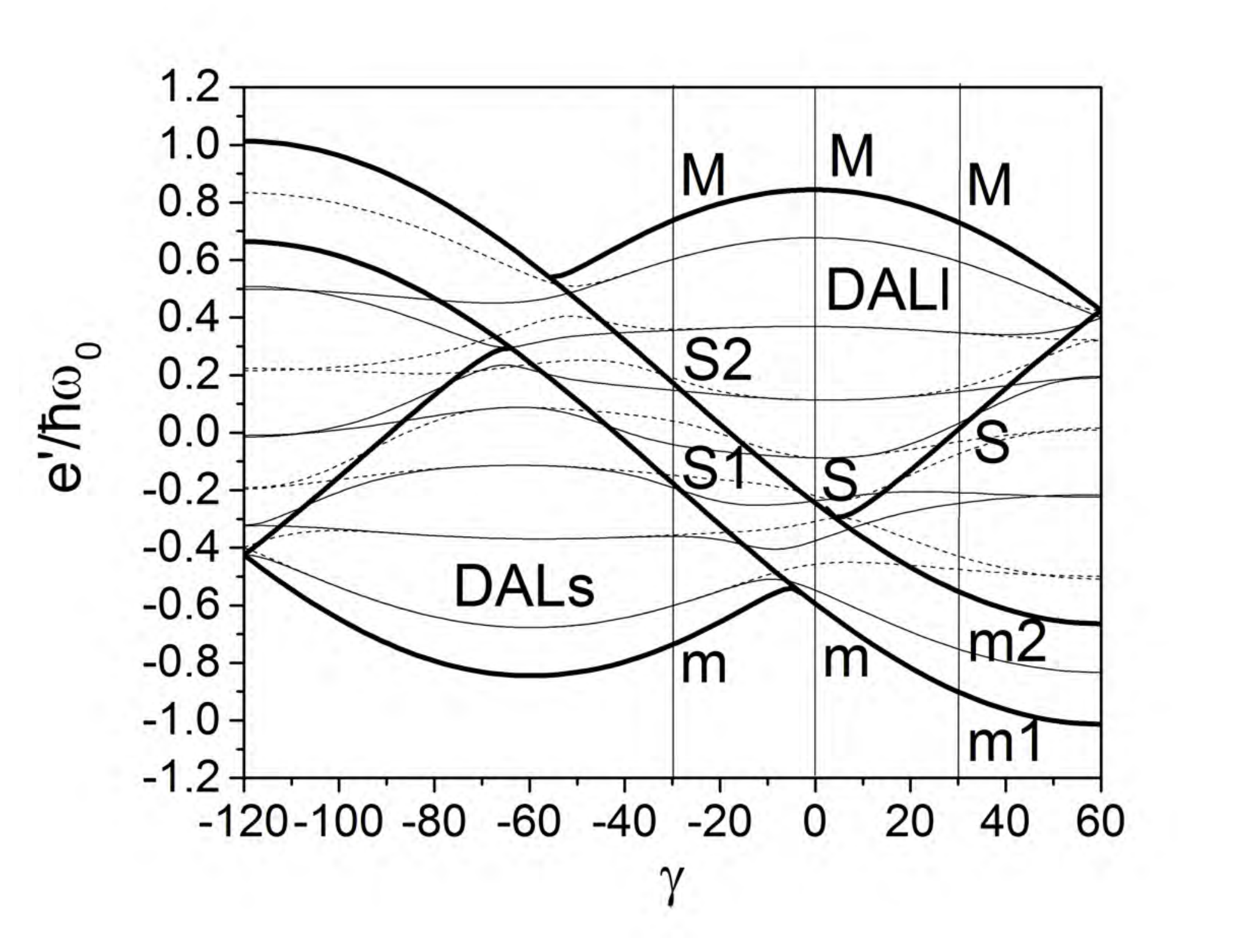} 
      \caption{\label{f:levelsom0ga}   Left: Quantal energies  of an i$_{13/2}$ \qp particle in the triaxial potential with $\varepsilon=0.26$ 
      as function of the triaxiality parameter $\ga$ for $\om=0$. Right: Like the left panel  $\om=0.025\om_0$.       
      The  cases $\ga=-30^\circ,~0^\circ,~30^\circ$  discussed below are indicated by vertical lines.  
        The separatrices and the maxima and minima of the classical energy
      are shown as thick  curves and labeled in accordance with the other figures.}
\end{center}
 \end{figure}

 \subsection{Triaxial shape, rotation about a principal axis}
 
The orbits are the intersection curves between the \am sphere and the triaxial energy ellipsoid.
Figs. \ref{f:orbitstrax}  illustrate    the topology  of the orbits  at $\om=0$, without pairing and $\ga=\pm30^o$.   In contrast to the axial potential, 
there are two classes of DAL orbits, the phase spaces of which
are delineated by the separatrix S.   The orbits  $e>e_S$ are aligned with (precess around) the long (l-) axis. They appear where the 
energy ellipsoid sticks out of the \am sphere. The $e<e_S$ orbits are aligned 
 with the short (s-) axis. They appear where the \am sphere sticks out of the energy ellipsoid.   The projections onto the 1-3-plane 
 are ellipses and hyperbolas, for $\ga=-30^\circ$ and $30^\circ$, respectively. The separatrix corresponds to the ellipse that touches the
\am sphere or to the asymptotes of the hyperbolas, respectively.   The ratio
\beq\label{eq:axratio}
\frac{a_1}{a_3}=\left[\frac{\sin \ga}{\cos\left(\ga+30^\circ\right)}\right]^{1/2}
 \eeq
  is the ratio of the half-axes of the ellipses 
for $-60^\circ\leq\ga\leq0^\circ$ or the slope of the asymptotes for $0^\circ\leq\ga\leq60^\circ$. The left panel of Fig. \ref{f:levelsom0ga}  shows how the 
topology changes with the shape.
For decreasing $\vert \ga \vert$ the phase space of the DAL orbits aligned with the l-axis (DALl) increases
at expense of the phase space of the DAL orbits aligned with the s-axis (DALs) and vice versa for 
increasing $\vert \ga \vert$. For the axial limits the  separatrix disappears and the orbits are circular 
in the medium-long projection (used in the left panel of Fig. \ref{f:orbitstrax}). There  are only DALl orbits for $\ga=0^\circ$, which are  horizontal lines
in the short-long projection (used in the right panel of Fig. \ref{f:orbitstrax}).
There are only DALs orbits for $\vert\ga\vert=60^\circ$, which are vertical lines in the short-long projection     
The  lowest quantal routhians are shifted above the classical minima by the zero-point energy, which is larger  at strong triaxility, where the orbits are well confined 
by the potential, than
 at axial shape, where they are more delocalized. 
    
      \begin{figure}[t]
  \begin{center}
  \includegraphics[width=0.48\linewidth]{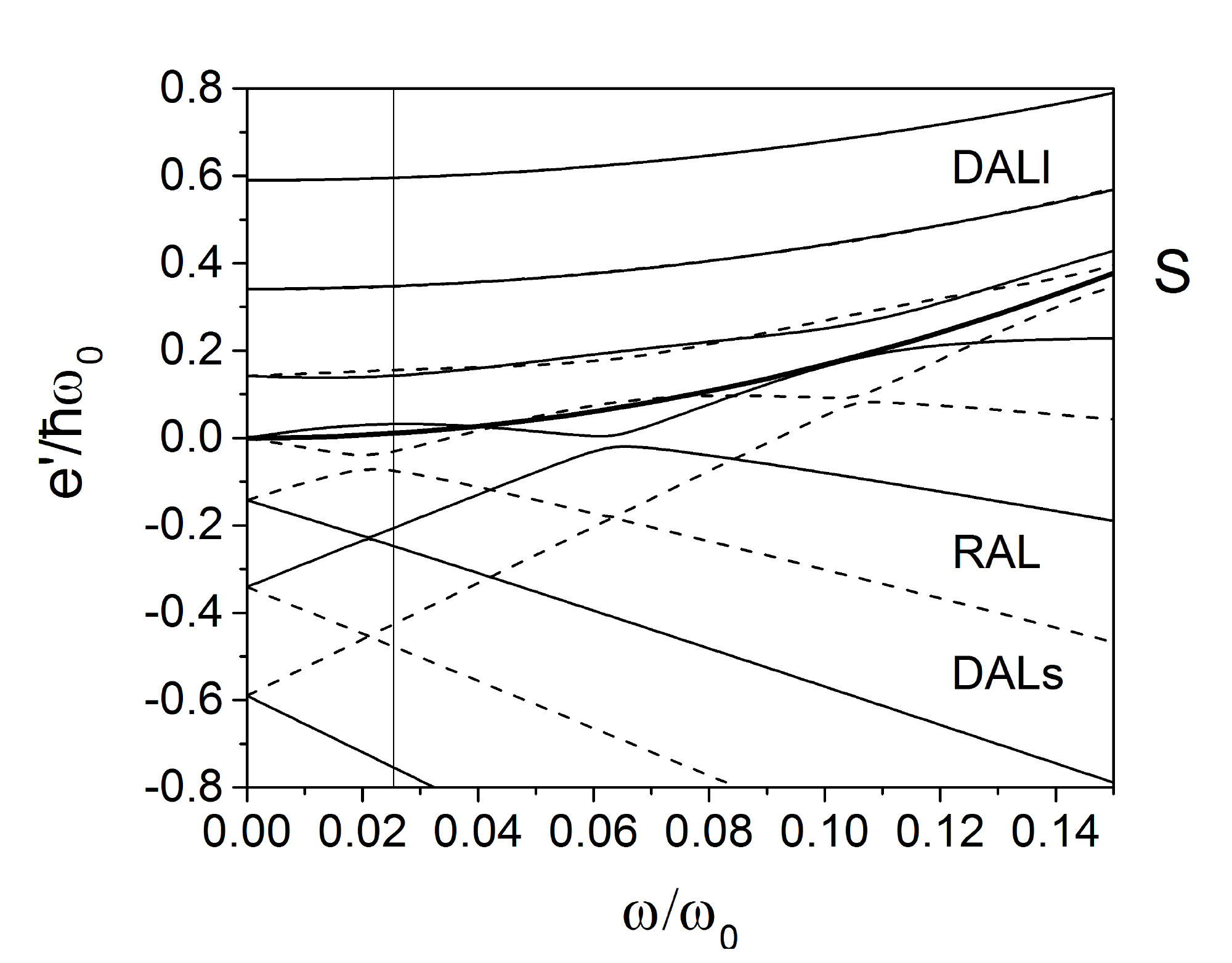} 
 \includegraphics[width=0.48\linewidth]{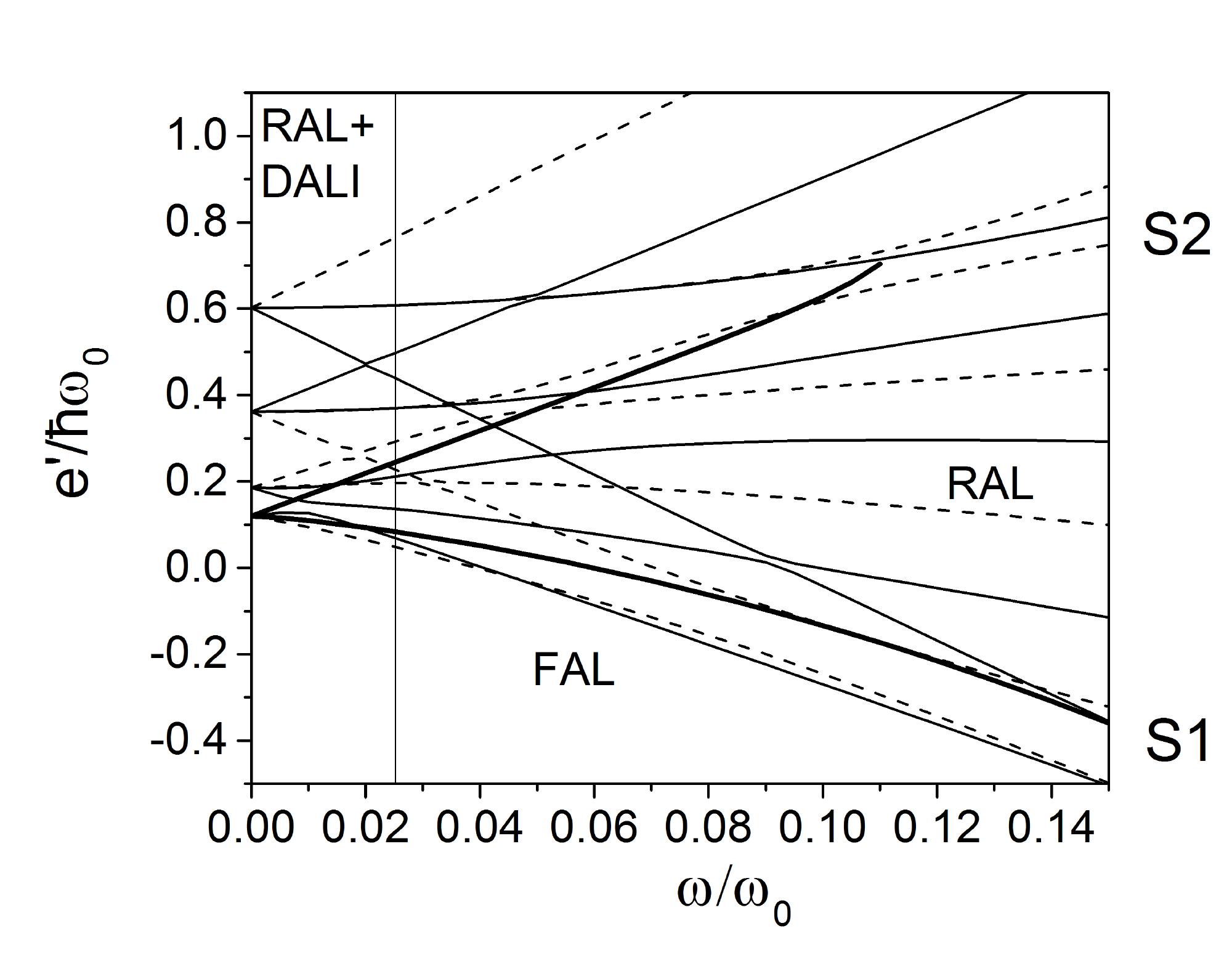}  
   \includegraphics[width=0.47\linewidth]{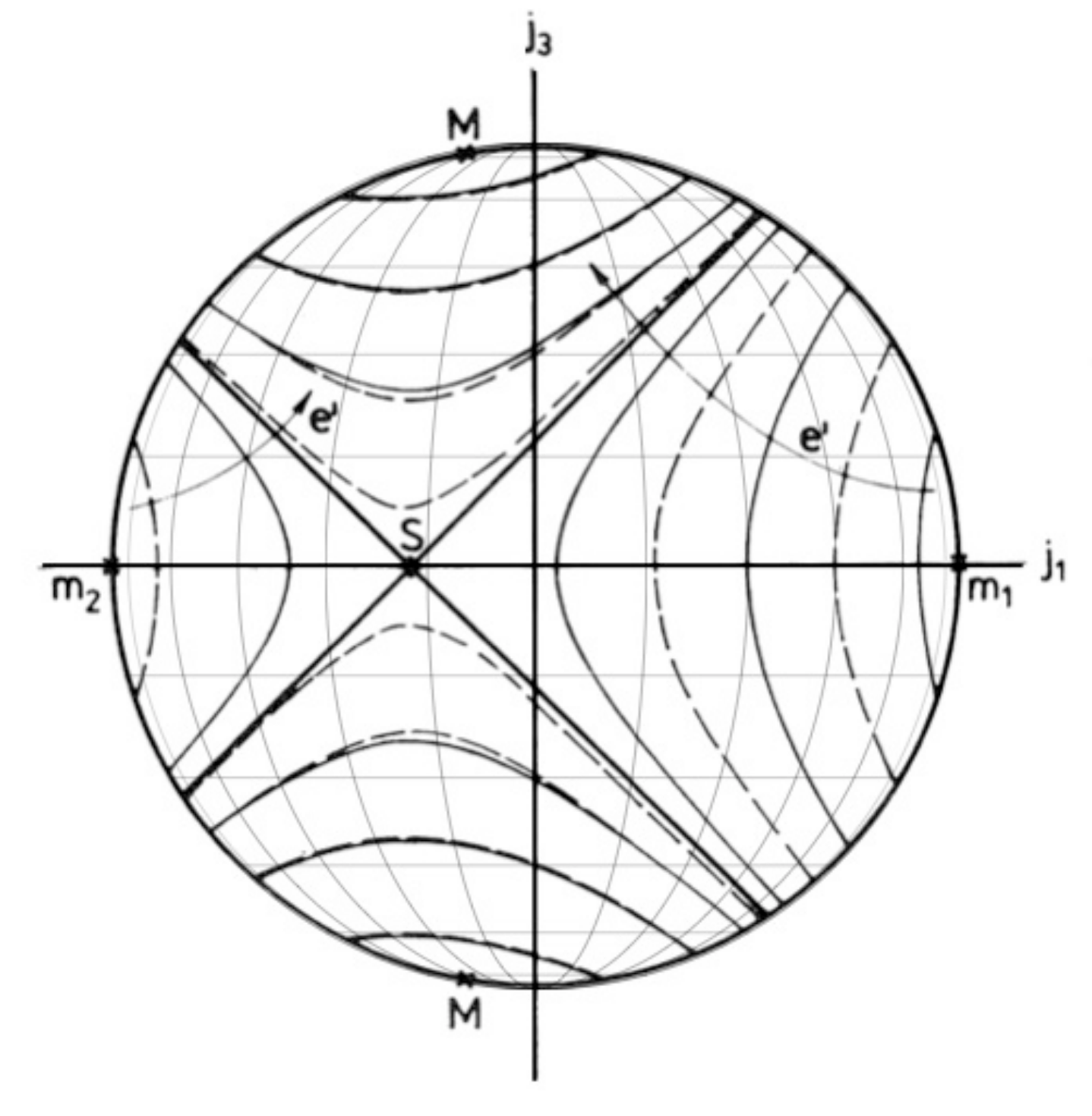} 
  \includegraphics[width=0.52\linewidth]{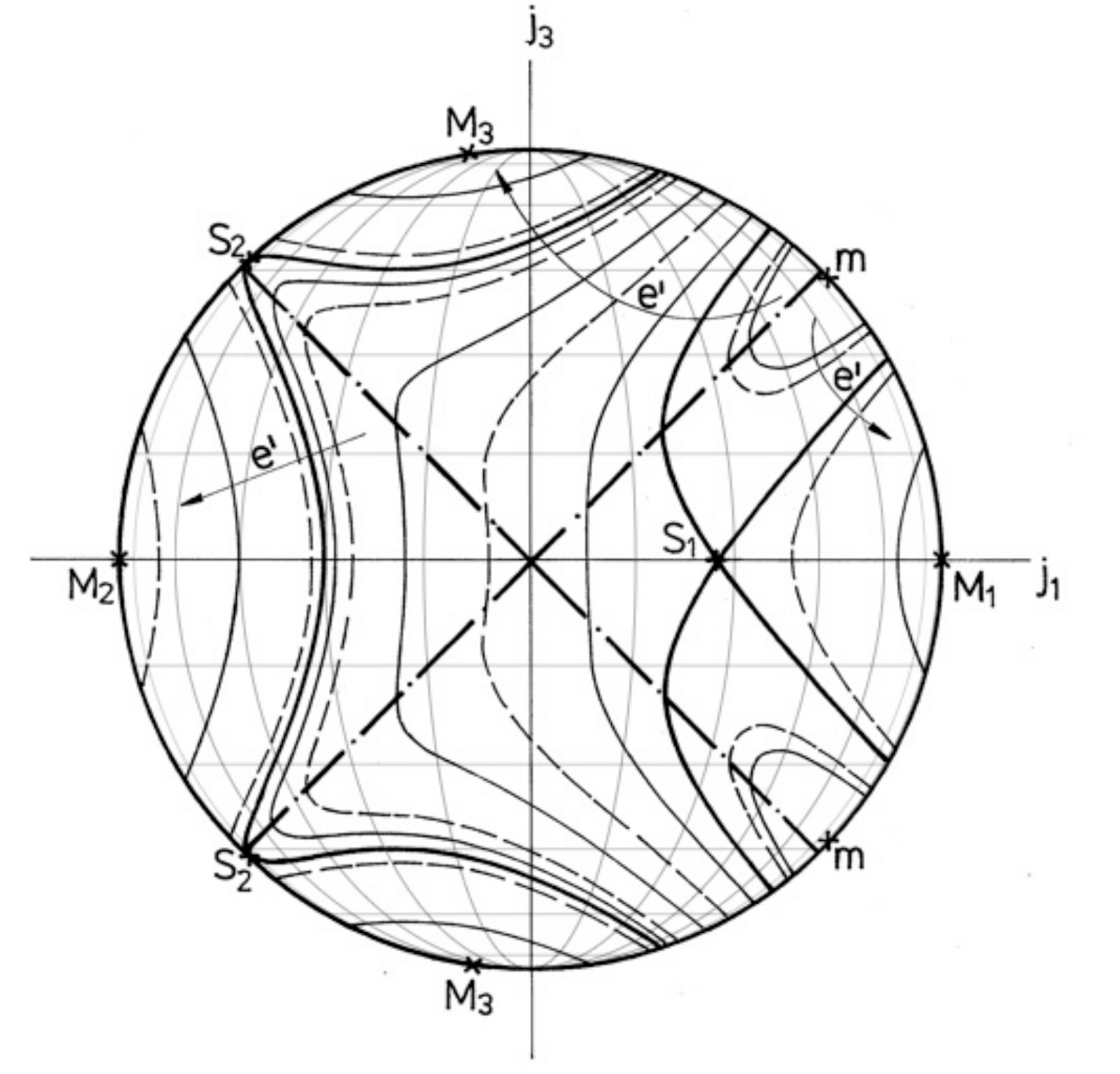} 
    \caption{\label{f:orbitstraxs}  Routhians and classical orbits of an i$_{13/2}$ \qp
          in the triaxial potential with $\varepsilon=0.26, \gamma=30^\circ$, rotating with the frequency $\om$ about the 1- axis, which is the short axis. 
          The layout is analog to Fig. \ref{f:orbitsax}. \\
          Upper left panel: Quantal routhians for zero pairing. \\Upper right panel:   Quantal routhians for $\Delta=0.12~\hbar\om_0$ and $\lambda=0$.  \\       
       Lower left panel:  Classical orbits  in \am space for zero pairing. The figure shows the 1 (short)-3 (long) -projection of the orbits, which  lie on the surface of the 
    sphere of fixed \amd. The frequency $\om=0.025\om_0$. The energy of the orbits increases in direction of the arrow. \\   
    Lower right panel: Analogous to the lower left panel for finite pairing. The dash-dotted line is the orbit with $e'(0)=\lambda=0$.}

  \end{center}
 \end{figure}

  \begin{figure}[t]
  \begin{center}
 \includegraphics[width=0.48\linewidth]{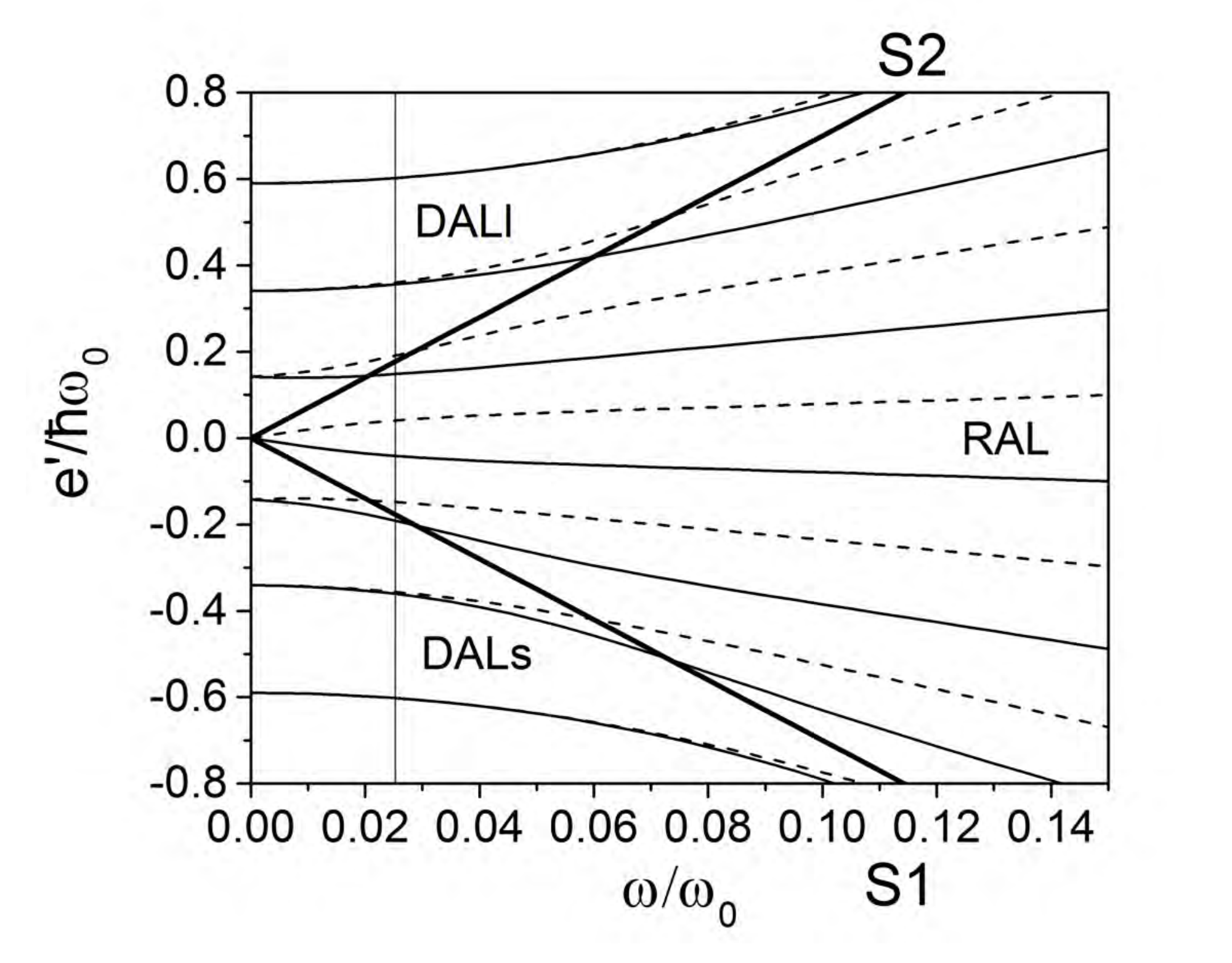} 
 \includegraphics[width=0.48\linewidth]{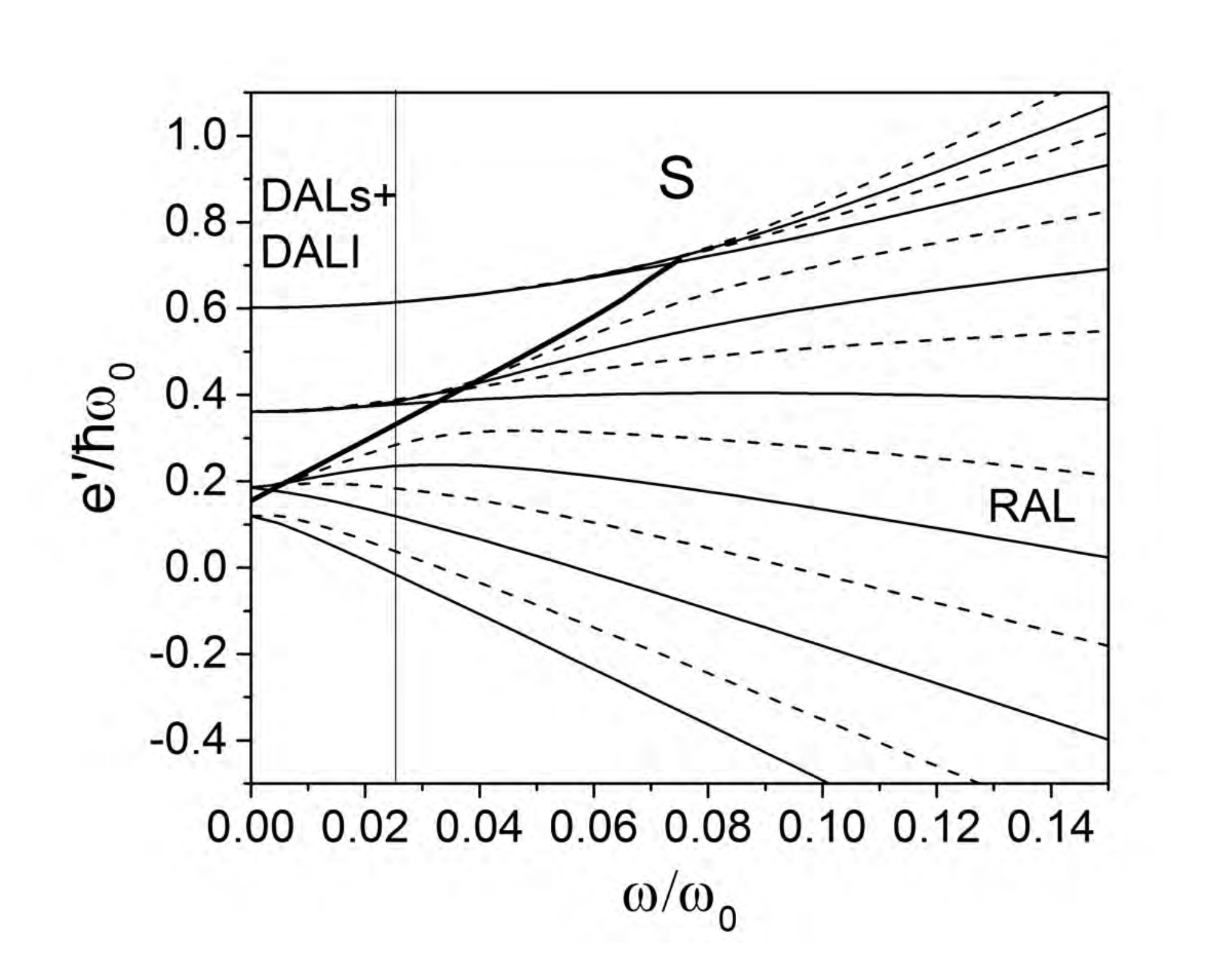}  
  \includegraphics[width=0.48\linewidth]{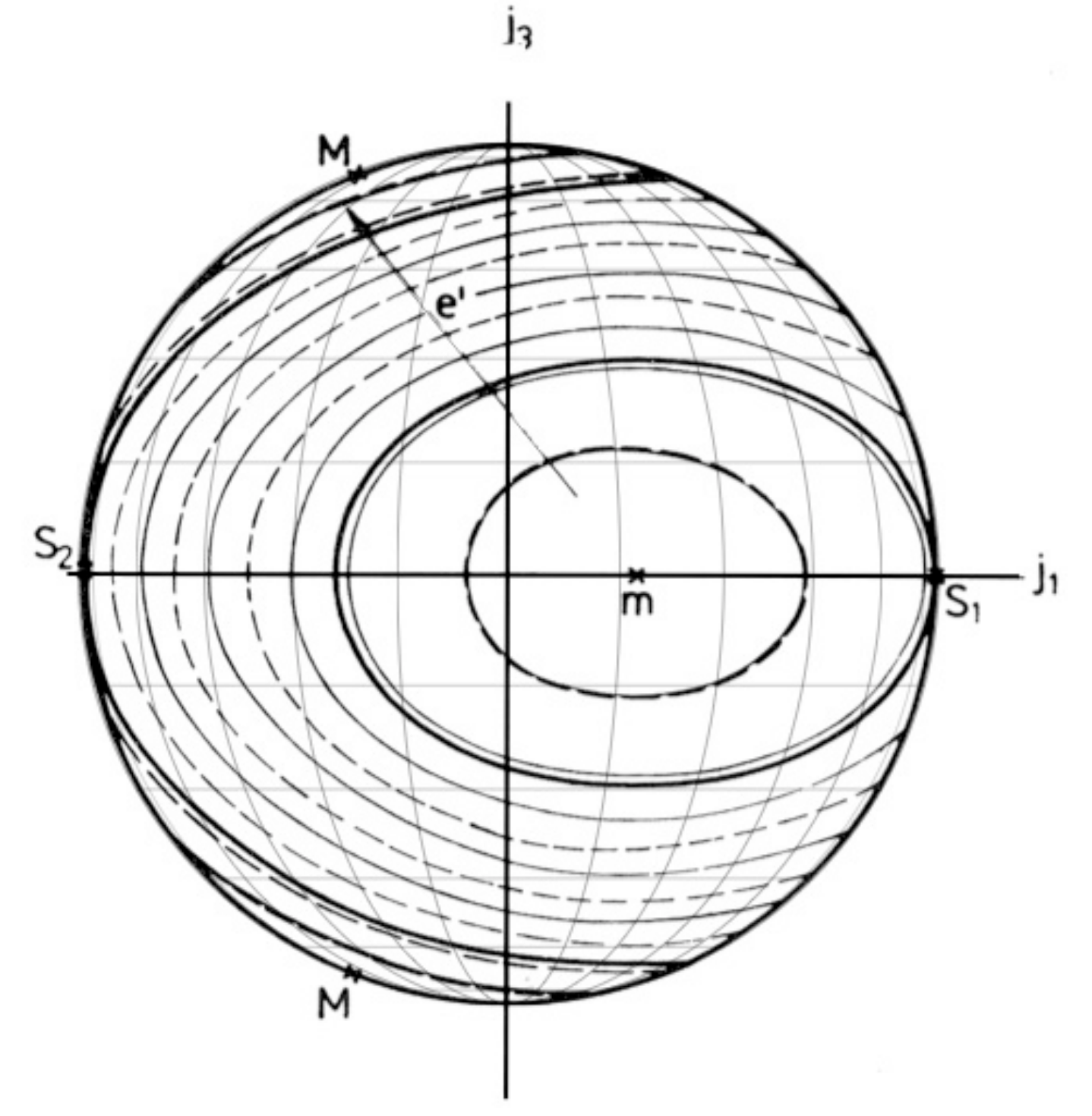} 
   \includegraphics[width=0.48\linewidth]{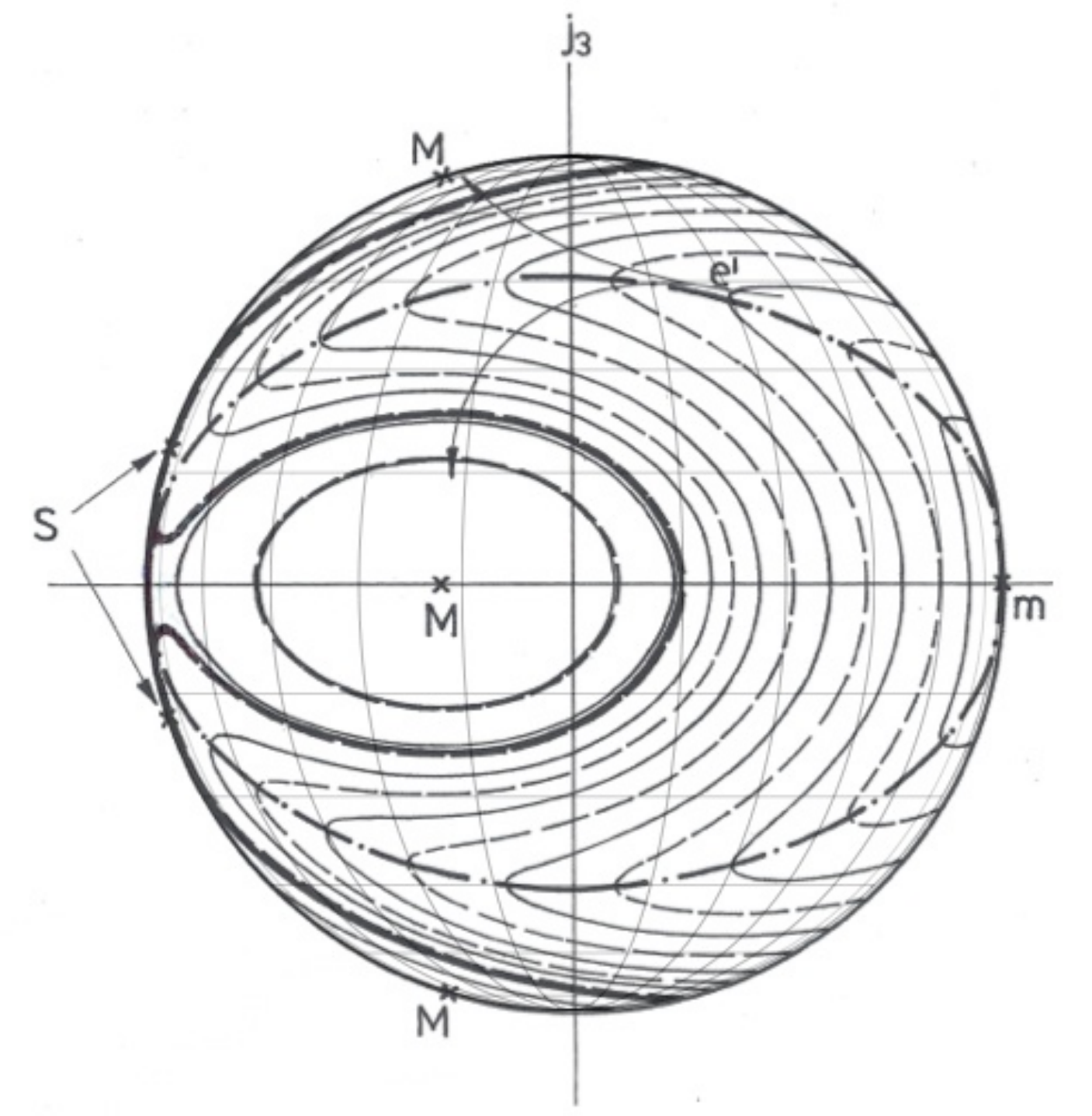}  

    \caption{\label{f:orbitstraxm}  Routhians and classical orbits of an i$_{13/2}$ \qp
          in the triaxial potential with $\varepsilon=0.26, \gamma=-30^\circ$, rotating with the frequency $\om$ about the 1- axis, which is the medium axis. 
          The layout is analog to Fig. \ref{f:orbitsax}. \\
          Upper left panel: Quantal routhians for zero pairing. \\Upper right panel:   Quantal routhians for $\Delta=0.12~\hbar\om_0$ and $\lambda=0$.  \\       
       Lower left panel:  Classical orbits  in \am space for zero pairing. The figure shows the medium-long-projection of the orbits, which  lie on the surface of the 
    sphere of fixed \amd. The frequency $\om=0.025\om_0$. The energy of the orbits increases in direction of the arrow.  \\   
    Lower right panel: Analogous to the  lower left panel for finite pairing. The dash-dotted line is the orbit with $e'(0)=\lambda=0$.}
  \end{center}
 \end{figure}
 
 In the following the 1-axis is chosen as the rotational axis, e.g $\vth=90^\circ$. 
 From Eqs (\ref{eq:ellipsoid},\ref{eq:ellipsoid13}) it is clear that increasing the rotational frequency $\om$ merely shifts  the centers of the ellipses and hyperbolas 
 along the 1-axis to  $j_{1c}$ given by Eq. (\ref{eq:center}).
 Showing $\ga=30^\circ$ as an example,  the left panels of Fig. \ref{f:orbitstraxs} illustrate  
  the case  $0^\circ\leq\ga\leq60^\circ$ of a particle in the triaxial potential that rotates about the short axis.
 The 1-3-projection of the orbits are hyperbolas. 
 The asymptotes are the separatrix. Their crossing  is the saddle point, 
 which is  the solution $j_2\not=0,~j_1=j_{1c},~j_3=j_{3c}=0$ of Eqs. (\ref{eq:qdot},\ref{eq:pdot}).
   It has the energy 
  \beq
  e'_S=2\ka\left[\sin\left(\ga-30^\circ\right) j^2+\left(\frac{\om}{\ka}\right)^2\frac{1}{16\sqrt{3}\sin\ga}\right]. 
  \eeq
  As a new feature compared to the axial case,
  one has "deformation assisted RAL". The DALs orbits with $e'_{m1}\leq e' \leq e'_s$ lie in the  sector right to the separatrix. They
    revolve the 1-axis (s), which is also the rotational axis, i. e. they are RAL as well. 
  The DALs orbits with $e'_{m2}\leq e' \leq e'_s$  lie in the sector left to the separatrix. 
  They also revolve the 1-axis (s), but they are anti-aligned with the s-axis.   
  This region exists only for $-j_{1c}<j$, i. e. only as long as  the center of the seperatrix lies inside the \am sphere.
  The sectors above and below the separatrix contain the DALl orbits, which are two-fold degenerate.
  The topology is reflected by the routhians in the upper left panel of Fig. \ref{f:orbitstraxs}. Below the separatrix the orbits  are aligned or anti-aligned with the s-axis. 
  For $ \om/\ka>j(4\sqrt{3}\sin\ga)$ the center of the asymptotes lies outside the \am sphere, and there are no anti-RAL orbits left. This is where the highest anti-aligned 
  routhian meets the separatrix. Now the two-fold degenerate DAL orbits are located above a new separatrix, which is the hyperbola that touches the \am sphere from inside. 
  Like for the parabolas in the axial case, the curvature of the hyperbolas decreases with $\om$. 
  For $\om\rightarrow\infty$ they approach vertical lines, and only RAL orbits survive. 
   For $\ga\rightarrow 0$ one has $j_c\rightarrow-\infty$, and the hyperbolas become parabolas.  For $\ga\rightarrow 60^\circ$, only DALs orbits remain, which are vertical lines and RAL. 
      
  The right panel of Fig. \ref{f:levelsom0ga} displays how the quantal routhians and the stationary  points of the classical routhian change 
   with the triaxiality parameter $\ga$  for a given frequency of $\om=0.025\om_0$.
   The  zero-point energy of the lowest quantal routhian is larger for the  DAL orbits at strong triaxility, which are well confined by the potential,
    than the RAL orbits at axial shape, which are less confined by the cranking term. 
    The curve m1 shows the classical minimum of the anti-aligned RAL orbits in the $0^\circ\leq\ga\leq60^\circ$ sector. 
     It is noted that m1 continues S of the axial case, both of which corresponding to the touch point of the orbit with the \am sphere at $j_1=-j,~j_3=0$. The separatrix S, 
   which corresponds to the center of the hyperbolas, branches off this curve at slightly positive $\ga$ values. Understanding this in terms of geometry requires a more detailed 
   analysis, which is omitted, because it is not particularly relevant.

    The above discussion for $\ga=0^\circ$  exemplifies the general relation between the quasiparticle and the particle orbits. 
    Applying it to the case $\ga=30^\circ$, consider the upper hemisphere $j_3>0$. 
   (The lower hemisphere is symmetric.)
    Take the particle orbits obtained for $\De=0$ and overlay the Fermi orbit $e'(\om=0,j_1,j_3)=\la$ (dash-dotted).   The branch of the quasiparticle orbit that lies 
    above the Fermi orbit is particle-like and remains about the same.
   The branch  that lies below the Fermi orbit is hole-like, i. e. it is  obtained by reflection through the 2-3- plane. In the 
   vicinity of the Fermi orbit the particle- and the hole-branches are 
   bent such that they smoothly match, where the larger $\De$ the smoother the bend. 
 
 The lower right panel of Fig. \ref{f:orbitstraxs} shows the quasiparticle orbits that emerge from the particle orbits in the lower left panel.
     For the displayed case of $\la=0$ the Fermi hyperbolas coincide with the dash-dotted asymptotes. 
    The reflected hole-branches close the particle branches in the right hemisphere, which generates new orbits of FAL nature. The FAL region is separated 
    by the separatrix S1 from the two disjunct RAL spaces, one on the right-  and one on the left-hand side, having large and small alignment $\langle j_1\rangle$,
    respectively. They are separated by S2 from the DALl regions above and below and  the anti-aligned RAL region to the left. The various regions are 
    indicated in the routhian diagram in the upper right panel.  The quasiparticle orbits for other values of $\lambda$  can be qualitatively constructed by reflecting 
     the particle orbits through the Fermi orbit in the non-rotating potential with the energy $e=\lambda$, which are the hyperbolas away from the 
     asymptotes in Fig. \ref{f:orbitstrax} right (see Fig. \ref{f:orbitstrax}).          
 
  Showing $\ga=-30^\circ$ as an example, the left panels of Fig. \ref{f:orbitstraxm} illustrate the case $-60^\circ \leq\ga\leq0^\circ$   
  of an i$_{13/2}$  particle in the triaxial potential that rotates about the medium axis.
 The 1-3-projected orbits are ellipses. 
 With increasing $\om$ their center moves to the right to $j_{1c}$ given by Eqs. (\ref{eq:center}). There are two separatices S1 and S2, which  are the ellipses 
 that touch the circle   right or left, respectively. The touching points are the solutions $j_2=j_3=0,~j_1=\pm j$ of Eqs. (\ref{eq:qdot},\ref{eq:pdot}).
 Their energies are
 \beq
  e'_{S1,S2}=-\frac{\ka}{2}\sin\left(\ga+30^\circ\right) j^2\pm\om j.
  \eeq
  Inside S1 ($e'<e'_{S1}$) lie the DALs orbits that revolve the s-axis and outside S2 
 ($e'>e'_{S2}$) lie the DALl orbits that
 revolve the l-axis. Between the separatrices a belt of RAL orbits is located. The two separatrices disappear for   $ \om/\ka>j\vert4\sqrt{3}\sin\ga\vert$, when
 $j_c>j$, and only the RAL orbits remain. 
 The topology is reflected by the routhians in the upper left panel of Fig. \ref{f:orbitstraxm}. Outside the separatrices one sees the doubly degenerate  the DAL orbits
 with some anti-alignment (cf. discussion of $\ga=0$),  inside there are the  signature-split RAL orbits.  As a new feature, the RAL orbits do not carry much 
 aligned \am $\langle j_1\rangle$ when the RAL phase space is still small. The contributions of the right and left turning points are comparable because 
 the orbits extend far to the left, where $j_2$ is small (see discussion for the axial case).  
  With increasing $\om$ the centers of the ellipses move to the right. The ellipse S1 shrinks
 and with it the phase space of the DALs orbits inside. RAL orbits with positive $\langle j_1\rangle$ emerge outside.  Likewise, the ellipse S2 grows with $\om$, 
 the phase space of the DALl orbits outside shrinks, and RAL orbits with negative $\langle j_1\rangle$ emerge inside. For $\om=-4\ka\sqrt{3} j\sin\ga$
 the center of S1 reaches the surface of the sphere, and no DALs orbits remain. For $\om=2\ka\sqrt{3} j\cos (\ga+30^\circ)$ the curvature of S2 
 equals the one of the circle, and no DALl orbits remain. In case of $\ga=-30^\circ$  the DAL orbits disappear at the common frequency $\om=2\ka\sqrt{3} j$.
 
 For $\ga\rightarrow0^\circ$ the center $j_c\rightarrow\infty$ and 
 $a_1/a_3\rightarrow0$. The separatrix S1 disappears, the  orbits become parabolas, and the above-discussed scenario for prolate shape is approached. 
 Fig. \ref{f:levelsom0ga} right demonstrates that slightly below $\ga=0$ the separatrix S1 merges with the m, the minimum of the classical routhian, which continues as m  to $\ga>0$.
 For $\ga\rightarrow-60^\circ$ the ellipses become circles, and the separatrix S2 disappears. The DALs orbits between m and  S1 have minimal energy for oblate shape, which
 reflects their good overlap with the potential. 
  In the RAL region the orbit that is maximally anti-aligned has the largest energy.
  With decreasing energy, the projection $\langle j_1 \rangle$ becomes less negative and eventual positive. As seen in Fig. \ref{f:levelsom0ga}, with increasing $\om$
  the RAL belt opens from the $\om=0$ - separatrix. The orbits nearby are weakly coupled to triaxial potential, because the orbits change from DALl to DALs at S. 
  As a consequence the cranking term dominates the energy order.  
  
  Oblate shape and rotation about the l-axis  is also realized for $\ga=-180^\circ$.    Compared to the prolate shape $\ga=0^\circ$, this choice just changes the sign
  of the $j_3^2$ term, which is equivalent with $j_1\rightarrow-j_1$ and $e'\rightarrow-e'$. This corresponds to a  reflection of orbits in the lower left panel of Fig. \ref{f:orbitsax} 
  through the 2-3-plane and a horizontal reflection 
  of the upper left panel. The two alternative pictures correspond to different projections of the three-dimensional orbits. For $\ga=-60^\circ$  the 1-3-projection is the
  l-l-projection, which gives shifted circles. 
  For $\ga=-180^\circ$ the 1-3-projection is  the s-l-projection. The l-l-projection  is obtained by eliminating $j_3^2$ for $j_2^2$ in
   Eq. (\ref{eq:ellipsoid}), which results in orbits that  are circles with centers on the 1-axis.    
    
 The lower right panel of Fig. \ref{f:orbitstraxm} shows the quasiparticle orbits that emerge from particle orbits in the lower left panel. 
 Qualitatively they are constructed by reflecting the particle orbits  through the Fermi ellipse, which  is shown by the dashed dotted curve, and connecting   
    the reflected hole-branches  inside with  the particle branches outside. The reflection increases strongly the aligned \am  $\langle j_1\rangle$ 
    of the lowest  RAL  orbits. The separatrix S is obtained by connecting the unpaired reflected separatrix S1 with S2 through the Fermi ellipse. 
 Its saddle points S are close to S2 and the reflected saddle point S1.      
  The DALl outside S2 retain their particle character while the DALs orbits inside S1 become holes.  For the considered special case
    $\ga=-30^\circ$ and $\la=0$ they are degenerate.       
  
   \begin{figure}[t]
  \begin{center}
  \includegraphics[width=0.48\linewidth]{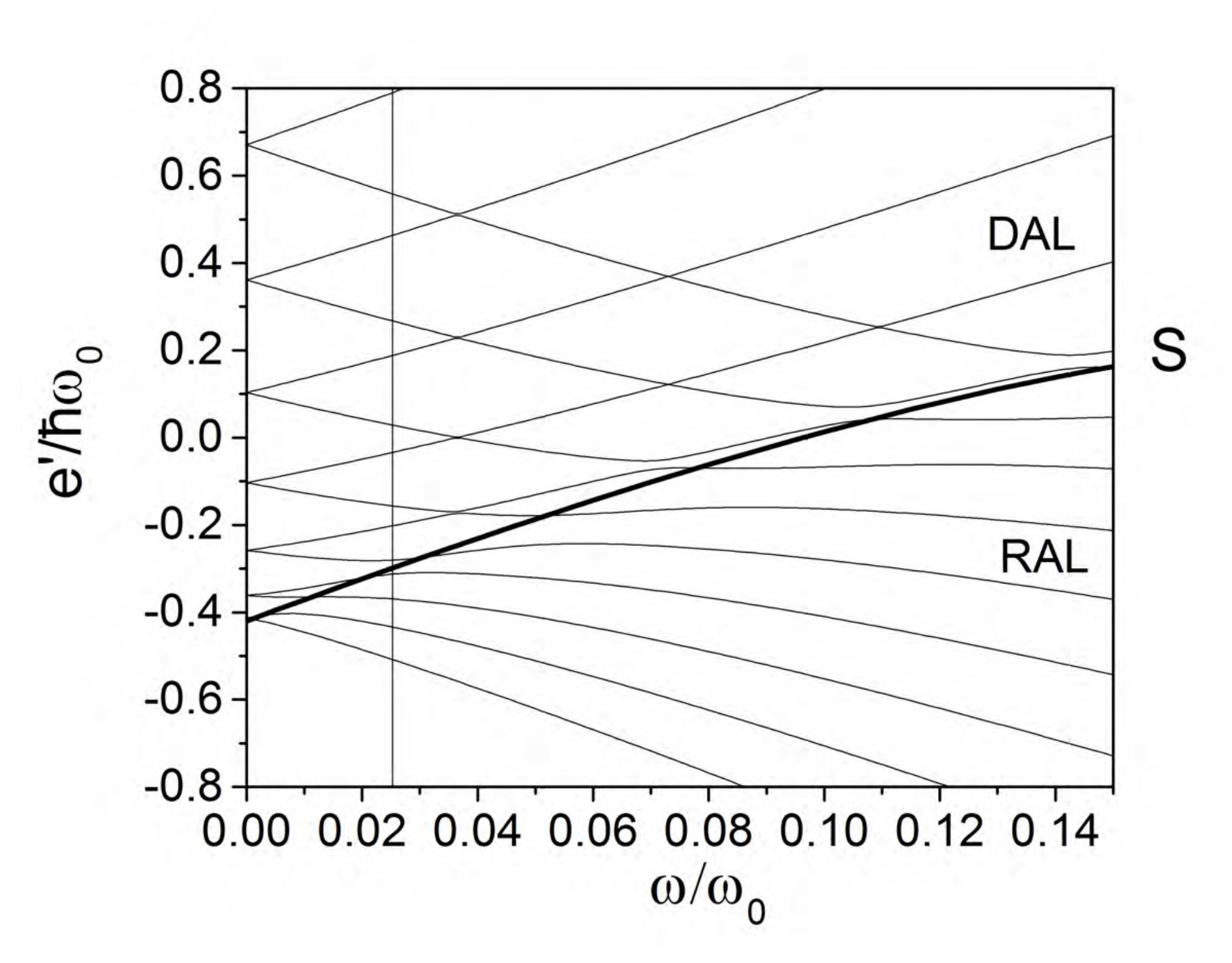} 
    \includegraphics[width=0.48\linewidth]{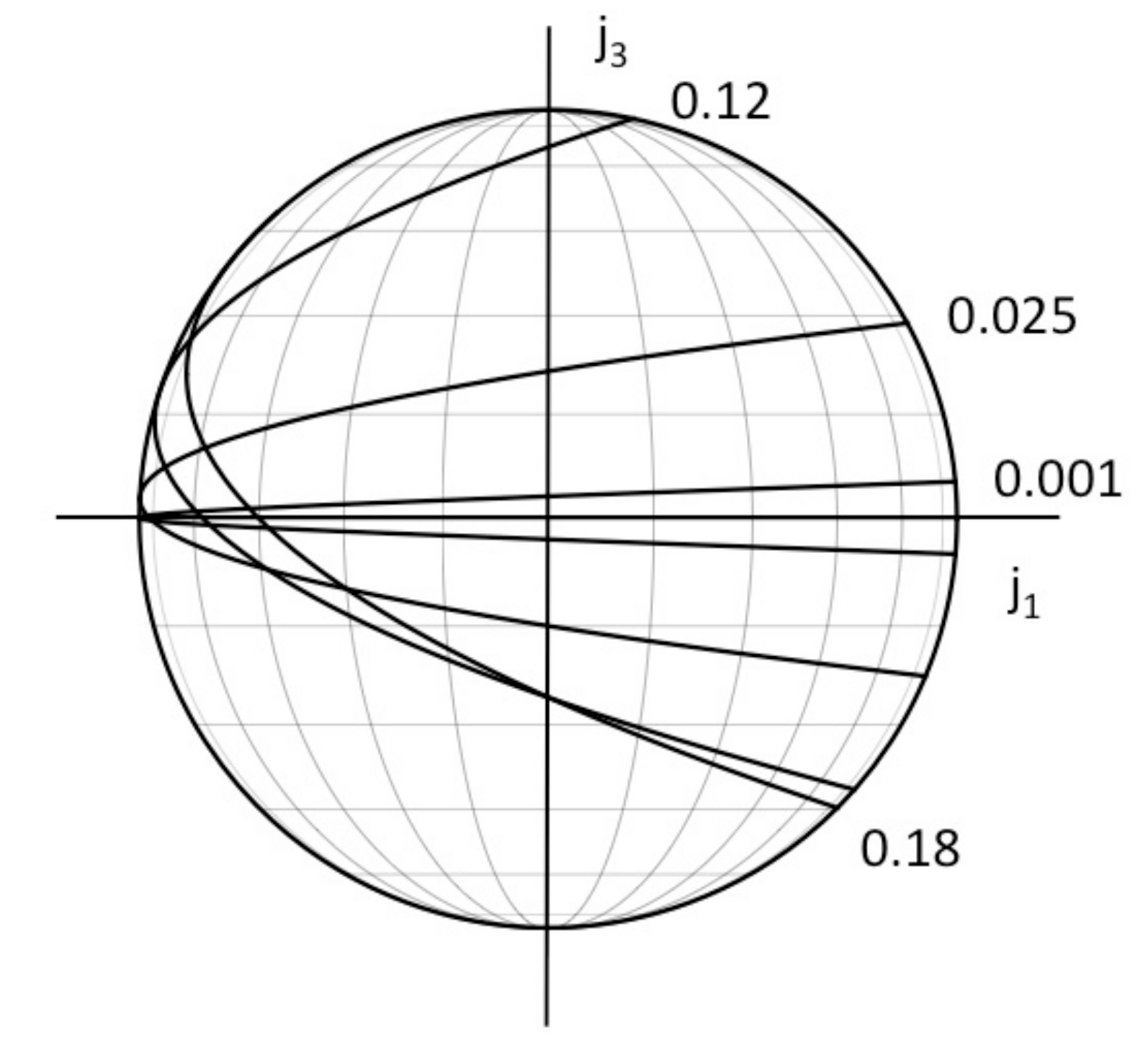} 
     \includegraphics[width=\linewidth]{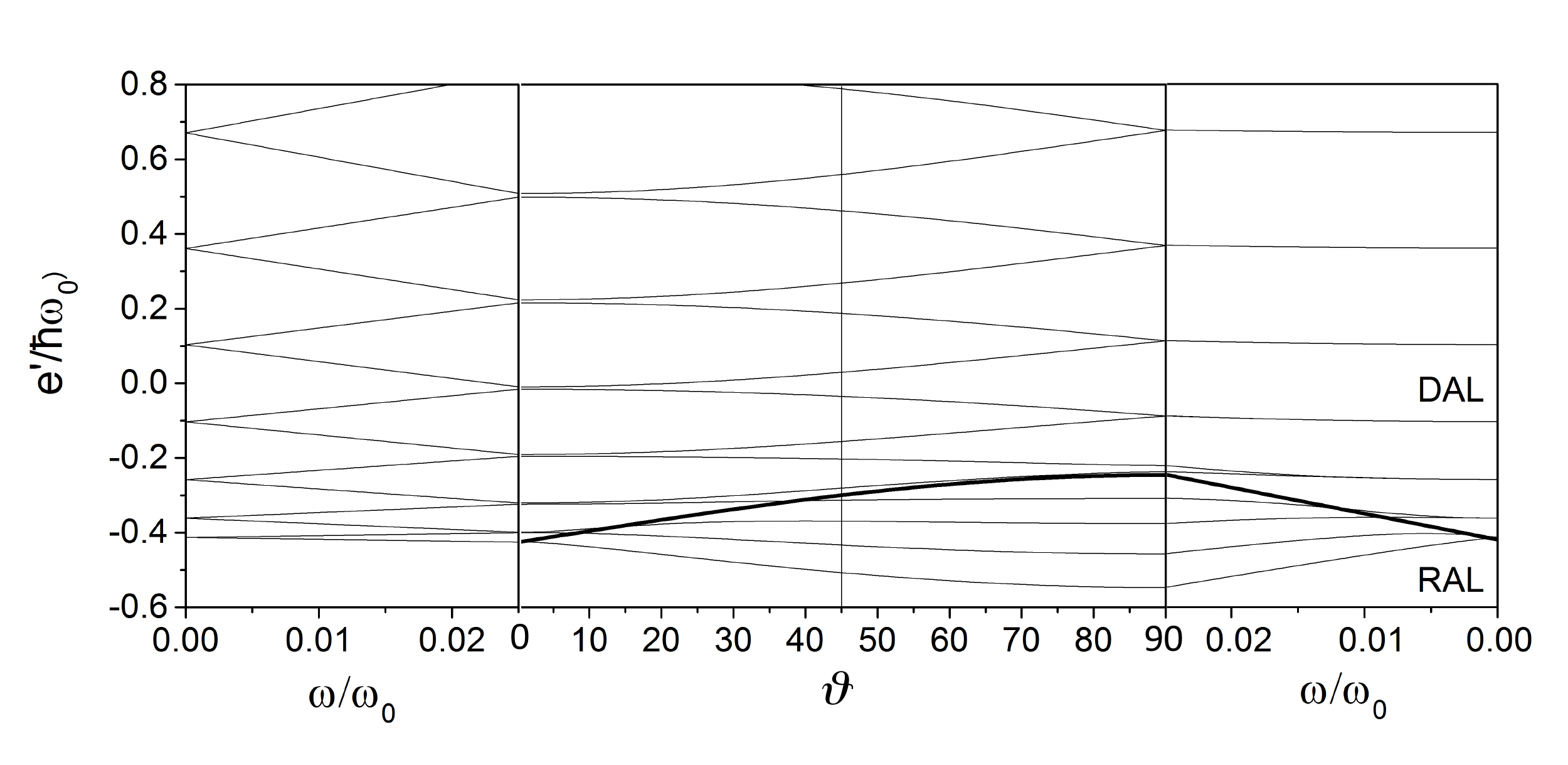}    
    \caption{\label{f:levelsaxth}  Upper left panel: Quantal routhians  of an i$_{13/2}$ particle
          in the prolate potential $\varepsilon=0.26$ that rotates with the frequency $\om$ about the  axis that is
          tilted by $45^\circ$ into the 1-3-plane. Upper right panel: Separatrices of the classical orbits that belong to the left panel. 
           Lower part middle panel:  Quantal routhians as functions of the tilt angle $\vth$ at $\om=0.025\om_0$. The left and right panels show the routhians 
           for rotation about the l- and s-axes, respectively. 
           The layout is analog to Figs. \ref{f:orbitsax}, \ref{f:sepax}. }
  \end{center}
 \end{figure}
        
  \subsection{Rotation about a tilted axis}     
  
   The upper left panel of Fig. \ref{f:levelsaxth}  illustrates the case $\ga=0$ of a prolate  potential that rotates about an  axis  tilted by $45^\circ$ into the 1-3-plane,
  which is the plane spanned by  the short axis and the long symmetry axis, respectively.   
  The tilt shifts the axis of the parabolas to $j_{3c}=\cos\vth\om/(6\ka)$ and reduces the curvature to $6\ka/(\sin\vth\om)$, 
  which can be seen  in the upper right panel that shows how the separatrix changes with $\om$.
 The DAL orbits outside the separatrix are no longer symmetric with respect to the 1-axis, and their routhians are no longer  
  degenerate. At low frequency they are split by  $-\om\langle j_3 \rangle \cos\vth$. 
  The RAL orbits inside S revolve an  axis that
  has the angle $\arcsin (j_{3c}/ j)$ with the 1- axis. 
  With increasing $\om$ the curvature of the parabolas decreases and their  axes move up. In this way, the upper DAL orbits 
  cross over into the RAL space. 
  The separatrix disappears for $\om\approx0.18\om_0$ when the curvature of the parabola
  equals  that of the circle at the touching point. There is no longer a topological difference between RAL and DAL. 
   Increasing  $\om$ further the orbits approach straight lines with the slope  $-\cot\vth$ (-1 for the displayed case), i. e. they approach the limit  
   of a fixed \am projection on the tilted axis. The lower panel of Fig. \ref{f:levelsaxth} 
   illustrates how the quantal routhians change with $\vth$.  Changing the tilt angle transforms the orbits 
    from parabolas centered around the 1-axis for  $\vth=90^\circ$  
    via parabolas with a reduced curvature centered around the $j_{3c}$ axis  to horizontal lines for 
      $\vth=0^\circ$. These  DAL orbits are circles revolving the 3-axis with $ j_3=\pm \Omega$ being the conserved \am projection
      on the symmetry axis. They are also RAL orbits because  the cranking axis agrees with the symmetry axis.

     \begin{figure}[t]
  \begin{center}
 \includegraphics[width=\linewidth]{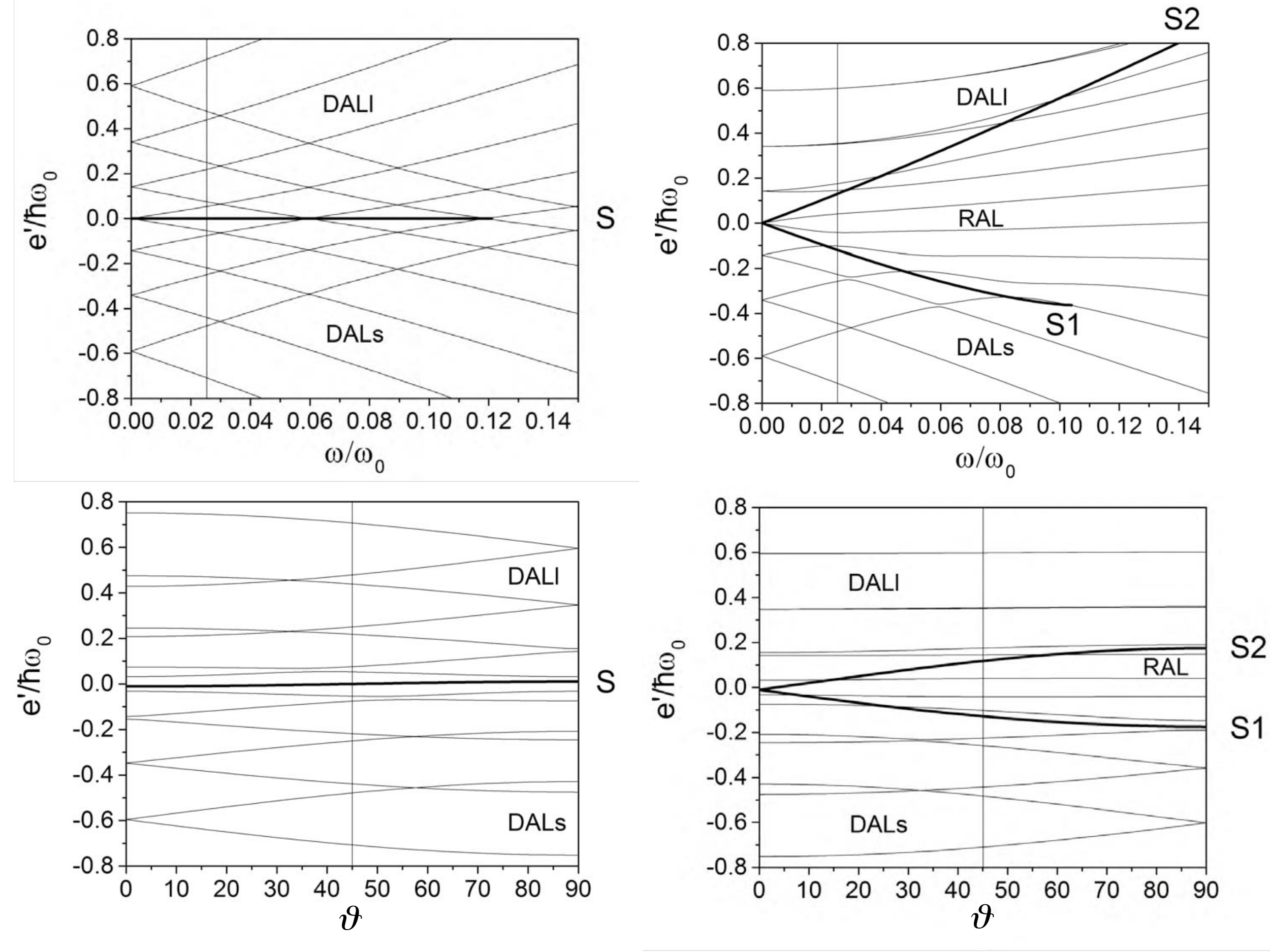}  
    \caption{\label{f:levelstraxth}  Quantal routhians  of an i$_{13/2}$ particle  
    in the triaxial potential with $\varepsilon=0.26, \vert \gamma\vert = 30$, rotating with the frequency $\om$ about a tilted axis.
   The layout is analog to Fig. \ref{f:orbitsax}. \\
    Upper left panel: $\ga=30^\circ,~\vth=45^\circ$, rotation axis tilted into the s-l plane. \\
    Upper right panel: $\ga=-210^\circ,~\vth=45^\circ$, rotation axis tilted into the m-s plane  \\       
    Lower  left panel: $\ga=30^\circ,~\om=0.025\om_0$.\\ The axis turns from the l- axis ($\vth=0$) to the s-axis ($\vth=\pi/2$).  \\   
    Lower right panel: $\ga=90^\circ,~\om=0.025\om_0$.\\ The axis turns from the m- axis ($\vth=0$) to the s-axis ($\vth=\pi/2$).}
  \end{center}
 \end{figure}

  For non-axial shape the centers of the particle orbits are shifted to $j_{1c}$ and $j_{3c}$ given by Eq. (\ref{eq:center}). 
    The upper left panel of Fig. \ref{f:levelstraxth} shows the routhians for $\ga=30^\circ$ and $\vth=45^\circ$ as an
  example for the rotation about an axis in the s-l-plane. The 1-3-projected orbits are hyperbolas. 
   The asymptotes are the two separatrix branches, which cross  at S located at   $j_{1c}$ and $j_{3c}$.  
    The lower left panel of  Fig. \ref{f:orbitstraxs}  illustrates the case $\vth=90^\circ$. For $\vth<90^\circ$ S is moved up by $j_{3c}$(=$-j_{1c}$ for $\vth=45^\circ$).   
     The DALs orbits right to S are also RAL orbits which revolve  an  axis that has the angle $\arcsin (j_{3c}/ j)$ with the 1- axis.   
With increasing $\om$ the center of S moves upward left on a straight line with the slope $-\cot\vth\sin\ga/\cos(\ga+30^\circ)$ (-1 for the displayed  case), which
     increases the phase space below  the asymptote with the slope +1 at the expense of the phase space above it.
    This means that DALl orbits with $\langle j_3 \rangle>0$ move into the DALs area and DALs orbits with $\langle j_1\rangle<0$ move into the DALl
    area. These are the routhians that cross the S line in Fig. \ref{f:levelstraxth} up-left.  When the center of 
    S reaches the \am sphere at $j_{1c}^2+j_{3c}^2=j^2$ (at $\om=0.12\om_0$ for the shown case), 
    the separatrix disappears, and all orbits become  RAL  revolving the tilted rotational axis. 
   Increasing  $\om$ further, they approach straight lines with  slope $-\cot\vth$ (-1 for the displayed case).
    The lower middle panel of  Fig. \ref{f:levelstraxth} shows the routhians for $\om=0.025\om_0$ as a function of the tilt angle $\vth$ of the rotational axis. 
    While the rotational axis is swept from the l- axis to the s- axis the center S of the asymptotes moves on a circle from the 3-axis to the 1- axis, where it is located in 
    the lower left panel of Fig. \ref{f:orbitstraxs}.  For $\vth=0$ the DALl orbits are also RALl   split by   $-\om\langle j_3 \rangle$, whereas the DALs orbits 
    are disjunct and degenerate. Increasing $\vth$   the doublets split by $-\om\sin\vth\langle j_1 \rangle$. Decreasing  $\vth$ from $90^\circ$ the analog  
    appears, where  DALl is exchanged with DALs.
          For $\ga\rightarrow0$, $j_{1c}$  moves far left and  the hyperbolas become the parabolas of the axial case.

  The 1-3-projection of the orbits corresponds to the m-l-projection for $-60^\circ\leq \ga \leq 0^\circ$ and to the m-s-projection for $-240^\circ\leq \ga \leq -180^\circ$ 
  (cf. Table \ref{t:axes}). As seen from Eqs. (\ref{eq:ellipsoid13}, \ref{eq:center}), rotation about an axis that is 
  tilted by $\vth$ with respect to the m- axis into the m-s- plane
  is related to rotation about an axis that is tilted by the same angle into the m-l-plane by the replacement $\{e',~j_{1c},~j_{3c} \}\rightarrow \{-e',~-j_{1c},~-j_{3c}\}$.  
   
    The upper right panel of Fig. \ref{f:levelstraxth} shows the routhians for $\ga=-210^\circ$ and $\vth=45^\circ$ as an
  example for the rotation about an axis in the m-s-plane.   
  The 1-3-projected orbits are ellipses  with the center $j_{1c}$ and $j_{3c}$,
  which moves downward left on a straight line with the slope $-\cot\vth\sin\ga/\cos(\ga+30^\circ)$ (1 for the shown case) with increasing $\om$.
 The lower  left panel of Fig.   \ref{f:orbitstraxm}  shows the m-l-projection  for rotation about the m-axis, i.e. $\vth=90^\circ,~\ga=-30^\circ$. 
  The m-s-projection is obtained  by reflecting through the 3-axis and exchanging maxima (M) and minima (m).  
   For $\vth<90^\circ$ the two separatrices are the ellipses that touch the circle
  at S1 and S2 below the 1-axis. Between them lie the RAL orbits which revolve  an  axis that has the angle $\arcsin (-j_{3c}/ j)$ with the 1-axis. While the
  points S1 and S2 move down on the circle with $\om$ the RAL region grows at expense of the DAL regions. The DALs region shrinks faster than DALl region. 
  
  The outer separatrix S1 
  disappears when the curvature at the touching point is equal to the curvature  of the circle, which occurs at
   $\om=0.104\om_0$ for the illustrated case (similar to the disappearance of the parabolic 
  separatrix in the right panel of Fig. \ref{f:levelsaxth}). 
  For larger $\om$ the RAL orbits approach straight lines with the slope $\cot\vth$. 
  When the center reaches the \am sphere at $j_{1c}^2+j_{3c}^2=j^2$ and the touching ellipse becomes
  a point, which occurs at   $\om=0.208\om_0$ for the illustrated case, the inner separatix S2  disappears, and only RAL orbits remain.  
   The lower right panel of Fig. 
  \ref{f:levelstraxth} shows the routhians for $\ga=-210^\circ$ and $\om=0.025\om_0$ as functions of the tilt angle,
   i. e. how they change from rotation about the s-axis to rotation about the m-axis.
  The development of the orbits can be anticipated from the lower left panel of Fig. \ref{f:orbitstraxm}. 
  The center of the ellipses moves on a circle from the 1-axis to the 3-axis when  $\vth$ 
  changes from $90^\circ$ to  $0^\circ$. The RAL region between the two ellipses shrinks on this path, which is reflected by the quantal routhians.
   For $\vth\rightarrow0^\circ$ the two separatices merge into one ellipse that touches the circle on both sides, which is the 2-3-projection of the asymptotes
  in the lower left panel of Fig. \ref{f:orbitstraxs} (rotated by $90^\circ$, M$\leftrightarrow$m).       
  
  As for rotation about the principal axes, the quasiparticle orbits can be qualitatively  constructed from the particle orbits  by reflection through the Fermi orbits.
  
  \subsection{Induced triaxiality and tilt angles}
  
  \begin{figure}[t]
  \begin{center}
  \includegraphics[width=0.48\linewidth]{levelsom025ga.pdf} 
 \includegraphics[width=0.48\linewidth]{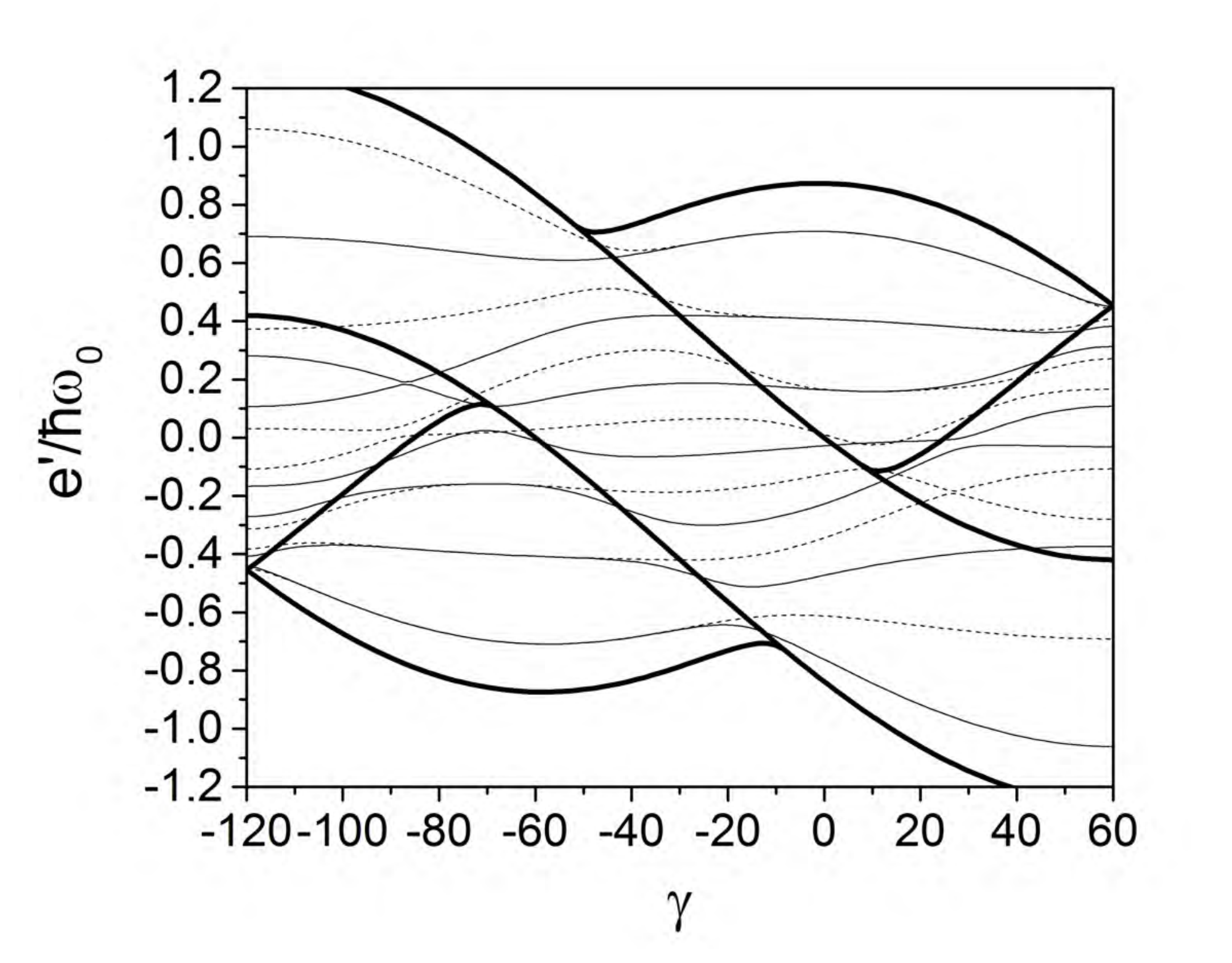}  
      \caption{\label{f:levelsga}  Routhians  of an i$_{13/2}$ \qp particle in the triaxial potential with $\varepsilon=0.26$ as function of  the triaxiality parameter $\ga$ 
      rotating about the 1-axis with the frequency $\om=0.025\om_0$  (left panel) and
      $\om=0.060\om_0$  (right panel). The layout is analog to Fig. \ref{f:orbitsax}. The  cases $\ga=-30^\circ,~0^\circ,~30^\circ$
      shown in Figs. \ref{f:orbitsax} - \ref{f:orbitstraxm} are marked.  The separatrices and maxima and minima of the classical routhian
      are shown as thick  curves and labeled in accordance with the other figures.}

  \end{center}
 \end{figure}

  The triaxial deformation parameter $\ga$ is susceptible to the polarization by the excited quasiparticles. 
   The equilibrium value of $\ga$ can be  seen as the balance between the drive of the excited quasiparticles and the 
  restoring force of the rotating "core", which is the  configuration from which the quasiparticles or particles and holes are excited \cite{FM83}. 
  For a qualitative guess, which is often enough,
  it is sufficient to discuss the general trends.  The balance between the coupling to 
  the deformed potential, which favors a large overlap of the density distribution of the orbit with the deformed potential, and the 
  inertial forces, which favor the alignment of the particle \am with the rotation axis,   determines the preferred $\ga$ value and tilt angle $\vth$.  
  To make a picture yourself of the density distribution,  imagine the torodial  density distribution of the high-j orbital (see Fig. \ref{f:qpcouplings}) to be 
  averaged over the various orientations corresponding to the classical orbits of $\vec j$.  
  
  Figs. \ref{f:levelsom0ga} and \ref{f:levelsga} show  the single particle routhians as functions of $\ga$ for rotation about the principal 1-axis with two different frequencies.
    For $\om=0$ the minimum of the lowest state in the shell  lies at $\vert \ga \vert=60^\circ$, because $\vec j$ 
  of the classical orbit m is aligned with the symmetry s-axis, which makes best overlap with the potential (cf. Fig. \ref{f:levelsom0ga} and discussion above).  
    For finite $\om$, $\ga=+60^\circ$ is lower because m is aligned with the 1-axis (which is s),
  which gives the extra energy gain of $\om j_1 $. This is not the case for $\ga=-60^\circ$, where m is aligned with the  2-axis (which is s). 
  The rotational gain $\om j_1$ is small for this DAL orbit, which 
  corresponds to  M in the lower left panel of Fig. \ref{f:orbitsax}  with $j_1\rightarrow-j_1$. 
  Prolate $\rightarrow$ oblate is achieved by changing the sign of the $j_3^2$ term, which is equivalent with 
  $e'\rightarrow-e'$ and $j_1\rightarrow-j_1$. For $\om=0$ and triaxial shape the  m orbit has DALs character. Its energy is higher than for oblate shape because 
  the density torus overlaps better with the potential that has the two l axes  in the torus' plane instead of the axes m and l. 
  For finite $\om$, $\ga>0$ is preferred because the
   m orbit is both DALs and RAL giving
  the extra energy gain $\om j_1$ compared to $\ga<0$, where the gain is small because it is DALs but not RAL (cf. Figs. \ref{f:orbitstraxs} and \ref{f:orbitstraxm}).  
  
The highest state  appears as a hole in the inert filled  shell, which has \mbox{$e'_{hole}=-e'_{particle}$}, i. e. its minima are the maxima of the particle states. 
As seen in the left panel of Fig. \ref{f:levelsom0ga},   the maxima of the M orbit lie at
$\ga=0^\circ,~-120^\circ$ for $\om=0$. The orbit with maximal energy has the smallest overlap with the potential: The axes s and s lie in the torus' plane  for prolate shape. 
For $\om>0$,  M at $\ga=-120^\circ$ is shifted up by $\om j_1$ because the $j_1$ is aligned with the symmetry axis l,  whereas for  $\ga=0^\circ$ the term $\om j_1$ is small (cf.
Fig. \ref{f:orbitsax}). For triaxial shape the  M orbit has DALl character. Its overlap with the potential is larger then for prolate shape, which is clear from geometry:
  The density torus overlaps less with the potential that has the axes s and s in the torus' plane intend of the axes m and s. 
   For finite $\om$, $\ga<-60^\circ$ is preferred because the M orbit is both DALl and RAL giving
  the extra upward shift of $\om j_1$. For  $-60^\circ<\ga<0^\circ$, the shift is small because the orbit  is DALs but not RAL. 
  
  As seen in the left panel of  Fig. \ref{f:levelsom0ga},  the mid-shell states have a minimum at $\ga=0^\circ,~-120^\circ$ and
   a maximum at $\ga=\pm 60^\circ$ for $\om=0$. 
  A mid-shell orbit is a circle in the 1-2-plane  for axial shape, which means that 
  the plane of  the torus is rotated around the 3-axis. The resulting average density is strongly concentrated at the poles 
  of the \am sphere. Such dumbbell distribution has a good 
  overlap with the prolate potential when its symmetry axis is aligned with the symmetry l-axis of the potential ($\ga=0^\circ$). It has a poor overlap 
  when its symmetry axis is aligned with the symmetry s-axis of the oblate potential ($\ga=60^\circ$, in this case the symmetry axis is 1). 
  When the stripe between the separatrices opens up with increasing $\om$ the minima and maxima  are shifted  to triaxial shapes along with the separatrices (cf. Fig. \ref{f:levelsga}).
  
  The states that are DAL but not simultaneously RAL favor a tilt of the rotational axis away from the principal axis. 
  As seen in the lower left panels of Figs. \ref{f:orbitsax} - \ref{f:orbitstraxm}, 
  these orbits have a large $\pm j_3$ projection but a small negative $j_1$ projection. The cranking term
   $-\om(j_1\sin \vth+j_3\cos\vth)$ decreases with $\vth$ for  the $j_3>0$ orbits.
 The RAL states favor rotation about the principal axis 1 when they have a substantial positive $j_1$ component, because the induced 
 component $j_3=j_{3c}\propto\om$ remains small. The lower panel of  Fig. \ref{f:levelsaxth} illustrates this for axial shape. When $\vth$ changes from
   $90^\circ$ to $0^\circ$ the degenerate signature partners of the DAL orbits  split rapidly because of the  term 
   $\mp\om \cos\vth \vert j_3\vert$, where $\vert j_3\vert \approx \Omega$, which is the good \am projection 
   on the symmetry axis for $\vth=0^\circ$. The lower branch favors a finite tilt angle.
   The RAL orbits change to DAL orbits  with a small projection $j_3$. They favor  $\vth=90^\circ$ because the 
  energy loss of $-\om \sin\vth j_1$  is overcome by the gain in $-\om \cos\vth j_3$. 

 \begin{figure}[t]
  \begin{center}
 \includegraphics[width=\linewidth]{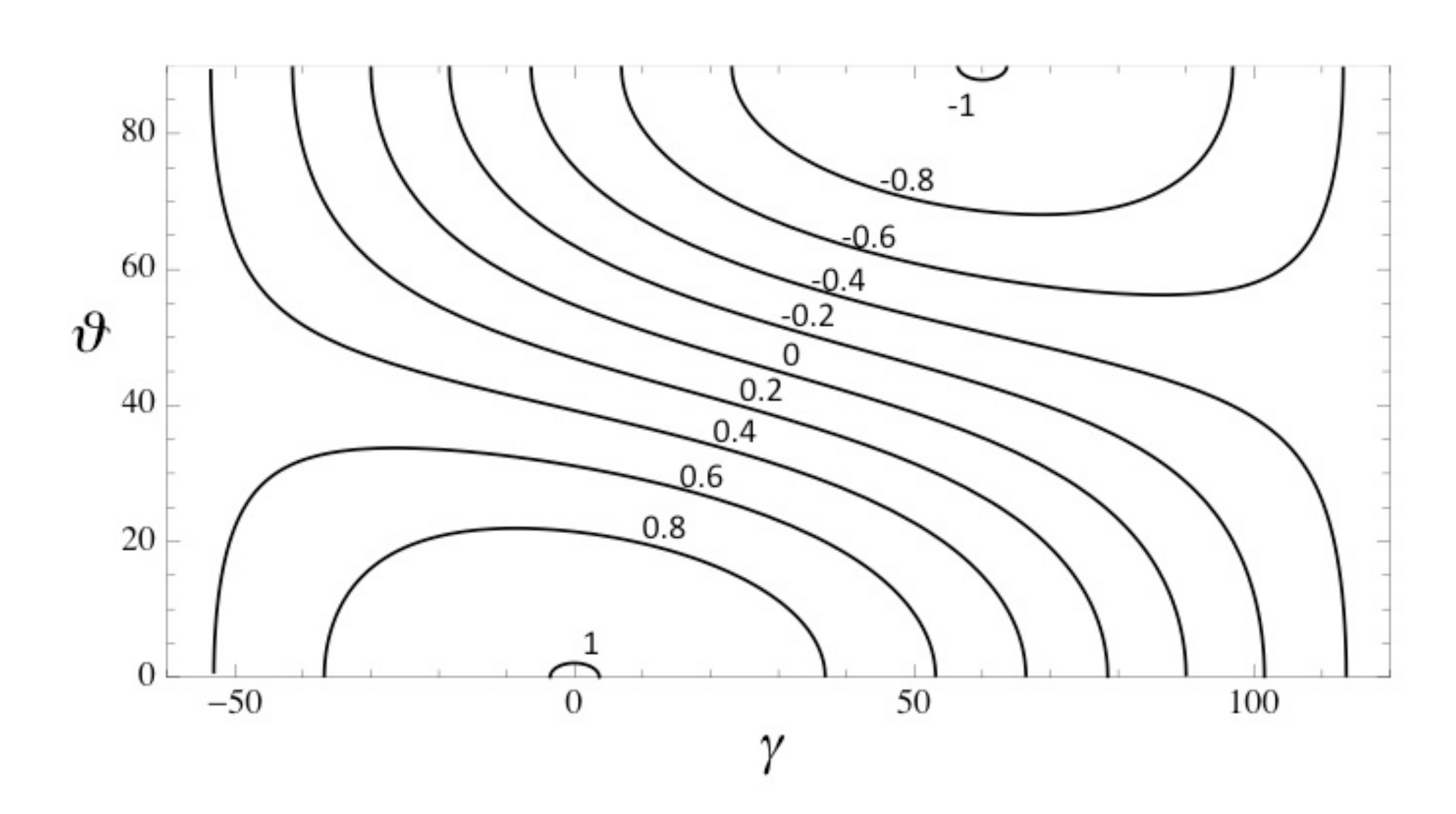}  

    \caption{\label{f:MinGamTh} 
    Location of the minimum of the lowest quasiparticle state with respect to the triaxiality parameter $\ga$ and the tilt angle $\vth$ within the 1-3-plane.
    The curves correspond to different occupancy of the j-shell. The numbers indicate the ratio $\lambda/(2\ka j^2)$, where $\pm2\ka j^2$ are the minimal and
    maximal energy of a particle in the non-rotating potential.   }
  \end{center}
 \end{figure}

For finite pairing, there is a simple relation between the energy minimum of the lowest quasiparticle with  respect to $\ga$ and $\vth$ and the chemical potential $\lambda$. 
For a given value of $\lambda$ the classical routhian (\ref{eq:orbitsp}) of the lowest quasiparticle  orbit touches the \am sphere. To good approximation the 
  touching point is located at $e\left(j_1,j_3=\sqrt{j^2-j_1^2}\right)=\lambda$, where
  \beq
  e'(\ga,\vth)=\Delta-\om\left(j_1(\ga)\sin\vth+j_3(\ga)\cos\vth\right).
  \eeq
   The routhian $e'((\ga,\vth)$ takes the lowest value when $\om\left(j_1(\ga)\sin\vth+j_3(\ga)\cos\vth\right)=\om j$, which according to Eq. (\ref{eq:ellipsoid})
   requires
   \beq\label{eq:MinGamTh}
   \lambda=\kappa j^2\left[\left(3\cos^2\vth-1\right)\cos\ga-\sqrt3\sin^2\vth\sin\ga\right].
       \eeq
 For all possible values of $\ga$, the classical energy takes values values between $-2\kappa j^2$ and $2\kappa j^2$, which restricts $\la$ to this interval and suggests 
 using the ratio $-1\leq\la/(2\ka j^2)\leq1$ as a parameter that indicates the occupancy of the j-shell.  
 
    Fig. \ref{f:MinGamTh} shows the solutions of Eq. (\ref{eq:MinGamTh}) 
 for various shell occupancies.  The region $-240^\circ\leq\ga\leq -60^\circ$ is related to the shown region by
  $\ga\rightarrow-60^\circ-\ga$ and $\vth\rightarrow90^\circ-\vth$. The upper edge  corresponds to rotation about the 1-axis and the lower to 
 rotation about the 3-axis. The solutions are  respectively
 \beq
\vth=90^\circ: \cos(\ga-60^\circ)=-\frac{\la}{2\ka j^2},~~~\vth=0^\circ:\cos\ga=\frac{\la}{2\ka j^2}.
  \eeq
  Depending on the $\ga$ sector, they represents rotation   about one of the principal axes. 
   For the  $\ga$ value at the minimum the  projection of $\vec j$ on 
  the rotation axis takes the maximal value, which is  $j_1=j$ or $j_3=j$ for $\vth=90^\circ$ and $\vth=0^\circ$,  respectively. 
  At bottom of the shell, where the quasiparticle orbit has the character of the lowest particle orbit, 
  it drives the shape toward $\ga=60^\circ$, i. e. 
  rotation about the symmetry s-axis of oblate shape. At the top 
  of the shell, where it has the character of a hole with the orbit of the highest particle, it drives the shape toward $\ga=0^\circ$, i. e. rotation about the symmetry 
  l-axis of prolate shape. A mid-shell  quasiparticle  drives toward $\ga=-30^\circ$ for $\vth=90^\circ$, which is triaxial shape rotating about 
  the m-axis ($\ga=90^\circ,~~\vth=0^\circ$ is the same solution).  
  
As seen in Fig. \ref{f:MinGamTh}, the tilt angle $\vth$ is the second degree of freedom to maximize the  
energy gain $-\vec \om \cdot\vec j$. Away from 
the optimal $\ga$ value the lowest \qp has FAL character (see the lower right panels of Figs. \ref{f:orbitsax} and \ref{f:orbitstraxm})). 
Aligning the rotational axis $\vec \om$ 
with the Fermi axis $\vec j_\la$  one achieves an energy gain of
$\om\left(\sin\vth j_{1\la}+\cos\vth j_{3\la}\right)$, where $j_{1\la}$ and $j_{3\la}$ can be estimated by the above-discussed reflection of the particle orbits
through the Fermi orbit. The gain in changing $\ga$ for $\vth=90^\circ$ is $\om j_{1\la}$, where $j_{1\la}$ is the value for the FAL orbit at the non-optimal $\ga$ value. 
In analogy, it is $\om j_{3\la}$ for $\vth=0^\circ$.
 For the partially filled shell there is a wide range of optimal $\ga-\vth$ combinations.    
Whether a change of $\ga$ or $\vth$ or a combination is preferred depends on  further excited quasiparticles and the 
quasiparticle  vacuum. 

           
\newpage    
\section{Table of content}\label{sec:content}

\vspace*{0.5cm}
\begin{tabular}{ll}
1. Introduction                                                        &        2     \\
2. The Unified Model: virtues  and limits              &   4      \\
 \hspace*{0.5cm} 2.1. The Bohr Hamiltonian 	&	5\\
 \hspace*{0.5cm} 2.2. The Deformed Shell Model    &  6\\
  \hspace*{0.5cm} 2.3. Deformed prolate nuclei		& 8\\
  \hspace*{1 cm}  2.3.1 The strong coupling limit	&  10\\
  \hspace*{1 cm}  2.3.2.  Comparison with experiment & 	13\\	
      \hspace*{0.5cm} 2.4. Spherical nuclei			& 21 \\
     \hspace*{0.5cm}  2.5. Microscopic basis of the Unified Model  &25 \\
   \hspace*{0.5cm}  2.6. Transitional nuclei 		& 30 \\
   \hspace*{0.5cm}  2.7. Quasiparticle triaxial rotor model  & 31 \\
   \hspace*{0.5cm}  2.8. Approaches beyond the Unified Model   &   32 \\
   3. Rotating mean field										&   34  \\
   \hspace*{0.5cm}  3.1. Selfconsistent cranking model				&   36   \\
   \hspace*{0.5cm} 3.2. Symmetries								&  45 \\
   \hspace*{1 cm}  3.2.1. Axis of rotation is a principal axis			&   47  \\
   \hspace*{1 cm}  3.2.2. Axis of rotation in a principal plane		&   47   \\
   \hspace*{1 cm}  3.2.3. Axis of rotation out of the principal planes	&    48 \\
   \hspace*{0.5cm} 3.3. Rotational frequency					&   49 \\
   \hspace*{0.5cm}  3.4. Calculation of the mean field shape		&   52  \\
 \hspace*{0.5cm} 3.5. Geometry and rotational response of quasiparticle orbitals & 53 \\
 \hspace*{0.5cm} 3.6. Cranked Shell Model					&    56  \\
 \hspace*{1 cm} 3.6.1. Bands as quasiparticle configurations		&    59  \\
 \hspace*{1 cm}  3.6.2. Bandcrossings						&     59\\
 \hspace*{1 cm}  3.6.3. Cranked Shell Model  classification of bands				&     63 \\
 \hspace*{0.5cm} 3.7. Cranked shell correction approach			&      67\\
 \hspace*{0.5cm}  3.8. Rotation about a tilted axis				&   69 \\
 \hspace*{1 cm}  3.8.1. The spinning clockwork picture			& 69\\
 \hspace*{1 cm} 3.8.2. Appearance of tilted rotation				&  72  \\
 \hspace*{1 cm}  3.8.3. Tilted axis cranking solutions of axial nuclei			&  73 \\
 \hspace*{1cm}  3.8.4. Approximate tilted axis cranking solutions	&  78\\
 \hspace*{1 cm}  3.8.5. Change of symmetry					&  80 \\
 \hspace*{1 cm}  3.8.6. Tilted axis cranking solutions 	of triaxial nuclei	& 84\\
 \hspace*{1.5cm} Wobbling								&  85\\
 \hspace*{1.5cm}  Transverse wobbling						& 86\\
 \hspace*{1.5cm} Chirality									& 90
  \end{tabular}
 \newpage
 \begin{tabular}{ll}
 4. Emergence and disappearance of collective degrees of freedom	& 96\\
 \hspace*{0.5cm} 4.1. Band termination 						& 97 \\
 \hspace*{0.5cm}  4.2. Magnetic rotation 						&  102 \\
 \hspace*{0.5cm} 4.3. Shears geometry 						& 110\\
 \hspace*{0.5cm} 4.4. Emergence of rotational bands			& 113\\
  \hspace*{1 cm} 4.4.1. Coherence length						& 114\\
 \hspace*{1 cm} 4.4.2. Regularity							&  116\\
 \hspace*{1 cm} 4.4.3. Three examples						&  118 \\
 \hspace*{0.5cm} 4.5. Tidal Waves							& 121\\
 \hspace*{0.5cm} 4.6. Coherence of the deformation degrees of freedom &123\\
 5. Outlook											&  125\\
 6. References											& 126\\
 Appendix A. Semiclassical analysis of high-j quasiparticle states 	& 132\\
 \hspace*{0.5cm}  A.1. Axial shape, rotation about the short axis	& 135\\
 \hspace*{0.5cm}  A.2. Triaxial shape, rotation about a principal axis	&  139\\
 \hspace*{0.5cm}  A.3. Rotation about a tilted axis				& 145\\
  \hspace*{0.5cm} A.4. Induced triaxiality and tilt angles			& 149
  \end{tabular}


\end{document}